%% file: main.tex
\documentclass[a4paper,10pt]{article}

\usepackage[top=1in, left=0.9in, bottom=1in, right=0.9in]{geometry}

\usepackage{amsmath,amssymb,amsfonts}
\allowdisplaybreaks[4]

\usepackage[utf8]{inputenc}
\usepackage{cite}

\usepackage{graphicx}
\usepackage{xcolor}
\usepackage{bm}

\usepackage[export]{adjustbox}
\usepackage{slashed}

\usepackage{booktabs}

\usepackage[colorlinks=true,linkcolor=blue,citecolor=blue,urlcolor=blue]{hyperref}

\let\counterwithin\relax
\usepackage{chngcntr}

\usepackage{longtable}
\usepackage{multirow}
\usepackage{makecell}
\usepackage{diagbox}
\usepackage{adjustbox}
\usepackage[vcentermath]{youngtab}
\usepackage{extarrows}
\usepackage{lscape}
\usepackage{bbm}
\usepackage[bb=boondox]{mathalpha}

\usepackage{subfig}

\usepackage{ytableau}
\usepackage[vcentermath]{youngtab}

\usepackage{tikz}
\usepackage[compat=1.1.0]{tikz-feynhand}
\usetikzlibrary{mindmap,chains,trees, shapes.geometric, arrows, positioning, shadows}

\usepackage{mathtools}
\usepackage{cancel}

\newcommand{\eq}[1]{\begin{equation}\begin{split} #1 \end{split}\end{equation}}

\newcommand{\eqs}[1]{\begin{eqnarray} #1 \end{eqnarray}}

\newcommand{\lra}[1]{\langle #1 \rangle}

\tikzset{thick bos/.style={very thick,decorate,decoration={snake,amplitude=1pt,segment length=4pt}}}
\tikzset{thick sca/.style={very thick,dash=on 4pt off 2pt phase 0pt}}

\newcommand{\fer}[3]{\propag[#1,fer] (#2)--(#3);
\draw[#1,style=very thick] (#2)--(#3);}

\newcommand{\antfer}[3]{\propag[#1,antfer] (#2)--(#3);
\draw[#1,style=very thick] (#2)--(#3);}

\tikzset{thick bos/.style={very thick,decorate,decoration={snake,amplitude=1pt,segment length=4pt}}}
\tikzset{thick sca/.style={very thick,dash=on 4pt off 2pt phase 0pt}}

\newcommand{\sca}[1] {\draw[brown,thick sca] (#1)--(v1)} 
\newcommand{\ferflip}[4]{
\draw[#3,very thick,rotate=#2] (0.7*#1,0)--(0.46*#1,0);
\draw[#4,very thick,decoration={markings,mark=at position 0.56 with {\arrow{Triangle[length=4pt,width=4pt]}}},postaction={decorate},rotate=#2] (0.46*#1,0)--(v1);
\draw[very thick,rotate=#2] plot[mark=x,mark size=2.5] coordinates {(0.46*#1,0)};
\draw[rotate=#2] (0.1*#1,-0.08) -- (0.16*#1,0.08);} 
\newcommand{\antferflip}[4]{
\draw[#3,very thick,rotate=#2] (0.7*#1,0)--(0.46*#1,0);
\draw[#4,very thick,decoration={markings,mark=at position 0.82 with {\arrow{Triangle[length=4pt,width=4pt]}}},postaction={decorate},rotate=#2] (v1)--(0.46*#1,0);
\draw[very thick,rotate=#2] plot[mark=x,mark size=2.5] coordinates {(0.46*#1,0)};
\draw[rotate=#2] (0.1*#1,-0.08) -- (0.16*#1,+0.08);
} 
\newcommand{\bos}[2] {\draw[#2,thick bos] (#1)--(v1)} 
\newcommand{\bosflip}[4] {
\begin{scope}[rotate=#2]
\clip (0,-0.1) rectangle (0.44*#1,0.1); 
\draw[#3,thick bos] (0,0)--(#1,0);
\end{scope}
\begin{scope}[rotate=#2]
\clip (0.44*#1,-0.1) rectangle (0.7*#1,0.1); 
\draw[#4,thick bos] (0,0)--(#1,0);
\end{scope}
\draw[very thick,rotate=#2] plot[mark=x,mark size=2.5] coordinates {(0.43*#1,0)};
\draw[rotate=#2] (0.17*#1,-0.08) -- (0.23*#1,+0.08);
} 
\newcommand{\bosflipflip}[5] {
\begin{scope}[rotate=#2]
\clip (0,-0.1) rectangle (0.35*#1,0.1); 
\draw[#3,thick bos] (0,0)--(#1,0);
\end{scope}
\begin{scope}[rotate=#2]
\clip (0.35*#1,-0.1) rectangle (0.55*#1,0.1); 
\draw[#4,thick bos] (0,0)--(#1,0);
\end{scope}
\begin{scope}[rotate=#2]
\clip (0.55*#1,-0.1) rectangle (0.7*#1,0.1); 
\draw[#5,thick bos] (0,0)--(#1,0);
\end{scope}
\draw[very thick,rotate=#2] plot[mark=x,mark size=2.5] coordinates {(0.55*#1,0)};
\draw[very thick,rotate=#2] plot[mark=x,mark size=2.5] coordinates {(0.35*#1,0)};
\draw[rotate=#2] (0.17*#1,-0.08) -- (0.23*#1,+0.08);
\draw[rotate=#2] (0.07*#1,-0.08) -- (0.13*#1,+0.08);
} 

\newcommand{\Ampone}[3] {
\begin{tikzpicture}[baseline=-0.1cm] \begin{feynhand}
\setlength{\feynhandarrowsize}{4pt}
\vertex [particle] (i1) at (#1,0) {$#2$}; 
\vertex (v1) at (0,0);
#3;
\vertex[dot] (v1) at (0,0) {};
\end{feynhand} \end{tikzpicture}
}

\newcommand{\Ampthree}[6] {
\begin{tikzpicture}[baseline=-0.1cm] \begin{feynhand}
\setlength{\feynhandarrowsize}{4pt}
\vertex [particle] (i1) at (-1.01,0) {$#1$}; 
\vertex [particle] (i2) at (0.579,0.827) {$#2$}; 
\vertex [particle] (i3) at (0.579,-0.827) {$#3$}; 
\vertex (v1) at (0,0);
#4;
#5;
#6;
\end{feynhand} \end{tikzpicture}
}
\newcommand{\Ampfour}[8] {
\begin{tikzpicture}[baseline=-0.1cm] \begin{feynhand}
\setlength{\feynhandarrowsize}{4pt}
\vertex [particle] (i1) at (-0.714,0.714) {$#1$}; 
\vertex [particle] (i2) at (0.714,0.714) {$#2$}; 
\vertex [particle] (i3) at (0.714,-0.714) {$#3$}; 
\vertex [particle] (i4) at (-0.714,-0.714) {$#4$}; 
\vertex (v1) at (0,0);
#5;
#6;
#7;
#8;
\end{feynhand} \end{tikzpicture}
}

\newcommand{\hi}[1]{\hat{\mathbf{i}}_{#1}}

\newcommand{\beq}{\begin {equation}}
\newcommand{\eeq}{\end   {equation}}
\newcommand{\bea}{\begin {eqnarray}}
\newcommand{\eea}{\end   {eqnarray}}
\newcommand{\baa}{\begin {array}   }
\newcommand{\eaa}{\end   {array}   }
\newcommand{\bit}{\begin {itemize} }
\newcommand{\eit}{\end   {itemize} }
\newcommand{\be }{\begin {equation}}
\newcommand{\ee }{\end   {equation}}

\begin{document}


\begin{center}


{\Large \textbf  {Massless-Massive Amplitude Correspondence III: \\ Massive Amplitude Bases in the SMEFT}}\\[10mm]

Yu-Han Ni$^{a, b}$\footnote{niyuhan@cuhk.edu.cn}, Hao Sun$^{e, f}$\footnote{hao.sun@desy.de}, 
Yi-Ning Wang$^{a, b}$\footnote{wangyining@itp.ac.cn}, Chao Wu$^{a, b}$\footnote{wuch7@itp.ac.cn}, Jiang-Hao Yu$^{a, b, c, d}$\footnote{jhyu@itp.ac.cn}\\[10mm]

\noindent
$^a${\em \small Institute of Theoretical Physics, Chinese Academy of Sciences, Beijing 100190, P. R. China}  \\
$^b${\em \small School of Physical Sciences, University of Chinese Academy of Sciences, Beijing 100049, P.R. China}   \\
$^c${\em \small School of Fundamental Physics and Mathematical Sciences, Hangzhou Institute for Advanced Study, \\ UCAS, Hangzhou 310024, China} \\
$^d${\em \small International Centre for Theoretical Physics Asia-Pacific, Beijing/Hangzhou, China}\\
$^e${\em \small Institute of High Energy Physics, Chinese Academy of Sciences, Beijing 100049, China}\\
$^f${\em \small Deutsches Elektronen-Synchrotron DESY, Notkestr. 85, 22607 Hamburg, Germany} \\[10mm]

\date{\today}

\end{center}

\begin{abstract}
\noindent

We develop a systematic correspondence between massless contact amplitudes in an unbroken theory and massive contact amplitudes after spontaneous symmetry breaking.  Our construction employs the spin-transversality (ST) massive amplitude basis, with the systematic high energy expansion through minimal-helicity-chirality (MHC) amplitudes. The resulting $U(2)=SU(2)\times U(1)_t$ description of a massive particle makes the semi-standard Young-tableau construction of massless Lorentz structures directly applicable to massive amplitudes. When the leading-order MHC component has a massless contact limit, it is one-to-one matched directly to its UV amplitude. Otherwise, five exceptional classes of ST amplitudes are identified, their first non-zero descendant components are matched through conserved current couplings to the massless contact amplitude. We apply the framework to the one-flavor electroweak sector of the Standard Model Effective Field Theory (SMEFT) through dimension eight, obtaining explicit relations between unbroken-phase Wilson coefficients and broken-phase ST amplitude coefficients for amplitudes with three to eight external particles.

\end{abstract}

\newpage

\tableofcontents

\setcounter{footnote}{0}
\counterwithin{equation}{section}

\newpage

\input{sec1-intro}

\input{sec2-spinor}

\input{sec3-matching}

\input{sec4-3pt4pt}

\input{sec5-smeft}

\input{sec6-con}

\section*{Acknowledgments}
This work is supported by the National Science Foundation of China under Grants No. 12347105, No. 12375099 and No. 12047503, and the National Key Research and Development Program of China Grant No. 2020YFC2201501, No. 2021YFA0718304.
H.S. is supported by the Helmholtz-OCPC International
Postdoctoral Exchange Fellowship Program.

\appendix

\input{sec7-app}

\bibliographystyle{JHEP}
\bibliography{ref}

\end{document}

%% file: sec1-intro.tex
\section{Introduction}

The Standard Model Effective Field Theory (SMEFT) parametrizes the effects of heavy new physics through a tower of local operators built from Standard Model (SM) fields and invariant under the unbroken gauge group. Above the electroweak scale, the effective Lagrangian is organized as
\begin{equation}
\mathcal L_{\rm SMEFT}=\mathcal L_{\rm SM}
+\sum_{d\geq5}\sum_i\frac{C_i^{(d)}}{\Lambda^{d-4}} \mathcal O_i^{(d)},
\end{equation}
where $\mathcal O_i^{(d)}$ denotes a gauge-invariant operator of mass dimension $d$, $C_i^{(d)}$ its dimensionless Wilson coefficient, and $\Lambda$ the scale of new physics. Two practical problems must be solved before this effective Lagrangian can be confronted with data: finding an independent operator basis at each dimension $d$, and translating that basis, after electroweak symmetry breaking (EWSB), into the massive interactions that actually enter scattering amplitudes of the physical $W^\pm$, $Z$, $h$, and fermion states.

On-shell methods address both problems at once by working directly with contact scattering amplitudes rather than with off-shell operators: a local operator is equivalent to a polynomial contact amplitude, while integration by parts (IBP) and the equations of motion (EOM) become on-shell, momentum-conservation redundancies. For massless particles, the spinor-helicity formalism~\cite{Parke:1986gb,Bern:1996je,Dixon:1996wi,Elvang:2013cua,Cheung:2017pzi,Travaglini:2022uwo,Badger:2023eqz} removes gauge redundancy and makes this correspondence explicit. Its natural generalization to an arbitrary number of external legs is furnished by conformal-helicity duality: an $N$-particle massless amplitude is simultaneously a representation of the conformal group $SU(2,2)$ and of a dual $U(N)$ group, and the latter lets one enumerate independent Lorentz structures systematically through semi-standard Young tableaux (SSYT)~\cite{Henning:2019enq,Henning:2019mcv,Li:2020gnx,Li:2022tec}. Combined with the group-theoretic construction of flavor and gauge tensors, this program has produced complete massless operator bases for the SMEFT and other effective theories to high mass dimension~\cite{Shadmi:2018xan,Aoude:2019tzn,Ma:2019gtx,Durieux:2019siw,Li:2020gnx,Li:2020xlh,Li:2020tsi,Li:2020zfq,Li:2021tsq,AccettulliHuber:2021uoa,Li:2022tec,Ren:2022tvi}.

The corresponding construction in the broken phase is considerably more subtle, because a massive one-particle state carries an $SU(2)$ little-group index rather than a $U(1)$ phase. Arkani-Hamed, Huang, and Huang (AHH) introduced massive spinor-helicity variables that make this little-group covariance manifest~\cite{Arkani-Hamed:2017jhn}, a formalism since applied widely to electroweak physics~\cite{Franken:2019wqr, Bachu:2019ehv, Ballav:2020ese, Wu:2021nmq, Ballav:2021ahg, Liu:2022alx, Bachu:2023fjn, Ema:2024rss}, effective field theory~\cite{Durieux:2019eor,Durieux:2020gip, Dong:2021vxo,Balkin:2021dko, DeAngelis:2022qco, Dong:2022mcv,Sun:2022ssa,Sun:2022snw, Liu:2023jbq,Song:2023lxf,Song:2023jqm,Goldberg:2024eot,Dong:2024dce}, and gravitational-wave physics~\cite{Guevara:2018wpp, Chung:2018kqs, Guevara:2019fsj,  Maybee:2019jus, Arkani-Hamed:2019ymq}. 
Existing massive amplitude bases~\cite{Durieux:2020gip,Dong:2021vxo,Balkin:2021dko,DeAngelis:2022qco,Dong:2022mcv} are well suited to constructing independent infrared (IR) massive structures. Their connection to ultraviolet (UV) operator dimension and to unbroken-phase Wilson coefficients, however, can be obscured. For example, a chiral basis leaves the mass dimension implicit, whereas relations used to obtain minimal stripped-contact-term or spinor-structure bases can introduce external-mass ratios that have no direct UV interpretation.

The spin-transversality (ST) formalism~\cite{Ni:2024yrr,Ni:2025xkg,Ni:2026wiz,Ni:2026mia} provides a useful alternative when the UV--IR correspondence is the central objective. The ST formalism was motivated by introducing an additional $U(1)$ symmetry, which is associated with a transversality quantum number and is closely related to chirality. In the present work, to maintain consistency with the SSYT method for massless amplitudes, we adopt an alternative description for the ST formalism. A massive momentum can be represented by two null momenta and hence a dual $U(2)$ group can be identified from conformal-helicity duality. This permits a single massive particle to be described by the $U(2)=SU(2)\times U(1)_t$ symmetry of an associated two-particle massless system: $SU(2)$ is the massive little group and $U(1)_t$ defines a transversality quantum number. In the minimal-helicity-chirality (MHC) expansion, an ST spinor is resolved into massless spinors of definite helicity and transversality~\cite{Ni:2025xkg}. The high-energy expansion in $\mathbf m/E$ then supplies a graded bridge between massive and massless amplitudes. Schematically,
\begin{equation}
\begin{array}{c|ccc}
\hline
 & \text{massless} & \text{MHC} & \text{ST} \\
\hline
\text{symmetry} & U(1)_h & U(1)_h\times U(1)_t & U(2) \\
\text{description} & \text{two massless particles} & \text{one massive particle} & \text{one massive particle} \\
\hline
\end{array}
\end{equation}
For an amplitude with $N$ massless and $N'$ massive external particles, this construction is naturally organized by $U(N+2N')$. It therefore allows the SSYT machinery for massless amplitudes to be applied directly to massive Lorentz structures.

This correspondence also identifies a necessary qualification to a naive high-energy matching. In generic cases, the first non-vanishing MHC component of an ST amplitude is matched directly to a massless contact amplitude. We call this \textit{direct matching}. There are, however, structures whose leading high-energy component vanishes because it is a conserved current. Their first non-zero massive contribution is a descendant and must be matched through the covariant-derivative coupling $A_\mu J^\mu$, together with possible operator contributions implied by the Goldstone equivalence theorem. We call this \textit{exceptional matching}. Thus, direct matching is controlled by unbroken-phase operator coefficients, whereas exceptional matching also involves renormalizable gauge couplings.

In this paper, we formulate the ST basis in the $U(2)$ language compatible with the SSYT construction and use the MHC expansion to derive a systematic massless--massive matching procedure. We classify the exceptional structures of a generic massive theory into five conserved-current classes, $VVS$, $VSS$, $ffV$, $VVV$, and $VVVV$. For the SM particle content, only $VVS$, $ffV$, and $VVV$ are realized: $VSS$ and $VVVV$ require, respectively, two distinct massive scalars or four distinct massive vectors. We then apply the method to the one-flavor electroweak sector of the SMEFT through dimension eight. Our explicit results cover amplitudes with three to eight external particles. Quark and gluon fields may be retained through generic fermion labels, while their explicit $SU(3)_C$ color tensors are suppressed because color factorizes from the electroweak matching. The resulting relations express broken-phase amplitude coefficients in terms of unbroken-phase Wilson coefficients and, where required, SM couplings.

\begin{figure}[htb!]
\centering
\includegraphics[width=0.95\linewidth]{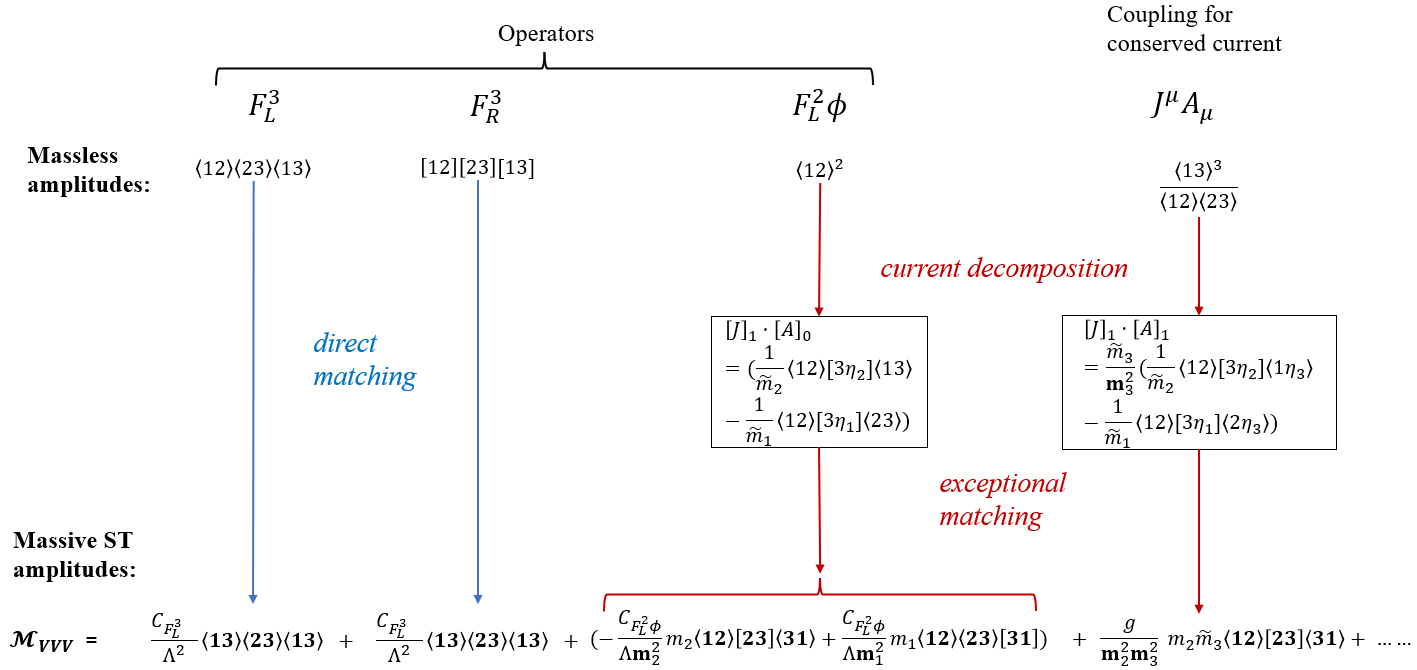}
\caption{A $VVV$ ST basis and its correspondence to massless amplitudes. The detailed coefficient relations are given in Eq.~\eqref{eq:VVV_result}.}
\label{fig:massive_diagram}
\end{figure}

Figure~\ref{fig:massive_diagram} illustrates the two branches for a $VVV$ amplitude. A direct contribution arises from the massless operator $F_{L/R}^3$ and matches a primary ST structure without an explicit mass factor. Exceptional contributions instead relate descendant ST structures to $F_{L/R}^2\phi$ operators and to the gauge coupling of the conserved current. We use the massless operator conventions and coefficients of Ref.~\cite{Li:2020gnx} throughout.

The remainder of this paper is organized as follows. Section~\ref{sec:2} reviews massless spinor-helicity and conformal-helicity duality and develops the ST description of massive states. Section~\ref{sec:matching} introduces the MHC expansion and the direct and exceptional matching branches. In section~\ref{sec:3}, we construct the lowest-dimension three- and four-point massive EFT bases and analyze all conserved-current classes. Section~\ref{sec:SMEFT} incorporates electroweak gauge tensors and presents the SMEFT matching results through dimension eight. We conclude in section~\ref{sec:con}.

%% file: sec2-spinor.tex
\section{Spin-Transversality Basis of Massive Amplitudes}
\label{sec:2}

There are two commonly used perspectives on the on-shell particle states in four-dimensional spacetime, each offering distinct advantages depending on the context of the amplitude under consideration: 
\begin{itemize}
    \item The first~\cite{Wigner:1939cj,Arkani-Hamed:2017jhn} is the Wigner particle state $|p,\sigma\rangle$, which furnishes a representation of the Poincar\'e group $ISO(3,1)$ and is characterized by momentum $p$, with $\sigma$ denoting the spin or helicity of the particle. In this framework, the one-particle state transforms under a local representation. This description is well suited for the 3-point amplitudes (particularly in the Standard Model), as they can be treated simply as a contraction of three-particle states. However, extending this approach to the $N$-particle case is not easy, because additional momentum structures not present in the particle state $|p,s\rangle$ must be taken into account.

    \item  An alternative approach~\cite{Li:2020gnx,Li:2022tec} involves the representation of the conformal group and the helicity group $SU(2,2) \times U(1)$.~\footnote{Although the conformal group is often presented as $SO(4,2)$, in this article we work with its double cover $SU(2,2)$. This choice is motivated by the presence of spinor fields, for which the spinorial representation is most naturally realized within $SU(2,2)$.} Although one cannot define an asymptotic particle state with fixed momentum in this case, since momentum can be rescaled under conformal symmetry, the momentum structure is instead incorporated into the definition of the particle states themselves. As a result, this formulation offers a natural pathway for generalizing to the $N$-particle case, which avoids limitations of the Wigner representation.
\end{itemize}
We adopted the first perspective in our previous work~\cite{Ni:2024yrr,Ni:2025xkg,Ni:2026wiz,Ni:2026mia}, which focuses on 3-point contact amplitudes and on higher-point amplitudes constructed from them. In this work, however, our aim is to investigate contact amplitudes beyond the 3-point level. Therefore, we take the second perspective to study the amplitude-operator correspondence. 
In this section, we review the basics of the spinor-helicity formalism firstly, and discuss the conformal symmetry of the massless amplitudes. Then we extend the argument to the massive case, which introduces a new quantum number, transversality $t$, and leads to the spin-transversality (ST) spinors.


\subsection{Poincar\'e Spinor-Helicity Formalism}
\label{sec:Poincare}


Let us review the spinor-helicity formalism, which uses the representation of the Poincaré group to describe the particle state. 


Since the Poincaré representation also transforms under its subgroup, the Lorentz group, it is useful to begin with the on-shell variables in the Lorentz representation. 
Using the double cover of the Lorentz group, $SU(2)_l\times SU(2)_r$, the irreducible representations are labeled as $(j_l,j_r)$, where $j_l$ and $j_r$ are half integers, denoting the quantum number for $SU(2)_l$ and $SU(2)_r$, respectively. The basic variables are of the Lorentz representations $(\frac{1}{2},0)$ and $(0,\frac{1}{2})$, which are referred to as the left- and right-handed spinors, respectively,
\begin{equation}
    \lambda_\alpha \sim \left(\frac{1}{2},0\right)\,,\quad \tilde{\lambda}_{\dot\alpha} \sim \left(0,\frac{1}{2}\right)\,.
\end{equation}
We have used the Van der Waerden notation~\cite{Dittmaier:1998nn,Schwinn:2007ee}, in which the left-handed indices are undotted, while the right-handed indices are dotted. The left- and right-handed spinors are related by complex conjugates, which interchange the dotted and undotted indices, $\left(\lambda_\alpha\right)^\dagger = \tilde{\lambda}_{\dot{\alpha}}$.
The spinor indices can be raised or lowered by the antisymmetric tensors $\epsilon^{\alpha\beta},\epsilon_{\dot\alpha\dot\beta}$,
\begin{equation}
    \lambda^\alpha = \epsilon^{\alpha\beta}\lambda_\beta = -\lambda_\beta \epsilon^{\beta\alpha}\,,\quad \tilde{\lambda}_{\dot\alpha} = \epsilon_{\dot\alpha\dot\beta} \tilde{\lambda}^{\dot\beta} = - \tilde{\lambda}^{\dot\beta}\epsilon_{\dot\beta\dot\alpha}\,,
\end{equation}
whose matrix forms are
\begin{equation}
\label{eq:epsilonmatrix}
    \epsilon_{\alpha\beta} = \epsilon_{\dot\alpha\dot\beta} = \left(\begin{array}{cc}
0 & -1 \\ 1 & 0
    \end{array}\right)\,, \quad \epsilon^{\alpha\beta} = \epsilon^{\dot\alpha\dot\beta} = \left(\begin{array}{cc}
0 & 1 \\ -1 & 0
    \end{array}\right)\,,
\end{equation}
satisfying $\epsilon^{\alpha\beta}\epsilon_{\beta\gamma} = \delta^\alpha{}_\gamma\,, \epsilon^{\dot\alpha\dot\beta}\epsilon_{\dot\beta\dot\gamma} = \delta^{\dot\alpha}{}_{\dot\gamma}$. 
The massless momentum $p$ is of the $(\frac{1}{2},\frac{1}{2})$ representation, and its spinors expression is
\begin{equation}\label{eq:MomentaSpinorRelationMassless}
    p^{\dot\alpha\alpha} = p^\mu \overline{\sigma}_\mu^{\dot\alpha\alpha} = \tilde{\lambda}^{\dot\alpha}\lambda^\alpha = \left(\begin{array}{cc}
        p^0+p^3 & p^1-ip^2 \\
        p^1+ip^2 & p^0-p^3
    \end{array}\right)\,,
\end{equation}
where $\overline{\sigma}^\mu = (1,-\vec{\sigma})$ with the Pauli matrices $\vec{\sigma}$. 

To describe massless particle states by spinors, we need the spinor expressions of the Poincar\'e generators, including Lorentz generators and translation generators. In the spinor representation, the translation generator $p^{\dot\alpha\alpha}=\tilde\lambda^{\dot\alpha}\lambda^{\alpha}$ has already been given in Eq.~\eqref{eq:MomentaSpinorRelationMassless}. For the Lorentz generators,
their spinor expressions are 
\begin{equation}
    M_{\alpha\beta} = i \lambda_{(\alpha} \frac{\partial}{\partial \lambda^{\beta)}}\,,\quad \tilde{M}_{\dot\alpha\dot\beta} = \tilde{\lambda}_{(\dot\alpha} \frac{\partial}{\partial \tilde{\lambda}^{\dot\beta)}}\,,
\end{equation}
where $(\cdots)$ denotes a symmetrization of spinor indices. 
Besides, the momentum is also invariant under the application of $H$ that
\begin{equation}
    H =  \frac{1}{2} \left(-\lambda_{\alpha} \frac{\partial}{\partial \lambda_{\alpha}} +\tilde{\lambda}_{\dot{\alpha}} \frac{\partial}{\partial \tilde{\lambda}_{\dot\alpha}}\right)\,,
\end{equation}
which generates a $U(1)$. Under this $U(1)$ group, the massless spinors are charged as
\begin{equation}
    H\lambda_\alpha \rightarrow -\frac{1}{2} \lambda_\alpha\,,\quad H\tilde{\lambda}^{\dot{\alpha}} \rightarrow \frac{1}{2} \tilde{\lambda}^{\dot\alpha}\,.
\end{equation}
It is usually referred to as the little group of massless particles, whose charge is the helicity $h$.

The Wigner 1-particle states of massless particles are labeled by $|p,h\rangle$, where $p$ is momentum and $h$ is the helicity, satisfying $W^0|p,h\rangle = h |p,h\rangle$. Thus, the left- and right-hand spinors themselves are 1-particle states of  massless fermions, 
\eq{
\lambda_{\alpha} \sim |p,-\frac{1}{2}\rangle, \quad
\tilde{\lambda}^{\dot{\alpha}} \sim |p,+\frac{1}{2}\rangle\,,
}
where we use $\sim$ to imply the equivalence between the 1-particle states and their spinor expressions, 
Generally, the 1-particle states with helicity $h$ are given by the symmetric products of these spinors
\begin{equation}
|p,h\rangle\sim
\left\{\begin{aligned}
\lambda_{(\alpha_1}\lambda_{\alpha_2}\cdots\lambda_{\alpha_h)},\quad h<0,\\
\tilde{\lambda}^{(\dot{\alpha}_1}\tilde{\lambda}^{\dot{\alpha}_2}\cdots\tilde{\lambda}^{\dot{\alpha}_h)},\quad h\ge 0. 
\end{aligned}\right.
\end{equation}

For massive particles, we can similarly use momentum and the little group to describe their particle states. According to Wigner's classification, the little group for a massive particle is $SU(2)$. 
Thus, we need to collect the spinor $\lambda$ ($\tilde{\lambda}$) and the auxiliary spinor $\eta$ ($\tilde{\eta}$) into a $SU(2)$ doublet,
\begin{equation}  \label{eq:massive_doublet}
\lambda_{I\alpha} = (\lambda_{\alpha}\;\eta_{\alpha})\quad \tilde\lambda_{\dot\alpha}^I = (\tilde{\lambda}_{\dot\alpha}\;\tilde{\eta}_{\dot\alpha})\,,
\end{equation}
where $I=1,2$ is the little group index. The massive momentum $\mathbf{p}$ is also little-group invariant,
\begin{eqnarray}\label{eq:MomentaSpinorRelationMassive}
    \mathbf{p}^{\dot\alpha\alpha}=\mathbf{p}^\mu \overline{\sigma}_\mu^{\dot\alpha\alpha} = \lambda^{\alpha}_I\tilde{\lambda}^{I\dot\alpha}=\lambda^\alpha\tilde\lambda^{\dot\alpha}+\eta^\alpha\tilde\eta^{\dot\alpha}= \left(\begin{array}{cc}
        \mathbf{p}^0+\mathbf{p}^3 & \mathbf{p}^1-i\mathbf{p}^2 \\
        \mathbf{p}^1+i\mathbf{p}^2 & \mathbf{p}^0-\mathbf{p}^3
    \end{array}\right)\,,
\end{eqnarray}
where we use the bold notation for massive variables. 
The non-vanishing determinants characterize the massive property of the spinors,
\begin{equation}
    \mathbf m = \det(\lambda^I_\alpha) = \frac{1}{2}\lambda^I{}^\alpha\lambda^J_\alpha \epsilon_{IJ}\, = \det(\tilde{\lambda}^I_{\dot\alpha}) = \frac{1}{2}\tilde{\lambda}^I_{\dot\alpha}\tilde{\lambda}^{J\dot\alpha}\epsilon_{IJ}\,,
\end{equation}
which leads to the Dirac equations
\begin{eqnarray}
\label{eq:diracequation}
\lambda^{I\alpha}\mathbf{p}_{\alpha\dot\alpha}=-\mathbf m\tilde{\lambda}^I_{\dot\alpha},\quad \tilde{\lambda}^I_{\dot\alpha}\mathbf{p}^{\dot\alpha\alpha}=-\mathbf{m}\lambda^{I\alpha}\,.
\end{eqnarray}

There are two ways to treat this $SU(2)$ little group transformations:

\begin{itemize}
\item One approach is to treat the little group as a subgroup of the Lorentz group~\cite{Conde:2016izb}. In this approach, the momentum $\mathbf{p}$ is treated as a coordinate in the Hilbert space rather than an operator acting on it, and the Pauli–Lubanski vector $W$ becomes a linear combination of Lorentz generators.

\item Another approach treats the $SU(2)$ as a spin group acting on an internal spin space labeled by the index $I$.
In this approach, indices can be raised and lowered using the Levi-Civita tensor in the spin space:
\begin{equation}
\lambda^\alpha_I=\epsilon_{IJ}\lambda^{J \alpha},\quad 
\tilde\lambda^{I\dot\alpha}=\epsilon^{IJ}\tilde\lambda^{\dot\alpha}_J\,,
\end{equation}
where $\epsilon^{IJ}\,,\epsilon_{IJ}$ takes the matrix forms in Eq.~\eqref{eq:epsilonmatrix}.
The Pauli–Lubanski vector can then be projected into this space using the massive spinors $\lambda^I$ and $\tilde\lambda^I$, $W_{\alpha\dot{\alpha}} = J_{I}^{J} \lambda_{\alpha}^{I} \tilde{\lambda}_{\dot{\alpha} J}$, yielding the $SU(2)$ generators in the spin space:
\begin{equation} \label{eq:SU2_generator}
J_{(IJ)}=\frac{i}{2}\left(\lambda_{\alpha I}\frac{\partial}{\partial \lambda^{J}_{\alpha}} + \lambda_{\alpha J}\frac{\partial}{\partial \lambda^{I}_{\alpha}} + \tilde{\lambda}_{\dot{\alpha} I}\frac{\partial}{\partial \tilde{\lambda}^{J}_{\dot{\alpha}}} + \tilde{\lambda}_{\dot{\alpha} J}\frac{\partial}{\partial \tilde{\lambda}^{I}_{\dot{\alpha}}}\right)\,,
\end{equation}
where the two indices $I,J$ are symmetric.
In Ref.~\cite{Arkani-Hamed:2017jhn}, Arkani-Hamed, Huang and Huang (AHH) proposed a formalism for scattering amplitudes for particles of any spin and mass, adopting this way.
\end{itemize}


A helpful parameterization of the massive spinors is
\begin{eqnarray}
    \lambda^{\alpha}_1=\lambda^\alpha=\sqrt{\mathbf{p}^0+|\vec{\mathbf{p}
    }|}\begin{pmatrix}
        -s\\
        c
    \end{pmatrix},\quad
    \lambda^{\alpha}_2=\eta^\alpha=\sqrt{\mathbf{p}^0-|\vec{\mathbf{p}
    }|}\begin{pmatrix}
        -c\\
        -s^*
    \end{pmatrix},
\end{eqnarray}
\begin{eqnarray}
    \tilde\lambda^{1\dot\alpha}=\tilde{\lambda}^{\dot\alpha}=\sqrt{\mathbf{p}^0+|\vec{\mathbf{p}
    }|}\begin{pmatrix}
        -s^*\\
        c
    \end{pmatrix},\quad
    \tilde\lambda^{2\dot\alpha}=\tilde{\eta}^{\dot\alpha}=\sqrt{\mathbf{p}^0-|\vec{\mathbf{p}
    }|}\begin{pmatrix}
        -c\\
        -s
    \end{pmatrix}.
\end{eqnarray}
where $s=\sin\frac{\theta}{2}e^{i\varphi}, c=\cos\frac{\theta}{2}$ are given by the azimuth angle of $\vec{\mathbf{p}}$. 
This choice of spinor variables exhibits favorable high-energy (H.E.) behavior: $\eta,\tilde{\eta}\to0$. Thus, the spinors $\lambda/\tilde{\lambda} $ are said to be the large components, and the spinors $\eta/\tilde{\eta}$ are said to be the small components. Therefore, in the high-energy limit, the scattering amplitudes with higher powers of  $\eta,\tilde{\eta}$ are usually neglected.

We now turn to the 1-particle states for massive particles. The Dirac equation in Eq.~\eqref{eq:diracequation} implies that the Wigner particle state $|\mathbf p,s\rangle$ for the massive particle with given momentum $\mathbf p$ and spin $s$ is not unique. For fermion case $s=\frac12$, we have
\begin{equation}
|\mathbf p,\tfrac12\rangle \sim \lambda^I \; \text{or}\;\tilde{\lambda}^I,
\end{equation}
which are related to each other by the Dirac equations. 
For the field of spin $s$, the general 1-particle states are
\begin{equation}
   |\mathbf p,s\rangle \sim \lambda^{(I_1} \lambda^{I_2} \dots \lambda^{I_m} \tilde{\lambda}^{I_{m+1}} \dots \tilde{\lambda}^{I_{2s})}\,,\quad m=0,1,\dots, 2s\,. 
\end{equation}
Thus, there are $2s+1$ different spinor expressions for spin-$s$ massive particles.
For example, for the vector field of spin 1, there are 3 spinor expressions
\begin{equation}
|\mathbf p,1\rangle=\lambda^{(I}\lambda^{J)}\,,\quad \lambda^{(I}\tilde\lambda^{J)}\,,\quad \tilde\lambda^{(I}\tilde\lambda^{J)}\,.
\end{equation}
They are of the same little group representation, but different Lorentz representations $(1,0)$, $(\tfrac12,\tfrac12)$ and $(0,1)$. Such ambiguity makes the massive amplitudes complicated.


After considering the particle states, we are now able to describe scattering amplitudes. Since the scattering amplitudes should be Lorentz scalars, both massless and massive spinors should contract to each other in the spinor space. 
For convenience, we adopt the conventional notation that angle brackets \(\langle\cdot\rangle\) and square brackets \([\cdot]\) represent left- and right-handed spinor contractions, respectively. Specifically, angle brackets \(\langle i^Ij^J \rangle=\lambda^{I\alpha}_i\lambda_{j\alpha}^J\) denote the contraction of left-chiral spinors of massive particles $i$ and $j$, while square brackets \([i^Ij^J]=\tilde{\lambda}_{i\dot\alpha}^I\tilde{\lambda}^{J\dot\alpha}_j\) correspond to the contraction of right-chiral spinors.

Besides the Lorentz representation, we should also consider the little group representation for the scattering amplitudes. The massless scattering amplitude $\mathcal{A} = \mathcal{A}(\lambda^\alpha,\tilde{\lambda}_{\dot\alpha})$ is covariant under the little group $U(1)$, and the spinor expression is simple, since there are no indices from the $U(1)$. However, for the masssive amplitude $\mathcal{M}=\mathcal{M}(\lambda^{I\alpha},\tilde{\lambda}^I_{\dot\alpha})$, it is covariant under the little group $SU(2)$ and need to deal with the little group indices. In particular,
it can be seen that symmetrizing the little group indices of a particle yields the representation of a particle with spin \( s \). While any antisymmetric combination of indices such as $I_1I_2-I_2I_1$ will result in a \( SU(2) \)-invariant tensor via the antisymmetric tensor \( \epsilon^{I_1I_2} \) 
\begin{eqnarray}
\lambda^{I_1\alpha_1}\lambda^{I_2\alpha_2}-\lambda^{I_2\alpha_1}\lambda^{I_1\alpha_2}&=&\mathbf m\epsilon^{I_1I_2}\epsilon^{\alpha_1\alpha_2},\\
\tilde{\lambda}^{I_1\dot\alpha_1}\tilde{\lambda}^{I_2\dot\alpha_2}-\tilde{\lambda}^{I_2\dot\alpha_1}\tilde{\lambda}^{I_1\dot\alpha_2}&=&\mathbf{m}\epsilon^{I_1I_2}\epsilon^{\dot\alpha_1\dot\alpha_2},\\
\lambda^{I_1\alpha}\tilde{\lambda}^{I_2\dot\alpha}-\lambda^{I_2\alpha}\tilde{\lambda}^{I_1\dot\alpha}&=&-\epsilon^{I_1I_2}\mathbf{p}^{\dot\alpha\alpha}.
\end{eqnarray}
This implies that the spinors appearing in the amplitude, if they carry little group indices, can be assumed to be fully symmetric by default; there are a total of \( 2s \) such indices, which are used to describe a massive particle with spin \( s \). 
We adopt the \textbf{bolded} spinor notation to describe these spinors that are fully symmetric by default.
For example, a massive scattering amplitude of one spin-1 and two spin-1/2 particles takes the form
\begin{eqnarray}
    [1^{I}2^{(J_1}]\langle2^{J_2)}3^K\rangle\sim[\mathbf{12}]\langle\mathbf{23}\rangle.
\end{eqnarray}
Here, the left-hand side explicitly shows the symmetrized little group indices \( (J_1J_2) \), while the right-hand side uses bolded notation to denote the same full symmetry.


Based on these representations, considerable effort has been devoted to deriving a basis for massive scattering amplitudes. Due to the Dirac equation, the expression of particle states is not uniquely defined, and a choice of basis must be made:

\begin{itemize}
\item One common approach~\cite{Arkani-Hamed:2017jhn,Dong:2021vxo} chooses a chiral basis, in which the mass dimension is not fixed. Taking the three-point fermion-fermion-vector ($ffV$) case 
as an example, where particles 1 and 2 are fermions with spin $\tfrac12$, while particle 3 is a vector with spin $1$, we obtain four amplitudes: 
\begin{equation}
\mathbf{\langle23\rangle\langle31\rangle},\,
\mathbf{\langle2|p_2 p_3|3\rangle\langle31\rangle},\,
\mathbf{\langle23\rangle\langle3|p_3 p_1|1\rangle},\,
\mathbf{\langle2|p_2 p_3|3\rangle\langle3|p_3 p_1|1\rangle}\,.
\end{equation}
All the 3 particles are expressed solely by left-handed spinors, which leads to momentum insertions and makes it inconvenient to establish a direct relation to effective operators at a given mass dimension.
\item An alternative basis~\cite{Durieux:2019eor,Durieux:2020gip,Balkin:2021dko,DeAngelis:2022qco} is designed such that the momentum insertions in the amplitude are minimal. For the $ffV$ case, the four amplitudes take the form
\begin{equation}
\mathbf{\langle23\rangle\langle31\rangle},\,
\mathbf{[23]\langle31\rangle},\,
\mathbf{\langle23\rangle[31]},\,
\mathbf{[23][31]}\,.
\end{equation}
They are introduced as the stripped-connected-term (SCT) basis in Ref.~\cite{Durieux:2019eor}. Drops out the Lorentz scalars such as $\mathbf p_i\cdot \mathbf p_j$ and $\mathbf m$. Considering the linear relations, the SCT basis is further reduced to an independent spinor-structure basis. These linear relations may also be derived using numerical methods~\cite{AccettulliHuber:2021uoa,DeAngelis:2022qco}. 
\end{itemize}

Although these studies provide complete (or over-complete) bases for massive amplitudes, their correspondence to effective operators in the UV is not straightforward. For the first approach, the structures $\mathbf{\langle23\rangle\langle31\rangle}$ and $\mathbf{\langle2|23|3\rangle\langle3|31|1\rangle}$ have different mass dimensions, yet they actually correspond to the same-dimension operators $\bar{\psi}_L F^{\mu\nu}\sigma_{\mu\nu}\psi_L$ and $\bar{\psi}_R F^{\mu\nu}\sigma_{\mu\nu}\psi_R$, respectively. This conversion must be supplied manually. For the second approach, the linear relations are used to obtain a minimal basis, but they introduce factors of mass ratios. These mass ratios do not directly correspond to UV couplings, so their correspondence to the UV still requires conversion. A more detailed discussion is provided in appendix~\ref{app:basis}.
In the following subsections, we will introduce another massive amplitude basis, which has a more straightforward connection to the massless basis construction. 


\subsection{Conformal $N$-particle States and Amplitudes}

The Wigner particle states do not incorporate momentum structure. Therefore, when constructing a basis, one must contend with redundancies arising from the Dirac equation, especially the massive basis. To address this, we can describe particle states using a spacetime symmetry larger than the Poincaré group, namely the conformal group $SU(2,2)$. 
Interestingly, for massless $N$-particle states, a duality exists between the conformal group and a $U(N)$ group, which provides a useful guide for constructing a basis for massless amplitudes.

\subsubsection*{1-particle states described by $SU(2,2)\times U(1)$}

The 1-particle states can be characterized by a representation of $SU(2,2)$, the 4-dimensional conformal group. The basics of the conformal group can be found in references such as Refs.~\cite{Rychkov:2016iqz,Gillioz:2022yze}. Here we present its generators in terms of spinors as
\eqs{
M_{\alpha\beta} &=& i \lambda_{(\alpha} \frac{\partial}{\partial \lambda^{\beta)}},\quad \tilde{M}_{\dot{\alpha} \dot{\beta}} = i \tilde{\lambda}_{(\dot{\alpha}} \frac{\partial}{\partial \tilde{\lambda}^{\dot{\beta})}} , \\
P_{\alpha\dot{\alpha}} &=& \lambda_{\alpha} \tilde{\lambda}_{\dot{\alpha}} ,\quad K_{\alpha\dot{\alpha}} = -\frac{\partial}{\partial \lambda^{\alpha}} \frac{\partial}{\partial \tilde{\lambda}^{\dot{\alpha}}} , \\
D_{+} &=& \frac{1}{2}\left(\lambda_\alpha \frac{\partial}{\partial \lambda_\alpha} + \tilde{\lambda}_{\dot\alpha}\frac{\partial}{\partial\tilde{\lambda}_{\dot\alpha}}\right) +1, }
where $M$ and $\tilde M$ are Lorentz generators, $P$ is the translation generator, $K$ is the special conformal generator, and $D_+$ is the dilatation generator. A representation of the conformal group, which can be uniquely identified by its corresponding primary state (with lowest dimension $\Delta$), is denoted by $[(\Delta,j_l,j_r),\dots]$. In this notation, $(\Delta,j_l,j_r)$ characterises a state possessing scaling dimension $\Delta$ and transforming in the left‑ and right‑handed representations. The symbol $\dots$ signifies further quantum numbers that are not part of the conformal‑group quantum labels. The numbers $j_l$ and $j_r$ still correspond to Lorentz generators $M$ and $\tilde M$,
\begin{align}
M^2\circ(\Delta,j_l,j_r)&=j_l(j_l+1)\times (\Delta,j_l,j_r),\\
\tilde M^2\circ(\Delta,j_l,j_r)&=j_r(j_r+1)\times (\Delta,j_l,j_r),
\end{align}
where we use $\circ$ to denote the application of generators on representations. $\Delta$ is the scaling dimension, which is positive and related to the dilatation $D_+$,
\begin{equation}
D_+\circ(\Delta,j_l,j_r)=\Delta\times(\Delta,j_l,j_r)\,,
\end{equation}
With the new quantum number $\Delta$, the momenta are embedded in the 1-particle states. For example, the particle states of fermions with the minimal $\Delta=\frac32$ are free of momenta,
\eq{\label{eq:conformal_rep1}
\lambda_{\alpha} \sim \left(\frac{3}{2},\frac{1}{2},0\right), \quad \tilde{\lambda}_{\dot{\alpha}} \sim \left(\frac{3}{2},0,\frac{1}{2}\right).
}
As the dimension $\Delta$ increases, the particle state begins to incorporate momentum structures. For $\Delta=\frac52$, we have
\eq{
p_{(\beta\dot\gamma}\,\lambda_{\alpha)} \sim \left(\frac{5}{2},1,\frac{1}{2}\right), \quad 
p_{\beta(\dot\gamma}\,\tilde{\lambda}_{\dot{\alpha})} \sim \left(\frac{5}{2},\frac{1}{2},1\right).
}

Besides the conformal group $SU(2,2)$ acting on spinor space, there is an additional internal symmetry $U(1)$ that serves as the helicity group for massless particles. Since $SU(2,2)$ and this $U(1)$ commute, they give rise to a duality in this quantum system, known as \textit{conformal-helicity duality}~\cite{Henning:2019enq}. 
This $U(1)$ is nothing but the little group of massless particles, with the generator 
\eq{
D_- = \frac{1}{2} \left(-\lambda_{\alpha} \frac{\partial}{\partial \lambda_{\alpha}} +\tilde{\lambda}_{\dot{\alpha}} \frac{\partial}{\partial \tilde{\lambda}_{\dot\alpha}} \right). 
}
whose quantum number is helicity $h$.
Returning to the fermion case, the particle states of different $\Delta$ share the same helicity $\pm\frac12$,
\begin{equation}
\begin{aligned}
\lambda_{\alpha},p_{\beta\dot\gamma}\,\lambda_{\alpha},\dots\quad &\sim\quad h=-\frac12, \\
\tilde{\lambda}_{\dot{\alpha}},p_{\beta\dot\gamma}\,\tilde{\lambda}_{\dot{\alpha}},\dots\quad &\sim\quad h=+\frac12.
\end{aligned}
\end{equation}
It shows that the particle states with different dimensions are degenerate when considering only the helicity representation. To distinguish them, we must consider the full representation of the direct product group $SU(2,2)\times U(1)$. Therefore, a 1-point massless particle state is denoted as
\eq{
[(\Delta,j_l,j_r),h]. 
}
The spinors are in a one-to-one correspondence to the massless particle state,
\eq{
\lambda_{\alpha} \sim \left[\left(\frac{3}{2},\frac{1}{2},0\right),-\frac{1}{2}\right], \quad \tilde{\lambda}_{\dot{\alpha}} \sim \left[\left(\frac{3}{2},0,\frac{1}{2}\right),+\frac{1}{2}\right].
}
Note that the left‑hand side gives the primary states for the conformal representation on the right‑hand side. Since a conformal representation is totally determined by its primary states, this notation is reasonable.

\subsubsection*{$N$-particle states described by $SU(2,2)\times U(N)$}


The $N$-particle states are described by the conformal group $SU(2,2)$ and its dual group, which generalizes the helicity group $U(1)$ for 1-particle states. A natural candidate for this dual group is $U(1)^N$, the direct product of helicity groups for each particle. However, a larger group can help resolve degeneracies among particle states, making $U(N)$ a more suitable choice, as its maximal torus is $U(1)^N$. Therefore, a conformal-helicity duality emerges between $SU(2,2)$ and $U(N)$, and the $N$-particle states are well described by the product group $SU(2,2)\times U(N)$.

Similarly, a general representation of $SU(2,2)\times U(N)$ is given by $[(\Delta,j_l,j_r),R]$, where $R$ denotes a representation of $U(N)$. The massless spinor $\lambda_i$ and $\tilde\lambda^i$, with $i=1,2,\dots,N$, correspond to the fundamental representation $\mathbf N$ and the anti-fundamental representation $\bar{\mathbf N}$ of $U(N)$~\footnote{Strictly speaking, the $\mathbf{N}$ and $\bar{\mathbf{N}}$ representations are those of $SU(N)$, not the full $U(N)$. However, this does not affect the construction of a complete amplitude basis.}, 
\begin{equation}
\lambda_i\sim \left[\left(\frac32,\frac12,0\right),\mathbf N\right],\quad
\tilde\lambda^i\sim\left[\left(\frac32,0,\frac12\right),\bar{\mathbf N}\right].
\end{equation}
So the general particle states are derived by multiplying these massless spinors.

Among these particle states, we are particularly interested in those corresponding to Lorentz scalars, $[(\Delta,0,0),R]$, as they directly contribute to contact scattering amplitudes. For the purpose of constructing amplitudes, it is sufficient to consider the scaling dimension $\Delta$ and the representation of $U(N)$. For example, two massless spinors can combine into symmetric and anti-symmetric representations of $U(N)$,
\begin{align}
\mathbf N\times\mathbf N&=\frac{\mathbf N(\mathbf N+1)}{2}+\frac{\mathbf N(\mathbf N-1)}{2},\\
\bar{\mathbf N}\times \bar{\mathbf N}&=\frac{ \bar{\mathbf N}(\bar{\mathbf N}+1)}{2}+\frac{ \bar{\mathbf N}(\bar{\mathbf N}-1)}{2},.
\end{align}
Only the antisymmetric representation $\frac{\mathbf N(\mathbf N-1)}{2}$ and $\frac{ \bar{\mathbf N}(\bar{\mathbf N}-1)}{2}$ yield Lorentz scalars:
\begin{equation}
\langle ij\rangle\sim [(3,0,0),\frac{ \mathbf N(\mathbf N-1)}{2}],\quad
[ij]\sim [(3,0,0),\frac{ \bar{\mathbf N}(\bar{\mathbf N}-1)}{2}].
\end{equation}
When the particle labels $i$ and $j$ are fixed, this provides an amplitude basis for two fermions and $N-2$ scalars.

For more general cases, the massless amplitudes, corresponding to the $N$-particle states invariant under the Lorentz group, are difficult to obtain, since they suffer from the following redundancies,
\begin{description}
    \item[{\normalfont\itshape Momentum conservation}] The $N$-point amplitude $\mathcal{A}_{N}$ satisfies the momentum conservation that $\sum_{i=1}^N P_i = 0$. The induced relation of amplitudes is
    \begin{equation}
        \sum_{i=1}^N \langle ai\rangle [ib] = 0\,,
    \end{equation}
    where $P_i^{\dot\alpha\alpha} = \tilde\lambda_i^{\dot\alpha}\lambda_i^\alpha = |i]\langle i|$. 
    \item[{\normalfont\itshape Schouten identity}] Any three spinors are not independent, which is stated by the Schouten identity that
    \begin{align}
        \lra{ij}\lambda_k + \lra{jk}\lambda_i + \lra{ki}\lambda_j &= 0\,, \\
        [ij]\tilde\lambda_k + [jk]\tilde\lambda_i + [ki]\tilde\lambda_j &= 0\,. 
    \end{align}
\end{description}
The duality between the conformal group $SU(2,2)$ and $U(N)$ provides a systematic construction of the amplitude basis that specifies the $N$ particles and the dimension $\Delta$. 
This method utilizes the Young tensor formalism to construct the Lorentz invariants, which, due to the duality, correspond to a special kind of Yong diagrams of the $U(N)$ group, referred to as the semi-standard Young tensor (SSYT) method~\cite{Li:2020gnx,Li:2020xlh,Li:2022tec}.

\subsubsection*{SSYT method}

Here we review the SSYT method.
We begin by considering representations of the conformal group $SU(2,2)$. In constructing scattering amplitudes, we only need to focus on representations of its Lorentz subgroup $SU(2)_l\times SU(2)_r$. The two $SU(2)$ factors can be treated independently. Each $SU(2)$ irrep corresponds to a standard Young diagram with at most two boxes per column. For example,
\begin{equation}
\ytableausetup{aligntableaux = center}
    \frac{1}{2} \sim \ydiagram{1}\,,\quad 1\sim \ydiagram{2}\,,\quad 0 \sim \ydiagram{1,1}\,.
\end{equation}
The SSYTs of each diagram form a basis of the corresponding representation. In this basis, the Schouten identity has been removed due to the reduction rules of the SSYTs, such as the Fock condition.

Moving to the spinors, the left- and right-handed spinors correspond to two Young diagrams
\begin{equation}
    \lambda_\alpha \sim \ydiagram{1} \,,\quad \tilde{\lambda}^{\dot\alpha} \sim \ydiagram[*(cyan)]{1}\,,
\end{equation}
where we have used different colors to distinguish them. The Lorentz invariants correspond to the trivial Young diagrams,
\begin{equation}
    \lra{ij} \sim \ydiagram{1,1}\,,\quad [kl] \sim \ydiagram[*(cyan)]{1,1}\,.
\end{equation}
Thus, a contact amplitude takes the form of Young diagram corresponding to the Lorentz group that
\begin{equation}
\label{eq:lorentz1}
    \mathcal{A}_N = \prod^n\lra{ij}\prod^{\tilde{n}}[kl] \sim \underbrace{\ydiagram{2,2}\dots \ydiagram{1,1}}_n \;\times\;  \underbrace{\ydiagram[*(cyan)]{2,2}\dots \ydiagram[*(cyan)]{1,1}}_{\tilde{n}}\,,
\end{equation}
where $n/\tilde{n}$ is the half number of the left-/right-handed spinors.


On the other hand, the spinors are also of the fundamental representations of the $U(N)$ group, 
\begin{equation}
    \begin{aligned}
        (\lambda_1\,,\lambda_2\,,\dots\,,\lambda_N) \in \mathbf{N}\,, \quad (\tilde\lambda_1\,,\tilde\lambda_2\,,\dots\,,\tilde\lambda_N) \in \overline{\mathbf{N}}\,.
    \end{aligned}
\end{equation}
Thus the total momentum $P = \sum_{i=1}^N P_i = \sum_{i=1}^N \tilde{\lambda}_i\lambda_i$ is invariant. The Lorentz representation in Eq.~\eqref{eq:lorentz1} uniquely determine its $U(N)$ group representation, 
\begin{equation}
    \underbrace{\ydiagram{2,2}\dots \ydiagram{1,1}}_n \;\times\; \overline{\underbrace{\ydiagram[*(cyan)]{2,2}\dots \ydiagram[*(cyan)]{1,1}}_{\tilde{n}}}  \;=\; 
    \underbrace{\ydiagram{2,2}\dots \ydiagram{1,1}}_n \;\times\; 
    \underbrace{\left.\begin{ytableau}
        *(cyan) & *(cyan) & \none[\dots] & *(cyan) \\ 
        *(cyan) & *(cyan) & \none[\dots] & *(cyan) \\ 
        \none[\vdots] & \none[\vdots] & \none & \none[\vdots] \\ 
        *(cyan) & *(cyan) & \none[\dots] & *(cyan) \\
    \end{ytableau}\right\}}_{\tilde{n}} N-2\,,
\end{equation}
where we have used the $N$-rank antisymmetric tensor $\epsilon_{i_1i_2\dots i_N}$ to transfer the anti-fundamental boxes to the fundamental ones. According to the Littlewood-Richardson rule of the Young diagram outer product, the tensor product of these two diagrams can be decomposed as
\begin{equation}
\label{eq:ydiagram2}
\ytableausetup{aligntableaux=center}
    \begin{aligned}
        & \underbrace{\ydiagram{2,2}\dots \ydiagram{1,1}}_n \;\times\; 
    \underbrace{\left.\begin{ytableau}
        *(cyan) & *(cyan) & \none[\dots] & *(cyan) \\ 
        *(cyan) & *(cyan) & \none[\dots] & *(cyan) \\ 
        \none[\vdots] & \none[\vdots] & \none & \none[\vdots] \\ 
        *(cyan) & *(cyan) & \none[\dots] & *(cyan) \\
    \end{ytableau}\right\}}_{\tilde{n}} N-2 \;=\;& \ytableausetup{aligntableaux=top}
\underbrace{\begin{ytableau}
        *(cyan) & *(cyan) & \none[\dots] & *(cyan) \\ 
        *(cyan) & *(cyan) & \none[\dots] & *(cyan) \\ 
        \none[\vdots] & \none[\vdots] & \none & \none[\vdots] \\ 
        *(cyan) & *(cyan) & \none[\dots] & *(cyan) \\
    \end{ytableau}}_{\tilde{n}} \underbrace{\ydiagram{2,2}\dots \ydiagram{1,1}}_n \;+\; \dots
    \end{aligned}
\end{equation}
On the right-hand side, the first decomposed diagram contains columns with $N-2$ boxes, while the other diagrams (denoted by the dots) contain columns with $N-1$ or $N$ boxes. Essentially, it is shown that all the decomposed diagrams besides the first one correspond to amplitudes proportional to the total momentum, thus are redundant due to momentum conservation. In particular, the Young diagram shown on the right of Eq.~\eqref{eq:ydiagram2} is called the primary Young diagram, which is determined uniquely by the particle number $N$ and the number of the left- and right-handed spinors $n/\tilde{n}$. Therefore, the independent amplitudes correspond to the SSYTs of the primary Young diagrams. 
The assignment of the numbers to fill the primary diagrams requires the helicity information of all particles
Given this helicity data, the numbers to fill the SSYTs are determined as follows,
\begin{equation}
\label{eq:fill}
    \# i = \tilde{n} - 2h_i \,,\quad i =1,2,\dots N\,,
\end{equation}
where $h_i$ is the helicity of the $i$th particle.

For illustration, we consider an example of the amplitudes of 4 fermions. The helicity structures are classified in 3 different sectors, $(----),(++++),(--++)$. The primary diagrams are 
\begin{equation}
\renewcommand{\arraystretch}{1.4}
    \begin{array}{ccc}
\quad (----)\quad  & \quad (++++)\quad  & \quad (--++)\quad \\
\ydiagram{2,2} & \ydiagram[*(cyan)]{2,2} & \begin{ytableau}
        *(cyan) & \\ *(cyan) & 
    \end{ytableau}
    \end{array}
\end{equation}
According to Eq.~\eqref{eq:fill}, the filling numbers are determined as
\begin{equation}
    \begin{aligned}
        (----)/(++++):\quad & \#1 = \#2 = \#3 = \#4=1\,,\\
        (--++):\quad & \#1=\#2=2\,,\#3=\#4=0\,,
    \end{aligned}
\end{equation}
then the SSYTs and the corresponding amplitudes can be obtained, which is presented in Table~\ref {tab:4fermion}.
\begin{table}[]
\renewcommand{\arraystretch}{1.8}
\ytableausetup{aligntableaux=center,smalltableaux}
    \centering
    \begin{tabular}{|c|c|c|c|c|c|}
\hline
helicity & primary diagram & SSYTs & amplitudes & operators\\
\hline
\multirow{2}{*}{$(++++)$} & \multirow{2}{*}{\ydiagram[*(cyan)]{2,2}} & \begin{ytableau}
    *(cyan) 1 & *(cyan) 2 \\ 
    *(cyan) 3 & *(cyan) 4
\end{ytableau} & $[13][24]$ & $(\psi^\dagger_1\psi^\dagger_3)(\psi^\dagger_2\psi^\dagger_4)$\\
& & \begin{ytableau}
    *(cyan) 1 & *(cyan) 3 \\ 
    *(cyan) 2 & *(cyan) 4
\end{ytableau} & $[12][34]$ & $(\psi^\dagger_1\psi^\dagger_2)(\psi^\dagger_3\psi^\dagger_4)$\\
\hline
\multirow{2}{*}{$(----)$} & \multirow{2}{*}{\ydiagram{2,2}} & \begin{ytableau}
     1 &  2 \\ 
     3 &  4
\end{ytableau} & $\langle13\rangle\langle24\rangle$ & $(\psi_1\psi_3)(\psi_2\psi_4)$\\
& & \begin{ytableau}
     1 &  3 \\ 
     2 &  4
\end{ytableau} & $\langle12\rangle\langle34\rangle$ & $(\psi_1\psi_2)(\psi_3\psi_4)$\\
\hline
$(--++)$ & \begin{ytableau}
        *(cyan) & \\ *(cyan) & 
    \end{ytableau} & \begin{ytableau}
        *(cyan)1 &1 \\ *(cyan)2 & 2
    \end{ytableau} & $\langle12\rangle [34]$ & $(\psi_1\psi_2)(\psi_3^\dagger\psi_4^\dagger)$\\
\hline
    \end{tabular}
    \caption{The primary diagrams, SSYTs, amplitudes, and the corresponding operator basis of the 4-fermion sector.}
    \label{tab:4fermion}
\end{table}
More details and examples of the Young tensor method can be found in Refs.~\cite{Li:2020gnx,Li:2020xlh,Li:2022tec}.


\subsubsection*{Amplitude-operator correspondence}

After deriving the independent contact amplitudes, we can use the operator-amplitude correspondence~\cite{Ma:2019gtx,Li:2020gnx,Li:2020xlh,Li:2022tec} to identify the corresponding independent operators in effective field theories.
Considering an effective operator $\mathcal{O}^{(d)}_N(x)$ of dimension $d$ and $N$ fields, its corresponding amplitudes can be obtained as
\begin{equation} \label{eq:amp-operator}
    \int d^4x \langle 0|\mathcal{O}^{(d)}_N(x) |\Phi_i\dots\Phi_N\rangle = \delta^{(4)}\left(\sum_{i=1}^N p_i\right) \mathcal{A}^{(d)}_N \,,
\end{equation}
where we have dropped the internal structures for now. On the right, the delta function implies momentum conservation, and $\mathcal{A}_N^{(d)}$ is the corresponding amplitude, which can be expressed by the spinors, $\mathcal{A}^{(d)}_N(\lambda_1,\lambda_2,\dots,\lambda_N)$.
According to the previous discussion, the massless particles correspond to the spinor expressions as follows
\begin{equation} \label{eq:correspondence}
\begin{tabular}{c|c|c|c}
    field & helicity & Lorentz representation & spinor form \\
    \hline
    $\phi$ & 0 & (0,0) & $1$ \\
    $\psi$ & $-\tfrac12$ & $\left(\frac{1}{2},0\right)$ & $\lambda_\alpha$ \\
    $\psi^\dagger$ & $+\tfrac12$ & $\left(0,\frac{1}{2}\right)$ &  $\tilde{\lambda}^{\dot\alpha}$ \\
    $F_L$ & $-1$ & (1,0) & $\lambda_\alpha\lambda_\beta$ \\
    $F_R$ & $+1$ & (0,1) &  $\tilde{\lambda}^{\dot\alpha} \tilde{\lambda}^{\dot\beta}$ \\
    $D$ & $0$ & $\left(\frac{1}{2},\frac{1}{2}\right)$ & $\lambda_{\alpha} \tilde{\lambda}_{\dot{\alpha}}$
\end{tabular}
\end{equation}
where $F^{\mu\nu}{L/R} = (F^{\mu\nu}\mp i\tilde{F}^{\mu\nu})/2$ with $\tilde{F}^{\mu\nu} = \epsilon^{\mu\nu\rho\sigma}F_{\rho\sigma}/2$ being the dual field-strength tensor.

Thus, given the SSYTs, the amplitude-operator correspondence leads to an operator basis. For example, the correspondence for four-fermion operators with no derivatives is illustrated in Table~\ref{tab:4fermion}.

\subsection{Massive Spinor from Conformal-Helicity Duality}

In this section, we extend the conformal symmetry from massless particle states to massive ones. Although a massive theory does not genuinely possess conformal symmetry, it is possible to describe the massive system in terms of a massless system with additional particles and to use quantum numbers from the conformal representation to label massive particle states.

The key point is that a massive momentum can be decomposed into two light-like vectors. Thus, a one-particle massive state can be described by a two-particle massless state. As discussed above, an $N$-particle massless state can be described by the symmetry $SU(2,2)\times U(N)$. Here $U(N)$ is dual to the conformal group $SU(2,2)$; we refer to it as the dual group. It is therefore natural to attempt to describe a massive particle using a 2-particle state that carries a representation of $SU(2,2)\times U(2)$. In this formalism, the two linearly independent~\footnote{Here we require $\lambda^1$ and $\lambda^2$ to be non‑proportional, which is equivalent to $\langle \lambda^1 \lambda^2 \rangle \neq 0$. In fact, the mass of the massive particle is entirely given by this overlap: $\boldsymbol{m} = \bigl|\langle \lambda^1 \lambda^2 \rangle\bigr|$. 
Conversely, if the mass is fixed in advance, $\lambda^1$ and $\lambda^2$ possess a residual freedom of choice, which corresponds to the $U(2)$ invariance of $m$. Physically, $U(2)=U(1)\otimes SU(2)$ accounts for a phase choice (transversality) and a rotation, i.e., the selection of the spin axis. Once the overall phase and spin axis are specified, the two spinors are completely fixed.

} massless spinors are packaged together as $\lambda^I$ or $\tilde\lambda^I$, where $I=1,2$. This spinor doublet can describe a massive particle, where the little‑group index $I$ corresponds to the spin components of a spin‑$1/2$ massive‑particle state. We refer to this doublet as a \textit{massive spinor}. When we wish to treat particle states of higher spin $s$, it suffices to totally symmetrize the little‑group indices of $2s$ massive spinors.

We first use the representation labels $(\Delta,j_l,j_r)$ from $SU(2,2)$ to characterize the massive state. In this context, the scaling dimension $\Delta$ is now given by $D_+$, which is slightly different from the massless one,
\eq{
D_+ = \frac{1}{2}\left(\lambda^I_\alpha \frac{\partial}{\partial \lambda^I_\alpha} + \tilde{\lambda}_{\dot{\alpha} I}\frac{\partial}{\partial\tilde{\lambda}_{\dot{\alpha} I}}\right) +2 \quad \mbox{for massive particle},
}
because a massive $D_+$ is the sum of two massless $D_+$'s. Acting on the massive spinors, we obtain 
\eq{D_+ \lambda^I_\alpha = \frac{5}{2} \lambda^I_\alpha,, \quad D_+ \tilde{\lambda}^I_{\dot\alpha} = \frac{5}{2} \tilde{\lambda}^I_{\dot\alpha}.}
It shows that $\Delta=\frac52$, which is different from the massless case where $\Delta=\frac32$ for spinors, as given in Eq.~\eqref{eq:conformal_rep1}. The remaining quantum numbers $j_l$ and $j_r$ remain the same as those for massless spinors.

We next consider the representation of the dual group $U(2)$, which can be factorized as $U(2) = U(1) \times SU(2)$. Here $SU(2)$ is identified as the spin group. Its generators are given by
\eq{
J^{I}_{J} = -\lambda_{\alpha}^{I} \frac{\partial}{\partial\lambda_{\alpha}^J} +\tilde{\lambda}_{\dot{\alpha} J} \frac{\partial}{\partial \tilde{\lambda}_{\dot{\alpha}I}} +\frac{1}{2} \delta^{I}_{J} \left(\lambda_{\alpha}^{K} \frac{\partial}{\partial\lambda_{\alpha}^K} -\tilde{\lambda}_{\dot{\alpha}K} \frac{\partial}{\partial \tilde{\lambda}_{\dot{\alpha}K}}\right),
}
which determines the spin $s$ of the massive particle state. This group is the same as the $SU(2)$ little group in the AHH formalism. The massive spinors are doublets under this spin group, which is similar to the AHH spinors in Eq.~\eqref{eq:massive_doublet}. The additional structure not present in the AHH formalism is the $U(1)$ factor, whose generator is defined by the  spinors as
\eq{
D_- =& \frac{1}{2}\left(-\lambda^I_\alpha \frac{\partial}{\partial \lambda^I_\alpha} + \tilde{\lambda}_{\dot{\alpha} I}\frac{\partial}{\partial\tilde{\lambda}_{\dot{\alpha} I}}\right) \\
}
This $U(1)$ defines a charge $t$ called transversality, which takes values for the left- and right-handed spinors, satisfying $t=j_r-j_l$ only if $\Delta=s+2$.

Combining the representations of $SU(2,2)$ and $U(2)$, a massive particle state can be characterized by the labels $[(\Delta,j_l,j_r), t, s]$. As a result, we assign the massive spinors to the representation that
\begin{equation}
\label{eq:stspinor}
    \lambda^I_\alpha \sim \left[\left(\frac{5}{2},\frac{1}{2},0\right), -\frac{1}{2}, \frac{1}{2}\right]\,,\quad 
    \tilde{\lambda}^I{}^{\dot\alpha} \sim \left[ \left(\frac{5}{2},0,\frac{1}{2}\right), +\frac{1}{2},\frac{1}{2}\right]\,,
\end{equation}
This implies that both the chirality $j_r-j_l$ from the conformal group $SU(2,2)$, and the transversality $t$ from the dual group $U(2)$, can serve to distinguish between left- and right-handed spinors. 

Before considering representations with higher dimensions, we first examine the relation between particle states with $\Delta=s+2$. Although these states carry different values of the transversality quantum number $t$, they can be connected. To reach that, we introduce raising and lowering operators $T^\pm$. In general, for a particle state with arbitrary $\Delta$ and $t$, these operators do not change $\Delta$, but only shift $t$. They are defined as
\begin{equation}\label{eq:T_definition}
    T^+_{\alpha\dot\alpha} = \tilde\lambda^I_{\dot\alpha} \frac{\partial}{\partial\lambda^{I\alpha}}\,,\quad T^-_{\alpha\dot\alpha} = \lambda^I_\alpha \frac{\partial}{\partial\tilde{\lambda}^{I\dot\alpha}}\,.
\end{equation}
Acting on states with $\Delta=s+2$, these operators transform spinors of different $t$ into one another. For example,
\begin{equation}
\text{spin-}\frac{1}{2}:\quad 
    \begin{aligned}
    \tilde{\lambda}^I & & &\xleftrightharpoons[\quad T^-\quad ]{T^+}& &\lambda^I  \\
    \left[\left(\frac{5}{2},0,\frac{1}{2}\right), +\frac{1}{2}, \frac{1}{2}\right] & & &\xleftrightharpoons[\quad T^-\quad ]{T^+}& &\left[\left(\frac{5}{2},\frac{1}{2},0\right), -\frac{1}{2}, \frac{1}{2}\right] 
\end{aligned}
\end{equation}
\begin{equation}
\text{spin-}1:\quad 
 \begin{aligned}
        \tilde\lambda^{(I}\tilde\lambda^{J)} & & & \xleftrightharpoons[T^-]{T^+} &&\qquad \lambda^{(I}\tilde\lambda^{J)} &&\xleftrightharpoons[T^-]{T^+} &&\lambda^{(I}\lambda^{J)}  \\
    [(3,0,1),+1, 1] &&& \xleftrightharpoons[T^-]{T^+} &&\left[\left(3,\frac{1}{2},\frac{1}{2}\right), 0, 1\right] &&\xleftrightharpoons[T^-]{T^+} &&[(3,1,0),-1,1]
    \end{aligned}
\end{equation}
The $T^{\pm}$ operators generate all the possible spinor structures from a given one. Generally, the spin-$s$ representations under $\Delta=s+2$ can be generated by $T^-$ from the $t=s$ one,
\eq{
[(s+2,0,s),+s, s] \xrightarrow{T^-} \left[\left(s+2, \frac{1}{2}, s-\frac{1}{2}\right), s-1, s\right] \xrightarrow{T^-} \cdots \xrightarrow{T^-} [(s+2,s,0), -s, s]. \label{eq:spinorT-}
}
The $[(s+2,0,s),+s,s]$ representation is called the highest weight of spin-$s$, and all the representations above are called the primary representations because their dimension $\Delta=s+2$ is minimal. 
Recall that the transversality $t$ is related to the chirality $j_l-j_r$. The number of possible chirality values is constrained by the spin $s$: a fermion has two chiralities, $j_l-j_r=-1/2,+1/2$, while a vector has three $j_l-j_r=-1,0,+1$, and so on. This imposes the constraint on the transversality-$t$ of spin-$s$ representations to 
\eq{
-s\leq t\leq s, 
}
where $t$ is an integer or a half-integer.

We now turn to particle states with higher dimensions, which involve additional spinors. In the massless case, such additional spinors contribute to the momentum structure $p=\lambda\tilde\lambda$. For  massive particle states, however, we have further possibilities: in addition to forming the momentum $p=\tilde\lambda^I\lambda_I$, we can also contract the massive spinors to define mass spurions:
\begin{equation}
\label{eq:masses}
    m = \det(\lambda^I_\alpha) = \frac{1}{2}\lambda^I{}^\alpha\lambda^J_\alpha \epsilon_{IJ}\,,\quad \tilde{m} = \det(\tilde{\lambda}^I_{\dot\alpha}) = \frac{1}{2}\tilde{\lambda}^I_{\dot\alpha}\tilde{\lambda}^{J\dot\alpha}\epsilon_{IJ}\,,
\end{equation}
where we use $m$ and $\tilde{m}$ instead of $\mathbf{m}$ to indicate they are of different transversalities under the enlarged conformal symmetry.
Their charges under $D_-$ are 
\begin{equation}
D_-\circ m = -m,\quad D_- \circ \tilde m = +\tilde m\,,
\end{equation}
which implies that $m$ carries transversality $t=-1$, while $\tilde m$ carries $t=+1$. Promoting masses to spurions, we can write down the EOM for massive particles consistent with transversalities. For a massive fermion, the Dirac equations take the form
\begin{eqnarray}
\lambda^{I\alpha}\mathbf{p}_{\alpha\dot\alpha}=-m\tilde{\lambda}^I_{\dot\alpha},\quad \tilde{\lambda}^I_{\dot\alpha}\mathbf{p}^{\dot\alpha\alpha}=-\tilde{m}\lambda^{I\alpha},.
\end{eqnarray}
If we impose the constraint ($m = \tilde{m} = \mathbf{m}$), we recover the ordinary Dirac equations. We identify $\lambda^{I}$ and $\tilde{m}\lambda^{I}$ as left-handed, and $\tilde{\lambda}^{I}$ and $m\lambda^{I}$ as right-handed. Thus, the EOM maps one chirality into the other.

Higher-dimensional massive states involving momentum structures can be treated similarly to their massless counterparts. The new element here is the structure involving the mass spurions. To systematically extract such structures, we promote the mass spurions in Eq.~\eqref{eq:masses} to generators. These generators simultaneously change $t$ and $\Delta$ and keep the Lorentz representation. In this way, the mass spurions extend the $U(1)$ transversality group to $ISO(2)$, whose algebra satisfies
\begin{equation}
\label{eq:iso2algebra}
    [D_-,m] = -m\,,\quad [D_-,\tilde{m}] = \tilde{m}\,,\quad [m,\tilde{m}]=0\,.
\end{equation}
In principle, we can repeatedly apply the generators $m$ and $\tilde m$ to construct the representation with higher values of $|t|$. However, among these representations, we are primarily interested in those that are related to the EOMs. For example, the spin-$\frac{1}{2}$ representations induced by the mass flips are
\begin{equation} \label{eq:EOM2}
    \begin{aligned}
        \tilde{\lambda}_{\dot{\alpha}}^I& & &\xrightarrow{m} & & m\tilde{\lambda}_{\dot{\alpha}}^I = -\lambda^{\alpha I} \mathbf{p}_{\alpha\dot{\alpha}} \\
\left[\left(\frac{5}{2},0,\frac{1}{2}\right), +\frac{1}{2}, \frac{1}{2}\right] &&& \xrightarrow{m}&& \left[\left(\frac{7}{2},0,\frac{1}{2}\right),-\frac{1}{2}, \frac{1}{2}\right], \\
\lambda_{\alpha}^I &&& \xrightarrow{\tilde{m}}&& \tilde{m} \lambda_{\alpha}^I = \mathbf{p}_{\alpha\dot{\alpha}} \tilde{\lambda}^{\dot{\alpha} I} \\ 
\left[\left(\frac{5}{2},\frac{1}{2},0\right), -\frac{1}{2}, \frac{1}{2}\right] &&&\xrightarrow{\tilde{m}} &&\left[\left(\frac{7}{2},\frac{1}{2},0\right), +\frac{1}{2}, \frac{1}{2}\right].
    \end{aligned}
\end{equation}
Similarly, the number of chirality flips is also constrained by the spin $s$. Since at most $2s$ flips are possible, $\Delta$ is restricted to
\eq{
s+2 \leq \Delta \leq 3s+2. \label{eq:deltarange}
}
Further chirality flips beyond this range would return the state to the initial one and therefore do not yield independent structures.
We now list all particle states satisfying the above constraints. The spin-$\tfrac12$ representations include
\eq{\label{eq:table_f}
\begin{tabular}{c|cc}
\hline
\diagbox{$\Delta$}{$t$} & $-\frac12$ & $+\frac12$ \\
\hline
$5/2$ & $\lambda^{I}_{\alpha}$ & $ \tilde{\lambda}^{I}_{\dot{\alpha}}$  \\
$7/2$ & $m \tilde{\lambda}^{I}_{\dot{\alpha}}$ & $\tilde{m} \lambda^{I}_{\alpha}$ \\
\hline
\end{tabular}
}
As shown in Eq.~\eqref{eq:EOM2}, the states in the same row are related by the $T^{\pm}$ generators, while states in the same column are related by chirality flips from the EOM. Multiplication by the mass spurions $m,\tilde{m}$ corresponds to skew moves in the table.
Similarly, the spin-$1$ representations include
\eq{\label{eq:table_v}
\begin{tabular}{c|ccc}
\hline
\diagbox{$\Delta$}{$t$} & $-1$ & $0$ & $1$ \\
\hline
$3$ & $\lambda^{(I}_{\alpha} \lambda^{J)}_{\beta}$ & $\lambda^{(I}_{\alpha} \tilde{\lambda}^{J)}_{\dot{\beta}}$ & $\tilde{\lambda}^{(I}_{\dot{\alpha}} \tilde{\lambda}^{J)}_{\dot{\beta}}$ \\
$4$ & $m \lambda^{(I}_{\alpha} \tilde{\lambda}^{J)}_{\dot{\beta}}$ & $\tilde{m} \lambda^{(I}_{\alpha} \lambda^{J)}_{\beta}, \ m \tilde{\lambda}^{(I}_{\dot{\alpha}} \tilde{\lambda}^{J)}_{\dot{\beta}}$ & $\tilde{m} \lambda^{(I}_{\alpha} \tilde{\lambda}^{J)}_{\dot{\beta}}$ \\
$5$ & $m^2 \tilde{\lambda}^{(I}_{\dot{\alpha}} \tilde{\lambda}^{J)}_{\dot{\beta}}$ & ${\color{gray} \mathbf{m}^2 \lambda^{(I}_{\alpha} \tilde{\lambda}^{J)}_{\dot{\beta}}}$ & $\tilde{m}^2 \lambda^{(I}_{\alpha} \lambda^{J)}_{\beta}$ \\
\hline
\end{tabular}
}
The entry colored in gray, $\mathbf{m}^2 \lambda^{(I}_{\alpha} \tilde{\lambda}^{J)}_{\dot{\beta}}$, is considered equivalent to $\lambda^{(I}_{\alpha} \tilde{\lambda}^{J)}_{\dot{\beta}}$, as it corresponds to two chirality flips that return the state to the initial one: 
\eq{
\mathbf{m}^2 \lambda^{(I}_{\alpha} \tilde{\lambda}^{J)}_{\dot{\beta}}=\mathbf{p}^{\dot{\alpha}\alpha}\mathbf{p}^{\dot{\beta}\beta} \tilde{\lambda}^{(I}_{\dot{\alpha}} \lambda^{J)}_{\beta}. 
}
The chirality is unchanged on both sides of the equation.
This is a special case of the spin-1 fields, whose states of transversality $t=0$ do not saturate the range in Eq.~\eqref{eq:deltarange}.
We have defined the $\Delta=s+2$ representations to be primary, so the other ones flipped by the masses are descendants. Specifically, the $\Delta=s+2+l$ representations with $l$ mass flips (except for $\mathbf{m}^2$) are called the $l$-th descendant.

The massive amplitudes constructed from the particle states in eqs.~\eqref{eq:table_f} and \eqref{eq:table_v} are said to be the spinor-transversality (ST) amplitudes, as they carry both spin and transversality information. 
In analogy with single-particle states, ST amplitudes can be categorized by the number of mass flips into primary and $l$-th descendant orders. 
For example, one of the primary amplitudes in class $ffV$ is $\langle\mathbf{13}\rangle [\mathbf{32}]$, which does not involve any mass flip. Its descendant amplitudes are
\eq{
\mbox{primary: }& \langle\mathbf{13}\rangle [\mathbf{32}]; \\
\mbox{1st descendant: }& \tilde{m}_1 \langle\mathbf{13}\rangle [\mathbf{32}], m_2 \langle\mathbf{13}\rangle [\mathbf{32}], m_3 \langle\mathbf{13}\rangle [\mathbf{32}], \tilde{m}_3 \langle\mathbf{13}\rangle [\mathbf{32}]; \\
\mbox{2nd descendant: }& \tilde{m}_1 m_2 \langle\mathbf{13}\rangle [\mathbf{32}], \tilde{m}_1 m_3 \langle\mathbf{13}\rangle [\mathbf{32}], \tilde{m}_1 \tilde{m}_3 \langle\mathbf{13}\rangle [\mathbf{32}], m_2 m_3 \langle\mathbf{13}\rangle [\mathbf{32}], m_2 \tilde{m}_3 \langle\mathbf{13}\rangle [\mathbf{32}]; \\
\mbox{3rd descendant: }& \tilde{m}_1 m_2 m_3 \langle\mathbf{13}\rangle [\mathbf{32}], \tilde{m}_1 m_2 \tilde{m}_3 \langle\mathbf{13}\rangle [\mathbf{32}].
}
These ST amplitudes are characterized by an overall dimension $\Delta$, as well as the transversality $t$ and spin $s$ for each external particle. Therefore, we can classify ST amplitudes using the following three concepts: 
\begin{itemize}
\item Spin category $\mathcal{S}$: a list of spins of the particles involved in the scattering process. The example above is of spin category $\mathcal{S}=\{\frac{1}{2},\frac{1}{2},1\}$.

\item Scaling dimension $\Delta$: the total scaling dimension of the ST amplitude.  If the primary ST amplitude has dimension $\Delta_0$, then its $l$-th descendant has dimension  $\Delta=\Delta_0+l$. For the $ffV$ amplitude, we find $\Delta_0=8$, which follows from summing the be the dimensions of the individual particle states given in Eqs~\eqref{eq:table_f} and \eqref{eq:table_v}. 

\item Transversality category $\mathcal{T}$: a list of transversalities of the particles involved in the scattering process. The primary amplitude of the example above is of transversality category $\mathcal{T} = \{-\frac{1}{2},+\frac{1}{2},0\}$. The 1st descendant amplitudes include
\eq{
\begin{array}{cccc}
\{+\frac{1}{2},+\frac{1}{2},0\}, & \{-\frac{1}{2},-\frac{1}{2},0\}, & \{-\frac{1}{2},+\frac{1}{2},-1\}, & \{-\frac{1}{2},+\frac{1}{2},+1\} \\
(++0), & (--0), & (-+-), & (-++),
\end{array}
}
where the second line gives the abridged notation. 
\end{itemize}
For a specific spin category $\mathcal{S}$, the ST amplitudes span a space that is a direct sum,
\begin{equation}
    \mathbf{M}_\mathcal{S} = \bigoplus_{l=0}^{l_{max}} [\mathbf{M}_\mathcal{S}]_l\,,
\end{equation}
where $l_{max}$ is determined by the spin category.
The total space $\mathbf{M}_\mathcal{S}$ is said to be a spin class, which contains all the ST amplitudes of different transversalities and dimensions. The subspace $[\mathbf{M}_{\mathcal{S}}]_0$ is spanned by all the primary ST amplitudes without any mass. It also takes a direct-sum form that
\eq{
[\mathbf{M}_{\mathcal{S}}]_0 = \bigoplus_{\mathcal{T},\Delta} [\mathbf{M}_{\mathcal{S}}^{\mathcal{T},\Delta}]_0. 
}
For example, the primary ST space of the class $ffV$ is spanned by the amplitudes in four distinct transversality categories,
\eq{
\begin{array}{ccccccccc}
[\mathbf{M}_{ffV}]_0 & = & [\mathbf{M}^{---}_{ffV}]_0 & \oplus & [\mathbf{M}^{-+0}_{ffV}]_0 & \oplus & [\mathbf{M}^{+-0}_{ffV}]_0 & \oplus & [\mathbf{M}^{+++}_{ffV}]_0, \\
&& \langle\mathbf{13}\rangle \langle\mathbf{23}\rangle && \langle\mathbf{13}\rangle [\mathbf{32}] && \langle\mathbf{23}\rangle [\mathbf{31}] && [\mathbf{13}] [\mathbf{23}]
\end{array}
}
These primary amplitudes are  related through the action of the operators of the two $T^\pm$ generators as defined in eq.~\eqref{eq:T_definition}:
\begin{equation} \begin{tikzpicture}[baseline=-0.1cm]
\path(0,0) node(C1) [rectangle] {$[\mathbf{23}][\mathbf{31}]$}(3,1) node(C2) [rectangle] {$[\mathbf{23}]\langle\mathbf{31}\rangle$}(3,-1) node(C3) [rectangle] {$\langle\mathbf{23}\rangle[\mathbf{31}]$}(6,0) node(C4) [rectangle] {$\langle\mathbf{23}\rangle\langle\mathbf{31}\rangle$};
\draw [thick,->] (C1)--node[above]{\small $T_1^-\cdot T_3^-$} (C2);
\draw [thick,->] (C1)--node[above]{\small $T_2^-\cdot T_3^-$} (C3);
\draw [thick,->] (C2)--node[above]{\small $T_2^-\cdot T_3^-$} (C4);
\draw [thick,->] (C3)--node[above]{\small $T_1^-\cdot T_3^-$} (C4);
\end{tikzpicture} \end{equation}
Besides, the linear space $[\mathbf{M}_{\mathcal{S}}]_{l>0}$ is said to be the $l$-th descendant space, which is obtained from the primary ones by the action of $m/\tilde{m}$.

As we have treated the masses $m,\tilde{m}$ as spurions, the ST basis of a specific category always contains descendant amplitudes simply obtained by multiplying $m/\tilde{m}$ with the primary ones. These amplitudes are not necessarily independent in other formalisms.
Here we compare the ST basis and some other basis developed previously.
\begin{itemize}
\item 
All the amplitudes form a polynomial built from spinor contractions.
If we impose the Lorentz symmetry $SO(3,1)$ and the $ISO(2)$ symmetry defined in Eq.~\eqref{eq:iso2algebra}, and eliminate the EOM and IBP redundancies, the resultant polynomials form a quotient algebra, denoted as
\begin{eqnarray}
    M_{ST}\equiv\mathbb{P}^{\mathbb{SO}(3,1) \times ISO(2)}[\boldsymbol{\lambda}_i,\boldsymbol{\tilde{\lambda}}_i]/{\text{\{EOM,IBP\}}},
\end{eqnarray}
which is the ST basis we propose in this paper.

\item If we break the $ISO(2)$ symmetry, which is equivalent to recover the spurions $m,\tilde{m}$ to real coefficients,
\begin{eqnarray}
    m_i\sim \mathbf{m}_i,\quad \tilde{m}_i\sim \mathbf{m}_i\,.
\end{eqnarray}
Accordingly, any descendant amplitudes in the ST basis are not more independent. The resultant amplitude basis is denoted as
\begin{eqnarray}
    M_{ST/EM}\equiv M_{ST}\big/\mathbb{P}[m_i,\tilde{m}_i],
\end{eqnarray}
where $EM$ represents the external mass.


\item 
Furthermore, when searching for resonances in a specific channel experimentally, the resonance corresponds to a fixed value of the invariant mass $\boldsymbol{m}_{\mathrm{Inv}} = \sqrt{(p_i+\dots p_j)^2}$. For example, the s-channel resonance of a $2\to 2$ scattering process corresponds to the invariant mass $\boldsymbol{m}_{\mathrm{Inv}}=\sqrt{s} = \sqrt{(p_1+p_2)^2}$. fixing the invariant mass $\boldsymbol{m}_{\mathrm{Inv}}$, we can perform the partial-wave expansion. 
This defines a new equivalence between the amplitudes, $a_1\simeq a_2$ if $a_1 =a_2 + \boldsymbol{m}_{\mathrm{Inv}} \times a_3$. Eliminating this equivalence, we obtain the basis
\begin{eqnarray}
    M_{ST/IM}\equiv  M_{ST}/\mathbb{P}[m_i,\tilde{m}_i,\boldsymbol{m}_{\mathrm{Inv}}]\,,
\end{eqnarray}
where $IM$ represents the internal mass. This defines the space on which the partial-wave expansion is performed.

\item If we mod out any Lorentz scalars such as $s_{ij}=(\boldsymbol{p}_i+\boldsymbol{p}_j)^2$, $\epsilon_{ijkl}\equiv \epsilon^{\mu\nu\rho\tau}\boldsymbol{p}_{i\mu}\boldsymbol{p}_{j\nu}\boldsymbol{p}_{k\rho}\boldsymbol{p}_{l\tau}$. This additional equivalence arises when we examine the spin‑dependent structure of amplitudes, where all scalars can then be regarded as expansion coefficients for the amplitude. This basis is denoted as
\begin{eqnarray}
    M_{ST/S}\equiv M_{ST}/\mathbb{P}[m_i,\tilde{m}_i,s_{ij},\epsilon_{ijkl}]\,,
\end{eqnarray}
where $S$ represents the Lorentz scalar structure.  These are the stripped-contact terms proposed in Refs.~\cite{Durieux:2019eor,Durieux:2020gip}. Detailed discussion is in appendix.~\ref{app:basis}.
\item Lastly, consider replacing the polynomial ring $\mathbb{P}[m_i,\tilde{m}_i,s_{ij},\epsilon_{ijkl}]$ with its fraction field $\mathrm{Frac}(\mathbb{P})$. This extension allows us to invert scalar quantities such as $1/s_{ij}$, and under this replacement the amplitudes form a module over $\mathrm{Frac}(\mathbb{P})$. Such objects are referred to as spinor structures in Ref.~\cite{Durieux:2019eor,Durieux:2020gip}. For $n$-point massive amplitudes with $n\ge 4$, the number of independent spinor structures is given by $\prod_i (2s_i+1)$. 
Further details are in appendix.~\ref{app:basis}.

\end{itemize}
Compared with other bases, the ST basis does not thoroughly eliminate redundancies.
In this paper, we refer to such ST amplitudes as a basis, because of 2 parts: firstly, the ST amplitudes are independent in the IR if the conformal symmetry is assumed, where the transversality is conserved; secondly, the ST amplitudes of different transversalities have independent UV origins according to the matching in section~\ref{sec:3}.
Consequently, 
its real advantage, however, lies in the UV: it establishes a more transparent correspondence to massless amplitudes, and the basis naturally captures UV structures. 
This is so because the ST basis carries the transversality quantum number $t$, which allows a direct relation to helicity in the UV. We will turn to this topic in the next section.

%% file: sec3-matching.tex
\section{Massless-Massive Amplitudes Matching}
\label{sec:matching}

In this section, we aim to establish a correspondence between massless amplitudes and the ST amplitude. Using this matching technique, we will distinguish the UV origins from the effective operators. This can be divided into two cases: direct matching from massless operators of the same mass dimension, and exceptional matching associated with conserved currents. Based on this massless-massive correspondence, we will present a method for constructing the massive basis as well as the corresponding coefficients in the SMEFT.

\paragraph{UV origins from the effective operators.}
In the SM, to distinguish the underlying massless structures, we classify the SM Lagrangian into two types: kinetic terms and operator terms. Kinetic terms generate elementary conserved currents via minimal coupling, which correspond directly to gauge-invariant contact amplitudes. In contrast, higher-dimensional operators may also give rise to current structures, but their associated contact amplitudes are not gauge-invariant and therefore require a factorized contribution with a pole structure. Generalizing the matching procedure to the SMEFT, the Lagrangian term can be separated as 
\begin{equation}
\begin{aligned}
\mbox{kinetic term}:& \quad \bar{\psi}\slashed{D}\psi,\, D_{\mu}\phi (D^{\mu}\phi)^*,\, -\frac{1}{4} F_{\mu\nu} F^{\mu\nu},& \\
\mbox{operator}:& \quad Y \psi^2 \phi,\, Y^* \bar{\psi}^2 \phi,\, \lambda_3 \phi^3,\, \lambda_4 \phi^4, \quad \frac{C_i^{(d)}}{\Lambda^{d-4}}\mathcal{O}^{(d)},& \\
\end{aligned}
\end{equation}
where the effective operator $\mathcal{O}^{(d)}$, together with the renormalizable terms, falls into the same operator type. For convenience, we distinguish the amplitudes corresponding to these two types of Lagrangian terms:
\begin{equation} \begin{aligned}
\text{kinetic term}\quad &\to\quad \mathcal{K},\\
\text{operator}\quad &\to\quad \mathcal{A}.
\end{aligned} \end{equation}
The amplitude $\mathcal{K}$ related to the kinetic term can include spurious pole. This is because the kinetic terms can be written as covariant derivative $D_\mu\to igA_\mu$, yielding a coupling between a gauge boson $A_\mu$ and a conserved current $J_\mu$, namely $A_\mu J^\mu$. Since the gauge field $A_\mu$ carries the pole structure, so does $\mathcal{K}$. In contrast, the contact amplitude $\mathcal{A}$ corresponds to an operator and has no pole structures, since there is no current structure.

\paragraph{Matching dictionary.}
For later reference, we summarize the objects and conventions used in the matching.  An ST amplitude $\mathbf{M}$ is a massive, $SU(2)$-little-group covariant contact amplitude.  Its high-energy expansion is resolved into MHC amplitudes $[\mathcal{M}]_l$, which have definite helicity and transversality and are ordered by powers of $\mathbf m/E$.  The integer $l$ labels a descendant order, and $l_0$ denotes the first non-vanishing order for a fixed massive structure.  The corresponding UV input is a massless contact amplitude $\mathcal A_{N+r}^{(d)}$, where $d$ is the mass dimension of its local operator and $r$ counts additional Higgs fields whose vacuum expectation values can enter the broken-phase coefficient.  We restrict the explicit matching below to contact UV amplitudes and to the leading case $r=0$, except where Higgs-field insertions are required by the electroweak projection.

The matching takes the following form:
\begin{equation} \begin{aligned}
\mathcal A_N^{(d)}\quad &\longrightarrow\quad [\mathcal M]_{0}
\quad\longrightarrow\quad \mathbf M,\qquad
d=N+n_\lambda,\\
\mathcal{K}\quad &\longrightarrow\quad [\mathcal M]_{l_0>0}
\quad\longrightarrow\quad \mathbf M,\qquad
d=N+n_\lambda-2l_0,
\label{eq:matching_dictionary}
\end{aligned} \end{equation}
where the last step restores massive little-group covariance. Here $n_\lambda$ denotes half the number of spinors $\lambda,\tilde\lambda,\eta,\tilde\eta$. If $[\mathcal M]_{0}$ has a non-vanishing massless limit, this is \textit{direct matching}.  If its leading component vanishes because it is a conserved current, the first non-zero descendant $[\mathcal M]_{l_0>0}$ must instead be matched to the current coupling $A_\mu J^\mu$ and, when allowed, to local operator contributions; this is \textit{exceptional matching}.  Thus the inputs of the construction are a massless Lorentz basis, its gauge tensors and Wilson coefficients, and the electroweak projectors, while the outputs are the ST basis and the broken-phase amplitude coefficients $c^{\mathcal T}_{\mathrm{class}}$.  The detailed definitions and the power-counting derivation are given in the following subsections.

\subsection{Minimal-Helicity-Chiraliry Amplitudes}
\label{sec:MHC}

Since our goal is to relate massless amplitudes to the massive ST amplitude, an intermediate step is needed. This urges us to consider the H.E. behavior of the ST amplitudes. To this end, we decompose the ST spinors into different components as follows,
\begin{eqnarray}
\label{eq:ahh}
    \lambda^{I}_{\alpha}=-\lambda_\alpha \zeta^{+I}+\eta_\alpha\zeta^{-I}\,,\quad \tilde{\lambda}^{I}_{\dot\alpha}=\tilde{\eta}_{\dot\alpha} \zeta^{+I}+\tilde{\lambda}_{\dot\alpha}\zeta^{-I}\,.
\end{eqnarray}
Here, $\zeta^{\pm I}$ characterizes the spin polarization in the $U(2)$ space:
\begin{eqnarray}
    \zeta^{+I}=\begin{pmatrix}
        1\\
        0
    \end{pmatrix},\;\zeta^{-I}=\begin{pmatrix}
        0\\
        1
    \end{pmatrix}.
\end{eqnarray}
In the H.E. regime $E\gg \mathbf m$, $\lambda_{\alpha}$ and $\tilde\lambda_{\dot\alpha}$ are the large components, whereas $\lambda_{\alpha}$ and $\tilde\lambda_{\dot\alpha}$ are the small components. They scale as follows
\eq{
\lambda_{\alpha}, \tilde{\lambda}_{\dot{\alpha}} \sim \sqrt{2E}\,,\quad \eta_{\alpha}, \tilde{\eta}_{\dot{\alpha}} \sim \frac{\mathbf{m}}{\sqrt{2E}}. 
}

In this regime, it is more convenient to describe particle states using the decomposed spinors $\lambda,\eta$ rather than the original ST spinors. Therefore, the full $U(2)=SU(2)\times U(1)_t$ symmetry is no longer appropriate, as certain generators within $U(2)$ mix $\lambda$ and $\eta$. Instead, we focus on a subgroup that better characterizes the particle states in the H.E. limit.

We first consider the $SU(2)$ subgroup, whose generators $J^I_J$ are defined in Eq.~\eqref{eq:SU2_generator}. Among them, one generator corresponds to the helicity $h$ for a massive particle,
\begin{equation}
\begin{split}
H=\frac{1}{2}\left(J_1^1 -J_2^2\right)= \frac{1}{2} \left(-\lambda_{\alpha} \frac{\partial}{\partial\lambda_{\alpha}} + \eta_{\alpha} \frac{\partial}{\partial\eta_{\alpha}} + \tilde{\lambda}_{\dot{\alpha}} \frac{\partial}{\partial\tilde{\lambda}_{\dot{\alpha}}} - \tilde{\eta}_{\dot{\alpha}} \frac{\partial}{\partial\tilde{\eta}_{\dot{\alpha}}}\right)\,,
\end{split}
\end{equation}
The decomposed spinors are eigenstates of this generator, so helicity remains a good quantum number:
\begin{align} \label{eq:h_value}
    H \lambda = -\frac{1}{2}\lambda,\quad H\eta = \frac{1}{2}\eta\,,\quad H\tilde\lambda = \frac{1}{2}\tilde{\lambda}\,,\quad H\tilde{\eta} = -\frac{1}{2}\tilde{\eta}\,.
\end{align}
This shows that the two components of ST spinors $\lambda^I$ or $\tilde\lambda^I$ have opposite helicity. The other two generators in $SU(2)$ are denoted by $J^{\pm}$,
\eq{\label{eq:generator_J+-}
J^+ = -J^2_1 = -\eta_{\alpha} \frac{\partial}{\partial \lambda_{\alpha}} +\tilde{\lambda}_{\dot{\alpha}} \frac{\partial}{\partial\tilde{\eta}_{\dot{\alpha}}},\quad J^- = -J^1_2 = -\lambda_{\alpha} \frac{\partial}{\partial \eta_{\alpha}} +\tilde{\eta}_{\dot{\alpha}} \frac{\partial}{\partial \tilde{\lambda}_{\dot{\alpha}}}. 
}
These act as ladder operators that raise or lower the helicity quantum number. Then we turn to the $U(1)_t$ subgroup, whose generator can be rewritten as
\begin{equation} \label{eq:t_value}
D_-= \frac{1}{2} \left(-\lambda_{\alpha} \frac{\partial}{\partial\lambda_{\alpha}} - \eta_{\alpha} \frac{\partial}{\partial\eta_{\alpha}} + \tilde{\lambda}_{\dot{\alpha}} \frac{\partial}{\partial\tilde{\lambda}_{\dot{\alpha}}} + \tilde{\eta}_{\dot{\alpha}} \frac{\partial}{\partial\tilde{\eta}_{\dot{\alpha}}}\right)\,,
\end{equation}
Under its action, the decomposed spinors exhibit different transversality properties,
\begin{align}
    D_- \lambda = -\frac{1}{2}\lambda,\quad 
    D_- \eta = -\frac{1}{2}\eta\,,\quad 
    D_- \tilde\lambda = \frac{1}{2}\tilde{\lambda}\,,\quad 
    D_- \tilde{\eta} = \frac{1}{2}\tilde{\eta}\,.
\end{align}
This indicates that transversality also becomes a good quantum number in the H.E. limit.  

Therefore, particle states in this framework should be characterized by both helicity $h$ and transversality $t$. Since transversality is closely related to chirality, we refer to the decomposed spinors as helicity-chirality spinors. Among all possible particle states constructed from these spinors, only those satisfying the condition $h=t$ can have a direct correspondence with massless amplitudes at UV. This is because, for a massless particle, chirality must equal helicity. We refer to states satisfying $h=t$ as \textit{minimal-helicity-chirality} (MHC) states. 

Although transversality is related to chirality, the two concepts are not identical. Their differences manifest in two respects. First, chirality is defined by the Lorentz representation $(j_l,j_r)$ of a particle: $c=j_r-j_l$. The spurion masses $m,\tilde m$ carry transversality but do not affect chirality. Therefore, while primary states satisfy $c=t$, descendant states may have $c\neq t$. Second, chirality distinguishes particles from anti-particles. Compared to the MHC state for a particle, the chirality $c$ of the corresponding anti-particle needs to be multiplied by an additional factor of $-1$. The relationship between chirality and transversality of the primary fermion/anti-fermion particle state is illustrated below,
\eq{
\begin{array}{c|ccc}
\hline
\qquad & \mbox{particle} & & \mbox{anti-particle} \\
\hline
t & \pm\frac{1}{2} & & \pm\frac{1}{2} \\
c & \pm\frac{1}{2} & \xrightarrow{\times(-1)} & \mp\frac{1}{2} \\
\hline
\end{array}
}
This distinction motivates our emphasis on "chirality" instead of merely transversality in the naming of MHC states. 

Let us now construct MHC states that are independent of momentum, built solely from the helicity-chirality spinors and mass spurions. We first consider the MHC state for spin-$\frac12$ particles. Starting from an ST particle state with a given transversality $t$, we decompose the ST spinors into components of definite helicity. Among these, only the component that satisfies $h=t$ yields the MHC state. We obtain
\eq{
\begin{tabular}{c|cc}
\hline
\diagbox{$\Delta$}{$t$} & $-\frac12$ & $+\frac12$ \\
\hline
$5/2$ & $-\lambda_{\alpha}$ & $ \tilde{\lambda}_{\dot{\alpha}}$  \\
$7/2$ & $m \tilde{\eta}_{\dot{\alpha}}$ & $\tilde{m} \eta_{\alpha}$ \\
\hline
\end{tabular}
}
As shown in Eqs.~\eqref{eq:h_value} and \eqref{eq:t_value},  the spinors $\eta$ and $\tilde\eta$ individually yield $h\neq t$. They therefore appear in the above table only when multiplied by the spurions $m$ or $\tilde m$. For spin-1, however, the corresponding MHC states (shown in the table below) do not fully obey this rule:
\eq{\label{eq:v_MHC_table}
\begin{tabular}{c|ccc}
\hline
\diagbox{$\Delta$}{$t$} & $-1$ & $0$ & $1$ \\
\hline
$3$ & $\lambda_{\alpha} \lambda_{\beta}$ & $-\lambda_{\alpha} \tilde{\lambda}_{\dot{\beta}}$ {\color{gray} $+\eta_{\alpha} \tilde{\eta}_{\dot{\beta}}$} & $\tilde{\lambda}_{\dot{\alpha}} \tilde{\lambda}_{\dot{\beta}}$ \\
$4$ & $-m \lambda_{\alpha} \tilde{\eta}_{\dot{\beta}}$ & $-\tilde{m} \lambda_{(\alpha} \eta_{\beta)}, \ m \tilde{\eta}_{(\dot{\alpha}} \tilde{\lambda}_{\dot{\beta})}$ & $\tilde{m} \lambda_{\alpha} \tilde{\eta}_{\dot{\beta}}$ \\
$5$ & $m^2 \tilde{\eta}_{\dot{\alpha}} \tilde{\eta}_{\dot{\beta}}$ & {\color{gray} $-m\tilde m\lambda_{\alpha} \tilde{\lambda}_{\dot{\beta}}$} $+m\tilde m\eta_{\alpha} \tilde{\eta}_{\dot{\beta}}$ & $\tilde{m}^2 \eta_{\alpha} \eta_{\beta}$ \\
\hline
\end{tabular}
}
For states with $t=0$, both $-\lambda\tilde\lambda$ and $\eta\tilde\eta$ satisfy $h=t=0$, so their sum gives the complete MHC states. For convenience, we shade in gray those terms in which $\eta$ and a spurion do not appear together, such as $\eta\tilde\eta$. 
It is safe to ignore these terms since they give no more independent massive amplitudes when recovering the little group covariance, which will be discussed later.
By analogy with the classification of ST particle states, we refer to the MHC term shown in black with conformal dimension $\Delta=2+s$ as the primary MHC state, and the others as the descendant MHC states.

Recall that $J^+$ and $J^-$ flip the helicity, while the spurions $m$ and $\tilde m$ flip transversality. By combining these operations, specifically through $mJ^-$ or $\tilde m J^+$, we can flip the primary MHC particle states and the descendants while preserving the condition $h=t$. It is helpful to express these flips diagrammatically in the following. We distinguish the fermions into particles and anti-particles by the direction of the arrows,
\begin{equation}
\mbox{spin-} \frac{1}{2}:\quad 
\begin{aligned}
-\lambda_{\alpha} &= \Ampone{1.5}{-\frac{1}{2}}{\fer{red}{i1}{v1}}&
&\xrightarrow{\tilde{m} J^+}& \tilde{m} \eta_{\alpha} &= \Ampone{1.5}{+\frac{1}{2}}{\ferflip{1.5}{0}{cyan}{red}}, \\
\tilde{\lambda}_{\dot{\alpha}} &= \Ampone{1.5}{+\frac{1}{2}}{\fer{cyan}{i1}{v1}}& &\xrightarrow{m J^-}& m \tilde{\eta}_{\dot{\alpha}} &= \Ampone{1.5}{-\frac{1}{2}}{\ferflip{1.5}{0}{red}{cyan}}, \\
 -\lambda_{\alpha} &= \Ampone{1.5}{-\frac{1}{2}}{\antfer{cyan}{i1}{v1}}&
&\xrightarrow{\tilde{m} J^+}& \tilde{m} \eta_{\alpha} &= \Ampone{1.5}{+\frac{1}{2}}{\antferflip{1.5}{0}{red}{cyan}}, \\
\tilde{\lambda}_{\dot{\alpha}} &= \Ampone{1.5}{+\frac{1}{2}}{\antfer{red}{i1}{v1}}& &\xrightarrow{m J^-}& m \tilde{\eta}_{\dot{\alpha}} &= \Ampone{1.5}{-\frac{1}{2}}{\antferflip{1.5}{0}{cyan}{red}}. 
\end{aligned}
\end{equation}
The first two lines are the flips of particles, and the last two lines are of antiparticles. The color of each diagram shows the chirality, with red for $-\frac{1}{2}$ and cyan for $+\frac{1}{2}$; and the number on the right is the helicity.
We denote the chirality flip ($m,\tilde{m}$) by a bolded cross "\begin{tikzpicture}
\draw[very thick] plot[mark=x,mark size=2.5] coordinates {(0,0)};
\end{tikzpicture}", and the helicity flip ($J^\pm$) by "/". The flip scenario shows the chirality before and after the flip. Take the first line as an example. It can be read from the diagram $\Ampone{1.5}{+\frac{1}{2}}{\ferflip{1.5}{0}{cyan}{red}}$ that the chirality before the flip is $-\frac{1}{2}$, because the part closer to the dot is $\Ampone{1}{ }{\fer{red}{i1}{v1}}$; and the chirality after the flip is $+\frac{1}{2}$. 

For the flips of vector bosons, we use wavy lines to represent them and use brown for the chirality 0 states. Thus, the flips are
\begin{equation}
\mbox{spin-}1:  \ 
\begin{aligned}
\lambda_{(\alpha} \lambda_{\beta)} &= \Ampone{1}{-1}{\bos{i1}{red}} \hspace{0.1em} \xrightarrow{\tilde{m}J^+} \tilde{m} \lambda_{(\alpha} \eta_{\beta)}\hspace{0.2em} = \Ampone{1.2}{0}{\bosflip{1.2}{0}{red}{brown}} \hspace{0.3em}\xrightarrow{\tilde{m}J^+} \tilde{m}^2 \eta_{(\alpha}\eta_{\beta)} = \Ampone{1.2}{+1}{\bosflipflip{1.2}{0}{red}{brown}{cyan}}, \\
-\lambda_{\alpha} \tilde{\lambda}_{\dot{\beta}} &= \Ampone{0.8}{0}{\bos{i1}{brown}} 
\left\{\begin{aligned}
&\xrightarrow{mJ^-} -m \lambda_{\alpha} \tilde{\eta}_{\dot{\beta}} = \Ampone{1.2}{-1}{\bosflip{1.2}{0}{brown}{red}} \xrightarrow{\tilde{m}J^+} \\
&\xrightarrow{\tilde{m}J^+} \tilde{m} \eta_{\alpha} \tilde{\lambda}_{\dot{\beta}}\hspace{0.78em} = \Ampone{1.2}{+1}{\bosflip{1.2}{0}{brown}{cyan}} \xrightarrow{mJ^-}
\end{aligned}\right. \hspace{0.1em}
m\tilde{m} \eta_{\alpha} \tilde{\eta}_{\dot{\alpha}} = \Ampone{1.2}{0}{\bosflipflip{1.2}{0}{brown}{red}{brown}}, \\
\tilde{\lambda}_{(\dot{\alpha}} \tilde{\lambda}_{\dot{\beta})} &= \Ampone{1}{+1}{\bos{i1}{cyan}} \hspace{0.1em} \xrightarrow{mJ^-} m \tilde{\lambda}_{(\dot{\alpha}} \tilde{\eta}_{\dot{\beta})}\hspace{0.2em} = \Ampone{1.2}{0}{\bosflip{1.2}{0}{cyan}{brown}} \hspace{0.3em}\xrightarrow{mJ^-} m^2 \tilde{\eta}_{(\dot{\alpha}} \tilde{\eta}_{\dot{\beta})} = \Ampone{1.3}{-1}{\bosflipflip{1.3}{0}{cyan}{brown}{red}}.
\end{aligned} 
\end{equation}
In the second line, both the primary state $-\lambda_{\alpha} \tilde{\lambda}_{\dot{\beta}}$ and the 2th descendant state $m\tilde{m}\eta_{\alpha} \tilde{\eta}_{\dot{\beta}}$ carry helicity $0$, so they are not independent.   Constrained by $SU(2)$ covariance, their combination must take the form
\begin{equation} \label{eq:longitudinal}
-\lambda_{\alpha} \tilde{\lambda}_{\dot{\beta}}+\frac{1}{\mathbf m^2}\times m\tilde{m}\eta_{\alpha} \tilde{\eta}_{\dot{\beta}}
\end{equation}
which matches the complete state $-\lambda_{\alpha} \tilde{\lambda}_{\dot{\beta}}+\eta_{\alpha} \tilde{\eta}_{\dot{\beta}}$ appearing in eq.~\eqref{eq:v_MHC_table}. Lastly, scalar bosons are denoted by dashed lines and do not need flipping, 
\eq{\label{eq:scalar_state}
\mbox{spin-}0: \ 1 = \Ampone{1.2}{0}{\sca{i1}}. 
}

The states above do not yet incorporate momentum dependence. We use the spin-0 case to illustrate how the classification extends when momentum structures are included. The $\Delta=2$ state shown in Eq.~\eqref{eq:scalar_state} has no momentum structure, while momentum dependence first appears at $\Delta=3$. In this case, we refer to $\lambda_{\alpha} \tilde{\lambda}_{\dot{\alpha}}$ as the primary state and $m\tilde m\eta_{\alpha} \tilde{\eta}_{\dot{\alpha}}$ as the descendant state. As in Eq.~\eqref{eq:longitudinal}, these two are not independent. We can also restore $SU(2)$ covariance by restricting the coefficients of the two terms, which gives the complete massive momentum $\mathbf{p}_{\alpha\dot{\alpha}} = \lambda_{\alpha} \tilde{\lambda}_{\dot{\alpha}} +\eta_{\alpha} \tilde{\eta}_{\dot{\alpha}}$.

By contracting these MHC states, we obtain the MHC amplitudes, which correspond precisely to the $\mathcal{H}=\mathcal{T}$ components of ST amplitudes. These MHC amplitudes are also characterized by spin $\mathcal{S}$ and transversality $\mathcal{T}$.
The MHC amplitude space $\mathcal{M}_{\mathcal{S}}$ is isomorphic to the ST one $\mathbf{M}_{\mathcal{S}}$,
\eq{
\mathcal{M}_{\mathcal{S}} \sim \mathbf{M}_{\mathcal{S}}. 
}
Therefore, the MHC amplitudes can also be categorized into primary $[\mathcal{M}_{\mathcal{S}}]_0$ and $l$-th descendant $[\mathcal{M}_{\mathcal{S}}]_l$. Using mass spurions and $J^{\pm}$ defined in eq.~\eqref{eq:generator_J+-}, we can give a recursion relation,
\eq{
[\mathcal{M}]_{l+1} \subset \tilde{m} J^+ [\mathcal{M}]_l \oplus mJ^- [\mathcal{M}]_l, \label{eq:MHC-recur}
}
where $\subset$ means the recursive method may lead to a redundant space, but $[\mathcal{M}]_{l+1}$ considers the independent one. The primary MHC amplitudes are composed of $\lambda$ and $\tilde{\lambda}$, and the $l$-th descendant contain $l$ $\tilde{m}\eta$ and $m\tilde{\eta}$. We still take the $ffV$ example:
\eq{
\mbox{primary: }& \langle13\rangle [32]; \\
\mbox{1st descendant: }& -\tilde{m}_1 \langle\eta_13\rangle [32], m_2 \langle13\rangle [3\eta_2], -m_3 \langle1\eta_3\rangle [32], \tilde{m}_3 \langle13\rangle [\eta_32]; \\
\mbox{2nd descendant: }& -\tilde{m}_1 m_2 \langle\eta_13\rangle [3\eta_2], -\tilde{m}_1 m_3 \langle\eta_13\rangle [\eta_32], \tilde{m}_1 \tilde{m}_3 \langle\eta_1\eta_3\rangle [32], \\ & m_2 m_3 \langle13\rangle [\eta_3\eta_2], -m_2 \tilde{m}_3 \langle1\eta_3\rangle [3\eta_2], \\
& {\color{gray} -m_3\tilde{m}_3 \langle1\eta_3\rangle [\eta_32]}; \\
\mbox{3rd descendant: }& -\tilde{m}_1 m_2 m_3 \langle\eta_13\rangle [\eta_3\eta_2], \tilde{m}_1 m_2 \tilde{m}_3 \langle\eta_1\eta_3\rangle [3\eta_2], \\
&{\color{gray} \tilde{m}_1 m_3 \tilde{m}_3 \langle\eta_1\eta_3\rangle [\eta_32], -m_2 m_3 \tilde{m}_3 \langle1\eta_3\rangle [\eta_3\eta_2]}; \\
{\color{gray} \mbox{4th descendant: }}&{\color{gray} \tilde{m}_1 m_2 m_3 \tilde{m}_3 \langle\eta_1\eta_3\rangle [\eta_3\eta_2]}. 
}
The amplitudes colored in gray are ignorable when considering the $SU(2)$ covariance and modding out the mass shell $\mathbf{m}^2$. The redundancies can always be removed by restricting the coefficients of each amplitude. Suppose the primary amplitude is $c_1 \langle13\rangle [32]$, and the 2nd descendant amplitude is $- c_2 m_3\tilde{m}_3 \langle1\eta_3\rangle [\eta_32]$. The $SU(2)$ covariance and mass shell requires that $c_1=c_2/\mathbf m^2_3$, because the helicity-0 state in spin-1 is $-\lambda_{\alpha}\tilde{\lambda}_{\dot{\beta}} +\eta_{\alpha}\tilde{\eta}_{\dot{\beta}}$. The other amplitudes colored in gray are determined in the same way.


The above classification for MHC amplitude actually gives a good expansion of the massive amplitude in the H.E. regime $E\gg \mathbf m$, which can be shown by a power counting analysis involving energy scale $E$ and mass scale $\mathbf m$. Recall that the spinors exhibit different scaling behavior: $\lambda,\tilde{\lambda}\sim \sqrt{E}$ and $\eta,\tilde{\eta}\sim\mathbf m/\sqrt{E}$. To ensure the $N$-point MHC amplitude has specific power counting, we apply the separated momentum conservation when taking the H.E. limit,
\eq{
\begin{cases}
\sum_{i=1}^{N} p_i = 0 \\
\sum_{i=1}^{N} \eta_i =0.
\end{cases}
}
Therefore, the energy and mass scaling of the $ffV$ amplitudes is
\eq{\label{eq:ffV_scaling}
\mbox{primary: }& \langle13\rangle [32] \sim 0; \\
\mbox{1st descendant: }& -\tilde{m}_1 \langle\eta_13\rangle [32] \sim \mathbf{m}^2 E; \\
\mbox{2nd descendant: }& -\tilde{m}_1 m_2 \langle\eta_13\rangle [3\eta_2] \sim \mathbf{m}^4; \\
\mbox{3rd descendant: }& -\tilde{m}_1 m_2 m_3 \langle\eta_13\rangle [\eta_3\eta_2] \sim \frac{\mathbf{m}^6}{E}.
}
The primary amplitude vanishes due to the separated momentum conservation. The amplitudes in one order are scaled the same. The order of MHC amplitudes is highly relevant to the order of energy scaling in the H.E. limit.
It can be inferred that 
\eq{
[\mathcal{M}]_{l+1} \sim \frac{\mathbf{m}^2}{E} [\mathcal{M}]_l. 
}

More generally, the power counting of an MHC amplitude is determined by the number of spinors, denoted by
\begin{eqnarray}
    n_{\lambda}\equiv \text{half number of }\lambda,\tilde{\lambda},\eta,\tilde{\eta}.
\end{eqnarray}
Note that the spurion masses $m\sim \lambda\eta$ and $\tilde m\sim \tilde\lambda\tilde\eta$ are composed of spinors, so each counts as two. However, the product $m\tilde{m}$ is identified as the physical mass squared $\mathbf{m}^2$, and thus does not enter into this counting. The MHC amplitude then scales as
\begin{equation} \label{eq:MHC_scaling}
[\mathcal{M}]_l\sim E^{n_\lambda}\left(\frac{\mathbf m}{E}\right)^{2l} \sim E^{n_\lambda - 2l} {\mathbf m}^{2 l}
\end{equation}
For the $ffV$ example, the 1st descendant amplitude $-\tilde{m}_1 \langle\eta_13\rangle [32]$ has $n_\lambda=3$ and $l=1$, which reproduces the result in Eq.~\eqref{eq:ffV_scaling}.

The above scaling analysis does not account for the coefficients $c_l$ of the amplitude $[\mathcal M]_l$. It is important to note that the mass dimensions of MHC amplitudes $[\mathcal M]_l$ at different orders are not identical, since $m\tilde\eta$ has a higher mass dimension than $\lambda$. By dimensional analysis, an $N$-point object has mass dimension $4-N$. In a renormalizable theory, this requires the coefficient to scale as
\begin{equation}
c_l\sim \mathbf m^{4-N-n_\lambda}.
\end{equation}
Since different descendant orders $l$ correspond to different values of $n_\lambda$, the mass dimension of the coefficient varies with $l$.

The above introduction of the MHC amplitude, together with the power-counting analysis, will then help us build the massless-massive correspondence in the following subsections.

\subsection{Massless-Massive Correspondence}

With the MHC amplitude as a tool, we are now in a position to establish a massless-massive correspondence. Before doing so, it is important to distinguish between the three types of expansion orders that appear in this paper. In matching an $N$-point massive amplitude, we assume that the corresponding UV object could be a massless amplitude with more external particles. The three notions of order are as follows:
\begin{itemize}
\item The IR power counting order $l$, labeling the $l$-th MHC amplitude $[\mathcal{M}]_l$, with first non-vanishing order $l_0$.
\item The UV power counting order $r$, counting the number of additional Higgs legs in the massless amplitude $\mathcal{A}_{N+r}$.
\item Another UV power counting order $d$, denoting the mass dimension of the massless effective operator $\mathcal{O}^{(d)}$.
\end{itemize}
Note that $d$ also corresponds to the conformal dimension of the massless amplitude.

To relate these different orders, we perform a power-counting analysis. We first consider the power counting for MHC amplitudes. In the H.E. regime where $E\gg \mathbf{m} $, an $N$-point MHC amplitude can be expanded in powers of $\mathbf{m}/E$, as shown in eq.~\eqref{eq:MHC_scaling}. For our purpose, it is sufficient to focus on the energy scaling and neglect the mass scaling:
\begin{equation} \label{eq:MHC_E_scaling}
[\mathcal{M}]_l\sim  E^{n_{\lambda}-2l},
\end{equation}

It is important to note that the leading order (LO) term in this expansion is not necessarily $l=0$. To illustrate this, consider the following primary and descendant MHC amplitudes: 
\begin{equation} \label{eq:ffV_case}
\begin{aligned}
{[\mathcal{M}]_0}&=\langle13\rangle [32]& &\sim E^2,\\
[\mathcal{M}]_1&= -\tilde{m}_1 \langle\eta_13\rangle [32]& &\sim E^1,\\
[\mathcal{M}]_2&= -\tilde{m}_1 m_2 \langle\eta_13\rangle [3\eta_2]& &\sim E^0,\\
&\cdots
\end{aligned}
\end{equation}
They correspond to $n_\lambda=2,3,4,\cdots$, respectively. These amplitudes can be interpreted as a $ffV$ amplitude or as an amplitude involving additional scalar bosons (such as $ffVS$). Both interpretations yield the same power-counting behavior, but their leading orders differ. For the $ffVS$ case, the leading order is the primary amplitude $[\mathcal M]_0$. However, for the $ffV$ case, the leading order is the descendant amplitude $[\mathcal M]_1$ due to the constraint of momentum conservation, $\sum_i p_i=0$. In general, we denote the true leading order as $[\mathcal M]_{l_0}$.



\paragraph{Direct matching}
Let us first consider the case with leading order $l_0=0$. We refer to this as \textit{direct matching}. In this case, there are two types of UV origins: massless contact amplitudes $\mathcal{A}$ and factorized amplitudes $\mathcal{F}$. We consider the contact amplitude first. An $N$-point massless contact amplitude exhibits the energy scaling 
\begin{equation} \label{eq:massless_scaling1}
\mathcal{A}_{N}^{(d)}\sim E^{d-N},
\end{equation}
where the dimension $d$ is defined as 
\begin{equation}
d=N+n_\lambda^{\text{massless}},
\end{equation}
and $n_\lambda^{\text{massless}}$ counts half the number of $\lambda$ and $\tilde\lambda$ in the massless amplitude. This dimension $d$ corresponds to the mass dimension for an effective operator, as shown in Eq.~\eqref{eq:amp-operator}. Note that the massless amplitude can have $r$ additional particles. In a theory with spontaneous symmetry breaking, these additional particles should be Higgs bosons. Thus, increasing the number of additional particles does not change the energy scaling behavior:
\begin{equation} \label{eq:massless_scaling}
\mathcal{A}_{N+r}^{(d+r)}\sim E^{d-N},
\end{equation}
The dimension of the corresponding effective operator is enlarged to $(d+r)$. This corresponds to contributions from operators with more Higgs doublets $H$ or $H^\dagger$. In this case, the additional Higgs bosons are not inserted into external lines but rather appear at the effective operator vertex, changing the contact amplitude from $\mathcal{A}_{N}^{(d)}$ to $\mathcal{A}_{N+1}^{(d+1)}$. This can be represented diagrammatically as
\eq{
\begin{tikzpicture}[baseline=-0.1cm] \begin{feynhand}
\setlength{\feynhandarrowsize}{4pt}
\vertex [particle] (i1) at (1.5,0) {$-\frac{1}{2}$}; 
\vertex (v2) at (0.75,0);
\vertex [particle] (i2) at (0.75,0.75) {};
\propag [antfer] (i1) to (v1);
\propag[sca] (i2) to (v1);
\vertex[blob, fill=white, draw=black, minimum size=15pt] (v1) at (0,0) {$O$};
\end{feynhand} \end{tikzpicture}
\quad \to \quad 
\begin{tikzpicture}[baseline=-0.1cm] \begin{feynhand}
\setlength{\feynhandarrowsize}{4pt}
\vertex [particle] (i1) at (1.5,0) {$+\frac{1}{2}$}; 
\vertex (v1) at (0,0);
\antfer{red}{i1}{v1};
\vertex[blob, fill=white, draw=black, minimum size=15pt] (v1) at (0,0) {$O$};
\end{feynhand} \end{tikzpicture}
=\tilde{\lambda}_{\dot{\alpha}} \ ,
}
The additional Higgs bosons acquire vacuum expectation values (VEVs) and contribute to the MHC amplitude coefficients. This type of contribution will be considered in this paper.

\begin{figure}[htbp]
\centering
\includegraphics[width=0.8\linewidth]{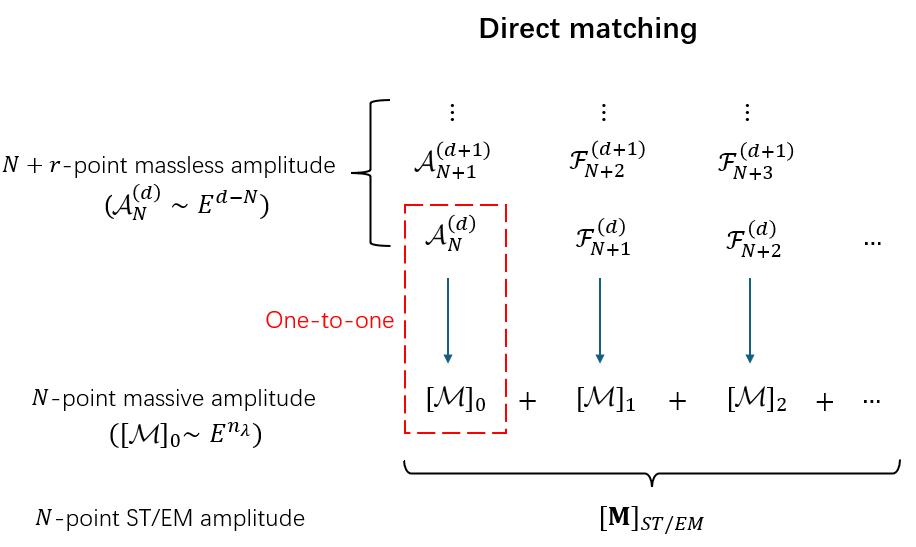}
\caption{Illustration of direct matching between massless contact amplitudes $\mathcal{A}$, massless factorized amplitude $\mathcal{F}$, and massive amplitudes $\mathcal{M}$. For the massless amplitude, $r$ denotes the number of additional Higgs bosons, $(d)$ denotes the mass dimension of the effective operator insertion. For the MHC amplitude, $l$ denotes its descendant order. The diagram shows a one-to-one correspondence between the leading-order massive amplitude $[\mathcal{M}]_0$ and the massless contact amplitude $\mathcal{A}_N^{d}$. Furthermore, the massive amplitudes $[\mathcal{M}]_l$ of different orders are unified into $\mathcal{M}_{ST/EM}$,  where the external masses have been modded out.
}
\label{fig:direct}
\end{figure}

We are now ready to build the massless-massive correspondence by matching these energy scalings. We find the relation $d=N+n_\lambda$, which is independent of $r$. Therefore, the massless contact amplitudes with different particle number can all match to the leading MHC amplitude $[\mathcal{M}]$. In this case, the MHC amplitude can receive contributions from UV operators of different mass dimensions. We can write the matching with the coefficient explicitly as
\begin{equation}\begin{aligned}
\text{LO matching}:\quad c_{0} [\mathcal{M}]_{0} \sim 
\sum_{r\ge 0} v^r \mathcal{A}_{N+r}^{(d+r)}.
\end{aligned} \end{equation}
Here we introduce the VEV $v$ to ensure that both sides of the equations have the same mass scaling, although we do not write it explicitly. The $r$ additional Higgs bosons contribute their VEVs and enter the coefficient $c_l$. From the power-counting analysis, we obtain
\begin{equation}
c_{0} \sim \frac{1}{\Lambda^{d-4}} \sum_r \left(\frac{v}{\Lambda}\right)^{r}.
\end{equation}
When we consider only the massless contact amplitude with the lower mass dimension, this matching will be a one-to-one correspondence, as shown in Figure~\ref{fig:direct}.

As an example, we consider the operator $\psi^{\dagger} (\slashed{D}\phi^*) \psi \phi$. It can match to a 4-point primary MHC amplitude of class $ffVS$ with helicity-$(-+00)$,
\eq{
\begin{array}{ccc}
\mbox{MHC} && \mbox{massless} \\
{[\mathcal{M}_{ffVS}^{-+00}]}_0 = \langle13\rangle [32] & \quad\xrightarrow{H.E.} & \quad  \psi^{\dagger}_2(\slashed{D}\phi^*_4) \psi_1 \phi_3\sim -\langle14\rangle [42],
\end{array}
}
where the subscript on the right-hand side denotes particle labels. The coefficient of this MHC amplitude is thus 
\eq{\label{eq:4pt_coef}
c_{ffVS}^{-+00} = -\frac{C}{\Lambda^2}. 
}
where $C$ is the Wilson coefficient of the operator $\psi^{\dagger} (\slashed{D}\phi^*) \psi \phi$.

The other UV origin is factorized massless amplitudes $\mathcal{F}$, which have at least one additional Higgs boson. For an insertion of a single operator $O^{(d)}$ with fixed dimension $d$, the scaling behavior will depend on the number of additional Higgs bosons:
\begin{equation}
\mathcal{F}^{(d)}_{N+r}\sim E^{d-N-r},\quad r>0.
\end{equation}
This factorized amplitude corresponds to fixing the effective operator while adding Standard Model vertices to the amplitude. This contribution is already captured by the on-shell Higgsing or Higgs insertion technique, where additional Higgs bosons are attached to external lines. For example, the attachment to an external fermion is represented by
\eq{
\begin{tikzpicture}[baseline=-0.1cm] \begin{feynhand}
\setlength{\feynhandarrowsize}{4pt}
\vertex [particle] (i1) at (1.5,0) {$-\frac{1}{2}$}; 
\vertex (v2) at (0.75,0);
\vertex [particle] (i2) at (0.75,0.75) {};
\propag [antfer] (i1) to (v1);
\propag[sca] (i2) to (v2);
\vertex[dot] (v1) at (0,0) {};
\end{feynhand} \end{tikzpicture}
\quad \to \quad 
 \Ampone{1.5}{-\frac{1}{2}}{\antferflip{1.5}{0}{cyan}{red}}=m\times \tilde{\eta}_{\dot{\alpha}}  \ ,
}
which corresponds to the subleading MHC state $m\tilde\eta$. However, such constructions will not be considered in this paper, as our primary focus is on contact UV amplitudes. A detailed discussion of Higgs insertion can be found in  Refs.~\cite{Ni:2026wiz, Ni:2026mia}. 

Comparing the energy scaling, we can match the massless factorized amplitude $\mathcal{F}$ to the higher-order MHC amplitude $[\mathcal{M}]_{r}$ 
\begin{equation} \begin{aligned}
&\text{higher-order matching}:& [\mathcal{M}]_{r}&\sim v^r\mathcal{F}_{N+r}^{(d)}.\\
\end{aligned} \end{equation}
More generally, the dimension $(d)$ can also increase in the higher-order matching, as shown in Figure~\ref{fig:direct}.In this paper, we are interested in the contact UV amplitude, so we can focus on the LO matching, and neglect the higher-order matching.






\begin{figure}[htbp]
\centering
\includegraphics[width=0.9\linewidth]{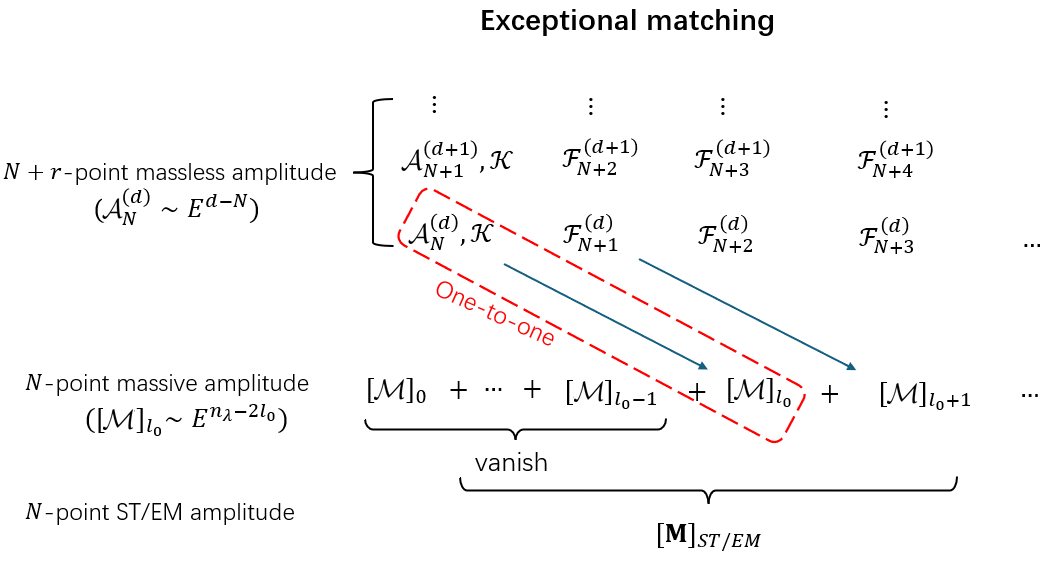}
\caption{Illustration of exceptional matching. In this case, the leading-order MHC amplitude is $[\mathcal{M}]_{l_0}$,  and the matching requires some deformations.
}
\label{fig:exceptional}
\end{figure}

\paragraph{Exceptional matching}
Next, we consider the case with leading MHC order $l_0>0$, which we refer as the \textit{exceptional matching}. The leading MHC amplitude $[\mathcal{M}]_{l_0}$ can correspond to the massless conatct amplitude $\mathcal{A}^{(d)}_N$, with the relation $d=N+n_\lambda-2l_0$. However, in this case, we need to include a new type of UV origin: the kinetic term $\mathcal{K}$. It will match to the massive amplitude  
\begin{equation}
[\mathcal{M}]_{l_0} \sim \mathcal{K}. 
\end{equation}
Generally, the higher-order matching in this case will also involve the factorized amplitude $\mathcal{F}$, which is illustrated in Figure~\ref{fig:exceptional}.

The kinetic structures include several terms $\bar{\psi}\slashed{D}\psi,\, D_{\mu}\phi (D^{\mu}\phi)^*,\, -\frac{1}{4} F_{\mu\nu} F^{\mu\nu}$.
As an example, we consider the kinetic term $\psi^{\dagger} \slashed{D} \psi$. It can be expanded as
\begin{equation} \begin{aligned}
\begin{array}{lccccc}
\text{kinetic term:} &  \bar{\psi}\slashed{D}\psi & = & \bar{\psi}\slashed{\partial}\psi & + & ig\bar{\psi}\slashed{A}\psi.
\end{array}
\end{aligned} \end{equation}
Here we use the expansion of the covariant derivative $D=\partial+igA$, where $g$ is the gauge coupling and $A$ is the gauge field. The second term $\psi^{\dagger} \slashed{A} \psi$ matches to a 3-point descendant MHC amplitude of class $ffV$  with helicity-$(-+-)$,
\begin{eqnarray}
    \begin{array}{ccc}
\mbox{MHC} && \mbox{massless}  \\
{[\mathcal{M}_{ffV}^{-+-}]}_{1} =  m_3 \langle13\rangle [\eta_32] &\quad\xrightarrow{H.E.} & \quad \psi^{\dagger}_2 \slashed{A}_3 \psi_1\sim \frac{\langle13\rangle [\eta_32]}{[\eta_33]}.
\end{array}
\end{eqnarray}
In this case, the massive coefficient contains only the gauge coupling
\eq{
c_{ffV}^{-+-} = \frac{ig}{\mathbf m_3^2}. 
}

In summary, we obtain a concrete correspondence between the massless amplitude basis and the MHC amplitudes. For MHC amplitudes, we focus on the nonvanishing leading descendant order $l_0$, whose matching we define as LO matching. The corresponding massless amplitudes may receive various UV origins, but we concentrate on two types: the massless leading operator with lowest dimension, and the kinetic term with gauge couplings. 

Note that, for the exceptional matching with leading order $l_0>0$, both massless contact and factorized amplitudes can match the leading-order massive amplitude. As shown in Figure~\ref{fig:exceptional}, this leading-order matching is a one-to-one correspondence, but we have not yet provided a criterion for determining which MHC amplitude relates to them. In the next subsection, we will turn to the massive amplitude itself and use the analysis of current decomposition for the massive amplitude to distinguish them.


\subsection{Current Decomposition for Massive Amplitudes}
\label{sec:current}

In the LO matching, when a vector boson is included, the primary MHC amplitude may vanish (e.g., the $ffV$ amplitude $\langle13\rangle[32]$). So the massless amplitude will not match the primary MHC amplitude but the descendant one. Hence, it is important to study the structure of MHC amplitudes involving vector bosons. When a vector boson carries the $(\tfrac12,\tfrac12)$ Lorentz representation, the MHC amplitudes can be decomposed into the contraction of a MHC current $\mathcal{J}^{\dot{\alpha} \alpha}$ and a massive vector boson $\mathbf{A}_{\alpha\dot{\alpha}}$,
\eq{
\mathcal{M} = \mathcal{J} \cdot \mathbf{A}.
}
We refer to this as the \textit{current decomposition}. The $N$-point MHC current is defined as 
\eq{
\mathcal{J}_{\alpha\dot{\alpha}}(1,2,...,N) = \langle\Omega|\mathcal{O}_{\alpha\dot{\alpha}}|\{1,2,...,N\}\rangle,
}
which is an $N$-point form factor in the $\left(\frac{1}{2}, \frac{1}{2}\right)$ Lorentz representation, characterized by the spin and helicity of each particle. 

The MHC currents are graded by descendant order in the same way as MHC amplitudes. For example, the $ffV$ MHC amplitudes contain the following $\mathcal{J} \cdot \mathbf{A}$ structures:
\eq{
\begin{array}{c|c|c|c}
\hline
\mbox{order} & \mbox{amplitude} & \mathcal{J} & \mathbf{A} \\
\hline
\mbox{primary} & \langle13\rangle [32] & |2]^{\dot{\alpha}} \langle1|^{\alpha} & |3\rangle_{\alpha} [3|_{\dot{\alpha}} \\
\hline
1\mbox{-st descendant} & -\tilde{m}_1 \langle\eta_13\rangle [32] & \tilde{m}_1 |2]^{\dot{\alpha}} \langle\eta_1|^{\alpha} & \multirow{2}{*}{ $-|3\rangle_{\alpha} [3|_{\dot{\alpha}}$ } \\
& m_2 \langle13\rangle [3\eta_2] & -m_2 |\eta_2]^{\dot{\alpha}} \langle1|^{\alpha} & \\
\hline
2\mbox{-nd descendant} & -\tilde{m}_1 m_3 \langle\eta_13\rangle [\eta_32] & \tilde{m}_1 |2]^{\dot{\alpha}} \langle\eta_1|^{\alpha} & -m_3 |3\rangle_{\alpha} [\eta_3|_{\dot{\alpha}} \\
& -\tilde{m}_1 m_2 \langle\eta_13\rangle [3\eta_2] & \tilde{m}_1 m_2 |\eta_2]^{\dot{\alpha}} \langle\eta_1|^{\alpha} & -|3\rangle_{\alpha} [3|_{\dot{\alpha}} \\
\hline
\end{array} \label{eq:ffV_JA}
}
Here we list the structures corresponding to the primary amplitude $\langle13\rangle [32]$ in the $ffV$ class. The $ffV$ class also contains another primary amplitude $[13]\langle32\rangle$, which can be decomposed in a similar way.

Similar to the MHC amplitudes, MHC currents are also classified into primary and descendant components,
\begin{equation}
\mathcal{J}=[\mathcal{J}]_0 \oplus [\mathcal{J}]_1 \oplus [\mathcal{J}]_2 \oplus\cdots
\end{equation}
where $\mathcal{J}$ denotes all the MHC currents under a given spin category, $[\mathcal{J}]_0$ is referred to as the primary current, and $[\mathcal{J}]_n$ the $n$-th descendant current. Similar to Eq.~\eqref{eq:MHC-recur}, they also satisfy the following recursion relation
\eq{
[\mathcal{J}]_{l+1} = mJ^+ [\mathcal{J}]_l \oplus \tilde{m} J^- [\mathcal{J}]_l,
}
where the mass spurions $m$ and $\tilde{m}$ run over the external legs on which the ladder operators act. In the $ffV$ example, the primary and 1st descendant currents are
\eqs{
{[\mathcal{J}]}_0 &=& -|2]^{\dot{\alpha}} \langle1|^{\alpha}, \\
{[\mathcal{J}]}_1
&=& \tilde{m}_1 J_1^- [\mathcal{J}]_0 \oplus m_2 J_2^+ [\mathcal{J}_0] \nonumber \\
&=& (\tilde{m}_1 |2]^{\dot{\alpha}} \langle\eta_1|^{\alpha}) \oplus ( -m_2 |\eta_2]^{\dot{\alpha}} \langle1|^{\alpha}), 
}
which is consistent with Eq.~\eqref{eq:ffV_JA}. 

\paragraph{Current-conservation condition}
These MHC currents can be classified by whether they satisfy
\begin{equation} \label{eq:conserved_condition}
\partial\cdot\mathcal{J}
\quad\xrightarrow{H.E.}\quad
\sum_{i=1}^N p_i\cdot[\mathcal{J}(1,\dots,N)]_0=0,
\end{equation}
where $\partial$ acts on the leading spinors. A current is conserved; otherwise it is non-conserved. The $ff$ and $ffs$ currents have the same spinor expression $[\mathcal{J}]_0=|2]\langle1|$, offer an interesting point of comparison: 
\begin{equation} \begin{aligned}
&\text{$ff$ current:}& (p_1+p_2)\cdot[\mathcal{J}]_0 
&=(p_1+p_2)_{\alpha\dot\alpha}\cdot|2]^{\dot\alpha}\langle1|^{\alpha}=0,\\
&\text{$ffs$ current:}& (p_1+p_2+p_3)\cdot[\mathcal{J}]_0
&=(p_1+p_2+p_3)_{\alpha\dot\alpha}\cdot |2]^{\dot\alpha}\langle1|^{\alpha}=[32]\langle21\rangle\neq 0.\\
\end{aligned} \end{equation}
Thus, the $ff$ current is conserved, whereas the $ffs$ current is not.

To see how these two cases relate to MHC amplitude, we can consider the H.E. limit for a massive vector boson. The H.E. limit of each term is naively
\begin{equation} \begin{aligned}
[\mathbf{A}(1^0)]_0 &= -|1\rangle_{\alpha} [1|_{\dot{\alpha}}& &\xrightarrow{H.E.}\quad \partial_{\alpha\dot{\alpha}} \phi, \\
[\mathbf{A}(1^-)]_1 &= -\frac{1}{\mathbf{m}_1^2} m_1 |1\rangle_{\alpha} [\eta_1|_{\dot{\alpha}}& &\xrightarrow{H.E.}\quad A^-_{\alpha\dot{\alpha}}, \\
[\mathbf{A}(1^+)]_1 &= \frac{1}{\mathbf{m}_1^2} \tilde{m}_1 |\eta_1\rangle_{\alpha} [1|_{\dot{\alpha}}& &\xrightarrow{H.E.}\quad A^+_{\alpha\dot{\alpha}}, \\
[\mathbf{A}(1^0)]_2 &= |\eta_1\rangle_{\alpha} [\eta_1|_{\dot{\alpha}}& &\xrightarrow{H.E.}\quad 0.
\end{aligned} \end{equation}
This correspondence may not always be explicit, due to IBP and EOM application in the amplitude matching. For a primary
MHC amplitude of the form $[\mathcal{J}]_0\cdot[\mathbf A]_0$, the H.E. limit gives
\begin{equation}
[\mathcal{J}]_0\cdot[\mathbf A]_0 \quad\xrightarrow{H.E.}\quad
[\mathcal{J}]_0\cdot \partial\phi \quad\overset{\text{IBP}}{\simeq}\quad
\phi \partial\cdot\mathcal{J}
\end{equation}
Therefore, conservation of the H.E. current can force the primary amplitude to vanish modulo IBP.
\begin{itemize}
\item  For an $(N+1)$-point amplitude, the corresponding $N$-point current may be conserved or non-conserved current. Modulo IBP, a conserved current corresponds to vanishing primary amplitudes, while a non-conserved current corresponds to a non-vanishing primary amplitude.

\item $N=2$-point current requires separate treatment.  In this case, all angle brackets or all square brackets vanish, due to three-point kinematics. Consequently, a primary amplitude constructed from a 2-particle current may vanish even when that current is not conserved. The $ff$ and $VS$ currents illustrate the distinction:
\begin{equation}
\begin{tabular}{c|c|c|c}
\hline
 & current $[\mathcal{J}]_0$ & H.E. limit for $\partial\cdot\mathcal{J}$ & primary amplitude \\
\hline
conserved & $ff$: $|2\rangle[1|$ & $[1|p_1+p_2|2\rangle=0$ & $[13]\langle32\rangle=0$ \\
\hline
non-conserved & $VS$: $|1\rangle[1|$ & $[1|p_1+p_2|1\rangle\neq0$ & $[13]\langle31\rangle=0$\\
\hline
\end{tabular}
\end{equation}
For the $VS$ current, $\partial\cdot\mathcal{J}\neq 0$, yet the associated primary amplitude vanishes. This difference arises because the three-particle kinematic relations are used in the latter case but not in the former.

\end{itemize}

The H.E. current-conservation condition is equivalently a Ward-identity test. After little-group covariance
is restored, components at different descendant orders are packaged into a massive current $\mathbf J$. The full massive
current need not be conserved:
\eq{
\partial \cdot \mathbf{J} \;\propto\; \mathbf{m} \mathbf{T}_\pi.
}
where $\mathbf{T}_\pi$ represents the Goldstone amplitude in L.E. and $\mathbf m$ denotes the mass of a vector boson. This is a specific case of the Goldstone Equivalence Theorem.
In these cases, the Ward identity in the H.E. and the Goldstone Equivalence Theorem at L.E. are combined. Thus, we examine the on-shell Higgs mechanism by studying the relationship between massive and massless currents for cases concerning Noether currents.

\paragraph{MHC classification with current decomposition}

Now we can use the current decomposition to classify contact MHC amplitudes and analyze their corresponding UV behavior. 
There are three types of MHC amplitudes classified by the decomposition and the H.E. limit of $\mathcal{J}$: 
\begin{description}
\item[type (i)] structures that cannot be decomposed into $\mathcal{J} \cdot \mathbf{A}$:

This includes classes without a vector, such as the $ffS$ amplitude $\langle12\rangle$, and structures in which no
vector is represented by a $(\tfrac12,\tfrac12)$ Lorentz representation, such as $\langle13\rangle\langle32\rangle$ in the $ffV$ class.

\item[type (ii)] structures that admit a current decomposition and whose primary current is non-conserved:

For example, the $ffVS$ structure $\langle13\rangle\langle32\rangle$ contains the non-conserved 3-point current $\mathcal{J}=|2]\langle1|$.

\item[type (iii)] structures that admit a current decomposition but whose primary amplitude vanishes:

This includes a conserved H.E. current and the special $VS$ current. An example is the $ffV$ structure], whose 2-point current $\mathcal{J}=|2]\langle1|$ is conserved.

\end{description}

A particle class can contain structures of more than one type. The types encountered in the present 3-point and 4-point analysis are summarized as follows:
\begin{equation}
\begin{tabular}{c|c|c|c|c}
\hline
 & \multirow{2}{*}{massive amplitude} & \multicolumn{2}{c|}{direct matching}  & exceptional matching \\
\cline{3-5}
 & & case (i) & case (ii) & case (iii) \\
\hline
3-point & $SSS$,$ffS$ & \checkmark & & \\
& $ffV$, $VSS$,  $VVS$, $VVV$ & \checkmark & & \checkmark \\
\hline
4-point & $SSSS$,$ffSS$,$ffff$ & \checkmark & & \\
 & $VSSS$,$ffVS$,$VVSS$,$ffVV$,$VVVS$ & \checkmark & \checkmark & \\
 & $VVVV$ & \checkmark & \checkmark & \checkmark \\
\hline
\end{tabular}
\end{equation}
Exceptional matching only occur in $ffV$, $VSS$, $VVS$, $VVV$ and $VVVV$ amplitude. For a more detailed discussion of classes with vanishing primary amplitude structures, see appendix~\ref{app:exception}.

\paragraph{Identify the UV origin} For types (i) and (ii), the primary MHC amplitude has a non-vanishing H.E. limit and can be matched directly. For type (iii), $[\mathcal J]_0\cdot[A]_0$ vanishes, so one must inspect descendants $[\mathcal J]_{l>0}$. We use $[\mathcal J]_l$ to denote a conserved current (and involve the $VS$ current), and $[\mathcal J']_l$ for a non-conserved current at descendant order $l$:
\begin{equation} \begin{aligned}
\partial\cdot[\mathcal J]_l
\quad&\to\quad
\sum_{i} p_i\cdot[\mathcal J]_l=0.\\
\partial\cdot[\mathcal J']_l
\quad&\to\quad
\sum_{i=1}^N p_i\cdot[\mathcal J']_l\neq 0.
\end{aligned} \end{equation}

The UV origin can be either a kinetic term or a local interaction operator. The above three types in the MHC classification give rise to the following:

\textbf{type (i):} For a structure without a current decomposition, the H.E. limit is matched directly to a leading local operator. For example, $\langle12\rangle$ in the $ffS$ class is associated with a Yukawa-type operator, while the chiral $ffV$ structure $\langle13\rangle\langle32\rangle$ is associated with a dipole operator containing a field strength. The corresponding UV operator is

\begin{equation} \begin{aligned}
&ffS:\langle12\rangle&
&\xrightarrow{H.E.}\quad \text{operator}: \bar\psi\psi\phi,\\
&ffV:\langle13\rangle\langle32\rangle&
&\xrightarrow{H.E.}\quad \text{operator}: ig \bar{\psi} (\sigma_{\mu\nu}F^{\mu\nu}) \psi.
\end{aligned} \end{equation}

\textbf{type (ii):} If the amplitude admits a current decomposition and the H.E. current is a non-conserved higher-point one $[\mathcal J']$, the amplitude in the H.E. limit $[\mathcal J']_0\cdot [\mathbf A]_0$ also corresponds to a operator. For example, the $ffVs$ amplitude,

\begin{equation}
ffVs:[\mathcal J']_0 \cdot [\mathbf{A}]_0=\langle13\rangle[32]
\quad\xrightarrow{H.E.}\quad \text{operator}: \bar{\psi} (\slashed{D} \phi^*) \psi \phi.
\end{equation}

\textbf{type (iii):} If the amplitude admits a current decomposition but the H.E. current is a conserved current $[\mathcal J]_0$ or $VS$ current, the amplitude vanishes in the H.E. limit. We first consider the conserved current $[\mathcal J]_0$. Taking the $ffV$ amplitude as an example, 
\begin{equation} \label{eq:vanish_primary}
ffV:[\mathcal J]_0 \cdot [\mathbf{A}]_0=\langle13\rangle[32]
\quad\xrightarrow{H.E.}\quad 0.
\end{equation}

In this case, we must inspect the descendant amplitude. Two candidate structures are $[\mathcal J]_0\cdot [\mathbf A]_1$ and $[\mathcal J']_1\cdot [\mathbf A]_0$. The first matches the kinetic term,
\begin{equation}
ffV:
[\mathcal J]_0 \cdot [\mathbf{A}]_1=\frac{\langle13\rangle[\eta_3 2]}{m_3}
\quad\xrightarrow{H.E.}\quad \text{kinetic term}: \bar{\psi} \slashed{A} \psi.
\end{equation}
This arises from the kinetic term $\bar{\psi} \slashed{D} \psi$ and thus corresponds to the conserved current $\bar{\psi} \gamma^\mu \psi$. 
This explains the vanishing result in eq.~\eqref{eq:vanish_primary}.
In the more general case, we can have a conserved current $[\mathcal J]_{l}$ at higher order $l>0$. This also corresponds to kinetic terms.

Then we consider $[\mathcal J']_1\cdot [\mathbf A]_0$. It corresponds to the operator,
\begin{equation}
ffV:[\mathcal J']_1 \cdot [\mathbf{A}]_0=\frac{\langle\eta_1 3\rangle[3 2]}{m_1}
\quad\xrightarrow{H.E.}\quad \text{operator}: \bar{\psi} \psi \phi.
\end{equation}
Therefore, the non-conserved current $[\mathcal J']_l$ relates to the UV operator.
This gives us a strategy to find the UV operators for the massive contact term, when the primary current is non-conserved. According to the LG covariance, we flip the non-conserved primary current $[\mathcal J]_0$ to the conserved descendant currents $[\mathcal J']_l$. These non-conserved descendant currents usually correspond to UV operators when contracting with $[\mathbf{A}]_0$,
\eq{
[\mathcal J']_l \cdot [\mathbf{A}]_0 \xrightarrow{H.E.} \mathcal{O}, \label{eq:J'A-O}
}
where $\mathcal{O}$ is the massless operator. The first non-conserved current need not occur at $l=1$. For a $VV$ current, the first descendant can remain conserved, requiring the second descendant. The $VS$ current is exceptional because its primary amplitude vanishes by three-particle kinematics even though the current itself is nonconserved.

The operator classes relevant to the type (iii) can be displayed most clearly by their low-energy projections:
\eq{
Y \psi^2 \phi, Y^* \bar{\psi}^2 \phi \xrightarrow{L.E.}& \begin{cases}
ffS \\ {\color{blue} ffV}
\end{cases} \\
\lambda_3 \phi^3 \xrightarrow{L.E.}& \begin{cases}
SSS \\ {\color{blue} VVS}
\end{cases} \\
\lambda_4\phi^4 \xrightarrow{L.E}& SSSS; \\
F_L^2 \phi, F_R^2 \phi \xrightarrow{L.E.}& \begin{cases}
VVS \\ {\color{blue} VVV}
\end{cases} \\
F_L D^2 \phi^3, F_R D^2 \phi^3 \xrightarrow{L.E.}& \begin{cases}
VVVS \\ VVSS \\ VSSS \\ {\color{blue} VVVV}
\end{cases}
}
where the first three lines are renormalizable operators, and the last two are EFT operators. Blue
entries mark classes in which at least one primary MHC structure vanishes in the HE limit. They need to perform exceptional matching, and the lowest
non-vanishing descendant is class dependent and can be first or second order. Detailed discussion will be given in the next section.

We summarize the situations of interest for the massless-massive correspondence as follows.
\begin{itemize}
\item Primary MHC amplitudes that cannot be decomposed into $[\mathcal J]_0\cdot[\mathbf{A}]_0$, or that can be decomposed but have a non-conserved H.E. current, belong to the direct matching. Taking their H.E. limit, they match to massless operators.

\item Primary MHC amplitudes that are decomposed into $[\mathcal J]_0\cdot[\mathbf{A}]_0$ vanish in the H.E. limit and belong to the exceptional matching. One must examine descendant currents to find two kinds of UV origins: a conserved current $[\mathcal J]_l$ corresponds to a gauge coupling between the conserved current and a gauge boson, while a non-conserved current $[\mathcal J]_l$ corresponds to an effective operator.

\end{itemize}



%% file: sec4-3pt4pt.tex
\section{Massive EFT Amplitude Basis}\label{sec:3}

The preceding discussion establishes a correspondence between massless amplitudes $\mathcal{A}$ and massive ST amplitudes $\mathbf{M}$. In this section, we apply this correspondence to construct the massive ST basis from the complete massless basis, and subsequently derive the corresponding massive operators. For clarity, we postpone discussion of the gauge tensor to the next section.

Our main focus here is on the matching for 3-point and 4-point contact amplitudes with spin $\le 1$. The matching procedure falls into two cases: 
\begin{itemize}
\item \textit{Direct matching}: This occurs when the primary MHC amplitudes remain non-vanishing in the H.E. limit. Here, a one-to-one correspondence exists between massless leading operators and primary MHC amplitudes. Thus, the number of independent primary MHC amplitudes in a massive class is equal to the number of independent massless leading operators when taking the H.E. limit.

\item \textit{Exceptional matching}: This occurs when the primary MHC amplitudes vanish in the H.E. limit. In generic massive theory, there are only five classes that belong to this category: $VVS$, $VSS$, $ffV$, $VVV$, and $VVVV$. Within the SMEFT, the classes $VSS$ and $VVVV$ are eliminated by the particle content. In this branch, the matching must instead be made between massless operators and descendant MHC amplitudes. 

\end{itemize}

We discuss these two matching branches separately and finally present the complete set of 3-point and 4-point results in lower dimensions, organized by transversality.

\subsection{3-point and 4-point Matching with Conserved Current}
\label{sec:exceptional_matching}

In this subsection, we carry out the exceptional matching in detail, applying the current decomposition introduced in Section~\ref{sec:current}. This type of matching occurs only for 3-point and 4-point amplitudes. Since the primary MHC amplitude vanishes in this case, we must instead consider the matching for the descendant MHC amplitude. In
the following, we first analyze the structure of the MHC amplitude, then find the corresponding massless structures.

Recall that the MHC amplitude $\mathcal{M}$ can be decomposed as
\begin{equation}
[\mathcal{M}]_l = \sum_i [\mathcal{J}]_{i} \cdot [\mathbf{A}]_{l-i},
\end{equation}
where $\mathcal{J}$ denotes the MHC current and $\mathbf{A}$ represent a vector boson. It can have two UV origins, depending on whether the current $\mathcal{J}$ is conserved. If it is conserved, the UV can come from the gauge coupling between a gauge boson and the conserved current. So the corresponding MHC amplitude should have the form:
\begin{equation}
[\mathcal{J}]_{i} \cdot [\mathbf{A}]_{1}.
\end{equation}
If the current is not conserved, the corresponding UV structure should be the operator. In this case, we require the vector boson $\mathbf{A}$ to have zero helicity, so it can only be $[\mathbf{A}]_0$ or $[\mathbf{A}]_2$. Thus we have
\begin{equation} \begin{aligned}
{}[\mathcal{M}]_0 &= [\mathcal{J}]_{0} \cdot [\mathbf{A}]_{0}, \\
[\mathcal{M}]_1 &= [\mathcal{J}]_{1} \cdot [\mathbf{A}]_{0}, \\
[\mathcal{M}]_2 &= [\mathcal{J}]_{2} \cdot [\mathbf{A}]_{0}+[\mathcal{J}]_{0} \cdot [\mathbf{A}]_{2}, \\
&\cdots
\end{aligned}
\end{equation}
Note that when there are multiple vector bosons $\mathbf{A}$, there can be more than one way to decompose $\mathcal{J}\cdot \mathbf{A}$. Thus, a term with $[\mathbf{A}]_2$ may be converted to one with $[\mathbf{A}]_2$ via a different current decomposition, such as $[\mathcal{J}]_0\cdot [\mathbf{A}]_2 \to [\mathcal{J}]_2\cdot [\mathbf{A}]_0$. We will see this mechanism at work in the following example of the $VVS$ amplitude.

Naively, one would have to scan over all the descendant currents to find the possible UV origins, but this task can be simplified considerably by helicity and dimensional analysis. From the previous section, we know that the massless operators containing at least one scalar boson can be derived from this mechanism, so we focus on them. For 3-point massless operators, angle and square brackets cannot appear simultaneously, as a result of 3-particle kinematics. Here we list the basis states with negative and zero helicity,
\eq{ \label{eq:3pt_except_UV}
\begin{array}{c|cc}
\hline
\text{helicity} & \mbox{massless basis} & \mbox{operator basis} \\
\hline
(0,0,0) & 1 & \phi^3 \\
(-\frac12, -\frac12, 0) & \langle12\rangle & \bar{\psi}\psi \phi \\
(-1, 0, 0) & \mbox{-} & \mbox{-} \\
(-1, -1, 0) & \langle12\rangle^2 & F_L^2 \phi \\
\hline
\end{array}
}
The positive-helicity structures can be obtained from the negative ones by converting the angle brackets into square ones.

The primary and descendant MHC amplitudes take the form:
\begin{equation}
[\mathcal{M}]_0=c_0 \times (\cdot)^{n},\qquad
[\mathcal{M}]_l=c_l \times m^l (\cdot)^{n},
\end{equation}
where $(\cdot)$ stands for an angle bracket $\langle\cdot\rangle$ or a square bracket $[\cdot]$. For a descendant MHC amplitude to match a massless operator, it must have the appropriate mass-power suppression. In the $ffV$ case discussed in the previous subsection, the coefficient behaves as $c_1\sim 1/m^2$, with each mass factor scaling like a bracket, $m\sim(\cdot)$. In general, the coefficient carries the mass power $c_l\sim 1/m^{2l}$. Since the mass can also be rewritten in bracket form, the $l$-th descendant amplitude can match a massless amplitude of the form
\begin{equation}
[\mathcal{M}]_l \quad\sim\quad \mathcal{A}=(\cdot)^{n-l}.
\end{equation}
Here the massless amplitude must be one of those in eq.~\eqref{eq:3pt_except_UV}. 

As shown in Appendix~\ref{app:exception}, there are five types of exceptional massive structures: $VVS$, $VSS$, $ffV$, $VVV$ and $VVVV$. In the following, we adopt the above analysis to identify the possible UV origins for these exceptional massive types.

\paragraph{1. $VVS$} The primary $SV$ current $[\mathcal J]_0$ is defined as
\begin{equation}
[\mathcal{J}(1^{0},2^{0})]_0=|2]^{\dot\alpha}\langle2|^{\alpha}.
\end{equation}
This primary current can be rewritten in the following form
\begin{equation}
|2]\langle2| = \frac12 (|2]\langle2|+|1]\langle1|) + \underbrace{\frac12 (|2]\langle2|-|1]\langle1|)}_{J^\mu}.
\end{equation}
The second term restores the Noether current for a charged scalar boson
\eq{
J^\mu=\phi^* (D^\mu \phi) - (D^\mu \phi) \phi.
}
which satisfies the conservation condition $\partial\cdot J = \frac{1}{2} (\langle1|p_1+p_2|1] -\langle 2|p_1+p_2|2]) =0$. The primary amplitude is generated by $[\mathcal J]_0\cdot [\mathbf{A}]_0$,
\eq{
(0,0,0): \quad [\mathcal{J}(1^{0},2^{0})]_0\cdot[\mathbf A(3^{0})]_0 = \langle 23\rangle [32]  
=\Ampthree{1^0}{2^0}{3^0}{\bos{i3}{brown}}{\bos{i2}{brown}}{\sca{i1}} \quad \xrightarrow{H.E.}\quad  0.
}

The descendant currents are given by
\begin{align}
[\mathcal{J}(1^{0},2^{-})]_1 &= m_2|\eta_2]^{\dot\alpha} \langle 2|^{\alpha}, \\
[\mathcal{J}(1^{0},2^{+})]_1 &= \tilde{m}_2|2]^{\dot\alpha} \langle \eta_2|^{\alpha}, \\
[\mathcal{J}(1^{0},2^{0})]_2 &= m_2\tilde{m}_2|\eta_2]^{\dot\alpha}\langle\eta_2|^{\alpha}.
\end{align}
The first two lines are the 1st descendant currents $[\mathcal J]_1$ and the last line is the 2nd descendant current $[\mathcal J]_2$. We do not need to consider $[\mathcal J]_1\cdot [\mathbf{A}]_0$, since it corresponds to helicity $(\pm 1,0,0)$ and such a corresponding massless operator does not exist, as shown in eq.~\eqref{eq:3pt_except_UV}. The 2nd descendant MHC $VVS$ amplitude is given by two terms
\begin{equation} \begin{aligned}
[\mathcal{J}(1^{0},2^{0})]_2 \cdot [\mathbf{A}(3^0)]_0 &=-c_{VVS,2}\times  m_2 \tilde{m}_2 \langle\eta_23\rangle [3\eta_2]=\Ampthree{1^0}{2^0}{3^0}{\bos{i3}{brown}}{\bosflipflip{1}{55}{brown}{red}{brown}}{\sca{i1}}, \\
[\mathcal{J}(1^{0},2^{0})]_0 \cdot [\mathbf{A}(3^0)]_2 &= -c_{VVS,1}\times m_3 \tilde{m}_3 \langle 2 \eta_3\rangle [\eta_3 2] =\Ampthree{1^0}{2^0}{3^0}{\bosflipflip{1}{-55}{brown}{red}{brown}}{\bos{i2}{brown}}{\sca{i1}}. 
\end{aligned} \end{equation}
where $c_{VVS,i}^{000}$ with $i=1,2$ are coefficients for different diagrams in helicity $(0,0,0)$. The spin group $SU(2)$ constrains the coefficients to satisfy $c_{VVS,1} = c_2$, so there is no need to consider the possibility of $c_1 \neq c_2$. Therefore, its H.E. limit is the $\phi^3$ operator,
\eq{(0,0,0):\quad
\Ampthree{1^0}{2^0}{3^0}{\bos{i3}{brown}}{\bosflipflip{1}{55}{brown}{red}{brown}}{\sca{i1}}
+\Ampthree{1^0}{2^0}{3^0}{\bosflipflip{1}{-55}{brown}{red}{brown}}{\bos{i2}{brown}}{\sca{i1}} 
\quad\xrightarrow{H.E.}\quad  \Ampthree{1^0}{2^0}{3^0}{\propag[gho] (i3) to (v1)}{\propag[gho] (i2) to (v1)}{\propag[sca] (i1) to (v1)} = \lambda_3,
}
where $\lambda_3$ is the coupling constant of the $\phi^3$ interaction in the UV. On the right-hand side, the dotted line represents the Goldstone boson in the UV, while the dashed line represents the scalar boson that will match onto the massive Higgs boson. This fixes the coefficients to be
\eq{
c_{VVS,1} = c_{VVS,2} = \frac{\lambda_3}{(\mathbf{m}_1^2-2\mathbf{m}_V^2)\mathbf{m}_V^2}.
}
where two vector masses are equal $\mathbf{m}_V=\mathbf{m}_2=\mathbf{m}_3$.

\paragraph{2. $VSS$} 
We now turn to the $VSS$ case, which directly corresponds to the Noether current
\begin{equation}
[\mathcal{J}(1^0,2^0)]_0=p_1^{\dot{\alpha}\alpha}-p_2^{\dot{\alpha}\alpha}.
\end{equation}
The corresponding primary $VSS$ amplitude also vanishes:
\begin{equation}
[\mathcal{J}(1^{0},2^{0})]_0\cdot[\mathbf A(3^{0})]_0 = \langle 3 |1-2|3]  \quad\xrightarrow{H.E.}\quad  0.
\end{equation}
However, we do not provide a diagrammatic representation here, since the momentum structure is not involved in the MHC particle states as defined in section~\ref{sec:MHC}, and additional labels would be required in the diagrams, which complicates the presentation.

The $VSS$ descendant amplitude does not correspond to a contact amplitude. The reason lies in the sign difference between the expansions $|\mathbf{p}]^{\dot{\alpha}}\langle\mathbf{p}|^{\alpha}=p^{\dot{\alpha}\alpha}-\eta^{\dot{\alpha}\alpha}$ and $\mathbf{p}^{\dot{\alpha}\alpha}=p^{\dot{\alpha}\alpha}+\eta^{\dot{\alpha}\alpha}$. In the $VVS$ case, the current takes the vector form $|\mathbf{p}]^{\dot{\alpha}}\langle\mathbf{p}|^{\alpha}$, whereas in the $VSS$ case it is momentum form $\mathbf{p}^{\dot{\alpha}\alpha}$. As a result, the $VVS$ descendant amplitude corresponds to the $\phi^3$ operator, but the $VSS$ case does not correspond to any operator due to this sign difference.

\paragraph{3. $ffV$}

In this case, the primary MHC amplitudes contain two $ff$ currents,
\eq{ \label{eq:primary_ff_current}
[\mathcal{J}(1^+,2^-)]_0 = |1]^{\dot{\alpha}} \langle2|^{\alpha}, \qquad
[\mathcal{J}(1^-,2^+)]_0 = |2]^{\dot{\alpha}} \langle1|^{\alpha}.
}
Their H.E. limit is the Noether current of a fermion field,
\eq{
J^{\mu} = \bar{\psi}\gamma^{\mu}\psi.
}
Therefore, the H.E. limits of the following primary MHC amplitudes $[\mathcal J]_0\cdot [\mathbf{A}]_0$ vanish:
\eqs{
(+\frac{1}{2},-\frac{1}{2},0): \quad  [\mathcal{J}(1^+,2^-)]_0 \cdot [\mathbf{A}(3^0)]_0 = \langle23\rangle [31] = \Ampthree{1^+}{2^-}{3^0}{\fer{cyan}{i1}{v1}}{\antfer{cyan}{i2}{v1}}{\bos{i3}{brown}}, \\
(-\frac{1}{2},+\frac{1}{2},0): \quad  [\mathcal{J}(1^-,2^+)]_0 \cdot [\mathbf{A}(3^0)]_0 = \langle13\rangle [32] = \Ampthree{1^-}{2^+}{3^0}{\fer{red}{i1}{v1}}{\antfer{red}{i2}{v1}}{\bos{i3}{brown}}. 
}

The 1st descendant current receives two contributions from the primary $ff$ currents in Eq.~\eqref{eq:primary_ff_current}:
\begin{equation}
[\mathcal{J}(1^-,2^-)]_1=
c_{ffV,1}\times m_1|\eta_1]^{\dot{\alpha}} \langle 2|^{\alpha}
+c_{ffV,2}\times m_2|\eta_2]^{\dot{\alpha}} \langle 1|^{\alpha},
\end{equation}
where $c_{ffV,i}$ are coefficients. Contracting with the vector boson to form $[\mathcal J]_1\cdot[\mathbf{A}]_0$, the two terms give
\begin{equation} \begin{aligned}
[\mathcal{J}(1^-,2^-)]_1 \cdot [\mathbf{A}(3^0)]_0 
&= c_{ffV,1}\times m_1 \langle23\rangle [3\eta_1] + c_{ffV,2}\times m_2 \langle13\rangle [3\eta_2] \\ 
&= \Ampthree{1^-}{2^-}{3^0}{\ferflip{1}{180}{red}{cyan}}{\antfer{cyan}{i2}{v1}}{\bos{i3}{brown}} +\Ampthree{1^-}{2^-}{3^0}{\fer{red}{i1}{v1}}{\antferflip{1}{55}{cyan}{red}}{\bos{i3}{brown}}.
\end{aligned}
\end{equation}
Taking the H.E. limit of $[\mathcal J]_1\cdot[\mathbf{A}]_0$, corresponding to the Yukawa coupling
\begin{equation}
\begin{aligned}
(-\frac{1}{2},-\frac{1}{2},0):\quad 
&\Ampthree{1^-}{2^-}{3^0}{\ferflip{1}{180}{red}{cyan}}{\antfer{cyan}{i2}{v1}}{\bos{i3}{brown}} +\Ampthree{1^-}{2^-}{3^0}{\fer{red}{i1}{v1}}{\antferflip{1}{55}{cyan}{red}}{\bos{i3}{brown}} \quad\xrightarrow{H.E.}\quad \Ampthree{1^-}{2^-}{0}{\propag[fer] (i1) to (v1)}{\propag[antfer] (i2) to (v1)}{\propag[gho] (i3) to (v1)}=Y\langle 12 \rangle. \label{eq:ffV-Yukawa2}
\end{aligned}
\end{equation}
There remains some freedom in the coefficients:
\eq{
- c_{ffV,1} \mathbf{m}_1^2 + c_{ffV,2} \mathbf{m}_2^2 = Y.
}
These will be fixed once the gauge group is introduced.

\paragraph{4. $VVV$} The primary current has four terms
\begin{equation} \begin{aligned} \label{eq:primary_VV_current}
{[\mathcal{J}(1^-,2^0)]}_0 &= -\langle12\rangle |2]^{\dot\alpha}\langle1|^{\alpha},&  
{[\mathcal{J}(1^0,2^-)]}_0 &= -\langle12\rangle |1]^{\dot\alpha}\langle2|^{\alpha},& \\
{[\mathcal{J}(1^+,2^0)]}_0 &= -[12] |1]^{\dot\alpha} \langle2|^{\alpha},&
{[\mathcal{J}(1^0,2^+)]}_0 &= -[12] |2]^{\dot\alpha} \langle1|^{\alpha}.&
\end{aligned} \end{equation}
In this case, these primary structures do not correspond to Noether currents in any renormalizable theory. Instead, they correspond to a topological current of the form
\begin{equation}
J_{\mu} = \partial^\nu (F_{\mu\nu}\phi) 
=  (\partial^\nu F_{\mu\nu})\phi +  F_{\mu\nu} (\partial^\nu \phi).
\end{equation}
where the second term reproduces the primary structures in eq.~\eqref{eq:primary_VV_current}. Current conservation is easily verified: $\partial_\mu J^{\mu}=\partial_\mu \partial^\nu (F_{\mu\nu}\phi)=0$, which vanishes due to the antisymmetry of the field-strength tensor $F_{\mu\nu}$. Since the above derivation of current conservation does not use the EOM, we refer to it as a topological current. 

The corresponding primary MHC amplitudes for these currents vanish in the H.E. limit,
\begin{align}
(-1,0,0): \quad [\mathcal{J}(1^-,2^0)]_0 \cdot [\mathbf{A}(3^0)]_0 &= \langle12\rangle [23] \langle31\rangle  
= \Ampthree{1^-}{2^0}{3^0}{\bos{i1}{red}}{\bos{i2}{brown}}{\bos{i3}{brown}}, \\
(0,-1,0): \quad [\mathcal{J}(1^0,2^-)]_0 \cdot [\mathbf{A}(3^0)]_0 &= \langle12\rangle \langle23\rangle [31]  
= \Ampthree{1^0}{2^-}{3^0}{\bos{i1}{brown}}{\bos{i2}{red}}{\bos{i3}{brown}}, \\
(+1,0,0): \quad [\mathcal{J}(1^+,2^0)]_0 \cdot [\mathbf{A}(3^0)]_0 &= [12] \langle23\rangle [31]  
= \Ampthree{1^+}{2^0}{3^0}{\bos{i1}{cyan}}{\bos{i2}{brown}}{\bos{i3}{brown}}, \\
(0,+1,0): \quad [\mathcal{J}(1^0,2^+)]_0 \cdot [\mathbf{A}(3^0)]_0 &= [12] [23] \langle31\rangle  
= \Ampthree{1^0}{2^+}{3^0}{\bos{i1}{brown}}{\bos{i2}{cyan}}{\bos{i3}{brown}}.
\end{align}
Here we list four primary amplitudes in which particle 3 is fixed to have helicity 0. The remaining two primary amplitudes, involving a helicity-$\pm1$ particle 3, can be obtained by permutations and are discussed in appendix~\ref{app:detail_MHC}. In total, there are six primary amplitudes, all of which vanish in the H.E. limit.

\begin{figure}[htbp]
\centering
\includegraphics[scale=1.5,valign=c]{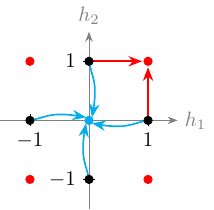}
\caption{Illustration of the ladder operators $mJ^-$ and $\tilde{m}J^+$ acting on the $VV$ current. Black dots represent the primary current $[\mathcal J]_0$. The arrow shows the flip from $[\mathcal J]_0$ to the 1st descendant current $[\mathcal J]_1$, indicated by red and cyan dots.} \label{fig:VV_flip}
\end{figure}

We now turn to the descendant currents. As illustrated in figure~\ref{fig:VV_flip}, the 1st descendant current $[\mathcal{J}]_1$ with nonzero helicity receives contributions from two primary current
\begin{align} 
\text{opposite sign}:&\quad {[\mathcal{J}(1^-,2^+)]}_1 = c^{-+}_{VVV,1} \tilde{m}_2 \langle1\eta_2\rangle |2]^{\dot{\alpha}} \langle1|^{\alpha} - c^{-+}_{VVV,2} m_1 [\eta_12] |2]^{\dot{\alpha}} \langle1|^{\alpha},  \\
\text{same sign}:&\quad [\mathcal{J}(1^-,2^-)]_1 = - c^{--}_{VVV,1} m_2 \langle12\rangle |\eta_2]^{\dot{\alpha}} \langle1|^{\alpha} -c^{--}_{VVV,2} m_1 \langle12\rangle |\eta_1]^{\dot{\alpha}} \langle2|^{\alpha}. \label{eq:J--1} 
\end{align}
The helicity-$(-+)$ MHC currents in the first line are conserved in the H.E. limit regardless of the coefficients  $c^{-+}_{VVV,1}$ and $c^{-+}_{VVV,2}$. For the helicity-$(--)$ MHC currents, we have
\begin{equation} \begin{aligned}
(-1,-1,0): [\mathcal{J}(1^-,2^-)]_1 \cdot [\mathbf{A}(3^0)]_0
&=-c^{--}_{VVV,1} m_2\langle12\rangle [\eta_23] \langle31\rangle-c^{--}_{VVV,2} m_1 \langle12\rangle \langle23\rangle [3\eta_1] \\
&=\Ampthree{1^-}{2^-}{3^0}{\bos{i1}{red}}{\bosflip{1}{55}{brown}{red}}{\bos{i3}{brown}}+\Ampthree{1^-}{2^-}{3^0}{\bosflip{1}{180}{brown}{red}}{\bos{i2}{red}}{\bos{i3}{brown}}.
\end{aligned} \end{equation}
Conservation here depends on the relation between the coefficients $c_{VVV,1}^{--}$ and $c_{VVV,2}^{--}$. There are two cases:

\begin{itemize}
\item If the coefficients satisfy $c_1^{--} \mathbf{m}_{VVV,1}^2 = c_2^{--} \mathbf{m}_{VVV,2}^2= c^\prime$, the current is conserved:
\eq{
[\mathcal{J}_c(1^-,2^-)]_1 = c^\prime\left(-\frac{1}{\tilde{m}_2} \langle12\rangle |\eta_2]^{\dot{\alpha}} \langle1|^{\alpha} -\frac{1}{\tilde{m}_1} \langle12\rangle |\eta_1]^{\dot{\alpha}} \langle2|^{\alpha}\right), \\
}
where the subscript $c$ denotes "conserved". The two particles in such a current are anti-symmetric, which requires a non-abelian gauge symmetry. One would expect an $f^{IJK}$ gauge structure to be attached to these currents, so they can satisfy the spin-statistics theorem. This leads to a Yang-Mills current of the form
\begin{equation}
J^\mu = f^{IJK} F^{\mu\nu} A_{\nu}.
\end{equation}

\item If, instead, the coefficients satisfy  $c_{VVV,1}^{--} \mathbf{m}_2^2 = -c_{VVV,2}^{--} \mathbf{m}_1^2= -c$, we obtain a non-conserved current:
\eq{
[\mathcal{J}_n(1^-,2^-)]_1 = c \left(\frac{1}{\tilde{m}_2} \langle12\rangle |\eta_2]^{\dot{\alpha}} \langle1|^{\alpha} -\frac{1}{\tilde{m}_1} \langle12\rangle |\eta_1]^{\dot{\alpha}} \langle2|^{\alpha}\right), \\
}
where the subscript $n$ denotes "non-conserved". The H.E. limit of $[\mathcal J_n]_1\cdot[\mathbf{A}]_0$ would be an EFT operator,
\eq{
[\mathcal{J}_n(1^-,2^-)]_1 \cdot [\mathbf{A}(3^0)]_0 
\quad \xrightarrow{H.E.} \quad 
\Ampthree{1^-}{2^-}{3^0}{\propag[bos] (i1) to (v1)}{ \propag[bos] (i2) to (v1)}{\propag[gho] (i3) to (v1); \vertex[dot] (v) at (0,0) {}} = \frac{C_{F_L^2 \phi} }{\Lambda^2} \langle12\rangle^2,
}
where $C_{F_L^2 \phi}$ is the Wilson coefficient. This fixes the coefficient to be $c=\frac{C_{F_L^2 \phi}}{\Lambda^2}$ and obtain
\begin{equation}
c_{VVV,1}^{--} = -\frac{C_{F_L^2 \phi}}{\Lambda^2 \mathbf{m}_1^2},\quad
c_{VVV,2}^{--} = \frac{C_{F_L^2 \phi}}{\Lambda^2 \mathbf{m}_2^2},
\end{equation}
Exchange symmetry does not require a non-abelian gauge group. If such a group were present, the gauge structure would simply be $\delta^{IJ}$. This is quite different from the "conserved" currents $[\mathcal J_c]_1$, which correspond to the massless currents of the SM.

\end{itemize}

The remaining 1st descendant MHC currents are the helicity-$(00)$ ones:
\eq{
[\mathcal{J}(1^0,2^0)]_1 = c_{1}^{00} \tilde{m}_1 (\langle\eta_12\rangle |2]^{\dot{\alpha}} \langle1|^{\alpha} +\langle12\rangle |2]^{\dot{\alpha}} \langle\eta_1|^{\alpha}) +c_{2}^{00} \tilde{m}_2 (\langle1\eta_2\rangle |1]^{\dot{\alpha}} \langle2|^{\alpha} +\langle12\rangle |1]^{\dot{\alpha}} \langle\eta_2|^{\alpha}) \\ -c_{3}^{00} m_1 ([\eta_12] |1]^{\dot{\alpha}} \langle2|^{\alpha} +[12] |\eta_1]^{\dot{\alpha}} \langle2|^{\alpha}) -c_{4}^{00} m_2 ([1\eta_2] |2]^{\dot{\alpha}} \langle1|^{\alpha} +[12] |\eta_2]^{\dot{\alpha}} \langle1|^{\alpha}), 
}
As shown in figure~\ref{fig:VV_flip}, this current receives contributions from all four primary currents, so there are four independent coefficient choices. The MHC amplitude $[\mathcal{J}(1^0,2^0)]_1\cdot [\mathbf{A}(3^0)]_0$ still vanishes in the H.E. limit. This is because both $\langle12\rangle$ and $[12]$ appear after contraction, and the result must vanish due to the 3-particle kinematics in the H.E. limit. 

This vanishing result at the 1st descendant level can also be understood from dimensional analysis. The $VVV$ amplitude has three brackets $(\cdot)^3$, whereas the helicity $(0,0,0)$ operator has zero brackets, as shown in eq.~\eqref{eq:3pt_except_UV}. This implies that we need to descend three mass orders (i.e., to the 3rd descendant) before we can potentially match the $\phi^3$ operator. However, as shown in appendix~\ref{app:detail_MHC}, a more careful calculation reveals that even the third descendant amplitude fails to reproduce this $\phi^3$ operator.

\paragraph{5. $VVVV$} We now consider the 4-point structures. The only exceptional matching occurs in the $VVVV$ amplitude. The corresponding primary current is
\eq{
[\mathcal{J}(1^0,2^0,3^0)]_0=(|1]\langle2|[23]\langle13\rangle-|2]\langle1|[13]\langle23\rangle).
}
It corresponds to a different topological current
\eq{\label{eq:baryon_current}
J_\mu=\varepsilon_{\mu\nu\rho\sigma} \partial^{\nu}\phi \partial^{\rho}\phi \partial^{\sigma}\phi.
}
Current conservation $\partial^\mu J_\mu=0$ follows from the antisymmetry of the Levi-Civita tensor, $\varepsilon_{\mu\nu\rho\sigma} \partial^\mu\partial^\nu \phi=0$. This topological current is related to the baryon current~\footnote{In the Skyrme model, the system is described by the left-current $L_\mu=U^\dagger\partial_\mu U$, where $U=\exp(i \tau\cdot\sigma/f_\pi)$ is the scalar field in the nonlinear $\sigma$ model. The baryon current is then defined as 
\begin{equation*}
B_\mu=\varepsilon_{\mu\nu\rho\sigma} \text{tr}(L^{\nu} L^{\rho} L^{\sigma}).
\end{equation*}
At leading order $L^\mu\to -\frac{i}{f_\pi} \tau\cdot \partial\sigma$, which reproduces eq.~\eqref{eq:baryon_current}.
} in the Skyrme model~\cite{Skyrme:1961vq,Holzwarth:1985rb}. The corresponding primary amplitude is given by
\begin{equation}
[\mathcal{J}(1^0,2^0,3^0)]_0\cdot[\mathbf{A}(4^0)]_0
=[41]\langle24\rangle[23]\langle13\rangle-[42]\langle14\rangle[13]\langle23\rangle
=\Ampfour{1^0}{2^0}{3^0}{4^0}{\bos{i1}{brown}}{\bos{i2}{brown}}{\bos{i3}{brown}}{\bos{i4}{brown}}.
\end{equation}
This amplitude also vanishes in the H.E. limit.

We then consider the possible UV origins among 4-point bosonic operators containing at least
one scalar. Since the primary MHC amplitude is of the form $(\cdot)^4$, the possible UV structure must have fewer than four brackets. For negative helicity, the candidate structures are
\eq{\label{eq:VVVV_UV}
\begin{array}{c|cc}
\hline
\text{helicity} & \mbox{massless basis} & \mbox{operator basis} \\
\hline
(0000) & 1, s_{ij} &  \phi^4, D^2\phi^4 \\
\hline
(-000) & \langle21\rangle[23]\langle13\rangle & F_L D^2 \phi^3 \\
\hline
(--00) & \langle12\rangle^2 & F_L^2 \phi^2 \\
\hline
(---0) & \langle12\rangle\langle23\rangle\langle31\rangle &  F_L^3 \phi \\
\hline
\end{array}
}

This current contributes to the massless amplitude at the 1st descendant level via $mJ^-$, taking the form:
\begin{eqnarray}
    [\mathcal{J}_0(1^-,2^0,3^0)]_1=c_{VVVV} m_1 \left(|\eta_1]\langle2|[23]\langle13\rangle-|2]\langle1|[\eta_13]\langle23\rangle\right),
\end{eqnarray}
These currents directly contribute to massless amplitudes as follows:
\eq{
[\mathcal{J}_0(1^-,2^0,3^0)]_1\cdot[\mathbf{A}(4^0)]_0
&=c_{VVVV} m_1 \left([4\eta_1]\langle24\rangle[23]\langle13\rangle-[42]\langle14\rangle[\eta_13]\langle23\rangle\right) =\Ampfour{1^-}{2^0}{3^0}{4^0}{\bosflip{1}{135}{brown}{red}}{\bos{i2}{brown}}{\bos{i3}{brown}}{\bos{i4}{brown}}\\
&\xrightarrow{H.E.}\frac{C_{F_L D^2 \phi^3 }}{\Lambda^3}\langle21\rangle[23]\langle13\rangle=\Ampfour{1^-}{2^0}{3^0}{4^0}{\propag[bos](i1) to (v1)}{\propag[sca](i2) to (v1)}{\propag[sca](i3) to (v1)}{\propag[sca](i4) to (v1)}.
}
Here, we have defined the massless operators $F_L D^2 \phi^3=F_L^{\mu\nu}(D_\mu\phi_2 D_\nu\phi_3-D_\mu\phi_3 D_\nu\phi_2)\phi_4$. The above H.E. limits yield the final matching relations:
\eq{
c_{VVVV}=-\frac{C_{F_L D^2 \phi^3 }}{\Lambda^3 \mathbf{m}^2_1}.
}
One can verify that the higher-order descendants will not match the operators in eq.~\eqref{eq:VVVV_UV}. Therefore, the only UV massless operator participating in the exceptional matching case for $VVVV$ is $F_{L/R} D^2 \phi^3$.

\paragraph{Summary} We now summarize the UV origins of the exceptional matching cases. The UV origins fall into two categories: the UV operators correspond to the non-conserved currents, and gauge couplings associated with conserved currents.

We first consider the UV operator contributions. For the five types of exceptional matching, we have identified all possible UV origins of the operators and their corresponding MHC amplitudes. By bolding the spinors in the MHC amplitudes (i.e., $\lambda,\eta\to \boldsymbol{\lambda}$), we can derive the descendant ST amplitude, which in turn yields the corresponding broken-phase operator structures. The results are summarized in the following table:
\eq{
\begin{array}{c|c|cc}
\hline
\text{class} & \mbox{massless basis} & \text{descendant ST amplitude} & \text{current decomposition} \\
\hline
VVS  & 1 & \makecell{c_{VVS,1}\times m_1\tilde{m}_1 \langle\mathbf{12}\rangle[\mathbf{12}]\\
+c_{VVS,2}\times m_2\tilde{m}_2 \langle\mathbf{12}\rangle[\mathbf{12}]} & [\mathcal{J}]_2 \cdot [\mathbf{A}]_0 \\
\hline
VSS  & \mbox{-} & \mbox{-} & \mbox{-} \\
\hline
ffV & \langle12\rangle & \makecell{c_{ffV,1} \times m_2 \langle\mathbf{13}\rangle[\mathbf{23}]\\
+c_{ffV,2} \times m_1[\mathbf{13}]\langle\mathbf{23}\rangle} & [\mathcal{J}]_1 \cdot [\mathbf{A}]_0 \\
\hline
VVV & \langle12\rangle^2 & 
\makecell{ c_{VVV,1} \times m_2\langle\mathbf{12}\rangle[\mathbf{23}] \langle\mathbf{31}\rangle  \\ + c_{VVV,2} \times m_1\langle\mathbf{12}\rangle\langle\mathbf{23}\rangle [\mathbf{31}] } & 
[\mathcal{J}]_1 \cdot [\mathbf{A}]_0 \\
\hline
VVVV & \langle21\rangle[23]\langle13\rangle & 
\makecell{ c_{VVVV} \times m_1 (\langle\mathbf{14}\rangle\langle\mathbf{23}\rangle[\mathbf{13}][\mathbf{24}]\\
-\langle\mathbf{13}\rangle\langle\mathbf{24}\rangle[\mathbf{14}][\mathbf{23}])  } & 
[\mathcal{J}]_1 \cdot [\mathbf{A}]_0  \\
\hline
\end{array}
}
Here, we have suppressed the helicity labels in the coefficients $c$. Apart from the $VSS$ type, all other types can receive contributions from massless operators in a generic massive theory. However, if we restrict our attention to the SMEFT, the particle content is constrained: there is only one type of massive scalar boson, the Higgs boson $h$, and three types of massive vector bosons, namely the $W^{\pm}$ and $Z$ bosons. These constraints eliminate the existence of $VSS$ and $VVVV$ exceptional matchings, as they would require two distinct massive scalar bosons and four distinct massive vector bosons, respectively.

For the remaining three classes ($VVS$, $ffV$ and $VVV$), we can also consider UV origins arising from gauge couplings between currents and gauge bosons. The typical matching takes the form
\eq{
\begin{array}{c|c|cc}
\hline
\text{class}  & \mbox{massless amplitude} & \text{descendant ST amplitude} & \text{current decomposition} \\
\hline
VVS  & \frac{\langle 12 \rangle\langle 31 \rangle}{\langle 23 \rangle} & c_{VVS}^\prime \times m_1\langle\mathbf{12}\rangle[\mathbf{12}]  & [\mathcal{J}]_0 \cdot [\mathbf{A}]_1 \\
\hline
ffV  & \frac{\langle 13 \rangle^2}{\langle 12 \rangle} & c_{ffV}^\prime \times m_3 \langle\mathbf{13}\rangle[\mathbf{23}] & [\mathcal{J}]_0 \cdot [\mathbf{A}]_1 \\
\hline
VVV  & \frac{\langle 13 \rangle^3}{\langle 12 \rangle\langle 23 \rangle} & 
\makecell{ c_{VVV,1}^\prime \times \tilde{m}_2 m_3 \langle\mathbf{12}\rangle[\mathbf{23}] \langle\mathbf{31}\rangle  \\ 
+ c_{VVV,2}^\prime \times m_1 \tilde{m}_2 [\mathbf{12}]\langle\mathbf{23}\rangle \langle\mathbf{31}\rangle \\ 
+ c_{VVV,3}^\prime \times m_1 m_3 [\mathbf{12}] [\mathbf{23}] \langle\mathbf{31}\rangle } & 
[\mathcal{J}]_1 \cdot [\mathbf{A}]_1 \\
\hline
\end{array}
}
where the coefficients $c^\prime$ are determined by the gauge couplings.

\subsection{Direct $N$-point Matching}

In this section, we discuss the direct matching, which can be applied to most of the $N$-point massive amplitudes.

In the previous discussion, we adopted the strategy of starting with a massive amplitude and finding the corresponding massless amplitude structures. Specifically, for an MHC amplitude structure with a given helicity category $\mathcal{H}=(h_1,h_2,\dots, h_N)$, we can find the corresponding massless amplitude through the helicity information:
\begin{equation}
[\mathcal M(1^{h_1=t_1},\dots,N^{h_N=t_N})]_{0} \quad=\quad \mathcal{A}^{(d)}(1^{h_1},\dots,N^{h_N}),
\end{equation}
where the superscript denotes the helicity of each particle. For 3-point massive amplitudes, this works because in each helicity category there is only one amplitude structure. We list the 3-point matching for all possible helicity categories in Table~\ref{tab:3pt}.

However, for the general $N$-point case, there will be more than one amplitude in the MHC basis; for example, the ST basis of the $ffVS$ with $\mathcal{T}=(-\frac12,-\frac12,0,0)$ is
\begin{equation}
\label{eq:exampleffvs}
[\mathbf{M}]_0 = \begin{pmatrix} 
  -\langle\mathbf{12}\rangle \langle\mathbf{3}|\mathbf{p}_4|\mathbf{3}] \\ 
  -\langle\mathbf{13}\rangle \langle\mathbf{2}|\mathbf{p}_4|\mathbf{3}] \end{pmatrix}\,,
\end{equation}
which corresponds to dimension-2 MHC-amplitude space. In general case, we must first ensure that these massive amplitudes form an independent basis, and then find the corresponding massless amplitudes.
However, the reduction of the massive amplitudes for $N>3$ is difficult, 
so we turn to the massless amplitudes and their matching for simplification, as the independent massless amplitudes correspond to the MHC basis one-by-one. 

The independent massless amplitudes can be constructed by the SSYT method~\cite{Li:2020gnx,Li:2020xlh,Li:2022tec}.
So we should change our strategy: start with the massless amplitude in a given helicity category and match it to the massive ST basis.
\begin{equation}
\mathcal{B}=
\begin{pmatrix}
\mathcal{A}_1 \\
\mathcal{A}_2 \\
\cdots
\end{pmatrix} \to [\mathbf{M}]_0\,.
\end{equation}
As an example, we consider the 4-point amplitudes of helicity category $\mathcal{H}=(-\frac12,-\frac12,0,0)$. In dimension $d=6$, two massless amplitudes form a basis:
\begin{equation}
\mathcal{B} = \begin{pmatrix}
-\langle12\rangle s_{34} \\ \langle13\rangle \langle24\rangle [34]
\end{pmatrix}\,,
\end{equation}
whose corresponding ST amplitudes are exactly in Eq.~\eqref{eq:exampleffvs}.
For all the four-point amplitude cases, the possible helicity categories and the lowest dimension are listed in Table~\ref{tab:4pt}.

In the general case, for given massive particle types, we know each particle spin $s_i$. So we can find all massless amplitudes with helicity categories $\mathcal{H}=(h_1,\cdots,h_n)$, in which $|h_i|\le s_i$. We streamline our matching method as follows:
\begin{itemize}
\item Given a class of massless amplitudes, specifying its helicity $\mathcal{H}$ and dimension $d$, we can construct its basic amplitudes $\mathcal{B}$, which can be done by the SSYT method;
\item Given the massless amplitudes, their quantum numbers $\mathcal{H}$ and transversality $\mathcal{T}$ are equal, which is also true for the MHC amplitudes.
According to the previous discussion, their direct counterparts of the massless amplitudes are the MHC amplitudes. The correspondence is dominated by the relation between $d$ and $n_\lambda$, with relation $d=n_\lambda + 2N$;
\item Further matching to ST amplitudes, the quantum number $\mathcal{H}$ becomes implicit. The MHC amplitudes match to the ST ones by direct bolding,
\begin{equation}
    i\, \rightarrow \boldsymbol{i}\,.
\end{equation}
\end{itemize}
By matching, we can identify both the ST basis of the massive amplitudes and the relations of the coefficients between the massive and massless amplitudes.

\begin{table}[t]
    \centering
    $$\begin{array}{c|c|c|c}
\hline
\mbox{spin} & n_\lambda & d & \mbox{helicity category} \\
\hline
SSS & 0 & 3 & (0,0,0) \\
\hline
ffS & 1 & 4 & (\pm\frac{1}{2},\pm\frac{1}{2},0) \\
\hline
ffV & 2 & 5 & (\pm\frac{1}{2},\pm\frac{1}{2},\pm1) \\
\hline
VVS & 2 & 5 & (\pm1,\pm1,0) \\
\hline
VVV & 3 & 6 & (\pm1,\pm1,\pm1) \\
\hline
\end{array}
$$
\caption{Three-point ST amplitudes and their direct matching across all possible helicity categories. Here, $d$ is the mass dimension of the massless operator, and $n_\lambda$ is half the number of spinors, satisfying the relation $d = n_\lambda + 3$. }
\label{tab:3pt}
\end{table}

\begin{table}[ht!]
    \centering
    \small
    {$$
\begin{array}{c|c|c|c}
\hline
\mbox{spin} & n_\lambda & d & \mbox{helicity category} \\
\hline
ffSS & 1 & 5 & (\pm\frac{1}{2},\pm\frac{1}{2},0,0) \\
& 2 & 6 & (\pm\frac{1}{2},\mp\frac{1}{2},0,0) \\
\hline
VSSS & 2 & 6 & (0,0,0,0) \\
& 3 & 7 & (\pm1,0,0,0) \\
\hline
ffff & 2 & 6 & \makecell{(\pm\frac{1}{2},\pm\frac{1}{2},\pm\frac{1}{2},\pm\frac{1}{2}), (\pm\frac{1}{2},\pm\frac{1}{2},\mp\frac{1}{2},\mp\frac{1}{2}), \\ (\pm\frac{1}{2},\mp\frac{1}{2},\pm\frac{1}{2},\mp\frac{1}{2}), (\pm\frac{1}{2},\mp\frac{1}{2},\mp\frac{1}{2},\pm\frac{1}{2})} \\
\cline{3-4}
& 3 & 7 & \makecell{(\pm\frac{1}{2},\pm\frac{1}{2},\pm\frac{1}{2},\mp\frac{1}{2}), (\pm\frac{1}{2},\pm\frac{1}{2},\mp\frac{1}{2},\pm\frac{1}{2}), \\ (\pm\frac{1}{2},\mp\frac{1}{2},\pm\frac{1}{2},\pm\frac{1}{2}), (\mp\frac{1}{2},\pm\frac{1}{2},\pm\frac{1}{2},\pm\frac{1}{2})} \\
\hline
ffVS & 2 & 6 & (\pm\frac{1}{2},\pm\frac{1}{2},\pm1,0), (\pm\frac{1}{2},\mp\frac{1}{2},0,0) \\
& 3 & 7 & (\pm\frac{1}{2},\pm\frac{1}{2},0,0), (\pm\frac{1}{2},\mp\frac{1}{2},\pm1,0), (\pm\frac{1}{2},\mp\frac{1}{2},\mp1,0) \\
& 4 & 8 & (\pm\frac{1}{2},\pm\frac{1}{2},\mp1,0) \\
\hline
VVSS & 2 & 6 & (\pm1,\pm1,0,0), (0,0,0,0) \\
& 3 & 7 & (\pm1,0,0,0), (0,\pm1,0,0) \\
& 4 & 8 & (\pm1,\mp1,0,0) \\
\hline
ffVV & 3 & 7 & \makecell{ (\pm\frac{1}{2},\pm\frac{1}{2},\pm1,\pm1), (\pm\frac{1}{2},\pm\frac{1}{2},0,0), (\pm\frac{1}{2},\pm\frac{1}{2},\mp1,\mp1), \\ (\pm\frac{1}{2},\mp\frac{1}{2},\pm1,0), (\pm\frac{1}{2},\mp\frac{1}{2},\mp1,0), \\ (\pm\frac{1}{2},\mp\frac{1}{2},0,\pm1), (\pm\frac{1}{2},\mp\frac{1}{2},0,\mp1) } \\
\cline{3-4}
& 4 & 8 & \makecell{ (\pm\frac{1}{2},\mp\frac{1}{2},\pm1,\pm1), (\mp\frac{1}{2},\pm\frac{1}{2},\pm1,\pm1), (\pm\frac{1}{2},\mp\frac{1}{2},0,0), \\ (\pm\frac{1}{2},\pm\frac{1}{2},\pm1,0), (\pm\frac{1}{2},\pm\frac{1}{2},0,\pm1), \\ (\pm\frac{1}{2},\pm\frac{1}{2},\mp1,0), (\pm\frac{1}{2},\pm\frac{1}{2},0,\mp1) } \\
\cline{3-4}
& 5 & 9 & (\pm\frac{1}{2},\pm\frac{1}{2},\pm1,\mp1), (\pm\frac{1}{2},\pm\frac{1}{2},\mp1,\pm1) \\
\hline
VVVS & 3 & 7 & (\pm1,\pm1,\pm1,0), (\pm1,0,0,0), (0,\pm1,0,0), (0,0,\pm1,0) \\
\cline{3-4}
& 4 & 8 & \makecell{ (\pm1,\pm1,0,0), (\pm1,0,\pm1,0), (0,\pm1,\pm1,0), \\ (\pm1,\mp1,0,0), (\pm1,0,\mp1,0), \\ (0,\pm1,\mp1,0), (0,0,0,0) } \\
\cline{3-4}
& 5 & 9 & (\pm1,\pm1,\mp1,0), (\pm1,\mp1,\pm1,0), (\mp1,\pm1,\pm1,0) \\
\hline
VVVV & 4 & 8 & \makecell{ (\pm1,\pm1,\pm1,\pm1), (\pm1,\pm1,0,0), (\pm1,0,\pm1,0), \\ (\pm1,0,0,\pm1), (0,\pm1,\pm1,0), (0,\pm1,0,\pm1), \\ (0,0,\pm1,\pm1), (\pm1,\pm1,\mp1,\mp1), (\pm1,\mp1,\pm1,\mp1), \\ (\pm1,\mp1,\mp1,\pm1), (\pm1,\mp1,0,0), (\pm1,0,\mp1,0), \\ (\pm1,0,0,\mp1), (0,\pm1,\mp1,0), (0,\pm1,0,\mp1), \\ (0,0,\pm1,\mp1), (0,0,0,0) } \\
\cline{3-4}
& 5 & 9 & \makecell{ (\pm1,\pm1,\pm1,0), (\pm1,\pm1,0,\pm1), (\pm1,0,\pm1,\pm1), \\ (0,\pm1,\pm1,\pm1), (\pm1,\pm1,\mp1,0), (\pm1,\mp1,\pm1,0), \\ (\mp1,\pm1,\pm1,0), (\pm1,\pm1,0,\mp1), (\pm1,\mp1,0,\pm1), \\ (\mp1,\pm1,0,\pm1), (\pm1,0,\pm1,\mp1), (\pm1,0,\mp1,\pm1), \\ (\mp1,0,\pm1,\pm1), (0,\pm1,\pm1,\mp1), (0,\pm1,\mp1,\pm1), \\ (0,\mp1,\pm1,\pm1), (\pm1,0,0,0), (0,\pm1,0,0), \\ (0,0,\pm1,0), (0,0,0,\pm1) } \\
\cline{3-4}
& 6 & 10 & \makecell{ (\pm1,\pm1,\pm1,\mp1), (\pm1,\pm1,\mp1,\pm1), \\ (\pm1,\mp1,\pm1,\pm1), (\mp1,\pm1,\pm1,\pm1) } \\
\hline
\end{array}$$
}
    \caption{Four-point ST amplitudes and their direct matching across all possible helicity categories. Here, $d$ is the mass dimension of a massless operator, and $n_\lambda$ counts the number of spinors in the ST basis, satisfying the relation $d = n_\lambda + 4$.}
    \label{tab:4pt}
\end{table}

For example, we illustrate our method with the 5-point class 
\begin{equation}
    Vf^4,\quad n_\lambda=3, \quad \mathcal{T}=\{0,-\frac{1}{2},-\frac{1}{2},-\frac{1}{2},+\frac{1}{2}\}\,.
\end{equation}
We neglect the gauge basis for simplicity and consider only the LO matching.
The corresponding massless class is $D\phi \psi_L^3\psi_R$, with quantum numbers $d=8, \mathcal{H} =  \mathcal{T}=\{0,-\frac{1}{2},-\frac{1}{2},-\frac{1}{2},+\frac{1}{2}\}$. 
Because the scalar field matches the massive vector field, 
in the ST basis, the massless basis $\mathcal{B}$ must satisfy Adler's zero condition:
\begin{equation}
\lim_{p_1\to 0} \mathcal{B}= 0.
\end{equation}
By SSYT, we find that the basis satisfying Adler's zero is $\mathcal{B}$,
\eq{
\mathcal{B} = \begin{pmatrix}
\langle13\rangle \langle24\rangle [15] \\
\langle12\rangle \langle34\rangle [15]
\end{pmatrix}\,,
}
which matches the primary MHC basis, and then matches the ST basis as follows,
\begin{equation}
    \mathcal{B} \xrightarrow{\text{matching}} [\mathcal{M}]_0 = \begin{pmatrix}
\langle13\rangle \langle24\rangle [15] \\
\langle12\rangle \langle34\rangle [15]
\end{pmatrix} \xrightarrow{\text{matching}} [\mathbf{M}]_0 = \begin{pmatrix}
\langle\mathbf{13}\rangle \langle\mathbf{24}\rangle [\mathbf{15}] \\
\langle\mathbf{12}\rangle \langle\mathbf{34}\rangle [\mathbf{15}]
\end{pmatrix}.
\end{equation}
We have simplified the notation by omitting the subscripts and superscripts of $[\mathcal{M}]_0$ and $[\mathbf{M}]_0$. These two amplitudes do not span the full ST basis of $Vf^4$, to obtain which we just need to run over all the allowed transversalities $\mathcal{T}$. Suppose the two massless coefficients are $C_{D\phi \psi_L^3\psi_R,n}$ with $n=1,2$, the matching massive coefficients are thus
\begin{equation}
    c_{Vf^4,n}^{0---+} = \frac{C_{D\phi \psi_L^3\psi_R,n}}{\Lambda^4}\,,\quad n=1,2\,.
\end{equation}
where $\Lambda$ is the cutoff of the SMEFT.

This method can be applied to higher-dimensional constructions that include more derivatives. For example, we consider the $VVVV$ amplitude in helicity $(0,0,0,0)$. The massless basis in general dimensions is given by
\eq{
\label{tab:massless_Amp_0000}
\begin{array}{c|c}
\hline
d & \mbox{massless basis} \\
\hline
8 & \begin{pmatrix}
\langle14\rangle\langle23\rangle[14][23] \\
\langle13\rangle\langle24\rangle[14][23] \\
\langle13\rangle\langle24\rangle[13][24] 
\end{pmatrix}\\
\hline
8+2n & \begin{pmatrix}
[34]
\langle34\rangle([12]\langle12\rangle)^{n+1}\\
[34]
\langle34\rangle([12]\langle12\rangle)^{n}[13]\langle13\rangle\\
[34]
\langle34\rangle([12]\langle12\rangle)^{n-1}([13]\langle13\rangle)^2\\
\dots
\\
[34]
\langle34\rangle[12]\langle12\rangle([13]\langle13\rangle)^n\\
[34]
\langle34\rangle([13]\langle13\rangle)^n[24]\langle24\rangle\\
([13]\langle13\rangle)^{n+1}[24]\langle24\rangle
\end{pmatrix}\\
\hline
\end{array}
}
Applying the direct matching, we obtain the corresponding ST basis:
\eq{
\label{tab:massive_Amp_0000}
\begin{array}{c|c}
\hline
n_\lambda & \mbox{primary ST basis} \\
\hline
4 & \begin{pmatrix}
\langle\mathbf{14}\rangle\langle\mathbf{23}\rangle[\mathbf{14}][\mathbf{23}] \\
\langle\mathbf{13}\rangle\langle\mathbf{24}\rangle[\mathbf{14}][\mathbf{23}] 
\\
\langle\mathbf{13}\rangle\langle\mathbf{24}\rangle[\mathbf{13}][\mathbf{24}]    
\end{pmatrix}\\
\hline
4+2n &\begin{pmatrix}
[\mathbf{34}]
\langle\mathbf{34}\rangle[\mathbf{12}]\langle\mathbf{12}\rangle(\mathbf{p}_1\cdot\mathbf{p}_2)^{n}\\
[\mathbf{34}]
\langle\mathbf{34}\rangle[\mathbf{12}]\langle\mathbf{12}\rangle(\mathbf{p}_1\cdot\mathbf{p}_2)^{n-1}\mathbf{p}_1\cdot\mathbf{p}_3\\
\dots
\\
[\mathbf{34}]
\langle\mathbf{34}\rangle[\mathbf{12}]\langle\mathbf{12}\rangle(\mathbf{p}_1\cdot\mathbf{p}_3)^n\\
\mathbf{p}_3\cdot\mathbf{p}_4(\mathbf{p}_1\cdot\mathbf{p}_3)^{n-1}[\mathbf{13}]\langle\mathbf{13}\rangle[\mathbf{24}]\langle\mathbf{24}\rangle\\
(\mathbf{p}_1\cdot\mathbf{p}_3)^{n}[\mathbf{13}]\langle\mathbf{13}\rangle[\mathbf{24}]\langle\mathbf{24}\rangle
\end{pmatrix}\\
\hline
\end{array}
}
The independence of these amplitude structures is discussed in appendix~\ref{app:exception}. The ST amplitudes take the form given by Eq.~\eqref{eq:VVVVSTamplitudes}, while the massless amplitudes correspond to their H.E. limits as shown in Eq.~\eqref{eq:VVVVmasslessamp}.

Finally, we can extend the amplitude-operator correspondence to the ST amplitudes $\mathbf{M}$ and operators in the broken phase. Considering an unbroken-phase operator $\mathbf{O}_N(x)$ with $N$ fields, its relation to the corresponding amplitudes is
\begin{equation}
    \int d^4x \langle 0|\mathbf{O}_N(x) |\Phi_i\dots\Phi_N\rangle = \delta^{(4)}\left(\sum_{i=1}^N \mathbf{p}_i\right) \mathbf{M} \,,
\end{equation}
where $\Phi_i$ represents a field in the broken phase. For a massive fermion, the left-handed and right-handed spinors are packaged into the Dirac field
\eq{
\begin{pmatrix}
|\boldsymbol{i}\rangle_{\alpha} \\ |\boldsymbol{i}]^{\dot{\alpha}}
\end{pmatrix} \sim f = \begin{pmatrix}
f_L \\ f_R
\end{pmatrix}, \ 
\begin{pmatrix}
[\boldsymbol{i}|_{\dot{\alpha}} & \langle\boldsymbol{i}|^{\alpha}
\end{pmatrix} \sim \bar{f} = \begin{pmatrix}
\bar{f}_R & \bar{f}_L
\end{pmatrix}\,,
}
where $f(\bar{f})$ is massive fermion (anti-fermion). 
For a massive vector boson $V$, different chirality components will not be unified into one field:
\eq{
|\boldsymbol{i}\rangle_{\alpha} |\boldsymbol{i}\rangle_{\beta} \sim V_{(\alpha\beta)}, \quad 
|\boldsymbol{i}\rangle_{\alpha} [\boldsymbol{i}|_{\dot{\alpha}} \sim V_{\alpha\dot{\alpha}}, \quad
[\boldsymbol{i}|_{\dot{\alpha}} [\boldsymbol{i}|_{\dot{\beta}} \sim V_{(\dot{\alpha}\dot{\beta})}.
}
Note that the component $V_{\alpha\dot{\alpha}}$ originates from the Goldstone amplitude in the UV and must satisfy Adler's zero condition. Another point to emphasize is that the above correspondence does not ensure equality of mass dimensions on both sides. The chiral components $V_{\alpha\beta}$ and $V_{\dot\alpha\dot\beta}$ can be identified with the field strength tensor, whereas if $V_{\alpha\dot{\alpha}}$ is identified as the non-chiral vector field, one should include a mass factor, i.e., $|\boldsymbol{i}\rangle_{\alpha} [\boldsymbol{i}|_{\dot{\alpha}} \sim \mathbf{m} V_{\alpha\dot{\alpha}}$. The correspondence for a massive scalar is trivial, $1\sim S$, and the massive momentum corresponds to the derivative $\mathbf{p}_{\alpha\dot{\alpha}} \sim D_{\alpha\dot{\alpha}}$.

The method above can also be applied to derive the $N$-point primary bases, not only 4-point ones. The required input is the massless basis for $N$-point amplitudes.

\subsection{List of 3-point and 4-point Massive EFT Amplitude Basis}

Using the matching procedure described above, we derive the 3-point and 4-point ST amplitude basis, along with the corresponding operator bases. For $N$-point direct matching, a massless amplitude of dimension $d$ corresponds to a massive amplitude with $n_\lambda$, satisfying the relation $d = n_\lambda + N$.

\paragraph{3-point basis} We begin with the 3-point basis. Both direct and exceptional matching contribute.

\begin{itemize}

\item \textit{Direct matching}: We list the negative helicity basis:
\eq{\label{eq:3pt_direct}
\begin{array}{c|c|cc|c|cc}
\hline
\text{class} & d & \text{helicity} & \mbox{massless basis} & n_\lambda & \text{primary ST basis} & \text{operator} \\
\hline
ffS & 4 & (--0) & \langle12\rangle & 1 & \langle\mathbf{12}\rangle & \bar{f}f S \\
\hline
ffV & 5 & (---) & \langle13\rangle\langle 23\rangle & 2 & \langle\mathbf{13}\rangle\langle \mathbf{23}\rangle & \bar{f}\sigma^{\mu\nu} f {V_L}_{\mu\nu} \\
\hline
VVS & 5 & (--0) & \langle12\rangle^2 & 2 & \langle\mathbf{12}\rangle^2 & {V_{1L}}_{\mu\nu} {V_{2L}}^{\mu\nu} S \\
\hline
VVV & 6 & (---) & \langle12\rangle\langle23\rangle\langle31\rangle & 3 & \langle\mathbf{12}\rangle\langle\mathbf{23}\rangle\langle\mathbf{31}\rangle & {V_{1L}}_{\mu\nu} {V_{2L}}^{\nu\rho} {V_{2L}}_{\rho}^{\mu} \\
\hline
\end{array}
}
The positive-helicity amplitudes can be obtained by taking the Hermitian conjugate of the negative-helicity ones, which replaces the angle bracket $|\mathbf{i}\rangle$ with the square bracket $|\mathbf{i}]$. Note that the $VSS$ class does not appear here, as it is absent in a theory with only one species of massive scalar boson.

\item \textit{Exceptional matching}: We first list the contribution from the UV operator,
\eq{
\begin{array}{c|c|cc|c|cc}
\hline
\text{class} & d & \text{helicity} & \mbox{massless basis} & n_\lambda & \text{descendant ST amplitude} & \text{operator} \\
\hline
ffV & 4 & (--0) & \langle12\rangle & 3 & \makecell{c_{ffV,1}\times m_2 \langle\mathbf{13}\rangle[\mathbf{23}]\\
+c_{VVV,2}\times m_1[\mathbf{13}]\langle\mathbf{23}\rangle} & \bar{f}\gamma^\mu f V_{\mu} \\
\hline
VVS & 3 & (000) & 1 & 4 & \makecell{c_{VVS,1}\times m_1^2\langle\mathbf{12}\rangle[\mathbf{12}]\\
+c_{VVS,2}\times m_2^2 \langle\mathbf{12}\rangle[\mathbf{12}]} & {V_1}_{\mu} {V_2}^{\mu} S \\
\hline
VVV & 5 & (--0) & \langle12\rangle^2 & 4 & 
\makecell{ c_{VVV,1}\times m_1\langle\mathbf{12}\rangle\langle\mathbf{23}\rangle [\mathbf{31}] \\ + c_{VVV,2}\times m_2\langle\mathbf{12}\rangle[\mathbf{23}] \langle\mathbf{31}\rangle } & 
{V_{1L}}_{\mu\nu} {V_{2L}}^{\nu\rho} {V_{2L}}_{\rho}^{\mu} \\
\hline
\end{array}
}
Here the coefficients $c$ are defined in section~\ref{sec:exceptional_matching}. They typically include factors of $1/\mathbf{m}^2$, which are accounted for in the mass dimension $d$ for the exceptional matching. This explains why the dimensions in the exceptional case are lower than those in the direct matching of eq.~\eqref{eq:3pt_direct} for the same class. Other helicity categories can be derived by permutations.

We next list the contributions arising from gauge-boson couplings to conserved currents:
\eq{
\begin{array}{c|c|cc|c|c}
\hline
\text{class} & d & \text{helicity} & \mbox{massless amplitude} & n_\lambda & \text{descendant ST amplitude} \\
\hline
ffV & 4 & (-+-) & \frac{\langle 12\rangle\langle 31\rangle}{\langle 23\rangle} & 3 & c_{ffV}^\prime \times m_3 \langle\mathbf{13}\rangle[\mathbf{23}] \\
\hline
VVS & 4 & (-0-) & \frac{\langle 13\rangle^2}{\langle 12\rangle} & 3 & c_{VVS,}^\prime  \times m_3\langle\mathbf{12}\rangle[\mathbf{12}]  \\
\hline
VVV & 4 & (-+-) & \frac{\langle 13\rangle^2}{\langle 12\rangle\langle 23\rangle} & 5 & 
\makecell{ c_{VVV,1}^\prime \times \tilde{m}_2 m_3 \langle\mathbf{12}\rangle[\mathbf{23}] \langle\mathbf{31}\rangle  \\ 
+ c_{VVV,2}^\prime \times m_1 \tilde{m}_2 [\mathbf{12}]\langle\mathbf{23}\rangle \langle\mathbf{31}\rangle \\ 
+ c_{VVV,3}^\prime \times m_1 m_3 [\mathbf{12}] [\mathbf{23}] \langle\mathbf{31}\rangle }  \\
\hline
\end{array}
}

\end{itemize}

\paragraph{4-point basis} We then consider the 4-point basis. Most 4-point bases arise solely from direct matching, with the exception of $VVVV$, which is listed at the end of this subsection.

\begin{itemize}
  
\item {$ffSS$} For convenience, we present the massless and massive ST bases separately. For the $ffSS$ class, the massless basis is simple and can be classified by the total helicity $h=h_1+h_2+h_3+h_4$:
\eq{
\begin{array}{c|c|cc}
\hline
d & \text{total helicity }h & \multicolumn{2}{c}{\mbox{massless basis}} \\
\hline
5 & h\neq 0 & \langle12\rangle & h.c. \\
\hline
6 & h=0 & \langle14\rangle [24] & perm. \\
\hline
\end{array}
}
Here, ``h.c.'' denotes Hermitian conjugation, which yields the opposite-helicity basis (e.g., $\langle12\rangle\to [12]$). ``Perm.'' indicates permutations of particles, which act on particles of the same spin and convert helicity labels accordingly {e.g., $\langle 14\rangle[24]\to \langle 24\rangle[14]$}.

After matching, the helicity structure maps onto the transversality of the ST amplitude. The corresponding massive ST basis is therefore organized by the total transversality $t=t_1+t_2+t_3+t_4$:
\begin{equation}
\begin{array}{c|c|c|c|c|c}
\hline
\multirow{2}{*}{$n_\lambda$} & 
\multicolumn{2}{c|}{\text{transversality}} & 
\multirow{2}{*}{\text{primary ST basis}} & 
\multirow{2}{*}{\text{operator basis}} & \text{\# of} \\
\cline{2-3}
 & \text{total} & \text{category} & & & perm. \\
\hline
  1 & t=-1 & (--00) & \langle\mathbf{12}\rangle & 
    \bar{f}_1 f_2 S_3 S_4 & 1 \\ 
    \hline
  2 & t=0 &  (-+00) & -\langle\mathbf{1}|\mathbf{p}_4|\mathbf{2}] & \bar{f}_1 (\slashed{D} S_4) f_2 S_3 & 2 \\
  \hline
\end{array}
\end{equation}
The last column indicates the number of independent bases obtained by permutations that preserve the total transversality. For instance, in the last row, permutation gives $-\langle\mathbf{1}|\mathbf{p}_4|\mathbf{2}]\to -\langle\mathbf{2}|\mathbf{p}_4|\mathbf{1}]$, yielding two terms. For simplicity, we list only bases with negative or zero total transversality ($t\le 0$). The $t>0$ bases can be obtained by Hermitian conjugation, e.g., $\langle\mathbf{12}\rangle\to [\mathbf{12}]$ for the $(++00)$ transversality case. 

\item {$VSSS$}

\eq{
\begin{array}{c|c|cc}
\hline
d & \text{total helicity }h & \multicolumn{2}{c}{\mbox{massless basis}} \\
\hline
6 & h=0 & \begin{pmatrix}
-s_{34} \\ -s_{24}
\end{pmatrix} & h.c. \\
\hline
7 & h\neq 0 & \langle13\rangle \langle14\rangle [34] &  \\
\hline
\end{array}
}
The basis enclosed in parentheses indicates that these amplitudes belong to the same helicity category.

\begin{equation}
\begin{array}{c|c|c|c|c|c}
\hline
\multirow{2}{*}{$n_\lambda$} & 
\multicolumn{2}{c|}{\text{transversality}} & 
\multirow{2}{*}{\text{primary ST basis}} & 
\multirow{2}{*}{\text{operator basis}} & \text{\# of} \\
\cline{2-3}
 & \text{total} & \text{category} & & & perm. \\
\hline
  2 & t=0 & (0000) & \begin{pmatrix}
    -\langle\mathbf{1}|\mathbf{p}_2|\mathbf{1}] \\
    -\langle\mathbf{1}|\mathbf{p}_3|\mathbf{1}]
    \end{pmatrix} & 
    \begin{pmatrix}
    (D_{\mu}S_2) V^{\mu} S_3 S_4 \\ (D_{\mu}S_3) V^{\mu} S_2 S_4
    \end{pmatrix} & 1 \\ 
    \hline
  3 & t=-1 &  (-000) & -\langle\mathbf{1}|\mathbf{p}_3\mathbf{p}_4|\mathbf{1}\rangle & S_2 (D_{\mu} S_3) (D_{\nu} S_4) {V_{L}}^{\mu\nu} & 1 \\
  \hline
\end{array}
\end{equation}

\item {$ffff$}

\eq{
\begin{array}{c|c|ccc}
\hline
d & \text{total helicity }h & \multicolumn{3}{c}{\mbox{massless basis}} \\
\hline
\multirow{2}{*}{6} & h\neq 0 & \langle12\rangle \langle34\rangle & perm., h.c. &  \\
\cline{2-5}
  & h=0 & \langle12\rangle [34] & perm. &  \\
\hline
7 & h\neq 0 & \langle13\rangle \langle23\rangle [34] & perm., h.c. & \\
\hline
\end{array}
}

\begin{equation}
\begin{array}{c|c|c|c|c|c}
\hline
\multirow{2}{*}{$n_\lambda$} & 
\multicolumn{2}{c|}{\text{transversality}} & 
\multirow{2}{*}{\text{primary ST basis}} & 
\multirow{2}{*}{\text{operator basis}} & \text{\# of} \\
\cline{2-3}
 & \text{total} & \text{category} & & & perm. \\
\hline
\multirow{2}{*}{2} & t=-2 & (----) & \langle\mathbf{12}\rangle \langle\mathbf{34}\rangle & \bar{f}_1 f_2 \bar{f}_3 f_4 & 1 \\
\cline{2-6}
& t=0 & (--++) & \langle\mathbf{12}\rangle [\mathbf{34}] & \bar{f}_1 f_2 \bar{f}_3 f_4 & 6 \\
\hline
3 & t=-1 & (---+) & \langle\mathbf{13}\rangle \langle\mathbf{2}|\mathbf{p}_3|\mathbf{4}] & \bar{f}_1 (D_{\mu} f_3) \bar{f}_2 \gamma^{\mu} f_4 & 4 \\
\hline
\end{array}
\end{equation}

\item {$ffVS$}

\eq{
\begin{array}{c|c|ccc}
\hline
d & \text{total helicity }h & \multicolumn{3}{c}{\mbox{massless basis}} \\
\hline
\multirow{2}{*}{6} & h\neq 0 & \langle13\rangle \langle23\rangle & h.c. &  \\
\cline{2-5}
  & h= 0 & \langle14\rangle [24] & perm. &  \\
\hline
7 & h\neq 0 & -\langle13\rangle \langle34\rangle [24] & \begin{pmatrix}
-\langle12\rangle s_{34} \\ \langle13\rangle \langle24\rangle [34]
\end{pmatrix} & perm., h.c. \\
\hline
8 & h=0 & \langle14\rangle \langle24\rangle [34]^2 &  \\
\hline
\end{array}
}

\begin{equation}
\begin{array}{c|c|c|c|c|c}
\hline
\multirow{2}{*}{$n_\lambda$} & 
\multicolumn{2}{c|}{\text{transversality}} & 
\multirow{2}{*}{\text{primary ST basis}} & 
\multirow{2}{*}{\text{operator basis}} & \text{\# of} \\
\cline{2-3}
 & \text{total} & \text{category} & & & perm. \\
\hline
\multirow{2}{*}{2} & t=-2 & (---0) & \langle\mathbf{13}\rangle \langle\mathbf{23}\rangle & \bar{f}_1 \sigma^{\mu\nu} f_2 {V_L}_{\mu\nu} S & 1 \\
\cline{2-6}
& t=0 & (-+00) & \langle\mathbf{13}\rangle [\mathbf{23}] & \bar{f}_1 \slashed{V} f_2 S & 2 \\
\hline
\multirow{2}{*}{3} & \multirow{2}{*}{$t=-1$} & (-+-0) & \langle\mathbf{13}\rangle \langle\mathbf{3}|\mathbf{p}_4|\mathbf{2}] & \bar{f}_1 \sigma^{\mu\nu} (\slashed{D} S) f_2 {V_L}_{\mu\nu} & 2 \\
&  & (--00) & \begin{pmatrix} -\langle\mathbf{12}\rangle \langle\mathbf{3}|\mathbf{p}_4|\mathbf{3}] \\ -\langle\mathbf{13}\rangle \langle\mathbf{2}|\mathbf{p}_4|\mathbf{3}] \end{pmatrix} & \begin{pmatrix} \bar{f}_1 f_2 (D_{\mu} S) V^{\mu} \\ \bar{f}_1 \slashed{V} (\slashed{D} S) f_2 \end{pmatrix} & 1 \\
\hline
5 & t=0 & (--+0) & \langle\mathbf{1}|\mathbf{p}_4|\mathbf{3}] \langle\mathbf{2}|\mathbf{p}_4|\mathbf{3}] & \bar{f}_1 \gamma^{\mu} \sigma^{\nu\rho} \gamma^{\sigma} f_2 (D_{\mu} D_{\sigma} S) {V_R}_{\nu\rho} & 1 \\
\hline
\end{array}
\end{equation}

\item {$VVSS$}

\eq{
\begin{array}{c|c|ccc}
\hline
d & \text{total helicity }h & \multicolumn{3}{c}{\mbox{massless basis}} \\
\hline
\multirow{2}{*}{6} & h\neq 0 & \langle12\rangle^2 & h.c. & \\
\cline{2-5}
  & h=0 & -s_{34} & &  \\
\hline
7 & h\neq 0 & \langle13\rangle \langle14\rangle [34] & perm.,h.c. & \\
\hline
8 & h=0 & \langle14\rangle^2 [24]^2 & perm. \\
\hline
\end{array}
}

\begin{equation}
\begin{array}{c|c|c|c|c|c}
\hline
\multirow{2}{*}{$n_\lambda$} & 
\multicolumn{2}{c|}{\text{transversality}} & 
\multirow{2}{*}{\text{primary ST basis}} & 
\multirow{2}{*}{\text{operator basis}} & \text{\# of} \\
\cline{2-3}
 & \text{total} & \text{category} & & & perm. \\
\hline
\multirow{2}{*}{2} & t=-2 & (--00) & \langle\mathbf{12}\rangle^2 & {V_{1L}}_{\mu\nu} {V_{2L}}^{\mu\nu} S_3 S_4 & 1 \\
\cline{2-6}
& t=0 & (0000) & \langle\mathbf{12}\rangle [\mathbf{12}] & {V_1}_{\mu} {V_2}^{\mu} S_3 S_4 & 1 \\
\hline
3 & t=-1 & (-000) & \langle\mathbf{12}\rangle \langle\mathbf{1}|\mathbf{p}_4|\mathbf{2}] & S_3 (D^{\nu} S_4) {V_2}^{\mu} {V_{1L}}_{\mu\nu} & 2 \\
\hline
4 & t=0 & (-+00) & \langle\mathbf{1}|\mathbf{p}_4|\mathbf{2}]^2 & S_3 (D_{\mu} D^{\nu} S_4) {V_{1L}}^{\mu\rho} {V_{2R}}_{\rho\nu} & 2 \\
\hline
\end{array}
\end{equation}

\item {$ffVV$}

\eq{
\begin{array}{c|c|ccc}
\hline
d & \text{total helicity }h & \multicolumn{3}{c}{\mbox{massless basis}} \\
\hline
\multirow{2}{*}{7} & \multirow{2}{*}{$h\neq0$} & \begin{pmatrix}
\langle{12}\rangle \langle{34}\rangle^2 \\
\langle{13}\rangle \langle{24}\rangle \langle{34}\rangle
\end{pmatrix} & \begin{pmatrix}
\langle{12}\rangle \langle{34}\rangle [{34}] \\ \langle{13}\rangle \langle{24}\rangle [{34}]
\end{pmatrix} & \langle{13}\rangle \langle{34}\rangle [{24}]  \\
& & [{12}] \langle{34}\rangle^2 & perm.,h.c. & \\
\hline
\multirow{2}{*}{8} & h\neq0 & -\langle{34}\rangle^2 \langle14\rangle [24] & \begin{pmatrix}
\langle14\rangle [24] s_{34} \\ \langle{14}\rangle [24] s_{24}
\end{pmatrix} & perm.,h.c. \\
\cline{2-5}
& h=0 & \begin{pmatrix}
\langle{13}\rangle \langle{23}\rangle s_{34} \\ -\langle{13}\rangle \langle23\rangle s_{24}
\end{pmatrix} & \makecell{-\langle{13}\rangle \langle23\rangle [24]^2\\ \langle{13}\rangle \langle23\rangle [{34}]^2} & perm. \\
\hline
9 & h\neq0 & -\langle{13}\rangle \langle23\rangle^2 [24] [34] & perm.,h.c. & \\
\hline
\end{array}
}

\begin{equation}
\begin{array}{c|c|c|c|c|c}
\hline
\multirow{2}{*}{$n_\lambda$} & 
\multicolumn{2}{c|}{\text{transversality}} & 
\multirow{2}{*}{\text{primary ST basis}} & 
\multirow{2}{*}{\text{operator basis}} & \text{\# of} \\
\cline{2-3}
 & \text{total} & \text{category} & & & perm. \\
\hline
\multirow{4}{*}{3} & t=-3 & (----) & \begin{pmatrix} \langle\mathbf{12}\rangle \langle\mathbf{34}\rangle^2 \\ \langle\mathbf{13}\rangle \langle\mathbf{24}\rangle \langle\mathbf{34}\rangle \end{pmatrix} & \begin{pmatrix} \bar{f}_1 f_2 {V_{3L}}_{\mu\nu} {V_{4L}}^{\mu\nu} \\ \bar{f}_1 \sigma^{\mu\nu} \sigma^{\rho\sigma} f_2 {V_{3L}}_{\mu\nu} {V_{4L}}_{\rho\sigma} \end{pmatrix} & 1 \\
\cline{2-6}
& \multirow{3}{*}{$t=-1$} & (--00) & \begin{pmatrix} \langle\mathbf{12}\rangle \langle\mathbf{34}\rangle [\mathbf{34}] \\ \langle\mathbf{13}\rangle \langle\mathbf{24}\rangle [\mathbf{34}] \end{pmatrix} & \begin{pmatrix} \bar{f}_1 f_2 {V_3}^{\mu} {V_4}_{\mu} \\ \bar{f}_1 \gamma^{\mu} \gamma^{\nu} f_2 {V_3}_{\mu} {V_4}_{\nu} \end{pmatrix} & 1 \\
& & (-+-0) & \langle\mathbf{13}\rangle \langle\mathbf{34}\rangle [\mathbf{24}] & \bar{f}_1 \sigma^{\mu\nu} \gamma^{\rho} f_2 {V_{3L}}_{\mu\nu} {V_4}_{\rho} & 4 \\
& & (++--) & [\mathbf{12}] \langle\mathbf{34}\rangle^2 & \bar{f}_1 f_2 {V_{3L}}_{\mu\nu} {V_{4L}}^{\mu\nu} & 1 \\
\hline
\multirow{4}{*}{4} & \multirow{2}{*}{$t=-2$} & (-+--) & \langle\mathbf{34}\rangle^2 \langle\mathbf{1}|\mathbf{p}_4|\mathbf{2}] & \bar{f}_1 \gamma^{\rho} f_2 {V_{3L}}^{\mu\nu} (D_{\rho} {V_{4L}}_{\mu\nu}) & 2 \\
& & (---0) & \begin{pmatrix} \langle\mathbf{13}\rangle \langle\mathbf{23}\rangle \langle\mathbf{4}|\mathbf{p}_3|\mathbf{4}] \\ \langle\mathbf{13}\rangle \langle\mathbf{24}\rangle \langle\mathbf{3}|\mathbf{p}_1|\mathbf{4}] \end{pmatrix} & \begin{pmatrix} \bar{f}_1 \sigma^{\mu\nu} f_2 (D_{\rho} {V_{3L}}_{\mu\nu}) {V_4}^{\rho} \\ (D_{\rho} \bar{f}_1) \sigma^{\mu\nu} \gamma^{\rho} \slashed{V}_4 f_2 {V_{3L}}_{\mu\nu} \end{pmatrix} & 2 \\
\cline{2-6}
& \multirow{2}{*}{$t=0$} & (-+00) & \begin{pmatrix} \langle\mathbf{34}\rangle [\mathbf{34}] \langle\mathbf{1}|\mathbf{p}_4|\mathbf{2}] \\ \langle\mathbf{13}\rangle [\mathbf{24}] \langle\mathbf{4}|\mathbf{p}_2|\mathbf{3}] \end{pmatrix} & \begin{pmatrix} \bar{f}_1 \gamma^{\nu} f_2 {V_3}^{\mu} (D_{\nu} {V_4}_{\mu}) \\ \bar{f}_1 \slashed{V}_3 \gamma^{\mu} \slashed{V}_4 (D_{\mu} f_2) \end{pmatrix} & 2 \\
& & (-+-+) & \langle\mathbf{13}\rangle [\mathbf{24}] \langle\mathbf{3}|\mathbf{p}_2|\mathbf{4}] & \bar{f}_1 \sigma^{\mu\nu} \gamma^{\delta} \sigma^{\rho\sigma} (D_{\delta} f_2) {V_{3L}}_{\mu\nu} {V_{4R}}_{\rho\sigma} & 4 \\
& & (--0+) & \langle\mathbf{13}\rangle [\mathbf{34}] \langle\mathbf{2}|\mathbf{p}_3|\mathbf{4}] & \bar{f}_1 (D_{\rho} \slashed{V}_3) \sigma^{\mu\nu} \gamma^{\rho} f_2 {V_{4R}}_{\mu\nu} & 2 \\
\hline
5 & t=-1 & (---+) & \langle\mathbf{13}\rangle \langle\mathbf{3}|\mathbf{p}_1|\mathbf{4}] \langle\mathbf{2}|\mathbf{p}_1|\mathbf{4}] & (D_{\kappa} D_{\tau} \bar{f}_1) \sigma^{\mu\nu} \gamma^{\kappa} \sigma^{\rho\sigma} \gamma^{\tau} f_2 {V_{3L}}_{\mu\nu} {V_{4R}}_{\rho\sigma} & 2 \\
\hline
\end{array}
\end{equation}

\item {$VVVS$}

\eq{\label{Tab:VVVSmassless}
\begin{array}{c|c|ccc}
\hline
d & \text{total helicity }h & \multicolumn{3}{c}{\mbox{massless basis}} \\
\hline
7 & h\neq 0 & \langle{12}\rangle \langle{23}\rangle \langle{13}\rangle & -\langle{12}\rangle \langle{14}\rangle [{24}] & perm.,h.c. \\
\hline
\multirow{2}{*}{8} & h\neq 0 & \begin{pmatrix}
-\langle{12}\rangle^2 s_{34} \\ -\langle{12}\rangle \langle{13}\rangle \langle24\rangle [34]
\end{pmatrix} & perm.,h.c. &  \\
\cline{2-5}
& h=0 & \begin{pmatrix}
-s_{24}^2 \\ -s_{34}^2 \\ -s_{34} s_{24}
\end{pmatrix} & -\langle{12}\rangle [{23}] \langle{1}4\rangle [34] & perm. \\
\hline
9 & h\neq0 & \langle{12}\rangle \langle{1}4\rangle \langle{2}4\rangle [34]^2 & perm.,h.c. &\\
\hline
\end{array}
}

\begin{equation}
\begin{array}{c|c|c|c|c|c}
\hline
\multirow{2}{*}{$n_\lambda$} & 
\multicolumn{2}{c|}{\text{transversality}} & 
\multirow{2}{*}{\text{primary ST basis}} & 
\multirow{2}{*}{\text{operator basis}} & \text{\# of} \\
\cline{2-3}
 & \text{total} & \text{category} & & & perm. \\
\hline
\multirow{2}{*}{3} & t=-3 & (---0) & \langle\mathbf{12}\rangle \langle\mathbf{23}\rangle \langle\mathbf{13}\rangle \quad  & {V_{1L}}_{\mu\nu} {V_{2L}}^{\nu\rho} {V_{3L}}_{\rho}^{\mu} S_4 \quad & 1 \\
\cline{2-6}
& t=-1 & (-000) & \langle\mathbf{12}\rangle \langle\mathbf{13}\rangle [\mathbf{23}] \quad  & {V_{1L}}_{\mu\nu} {V_2}^{\mu} {V_{3}}^{\nu} S_4 \quad & 3  \\
\hline
\multirow{3}{*}{4} & t=-2 & (--00) & \begin{pmatrix} -\langle\mathbf{12}\rangle^2 \langle\mathbf{3}|\mathbf{p}_4|\mathbf{3}] \\ \langle\mathbf{12}\rangle \langle\mathbf{13}\rangle \langle\mathbf{2}|\mathbf{p}_4|\mathbf{3}] \end{pmatrix} \quad  & \begin{pmatrix} {V_{1L}}_{\mu\nu} {V_{2L}}^{\mu\nu} {V_3}^{\rho} (D_{\rho} S_4) \\ {V_{1L}}_{\mu\nu} {V_{2L}}_{\rho\sigma} \end{pmatrix} \quad & 3 \\
\cline{2-6}
& \multirow{2}{*}{$t=0$} & (0000) & \begin{pmatrix} \langle\mathbf{13}\rangle [\mathbf{13}] \langle\mathbf{2}|\mathbf{p}_4|\mathbf{2}] \\ \langle\mathbf{12}\rangle [\mathbf{12}] \langle\mathbf{3}|\mathbf{p}_4|\mathbf{3}] \\ \langle\mathbf{12}\rangle [\mathbf{12}] \langle\mathbf{3}|\mathbf{p}_1|\mathbf{3}] \end{pmatrix} & \begin{pmatrix} {V_1}_\mu V_{2\nu} {V_3}^\mu (D^\nu S_4) \\ {V_1}_\mu {V_2}^\mu V_{3\nu} (D^\nu S_4) \\ {D^\nu V_1}_\mu {V_2}^\mu V_{3\nu} S_4 \end{pmatrix} & 1 \\
& & (-0+0) & \langle\mathbf{12}\rangle [\mathbf{23}] \langle\mathbf{1}|\mathbf{p}_4|\mathbf{3}] \quad  & {V_{1L}}_{\mu\nu} {V_2}^{\mu} {V_{3R}}^{\nu\rho} (D_\rho S_4) \quad  & 6 \\
\hline
5 & t=-1 & (--+0) & \langle\mathbf{12}\rangle \langle\mathbf{1}|\mathbf{p}_4|\mathbf{3}] \langle\mathbf{2}|\mathbf{p}_4|\mathbf{3}] \quad  & {V_{1L}}_{\mu\nu} {V_{2L}}^{\nu\rho} {V_{3R}}_{\rho\sigma} (D_\mu D^\sigma S_4) \quad & 3 \\
\hline
\end{array}
\end{equation}

\item {$VVVV$}:

\eq{\label{eq:4V_massless}
\begin{array}{c|c|cccc}
\hline
d & \text{total helicity }h & \multicolumn{4}{c}{\mbox{massless basis}} \\
\hline
6 & h\neq0 & \langle 12 \rangle^2 & h.c. \\
\hline
\multirow{2}{*}{8} & h\neq 0 & \begin{pmatrix}
\langle{12}\rangle^2 \langle{34}\rangle^2 \\ \langle{13}\rangle^2 \langle{24}\rangle^2 \\ \langle{12}\rangle \langle{34}\rangle \langle{13}\rangle \langle{24}\rangle
\end{pmatrix} & \begin{pmatrix}
\langle{12}\rangle^2 \langle{34}\rangle [{34}] \\ \langle{12}\rangle \langle{13}\rangle \langle{24}\rangle [{34}]
\end{pmatrix} & perm.,h.c. \\
\cline{2-6}
 & h=0 & \makecell{\langle{12}\rangle^2 [{34}]^2 \\ \langle{14}\rangle^2 [{24}]^2} & \begin{pmatrix}
\langle14\rangle\langle23\rangle[14][23] \\
\langle13\rangle\langle24\rangle[14][23] \\
\langle13\rangle\langle24\rangle[13][24] 
\end{pmatrix} & perm., \\
\hline
\multirow{2}{*}{9} & \multirow{2}{*}{$h\neq 0$} & \begin{pmatrix}
\langle{12}\rangle \langle{13}\rangle \langle{23}\rangle s_{34} \\ \langle{13}\rangle^2 \langle{24}\rangle \langle{2}3\rangle [3{4}]
\end{pmatrix} & \langle{12}\rangle \langle{13}\rangle \langle{2}3\rangle [3{4}]^2 & & \\
& & \begin{pmatrix}
-\langle13\rangle \langle14\rangle [34] s_{34} \\ \langle13\rangle \langle14\rangle [34] s_{24} 
\end{pmatrix} & perm.,h.c. & \\
\hline
10 & h\neq0 & \langle{13}\rangle^2 \langle{2}3\rangle^2 [3{4}]^2 & perm.,h.c. & \\
\hline
\end{array}
}
Each line corresponds to direct matching, except for the $d=6$ line, which arises from exceptional matching. For the direct matching, the primary ST basis is given as follows:
\begin{equation}
\begin{array}{c|c|c|c|c|c}
\hline
\multirow{2}{*}{$n_\lambda$} & 
\multicolumn{2}{c|}{\text{transversality}} & 
\multirow{2}{*}{\text{primary ST basis}} & 
\multirow{2}{*}{\text{operator basis}} & \text{\# of} \\
\cline{2-3}
 & \text{total} & \text{category} & & & perm. \\
\hline
\multirow{4}{*}{4} & t=-4 & (----) & 
\begin{pmatrix}
\langle\mathbf{12}\rangle^2 \langle\mathbf{34}\rangle^2 \\
\langle\mathbf{13}\rangle^2 \langle\mathbf{24}\rangle^2 \\
\langle\mathbf{12}\rangle \langle\mathbf{34}\rangle \langle\mathbf{13}\rangle \langle\mathbf{24}\rangle
\end{pmatrix} &
\begin{pmatrix}
{V_{1L}}_{\mu\nu} {V_{2L}}^{\mu\nu} {V_{3L}}_{\rho\sigma} {V_{4L}}^{\rho\sigma} \\
{V_{1L}}_{\mu\nu} {V_{2L}}_{\rho\sigma} {V_{3L}}^{\mu\nu} {V_{4L}}^{\rho\sigma} \\
{V_{1L}}_{\mu\nu} {V_{2L}}^{\nu\rho} {V_{3L}}^{\sigma\mu} {V_{4L}}_{\rho\sigma}
\end{pmatrix} & 1 \\
\cline{2-6}
& t=-2 & (--00) &
\begin{pmatrix}
\langle\mathbf{12}\rangle^2 \langle\mathbf{34}\rangle [\mathbf{34}] \\
\langle\mathbf{12}\rangle \langle\mathbf{13}\rangle \langle\mathbf{24}\rangle [\mathbf{34}]
\end{pmatrix} &
\begin{pmatrix}
{V_{1L}}_{\mu\nu} {V_{2L}}^{\mu\nu} {V_{3}}_{\rho} {V_{4}}^{\rho} \\
{V_{1L}}_{\mu\nu} {V_{2L}}^{\nu\rho} {V_{3}}^{\mu} {V_{4}}_{\rho}
\end{pmatrix} & 6 \\
\cline{2-6}
& \multirow{3}{*}{$t=0$} & (--++) &
\langle\mathbf{12}\rangle^2 [\mathbf{34}]^2 &
{V_{1L}}_{\mu\nu} {V_{2L}}^{\mu\nu} {V_{3R}}_{\rho\sigma} {V_{4R}}^{\rho\sigma} & 6 \\
& & (-+00) &
\langle\mathbf{13}\rangle \langle\mathbf{14}\rangle [\mathbf{23}] [\mathbf{24}] &
{V_{1L}}_{\mu\nu} {V_{2R}}^{\nu\rho} {V_{3}}^{\mu} {V_{4}}_{\rho} & 12 \\
& & (0000) &
\begin{pmatrix}
\langle\mathbf{14}\rangle\langle\mathbf{23}\rangle[\mathbf{14}][\mathbf{23}] \\
\langle\mathbf{13}\rangle\langle\mathbf{24}\rangle[\mathbf{14}][\mathbf{23}] 
\\
\langle\mathbf{13}\rangle\langle\mathbf{24}\rangle[\mathbf{13}][\mathbf{24}]  
\end{pmatrix} &
\begin{pmatrix}
{V_{1}}_{\mu} {V_{2}}_{\nu} {V_{3}}^{\nu} {V_{4}}^{\mu} \\
\text{tr}(\gamma^\mu\gamma^\nu\gamma^\rho\gamma^\sigma)
{V_{1}}_{\mu} {V_{2}}_{\rho} {V_{3}}_{\nu} {V_{4}}_{\sigma}
\\
{V_{1}}_{\mu} {V_{2}}_{\nu} {V_{3}}^{\mu} {V_{4}}^{\nu}
\end{pmatrix} & 1 \\
\hline
\multirow{3}{*}{5} & t=-3 & (---0) &
\begin{pmatrix}
\langle\mathbf{12}\rangle \langle\mathbf{13}\rangle \langle\mathbf{23}\rangle \langle\mathbf{4}|\mathbf{p}_3|\mathbf{4}] \\
\langle\mathbf{13}\rangle^2 \langle\mathbf{24}\rangle \langle\mathbf{2}|\mathbf{p}_3|\mathbf{4}]
\end{pmatrix} &
\begin{pmatrix}
{V_{1L}}_{\mu\nu} {V_{2L}}^{\nu\rho} (D_{\sigma} {V_{3L}}_{\rho}^{\mu}) {V_4}^{\sigma} \\
{V_{1L}}_{\mu\nu} {V_{2L}}_{\rho\sigma} (D^{\rho} {V_{3L}}^{\mu\nu}) {V_{4}}^\sigma
\end{pmatrix} & 4 \\
\cline{2-6}
& \multirow{2}{*}{$t=-1$} & (--0+) &
\langle\mathbf{12}\rangle \langle\mathbf{13}\rangle [\mathbf{34}] \langle\mathbf{2}|\mathbf{p}_3|\mathbf{4}] &
{V_{1L}}_{\mu\nu} {V_{2L}}^{\nu\rho} (D^{\sigma} {V_3}^{\mu}) {V_{4R}}_{\rho\sigma} & 12 \\
& & (-000) &
\begin{pmatrix}
-\langle\mathbf{12}\rangle \langle\mathbf{13}\rangle [\mathbf{23}] \langle\mathbf{4}|\mathbf{p}_3|\mathbf{4}] \\
\langle\mathbf{12}\rangle \langle\mathbf{13}\rangle [\mathbf{23}] \langle\mathbf{4}|\mathbf{p}_2|\mathbf{4}]
\end{pmatrix} &
\begin{pmatrix}
{V_{1L}}_{\mu\nu} {V_{2}}^{\mu} (D^{\sigma} {V_3}^{\nu}) {V_{4}}_{\sigma} \\
{V_{1L}}_{\mu\nu} (D^{\sigma}{V_{2}}^{\mu}) {V_3}^{\nu} {V_{4}}_{\sigma}
\end{pmatrix} & 4 \\
\hline
6 & t=-2 & (---+) &
\langle\mathbf{13}\rangle^2 \langle\mathbf{2}|\mathbf{p}_3|\mathbf{4}]^2 &
{V_{1L}}_{\mu\nu} {V_{2L}}^{\rho\sigma} (D_\rho D^\tau {V_{3L}}_{\mu\nu}) {V_{4R}}_{\sigma\tau} & 4 \\
\hline
\end{array}
\end{equation}

The $d=6$ line corresponds to exceptional matching. It matches onto the following descendant ST basis with transversality $(-000)$:
\begin{align}
c_{VVVV}\times m_1 (\langle\mathbf{14}\rangle\langle\mathbf{23}\rangle[\mathbf{13}][\mathbf{24}]-\langle\mathbf{13}\rangle\langle\mathbf{24}\rangle[\mathbf{14}][\mathbf{23}]) 
\quad\sim\quad
\varepsilon^{\mu\nu\rho\sigma} {V_{1}}_{\mu} {V_{2}}_{\rho} {V_{3}}_{\nu} {V_{4}}_{\sigma}.
\end{align}
However, this structure requires four distinct massive vector bosons, which are not present in the SMEFT.

\end{itemize}

%% file: sec5-smeft.tex
\newcommand{\calu}{\mathcal{U}}
\newcommand{\calm}{\mathcal{M}}

\section{Matched SMEFT Operators through Dimension Eight}
\label{sec:SMEFT}


\subsection{Scope and conventions}

We apply the matching framework to the one-flavor SMEFT through dimension eight, using the massless amplitude conventions of Ref.~\cite{Li:2020gnx}. Starting from the massless amplitudes, their bases are obtainable by the SSYT method, and according to the matching discussion in section~\ref{sec:2}, massless amplitudes correspond to the ST basis following a strict power-counting scheme.

Taking the gauge basis into account, the gauge symmetry of the massless amplitudes is, 
\begin{equation}
    \mathcal{G}_{\text{unbroken}} = SU(3)_C \times SU(2)_L\times U(1)_Y\,,
\end{equation}
and the gauge symmetry of massive amplitudes is the one in the broken phase,
\begin{equation}
    \mathcal{G}_{\text{broken}} = SU(3)_C \times U(1)_{\text{em}}\,,
\end{equation}
with $U(1)_{\text{em}}$ corresponding to electric charge conservation.
The spontaneous breaking of the electroweak gauge group $SU(2)_L\times U(1)_Y \rightarrow U(1)_{\text{em}}$ gives mass to the gauge bosons via the Higgs mechanism. 
Thus, the color symmetry and the electroweak symmetry behave differently in matching. Next, we will discuss them separately.

\paragraph{Electroweak symmetry}
Our emphasis is the electroweak projection across $SU(2)_L\times U(1)_Y\to U(1)_{\mathrm{em}}$.  
We retain generic fermion labels, including quark labels when useful for illustrating the electroweak projection, but do not list explicit $SU(3)_C$ color tensors.  
We consider leading matching with equal numbers of massless and massive external particles; higher-order matching involving additional soft Higgs fields is not included.

We reserve ``Wilson coefficient'' for an unbroken-phase coefficient $C_i^{(d)}$.  A coefficient $c_{\mathrm{class}}^{\mathcal T}$ instead multiplies a broken-phase ST basis element and can depend on $C_i^{(d)}$, SM couplings, vacuum expectation values, and the projectors defined below.  All displayed projectors use the field normalizations and electroweak mixing conventions specified in Eqs.~\eqref{eq:trans_Omega}--\eqref{eq:transo}.

Explicitly, we write the relations for the fields before and after the symmetry breaking as
\begin{align}
    (W^+,W^-,Z,A) \equiv V^{\mathbf{I}} &= O^{\mathbf{I} I}\left(W \oplus B\right)^I + \mathcal{U}^{Ii} H^\dagger_i + \mathcal{U}^{I}_i H^i\,,  \\
    (u,d,\nu,e) \equiv f^{\hat{\mathbf{i}}} &= \Omega^{\hat{\mathbf{i}}}_{\hat{i}} \left(Q\oplus d^\dagger_\mathbb{C} \oplus u^\dagger_\mathbb{C} \oplus L\oplus e^\dagger_\mathbb{C}\right)^{\hat{i}}\,,\\
    (\overline{u},\overline{d},\overline{\nu},\overline{e}) \equiv \overline{f}_{\hat{\mathbf{i}}} &= \Omega_{\hat{\mathbf{i}}}^{\hat{i}} \left(Q^\dagger\oplus d_\mathbb{C} \oplus u_\mathbb{C} \oplus L^\dagger\oplus e_\mathbb{C}\right)_{\hat{i}}\,,\\
    h &= \mathcal{U}^h_i H^i + \mathcal{U}^{hi} H^\dagger_i\,.
\end{align}
On the left-hand side of the equation, $V,f,\overline{f},h$ denote the massive vector bosons, fermions, antifermions, and scalars, respectively. Even though the symmetry is broken in this phase, the notation provides a bookkeeping device that groups related operator types under a common index structure, thereby simplifying the classification. For the massless fields, we use $I$ to range over the massless gauge bosons $W$ and $B$, including the $SU(2)_L$ adjoint space as a subspace; $i$ is the $SU(2)_L$ fundamental index, and $\hat{i}$ is the index among the massless fermion fields. The superscript denotes fermion fields, and the subscript denotes antifermion fields. 
For the massive fields, we use $\mathbf{I}$ ranging over the massive gauge bosons and $\hat{\mathbf{i}}$ ranging over the massive fermions, where the fermion and anti-fermion fields are also distinguished by superscripts and subscripts. The details of the two versions of notation are shown as follows,
\begin{eqnarray}
\begin{array}{c|c|c|c}
\hline
\text{broken phase} & \text{vector boson} & \text{fermion} & \text{scalar boson} \\
\hline
\text{index} & \mathbf{I} & \hat{\mathbf{i}} & {h} \\
\hline
\text{upper index value} & \{W^+,W^-,Z,A\} & \{u,d;\nu,e\} & \{h\} \\
\hline
\text{lower index value} & \mbox{-} & \{\bar{u},\bar{d};\bar{\nu},\bar{e}\} & \mbox{-} \\
\hline
\end{array}
\end{eqnarray}
\begin{equation}
\begin{array}{c|c|c|c}
\hline
\text{unbroken phase} & \text{gauge boson} & \text{fermion} & \text{scalar boson} \\
\hline
\text{index} & I & \hat{i} & i \\
\hline
\text{upper index value} & W,B & Q,L;u_{\mathbbm{C}},d_{\mathbbm{C}},e_{\mathbbm{C}} & H \\
\hline
\text{lower index value} & \mbox{-} & Q^{\dagger},L^{\dagger};u_{\mathbbm{C}}^{\dagger},d_{\mathbbm{C}}^{\dagger},e_{\mathbbm{C}}^{\dagger} & H^\dagger \\
\hline
\end{array} 
\end{equation}
The massless and massive fields are related by the projectors $O,\mathcal{U}$ and $\Omega$.
The $\Omega$ embeds fermion components, $O$ rotates gauge fields to vector mass eigenstates, and $\mathcal U$ projects Higgs-doublet components onto Goldstone modes or the physical Higgs. Their non-vanishing entries are
\begin{align}
\Omega^{\hat{\mathbf{i}}}_{\hat{i}}:&\qquad\qquad \Omega^{e}_{e_{\mathbbm{C}}} =\Omega^{u}_{u_{\mathbbm{C}}} =\Omega^{d}_{d_{\mathbbm{C}}} =\Omega^{e}_{L^2}= \Omega^{\nu}_{L^1}=\Omega^{u}_{Q^1}= \Omega^{d}_{Q^2}=1, \label{eq:trans_Omega}\\ 
\mathcal{U}^{ih},\mathcal{U}^{h}_i:&\qquad\quad
\mathcal{U}^{H^\dagger_2 h} = \mathcal{U}^{h}_{H^2} = \frac{1}{\sqrt{2}}, \\
\mathcal{U}^{i\mathbf{I}},\mathcal{U}^{\mathbf{I}}_i:&\qquad\quad
\mathcal{U}^{H^\dagger_1 W^-} = -\mathcal{U}^{W^+}_{H^1} = \sqrt{2}, \quad -\mathcal{U}^{Z}_{H^2} = \mathcal{U}^{H^\dagger_2 Z} = 1, \label{eq:trans_U_2}\\
O^{I \mathbf{I}}:&\quad
\left\{\begin{aligned}
O^{W^1 W^+} &= O^{W^1 W^-} =\frac{1}{\sqrt{2}},&  O^{W^3 Z} &= O^{B A} = \cos\theta_W\equiv\frac{g}{\sqrt{g^2+g^{\prime2}}},& \\ 
O^{W^2 W^+} &= -O^{W^2 W^-} = \frac{i}{\sqrt{2}},& 
-O^{B Z} &= O^{W^3 A} = \sin\theta_W\equiv\frac{g^\prime}{\sqrt{g^2+g^{\prime2}}}.&  
\end{aligned}\right. \label{eq:transo}
\end{align}
Here we list only the entries of $\Omega^{\hat{\mathbf{i}}}_{\hat{i}}$. The non-vanishing entries of $\Omega_{\hat{\mathbf{i}}}^{\hat{i}}$ follow by interchanging the two indices. In particular, we only consider the SM operators. Generally, the entries above can be changed by the higher-dimensional operators.

The advantage of this organization is to simplify the massive amplitudes sharing the same spinor structure but different electroweak structures. We encode electroweak information into the broken-phase amplitude coefficients, schematically written as
\eq{
({c_{\textrm{class}}^{\mathcal{T}}}^{\textrm{vec.}})^{\textrm{fer}.\; \textrm{sca}.},
}
where $\mathcal{T}$ denotes the transversality category, and the labels ``vec.'', ``fer.'', and ``sca.'' indicate vector, fermion, and scalar indices. The UV Wilson coefficients determine these broken-phase coefficients through the massless--massive correspondence.

\paragraph{Color gauge structure}
For the $SU(3)_C$ structures, a systematic construction of the gauge tensors is to utilize the Young tensor method.
Following the generalized Littlewood–Richardson rule employed in Refs.\cite{Li:2020gnx,Li:2020xlh}, we express the basis in terms of Levi-Civita tensors $\epsilon^{abc}$ carrying fundamental indices.  After converting anti-fundamental and adjoint indices into fundamental indices, we can obtain the tensor basis by outer product.
We illustrate this procedure with a concrete example $Z u d \bar{u} \bar{d} $, which is a subclass of the $Vf^4$ class.
It possesses non-trivial $SU(3)_C$ gauge structures, which do not change in matching. Quark $u$ and $d$ carried fundamental indices
\begin{equation}
u_{a_1} \sim \begin{ytableau}
a_1
\end{ytableau},\quad d_{a_2} \sim \begin{ytableau}
a_2
\end{ytableau}.
\end{equation}
For anti-quarks $\bar{u}$ and $\bar{d}$, we need the conversion from anti-fundamental indices to fundamental ones, which is realized by the $\epsilon_{abc}$ tensor,
\eq{
\bar{u}_{b_3 c_3} \equiv \epsilon_{a_3 b_3 c_4} \bar{u}^{a_3}\sim  \begin{ytableau}
b_3 \\ c_3
\end{ytableau}, \quad
\bar{d}_{b_4 c_4} \equiv \epsilon_{a_4 b_4 c_4} \bar{d}^{a_4} \sim  \begin{ytableau}
b_4 \\ c_4
\end{ytableau}.
}
Then, the gauge y-basis $T^y$, which is obtained by the outer product of the Young tableaux corresponding to the fields, is constructed from the generalized Littlewood-Richardson rule, 
\eq{
\ytableaushort[a_]{1} \xrightarrow{\ytableaushort[a_]{2}} \ytableaushort[a_]{12} \xrightarrow{\begin{ytableau}
b_3 \\ c_3
\end{ytableau}} \begin{ytableau}
a_1 & a_2 \\ b_3 \\ c_3 
\end{ytableau} \xrightarrow{\begin{ytableau}
b_4 \\ c_4
\end{ytableau}} \begin{ytableau}
a_1 & a_2 \\ b_3 & b_4 \\ c_3 & c_4
\end{ytableau} \to
\begin{aligned}[t]
&\epsilon^{a_1 b_3 c_3} \epsilon^{a_2 b_4 c_4}
+\epsilon^{a_1 b_4 c_3} \epsilon^{a_2 b_3 c_4} \\
&+\epsilon^{a_1 b_3 c_4} \epsilon^{a_2 b_4 c_3}
+\epsilon^{a_1 b_4 c_4} \epsilon^{a_2 b_3 c_3},
\end{aligned} \\
\ytableaushort[a_]{1} \xrightarrow{\ytableaushort[a_]{2}} \ytableaushort[a_]{1,2} \xrightarrow{\begin{ytableau}
b_3 \\ c_3
\end{ytableau}} \begin{ytableau}
a_1 & b_3 \\ a_2 \\ c_3
\end{ytableau} \xrightarrow{\begin{ytableau}
b_4 \\ c_4
\end{ytableau}} \begin{ytableau}
a_1 & b_3 \\ a_2 & b_4 \\ c_3 & c_4
\end{ytableau} \to
\begin{aligned}[t]
&\epsilon^{a_1 a_2 c_3} \epsilon^{b_3 b_4 c_4}
+\epsilon^{a_1 b_4 c_3} \epsilon^{b_3 a_2 c_4} \\
&+\epsilon^{a_1 a_2 c_4} \epsilon^{b_3 b_4 c_3}
+\epsilon^{a_1 b_4 c_4} \epsilon^{b_3 a_2 c_3}.
\end{aligned}
}
The next step is to contract $T^y$ with $\epsilon_{abc}$ for anti-particles, and obtain the gauge m-basis $T^m$,
\eq{
T^y \epsilon_{a_3 b_3 c_3} \epsilon_{a_4 b_4 c_4} = \begin{pmatrix}
6\delta_{a_4}^{a_1} \delta_{a_3}^{a_2} +6\delta_{a_3}^{a_1} \delta_{a_4}^{a_2} \\
-9 \delta_{a_4}^{a_1} \delta_{a_3}^{a_2} +3 \delta_{a_3}^{a_1} \delta_{a_4}^{a_2}
\end{pmatrix}
\quad\sim\quad T^m = \begin{pmatrix}
\delta_{a_4}^{a_1} \delta_{a_3}^{a_2} \\ \delta_{a_3}^{a_1} \delta_{a_4}^{a_2}
\end{pmatrix}. 
}
Thus, the two independent gauge tensors are $\delta_{a_4}^{a_1} \delta_{a_3}^{a_2} \,,\delta_{a_3}^{a_1} \delta_{a_4}^{a_2}$. Consequently, the matching massive amplitudes are extended to
\begin{equation}
\label{eq:vf4example}
c_{Vf^4,n}^{0---+} \times 
\begin{pmatrix}
\delta_{a_4}^{a_1} \delta_{a_3}^{a_2} \\ \delta_{a_3}^{a_1} \delta_{a_4}^{a_2}
\end{pmatrix} \times 
    \begin{pmatrix}
\langle\mathbf{13}\rangle \langle\mathbf{24}\rangle [\mathbf{15}] \\
\langle\mathbf{12}\rangle \langle\mathbf{34}\rangle [\mathbf{15}]
\end{pmatrix} \,,\quad n=1,2\,.
\end{equation}

\paragraph{Reproducible projection example.}
The following $Vf^4$ channel illustrates the procedure used throughout this section.  First select the massless operator tensor and its coefficient.  Next contract every unbroken index with the appropriate projector. Finally, fix the physical external-state labels; all terms with vanishing projector entries drop out.  For the transversality $\mathcal T=(0---+)$, the relevant leading-order operator types are $DHQ^2u_{\mathbbm C}Q^\dagger$ and $DH^\dagger Q^2d_{\mathbbm C}Q^\dagger$~\cite{Li:2020gnx}.  Before choosing external states, their contribution to the broken-phase amplitude coefficient is
\eq{
({c_{V f^4,n}}^{\mathbf{I}_1})_{\hat{\mathbf{i}}_2 \hat{\mathbf{i}}_3}^{\hat{\mathbf{i}}_4 \hat{\mathbf{i}}_5} \times \begin{pmatrix}
\delta_{a_4}^{a_1} \delta_{a_3}^{a_2} \\ \delta_{a_3}^{a_1} \delta_{a_4}^{a_2}
\end{pmatrix}\,,\quad n=1,2\,,
}
where $({c_{V f^4}}^{\mathbf{I}_1})_{\hat{\mathbf{i}}_2 \hat{\mathbf{i}}_3}^{\hat{\mathbf{i}}_4 \hat{\mathbf{i}}_5}$ is a broken-phase amplitude coefficient, not a Wilson coefficient. Considering the projectors, we find
\eq{
({c_{Vf^4}^{0---+}}^{\mathbf{I}_1})^{\hat{\mathbf{i}}_2 \hat{\mathbf{i}}_3}_{\hat{\mathbf{i}}_4 \hat{\mathbf{i}}_5}
= \left(-C_{D H Q^2 u_{\mathbbm{C}} Q^{\dagger},1} \delta^{\hat{i}_2}_{\hat{i}_5} \epsilon^{i_{1} \hat{i}_3} \delta^{u_\mathbbm{C}}_{\hat{i}_4} + C_{D H Q^2 u_{\mathbbm{C}} Q^{\dagger},2} \delta^{i_1}_{\hat{i}_5} \epsilon^{\hat{i}_2 \hat{i}_3} \delta^{u_\mathbbm{C}}_{\hat{i}_4} \right)  \mathcal{U}_{i_1}^{\mathbf{I}_1} \Omega_{\hat{i}_2}^{\hi{2}} \Omega_{\hat{i}_3}^{\hi{3}} \Omega^{\hat{i}_4}_{\hi{4}} \Omega^{\hat{i}_5}_{\hi{5}} \\
+\left( C_{D H^{\dagger} Q^2 d_{\mathbbm{C}} Q^{\dagger},1} \delta^{\hat{i}_2}_{i_1} \delta^{\hat{i}_3}_{\hat{i}_5} \delta^{d_\mathbbm{C}}_{\hat{i}_4} + C_{D H^{\dagger} Q^2 d_{\mathbbm{C}} Q^{\dagger},2} \delta^{\hat{i}_3}_{\hat{i}_1} \delta^{\hat{i}_2}_{\hat{i}_5} \delta^{d_\mathbbm{C}}_{i_4} \right) \mathcal{U}^{i_1 \mathbf{I}_1} \Omega_{\hat{i}_2}^{\hi{2}} \Omega_{\hat{i}_3}^{\hi{3}} \Omega^{\hat{i}_4}_{\hi{4}} \Omega^{\hat{i}_5}_{\hi{5}},
}
The tensors inside the bracket can be identified as $SU(2)_L$ group tensors by applying the particle labels in the Wilson coefficients. 
Once the particle labels in the broken phase are fixed, we can obtain the corresponding coefficient explicitly. For example, we choose
\begin{equation} \begin{aligned}
\mathbf{I}_1=W^+,\quad \hi{2}=\hi{3}=\hi{5}=d,\quad \hi{4}=u.
\end{aligned} \end{equation}
Only one term contributes to the broken-phase coefficient, yielding
\begin{equation} \begin{aligned}
({c_{Vf^4}^{0---+}}^{W^+})^{d d}_{u \bar{d}}=&-C_{D H Q^2 u_{\mathbbm{C}} Q^{\dagger},1} \delta^{i_2}_{i_5} \epsilon^{i_{1} i_3} \delta^{u_\mathbbm{C}}_{i_4}\times \mathcal{U}_{i_1}^{W^+} \Omega_{\hat{i}_2}^{d} \Omega_{\hat{i}_3}^{d} \Omega^{\hat{i}_4}_{u_{\mathbbm{C}}} \Omega^{\hat{i}_5}_{d} \\
=&-C_{D H Q^2 u_{\mathbbm{C}} Q^{\dagger},1} \delta^{\mathtt{i}_2}_{\mathtt{i}_5} \epsilon^{\mathtt{i}_{1} \mathtt{i}_3} \times \mathcal{U}_{H^{\mathtt{i}_{1}}}^{W^+} \Omega_{Q^{\mathtt{i}_2}}^{d} \Omega_{Q^{\mathtt{i}_3}}^{d} \Omega^{u_\mathbbm{C}}_{u} \Omega^{Q^{\dagger}_{\mathtt{i}_5}}_{d}\\
=&-C_{D H Q^2 u_{\mathbbm{C}} Q^{\dagger},1} \times \mathcal{U}_{H^{1}}^{W^+} \Omega_{Q^{2}}^{d} \Omega_{Q^{2}}^{d} \Omega^{u_\mathbbm{C}}_{u} \Omega^{Q^{\dagger}_{2}}_{d} \\
=& \sqrt{2} C_{D H Q^2 u_{\mathbbm{C}} Q^{\dagger},1}.
\end{aligned} \end{equation}
In the last line, we have used the transformation matrices from Eqs.~\eqref{eq:trans_Omega} and \eqref{eq:trans_U_2}. The same procedure can be applied to determine the unbroken-phase coefficients for other choices of particle labels.

In particular, the photon field $\mathbf I=A$ remains massless; thus the broken-phase field $V$ can be either massive or massless depending on the index value. For convenience, we denote the spinor indices of the vector field $V$ as 
\begin{equation}
    \mathbb{i} = \mathbb{1}\,,\mathbb{2}\,,\mathbb{3}\,,\dots\,,
\end{equation} 
where $\mathbb{i}$ takes the value $i$ for massless field and $\mathbf{i}$ for massive ones, as appropriate. Besides, the massive vector fields $W^\pm\,,Z$ are also matched from the derivatives on the Higgs fields due to the Higgs mechanism, $H^\dagger{}^i \partial H_i \rightarrow V$. In such cases, either one of the Higgs fields matches $V$ via the projector $\mathcal{U}$. We consider both cases. 
As specified above, we consider only LO matching with equal numbers of massless and massive external particles; higher-order matching involving soft scalars is excluded.

\paragraph{Scope}
We now apply the amplitude basis to EFT matching across EWSSB. For each process, we identify the gauge-invariant operators with specific helicity in the unbroken theory, and project the result onto the massive basis with corresponding transversality constructed. A broken-phase amplitude collects the contribution from different transversality categories $\mathcal{T}$ and can be written as
\begin{equation}
\label{eq:massivefor}
    \left(\mathcal{M}_{\text{massive class}}^{\mathbf{I}_1\dots\mathbf{I}_2\dots}\right){}^{\hi{3}\dots h\dots}_{\hi{4}\dots} = \sum_{\mathcal{T}} \left(c_{\text{massive class}}^{\mathcal{T}}\,{}^{\mathbf{I}_1\dots\mathbf{I}_2\dots}\right){}^{\hi{3}\dots h\dots}_{\hi{4}\dots} \mathcal{M}_{\text{massive class}}^{\mathcal{T}}\,.
\end{equation}
We work in the one-flavor electroweak sector and suppress explicit $SU(3)_C$ color tensors.  The color factors are unchanged by EWSSB and factorize from the amplitudes displayed below.  Consequently, the corresponding color-resolved quark and gluon results follow by attaching the appropriate color tensors to the electroweak coefficients.

In the following, we present the matching results of the massive amplitudes from the massless ones up to dimension 8 in the SMEFT. By convention, we organize the massive amplitudes by the number of external fields; in each class, we present the massive amplitude as in Eq.~\eqref{eq:massivefor}, then the Wilson coefficients. The massless amplitudes and the corresponding coefficients are presented in Ref.~\cite{Li:2020gnx}.
We keep the indices $\mathbf{I}\,,\hat{\mathbf{i}}$ of Wilson coefficients implicit, but their explicit values can be obtained directly once the numeric forms of the projectors $\Omega\,,\calu\,, O$ are substituted.
We summarize the matching results in Table~\ref{tab:smeft_navigation}, which lists the classes, the equations of matching, the numbers of UV operators, and the contributing UV operator families.



\begin{longtable}{c c c c p{9.5cm}}
\caption{Navigation table for the matched SMEFT amplitude classes of Section~\ref{sec:SMEFT}: number of external legs $N$, amplitude class, defining equation, number of contributing UV operators, and the SMEFT operator families entering its Wilson coefficient. A repeated family at the same operator dimension but with an independent flavor/tensor structure is indicated by a multiplicity $\times n$; the complete list of structures is given in the referenced equation. In particular, the renormalizable operators such as the Yukawas and dynamic terms are not counted.}
\label{tab:smeft_navigation} \\
\toprule
$N$ & class & eq. & \# ops & contributing UV operator families \\
\midrule
\endfirsthead
\multicolumn{5}{l}{\tablename\ \thetable{} -- continued} \\
\toprule
$N$ & class & eq. & \# ops & contributing UV operator families \\
\midrule
\endhead
\midrule
\multicolumn{5}{r}{continued on next page} \\
\endfoot
\bottomrule
\endlastfoot
\midrule
\multicolumn{5}{l}{\textit{3-point amplitudes}} \\
\midrule
3 & $f^2h$ & \eqref{eq:cls_f2h} & 1 & \scriptsize $L^2H^2$ \\
3 & $f\bar{f}V$ & \eqref{eq:cls_fbarV} & 2 & \scriptsize $Le_CB_L H^\dagger$, $Le_CW_LH^\dagger$ \\
3 & $f^2V$ & \eqref{eq:cls_f2V} & 2 & \scriptsize $L^2B_LH^2$, $L^2W_LH^2$ \\
3 & $V^3$ & \eqref{eq:VVV_result} & 4 & \scriptsize $W_L^3$, $B_L^2HH^\dagger$,$W_L^2HH^\dagger$, $W_LB_LHH^\dagger$ \\
3 & $V^2h$ & \eqref{eq:cls_V2h} & 3 & \scriptsize $B_L^2HH^\dagger$, $B_LW_LHH^\dagger$, $W_L^2HH^\dagger$ \\
3 & $h^3$ & \eqref{eq:cls_h3} & 4 & \scriptsize $D^2H^2H^\dagger {}^2$ ($\times2$), $H^3H^\dagger {}^3$, $H^2H^\dagger{}^2$\\
\midrule
\multicolumn{5}{l}{\textit{4-point amplitudes}} \\
\midrule
4 & $f^2\bar{f}{}^2$ & \eqref{eq:cls_f2bar2} & 13 & \scriptsize $L^2L^\dagger {}^2$, $e_C^2e_C^\dagger {}^2$, $Le_C^\dagger  e_C L^\dagger$, $D^2L^2L^\dagger {}^2$ ($\times2$), $D^2e_C^2e_C^\dagger {}^2$ ($\times2$), \dots\ (9 families total) \\
4 & $f^3\bar{f}$ & \eqref{eq:cls_f3bar} & 1 & \scriptsize $L^3e_CH$ \\
4 & $f^2h^2$ & \eqref{eq:cls_f2h2} & 4 & \scriptsize $L^2H^2$, $D^2L^2H^2$ ($\times2$), $DLe_C^\dagger  H^3$ \\
4 & $f\bar{f}h^2$ & \eqref{eq:cls_fbarh2} & 12 & \scriptsize $De_C^\dagger  e_C HH^\dagger$, $DL^\dagger  L HH^\dagger$ ($\times2$), $Le_CHH^\dagger {}^2$, $D^3LL^\dagger  HH^\dagger$ ($\times4$), $D^3e_C^\dagger  e_C HH^\dagger$ ($\times2$), $D^2Le_CHH^\dagger {}^2$ ($\times2$) \\
4 & $f\bar{f}Vh$ & \eqref{eq:cls_fbarVh} & 33 & \scriptsize $Le_CB_L H^\dagger$, $Le_CW_LH^\dagger$, $D^2Le_CB_LH^\dagger$ ($\times2$), $D^2Le_CW_LH^\dagger$ ($\times2$), $De_C^\dagger  e_C HH^\dagger$, \dots\ (15 families total) \\
4 & $f^2Vh$ & \eqref{eq:cls_f2Vh} & 2 & \scriptsize $L^2B_LH^2$, $L^2W_LH^2$ \\
4 & $f\bar{f}V^2$ & \eqref{eq:cls_fbarV2} & 34 & \scriptsize $DLL^\dagger  B_LW_L$, $DLL^\dagger  W_L^2$, $D^2Le_CB_LH^\dagger$ ($\times2$), $D^2Le_CW_LH^\dagger$ ($\times2$), $DLL^\dagger  B_LB_R$, \dots\ (24 families total) \\
4 & $V^4$ & \eqref{eq:cls_V4} & 22 & \scriptsize $B_L^4$, $B_L^2W_L^2$ ($\times2$), $W_L^4$ ($\times2$), $D^2B_L^2HH^\dagger$, $D^2B_LW_LHH^\dagger$ ($\times2$), \dots\ (14 families total) \\
4 & $V^3h$ & \eqref{eq:cls_V3h} & 17 & \scriptsize $D^2B_L^2HH^\dagger$, $D^2B_LW_LHH^\dagger$ ($\times2$), $D^2W_L^2HH^\dagger$ ($\times2$), $D^2B_LB_RHH^\dagger$, $D^2B_LW_RHH^\dagger$, \dots\ (12 families total) \\
4 & $V^2h^2$ & \eqref{eq:cls_V2h2} & 18 & \scriptsize $B_L^2HH^\dagger$, $B_LW_LHH^\dagger$, $W_L^2HH^\dagger$, $D^2B_L^2HH^\dagger$, $D^2B_LW_LHH^\dagger$ ($\times2$), \dots\ (12 families total) \\
4 & $Vh^3$ & \eqref{eq:cls_Vh3} & 6 & \scriptsize $D^2H^3H^\dagger {}^3$ ($\times3$), $D^2B_LH^2H^\dagger {}^2$, $D^2W_LH^2H^\dagger {}^2$ ($\times2$) \\
4 & $h^4$ & \eqref{eq:cls_h4} & 8 & \scriptsize $D^2H^2H^\dagger {}^2$ ($\times2$), $H^3H^\dagger {}^2$, $D^2H^3H^\dagger {}^3$ ($\times3$) \\
\midrule
\multicolumn{5}{l}{\textit{5-point amplitudes}} \\
\midrule
5 & $f^2\bar{f}{}^2h$ & \eqref{eq:cls_f2bar2h} & 13 & \scriptsize $DLe_C^\dagger  e_C^2H^\dagger$ ($\times3$), $DL^2L^\dagger  e_C H^\dagger$ ($\times3$), $L^2e_C^2H^\dagger {}^2$ ($\times2$), $e_C^\dagger {}^2e_C^2HH^\dagger$, $Le_C^\dagger  e_C L^\dagger  HH^\dagger$ ($\times2$), $L^2L^\dagger {}^2 HH^\dagger$ ($\times2$) \\
5 & $f^3\bar{f}h$ & \eqref{eq:cls_f3barh} & 1 & \scriptsize $L^3e_CH$ \\
5 & $f^2\bar{f}{}^2V$ & \eqref{eq:cls_f2bar2V} & 9 & \scriptsize $e_C^\dagger {}^2 e_C^2B_L$, $L^2L^\dagger {}^2B_L$, $L^2L^\dagger {}^2W_L$, $Le_C^\dagger  e_CL^\dagger  B_L$, $Le_C^\dagger  e_CL^\dagger  W_L$, \dots\ (7 families total) \\
5 & $f\bar{f}h^3$ & \eqref{eq:cls_fbarh3} & 12 & \scriptsize $Le_CHH^\dagger {}^2$, $D^2Le_CHH^\dagger {}^2$ ($\times6$), $De_C^\dagger  e_CH^2H^\dagger {}^2$, $DLL^\dagger  H^2H^\dagger {}^2$ ($\times4$) \\
5 & $f^2h^3$ & \eqref{eq:cls_f2h3} & 2 & \scriptsize $DLe_C^\dagger  H^3$, $L^2H^3H^\dagger$ \\
5 & $f^2Vh^2$ & \eqref{eq:cls_f2Vh2} & 2 & \scriptsize $L^2B_LH^2$, $L^2W_LH^2$ \\
5 & $f\bar{f}Vh^2$ & \eqref{eq:cls_fbarVh2} & 26 & \scriptsize $De_C^\dagger  e_{C}B_LHH^\dagger$ ($\times2$), $De_C^\dagger  e_{C}W_LHH^\dagger$ ($\times2$), $DLL^\dagger  B_LHH^\dagger$ ($\times4$), $DLL^\dagger  W_LHH^\dagger$ ($\times6$), $D^2Le_CHH^\dagger {}^2$ ($\times3$), \dots\ (9 families total) \\
5 & $f\bar{f}V^2h$ & \eqref{eq:cls_fbarV2h} & 24 & \scriptsize $Le_{C}B_L^2H^\dagger$, $Le_{C}W_LB_LH^\dagger$ ($\times2$), $Le_{C}W_L^2H^\dagger$ ($\times2$), $Le_{C}B_R^2H^\dagger$, $Le_{C}W_RB_RH^\dagger$, \dots\ (11 families total) \\
5 & $V^3h^2$ & \eqref{eq:cls_V3h2} & 5 & \scriptsize $B_LW_L^2HH^\dagger$, $W_L^3HH^\dagger$, $D^2B_LH^2H^\dagger {}^2$, $D^2W_LH^2H^\dagger {}^2$ ($\times2$) \\
5 & $V^2h^3$ & \eqref{eq:cls_V2h3} & 9 & \scriptsize $D^2B_LH^2H^\dagger {}^2$, $D^2W_LH^2H^\dagger {}^2$ ($\times2$), $B_L^2H^2H^\dagger {}^2$, $B_LW_LH^2H^\dagger {}^2$, $W_L^2H^2H^\dagger {}^2$ ($\times2$), $D^2H^3H^\dagger {}^3$ ($\times2$) \\
5 & $Vh^4$ & \eqref{eq:cls_Vh4} & 5 & \scriptsize $D^2B_LH^2H^\dagger {}^2$, $D^2W_LH^2H^\dagger {}^2$ ($\times2$), $D^2H^3H^\dagger {}^3$ ($\times2$) \\
5 & $h^5$ & \eqref{eq:cls_h5} & 3 & \scriptsize $H^3H^\dagger {}^2$, $D^2H^3H^\dagger {}^3$ ($\times2$) \\
\midrule
\multicolumn{5}{l}{\textit{6,7,8-point amplitudes}} \\
\midrule
6 & $f^2\bar{f}{}^2h^2$ & \eqref{eq:cls_f2bar2h2} & 7 & \scriptsize $L^2e_C^2H^\dagger {}^2$ ($\times2$), $e_C^\dagger {}^2e_C^2HH^\dagger$, $Le_C^\dagger  e_C L^\dagger  HH^\dagger$ ($\times2$), $L^2L^\dagger {}^2 HH^\dagger$ ($\times2$) \\
6 & $f^2h^4$ & \eqref{eq:cls_f2h4} & 1 & \scriptsize $L^2H^3H^\dagger$ \\
6 & $f\bar{f}h^4$ & \eqref{eq:cls_fbarh4} & 6 & \scriptsize $De_C^\dagger  e_CH^2H^\dagger {}^2$, $DLL^\dagger  H^2H^\dagger {}^2$ ($\times4$), $Le_CH^2H^\dagger {}^3$ \\
6 & $f\bar{f}Vh^3$ & \eqref{eq:cls_fbarVh3} & 6 & \scriptsize $Le_{C}B_LHH^\dagger {}^2$, $Le_{C}W_LHH^\dagger {}^2$ ($\times2$), $De_C^\dagger  e_CH^2H^\dagger {}^2$, $DLL^\dagger  H^2H^\dagger {}^2$ ($\times2$) \\
6 & $V^2h^4$ & \eqref{eq:cls_V2h4} & 6 & \scriptsize $B_L^2H^2H^\dagger {}^2$, $B_LW_LH^2H^\dagger {}^2$, $W_L^2H^2H^\dagger {}^2$ ($\times2$), $D^2H^3H^\dagger {}^3$ ($\times2$) \\
6 & $Vh^5$ & \eqref{eq:cls_Vh5} & 2 & \scriptsize $D^2H^3H^\dagger {}^3$ ($\times2$) \\
6 & $h^6$ & \eqref{eq:cls_h6} & 3 & \scriptsize $H^3H^\dagger {}^2$, $D^2H^3H^\dagger {}^3$ ($\times2$) \\
7 & $f\bar{f}h^5$ & \eqref{eq:cls_fbarh5} & 1 & \scriptsize $Le_CH^2H^\dagger {}^3$ \\
7 & $h^7$ & \eqref{eq:cls_h7} & 1 & \scriptsize $H^4H^\dagger {}^4$ \\
8 & $h^8$ & \eqref{eq:cls_h8} & 1 & \scriptsize $H^4H^\dagger {}^4$ \\
\end{longtable}

\subsection{3-point Matching Result}

Actually, the 3-point amplitudes are matched from both the SM and high-dimensional operators. Here we present both contributions.

\paragraph{$\boldsymbol{f^2h}$}
\begin{align}
    &\mathcal{M}^{\hi{1}\hi{2}h}_{f^2h}\left(\mathbf{\frac{1}{2},\frac{1}{2},0}\right) = 
    \left(c^{--0}_{f^2h}\right){}^{\hi{1}\hi{2}h}\lra{\mathbf{12}} 
    + \text{h.c.}\,,\label{eq:cls_f2h}
\end{align}

\begin{align}
    & \left(c^{--0}_{f^2h}\right){}^{\hi{1}\hi{2}h} = \left(\frac{C_{L^2H^2}}{\Lambda}\epsilon^{\hat{i}_1 i_3}\epsilon^{\hat{i}_2 i_4}\right)\Omega_{\hat{i}_1}^{\hi{1}}\Omega_{\hat{i}_2}^{\hi{2}}\mathcal{U}_{i_3}^{h} \mathcal{U}^h_{i_4}v\,.
\end{align}

\paragraph{\boldsymbol{$f\bar{f}V$}}
\begin{align}
    & \left(\mathcal{M}_{f\bar{f}V}^{\mathbf{I}_3}\right){}_{\hi{2}}^{\hi{1}}\left(\mathbf{\frac{1}{2},\frac{1}{2},1}\right) = 
    \left(c_{f\bar{f}V}^{---}{}^{\mathbf{I}_3}\right){}^{\hi{1}}_{\hi{2}} \lra{\mathbf{1}\mathbb{3}}\lra{\mathbf{2}\mathbb{3}} +\frac{1}{\mathbf{m}_3^2} \left( {c_{f\bar{f}V}^{+-0}}^{\mathbf{I}_3} \right)^{\hi{1}}_{\hi{2}} m_3 \langle\mathbf{2}\mathbb{3}\rangle [\mathbb{3}\mathbf{1}] 
    \notag \\
    &+ \left(c_{f\overline{f}V}^{+-0}{}^{\mathbf{I}_3}\right){}_{\hi{2}}^{\hi{1}} [\mathbf{13}]\lra{\mathbf{23}} \,,\label{eq:cls_fbarV}
\end{align}

\begin{align}
    & \left(c_{f\bar{f}V}^{---}{}^{\mathbf{I}_3}\right){}^{\hi{1}}_{\hi{2}} = \left(\frac{C_{Le_\mathbbm{C}B_L H^\dagger}\delta^{I_34} \delta^{\hat{i}_1}_{i_4}\delta^4_{\hat{i}_2}}{\Lambda^2} + \frac{C_{Le_\mathbbm{C}W_LH^\dagger}(\tau^{I_3})^{\hat{i}_1}_{i_4}\delta_{\hat{i}_2}^4}{\Lambda^2}\right)O^{\mathbf{I}_3}_{I_3}\Omega_{\hat{i}_1}^{\hi{1}}\Omega^{\hat{i}_2}_{\hi{2}} \mathcal{U}^{i_4 h} v \,,\label{eq:coefffv} \\
& \left( {c_{f\bar{f}V}^{+-0}}^{\mathbf{I}_3} \right)^{\hi{1}}_{\hi{2}} = -\sqrt{2} \left(\frac{1}{2} g (\tau^{I_3})_{\hat{i}_2}^{\hat{i}_1} +g' \delta^{I_3 4} \delta_{\hat{i}_2}^{\hat{i}_1}\right) O^{I_3 \mathbf{I}_3} \Omega_{\hat{i}_1}^{\hi{1}} \Omega^{\hat{i}_2}_{\hi{2}}\,,\\
& \left( {c_{f\bar{f}V}^{+-0}}^{\mathbf{I}_3} \right)^{\hi{1}}_{\hi{2}} = i\frac{Y}{2}\delta^{\hat{i}_1}_{i_3}\delta^{e_{\mathbb{C}}}_{\hat{i}_2} \mathcal{U}^{i_3\mathbf{I}_3} \Omega_{\hat{i}_1}^{\hi{1}} \Omega^{\hat{i}_2}_{\hi{2}}. 
\end{align}
Here the script $4$ appearing in the tensors $\delta^{4}_{\hat{i}_2}$ and $\delta^{I_3 4}$ denotes the right-handed lepton $e_\mathbbm{C}$ and the hypercharge gauge boson $B$, respectively.

\paragraph{\boldsymbol{$f^2V$}}
\begin{align}
    \left(\mathcal{M}_{f^2V}^{\mathbf{I}_3}\right){}^{\hi{1}\hi{2}} = \left(c_{f^2V}^{---}\,^{\mathbf{I}_3}\right){}^{\hi{1}\hi{2}}\lra{\mathbf{1}\mathbb{3}}\lra{\mathbf{2}\mathbb{3}}  + \text{h.c.}\,,\label{eq:cls_f2V}
\end{align}

\begin{align}
    \left(c_{f^2V}^{---}\,^{\mathbf{I}_3}\right){}^{\hi{1}\hi{2}} &= \left(\frac{C_{L^2B_LH^2}\epsilon^{\hat{i}_1 i_4}\epsilon^{\hat{i}_2 i_5}\delta^{I_43}}{\Lambda^3} + \frac{C_{L^2W_LH^2}\epsilon^{\hat{i}_1 i_4}(\tau^{I_3})^{\hat{i}_2 i_5}}{\Lambda^3}\right)\Omega_{\hat{i}_1}^{\hi{1}}\Omega_{\hat{i}_2}^{\hi{2}}O^{\mathbf{I}_3}_{I_3} \mathcal{U}_{i_4}^h\mathcal{U}_{i_5}^h v^2\,.
\end{align}

\paragraph{$\boldsymbol{V^3}$}
\begin{align}  \label{eq:VVV_result}
    \mathcal{M}^{\mathbf{I}_1\mathbf{I}_2\mathbf{I}_3}_{V^3} \left(\mathbf{1,1,1}\right) = & \left(c_{V^3}^{---}\right){}^{\mathbf{I}_1\mathbf{I}_2\mathbf{I}_3}\lra{\mathbb{12}} \lra{\mathbb{13}} \lra{\mathbb{23}}\, \notag \\ 
    &+\frac{1}{\mathbf{m}_1^2 \mathbf{m}_3^2} \left(c_{V^3}^{++0} \right)^{\mathbf{I}_1 \mathbf{I}_2 \mathbf{I}_3} m_1 m_3 [\mathbb{12}] [\mathbb{32}] \langle\mathbb{13}\rangle \\
    &+ \left(c_{V^3}^{--0} \right)^{\mathbf{I}_1 \mathbf{I}_2 \mathbf{I}_3} \left(\frac{1}{\mathbf{m_2}^2} m_2\langle\mathbb{12}\rangle [\mathbb{23}] \langle \mathbb{31}\rangle - \frac{1}{\mathbf{m_1}^2} m_1\langle \mathbb{12}\rangle \langle \mathbb{23} \rangle [\mathbb{31}]  \right)   
    + \text{h.c.}\, ,
\end{align}

\begin{align}
    \left(c_{V^3}^{---}\right){}^{\mathbf{I}_1\mathbf{I}_2\mathbf{I}_3} &= \left(\frac{C_{W_L^3}\epsilon^{I_1I_2I_3}}{\Lambda^2}\right) O^{\mathbf{I}_1}_{I_1} O^{\mathbf{I}_2}_{I_2} O^{\mathbf{I}_3}_{I_3}\,. \\
    \left(c_{V^3}^{++0} \right)^{\mathbf{I}_1 \mathbf{I}_2 \mathbf{I}_3} &= -g \epsilon^{I_1 I_2 I_3} O^{I_1 \mathbf{I}_1} O^{I_2 \mathbf{I}_2} O^{I_3 \mathbf{I}_3}\\
    \left(c_{V^3}^{--0} \right)^{\mathbf{I}_1 \mathbf{I}_2 \mathbf{I}_3} &= \left(\frac{C_{B_L^2HH^\dagger}\delta^{I_1 4}\delta^{I_2 4}\delta^{i_3}_{i_4}}{\Lambda^2} + \frac{C_{B_LW_LHH^\dagger}\delta^{I_1}_4(\tau^{I_2})^{i_3}_{i_4}}{\Lambda^2} + \frac{C_{W_L^2HH^\dagger}\delta^{I_1I_2}\delta^{i_3}_{i_4}}{\Lambda^2}\right) \notag \\
    & \times O^{\mathbf{I}_1}_{I_1} O^{\mathbf{I}_2}_{I_2}\calu^{\mathbf{I}_3}_{i_3} \calu^{i_4h}v \,.
\end{align}

\paragraph{$\boldsymbol{V^2h}$}
\begin{align}
    & \mathcal{M}^{\mathbf{I}_1\mathbf{I}_2h}_{V^2h}\left(\mathbf{1,1,0}\right) =\left(c_{V^2h}^{--0}\,{}^{\mathbf{I}_1\mathbf{I}_2}\right){}^{h} \lra{\mathbb{12}}^2 
    + \frac{1}{\mathbf{m}_1^2}\left(c_{V^2h}^{000}\,{}^{\mathbf{I}_1\mathbf{I}_2}\right){}^{h} m_1 \lra{\mathbb{12}}[\mathbb{12}]+ \text{h.c.}\,,\label{eq:cls_V2h}
\end{align}

\begin{align}
    \left(c_{V^2h}^{--0}\,{}^{\mathbf{I}_1\mathbf{I}_2}\right){}^{h} &= \left(\frac{C_{B_L^2HH^\dagger}\delta^{I_1 4}\delta^{I_2 4}\delta^{\hat{i}_3}_{i_4}}{\Lambda^2} + \frac{C_{B_LW_LHH^\dagger}\delta^{I_1}_4(\tau^{I_2})^{\hat{i}_3}_{i_4}}{\Lambda^2} + \frac{C_{W_L^2HH^\dagger}\delta^{I_1I_2}\delta^{\hat{i}_3}_{i_4}}{\Lambda^2}\right) \notag \\
    & \times O^{\mathbf{I}_1}_{I_1} O^{\mathbf{I}_2}_{I_2}\mathcal{U}^h_{\hat{i}_3} \mathcal{U}^h_{i_4} v\,, \\
    \left(c_{V^2h}^{000}\,{}^{\mathbf{I}_1\mathbf{I}_2}\right){}^{h} &= \sqrt{2}\left( \frac{1}{2}g (\tau^{I_1})^{i_2}_{i_3} + g^\prime \delta^{I_1 4} \delta^{i_2}_{i_3}   \right) O^{\mathbf{I}_1}_{I_1} (\mathcal{U}^{\mathbf{I}_2}_{i_2}  \mathcal{U}^{i_3 h}+\mathcal{U}^{i_3 \mathbf{I}_2}  \mathcal{U}^{h}_{i_2}).
\end{align}

\paragraph{$\boldsymbol{h^3}$}
\begin{align}
    & \mathcal{M}^{hhh}_{h^3}(\mathbf{0,0,0}) = \left(c_{h^3,1}^{000}\right){}^{hhh}s_{12} + \left(c_{h^3,2}^{000}\right){}^{hhh}\,,\label{eq:cls_h3}
\end{align}

\begin{align}
    \left(c_{h^3,1}^{000}\right){}^{hhh} &= \left(\frac{C_{D^2H^2H^\dagger{}^2,1}\delta^{{i}_1}_{i_3}\delta^{{i}_2}_{i_4}}{\Lambda^2} + \frac{C_{D^2H^2H^\dagger{}^2,2}\delta^{{i}_1}_{i_3}\delta^{{i}_2}_{i_4}}{\Lambda^2}\right)\mathcal{U}^{h}_{{i}_1}\mathcal{U}^{h}_{{i}_2}\mathcal{U}^{i_3 h} \mathcal{U}^{i_4 h} v\,,\\
    \left(c_{h^3,2}^{000}\right){}^{hhh} &= \left(\frac{C_{H^3H^\dagger{}^3}}{\Lambda^2} \delta^{{i}_1}_{i_4}\delta^{{i}_2}_{i_5}\delta^{{i}_3}_{i_6}\right)\mathcal{U}^h_{{i}_1}\mathcal{U}^h_{{i}_2}\mathcal{U}^h_{{i}_3}\mathcal{U}^{i_4 h}\mathcal{U}^{i_5 h}\mathcal{U}^{i_6 h}v^3 \notag \\
    & + 4\lambda \delta^{i_1}_{i_3}\delta^{i_2}_{i_4} \mathcal{U}_{i_1}^h \mathcal{U}_{i_2}^h \mathcal{U}^{i_3 h} \mathcal{U}^{i_4h} v\,.
\end{align}

All the massive amplitudes correspond to effective operators due to the massive amplitude-operator correspondence discussed previously. For example, the amplitudes $\lra{\mathbf{12}}$ of the class $f^2h$ corresponds to the operator
\begin{equation}
    \lra{\mathbf{12}} \sim (f_Lf_L)S\,,
\end{equation}
which violates lepton-number conservation. Another example is $\lra{\mathbb{12}}^2$ of the $V^2h$ class, the corresponding operator is
\begin{equation}
    \lra{\mathbb{12}}^2 \sim V_L^{\alpha\beta}V_L{}_{\alpha\beta} S \sim  V^{\mu\nu}V_{\mu\nu} S - V^{\mu\nu}\tilde{V}_{\mu\nu} S\,.
\end{equation}

Taking the numeric forms of the projectors in Eq.~\eqref{eq:trans_Omega}, Eq.~\eqref{eq:trans_U_2}, and Eq.~\eqref{eq:transo}, the Wilson coefficients take explicit values.
Considering $f\overline{f}V$, the Wilson coefficient is shown in Eq.~\eqref{eq:coefffv}. The index $\hi{2}$ is fixed to $e_R$, since the projector $\Omega$ behaves as a $\delta$-symbol. For the index $\hi{1}$, it takes $\nu_L$ or  $e_L$, then the electric-charge conservation determines the index $\mathbf{I}_3=Z,A$ or $\mathbf{I}_3=W^-$, respectively. In particular, as indicated by the projector $O$, the fields $A,Z$ are matched from the gauge fields $W,B$ in a mixing way, while the field $W^\pm$ is matched solely from the gauge field $W$. Thus, the non-vanishing Wilson coefficients are 
\begin{align}
    c_{e_L \bar{e}_R A}^{---}&\equiv \left(c_{f\bar{f}V}^{---}{}^{A}\right){}^{e_L}_{\bar{e}_R} = \frac{v}{\Lambda^2}\left(\cos\theta_W C_{Le_\mathbbm{C}B_L H^\dagger}+ \sin\theta_W C_{Le_\mathbbm{C}W_L H^\dagger}\right)\,,\\
    c_{e_L \bar{e}_R Z}^{---}&\equiv \left(c_{f\bar{f}V}^{---}{}^{Z}\right){}^{e_L}_{\bar{e}_R} = \frac{v}{\Lambda^2}\left(-\sin\theta_W C_{Le_\mathbbm{C}B_L H^\dagger} + \cos\theta_W C_{Le_\mathbbm{C}W_L H^\dagger}\right)\,,\\
    c_{\nu_L \bar{e}_R W^-}^{---}&\equiv \left(c_{f\bar{f}V}^{---}{}^{W^-}\right){}^{\nu_L}_{\bar{e}_R} = \frac{v}{\sqrt{2}\Lambda^2}C_{Le_\mathbbm{C}W_L H^\dagger}\,.
\end{align}
\begin{align}
    c_{e_L \bar{\nu}_L W^+}^{+-0} \equiv \left(c_{f\bar{f}V}^{+-0}{}^{W^+}\right){}^{e_L}_{\overline{\nu}_L} &= -\frac{1-i}{\sqrt{2}}g\,, \\
    c_{e_L \bar{e}_L Z}^{+-0} \equiv \left(c_{f\bar{f}V}^{+-0}{}^{Z}\right){}^{e_L}_{\overline{e}_L} &= -\sqrt{2}(\frac{g}{2}\cos\theta_W - g'\sin\theta_W)\,, \\
    c_{\nu_L \bar{\nu}_LZ}^{+-0} \equiv \left(c_{f\bar{f}V}^{+-0}{}^{Z}\right){}^{\nu_L}_{\overline{\nu}_L} &= \sqrt{2}(\frac{g}{2}\cos\theta_W + g'\sin\theta_W)\,, \\
    c_{e_L \bar{e}_L A}^{+-0} \equiv \left(c_{f\bar{f}V}^{+-0}{}^{A}\right){}^{e_L}_{\overline{e}_L} &= -\sqrt{2}(-\frac{g}{2}\sin\theta_W + g'\cos\theta_W)\,, \\
    c_{\nu_L \bar{\nu}_L A}^{+-0} \equiv \left(c_{f\bar{f}V}^{+-0}{}^{A}\right){}^{\nu_L}_{\overline{\nu}_L} &= -\sqrt{2}(\frac{g}{2}\sin\theta_W + g'\cos\theta_W)\,,\\
    c_{e_L \bar{e}_R Z}^{+-0} \equiv \left(c_{f\bar{f}V}^{+-0}{}^{Z}\right){}^{e_L}_{\overline{e}_R} &= i\frac{Y}{2}\,.
\end{align}
For completeness, we present all the non-vanishing Wilson coefficients of the 3-point amplitudes.

\subsubsection*{Explicit Wilson Coefficients of 3-point Matching}

\paragraph{$\boldsymbol{f^2h}$}

\begin{align}
    \left(c^{--0}_{f^2h}\right){}^{\nu_L\nu_Lh} = \frac{C_{L^2H^2}}{\sqrt{2}\Lambda} v\,,
\end{align}

\paragraph{\boldsymbol{$f\bar{f}V$}}

\begin{align}
    \left(c_{f\bar{f}V}^{---}{}^{A}\right){}^{e_R}_{\bar{e}_R} &= \frac{v}{\Lambda^2}\left(\cos\theta_W C_{Le_\mathbbm{C}B_L H^\dagger}+ \sin\theta_W C_{Le_\mathbbm{C}W_L H^\dagger}\right)\,,\\
    \left(c_{f\bar{f}V}^{---}{}^{Z}\right){}^{e_R}_{\bar{e}_R} &= \frac{v}{\Lambda^2}\left(-\sin\theta_W C_{Le_\mathbbm{C}B_L H^\dagger} + \cos\theta_W C_{Le_\mathbbm{C}W_L H^\dagger}\right)\,,\\
    \left(c_{f\bar{f}V}^{---}{}^{W^-}\right){}^{\nu_L}_{\bar{e}_R} &= \frac{v}{\sqrt{2}\Lambda^2}C_{Le_\mathbbm{C}W_L H^\dagger}\,,
\end{align}
\begin{align}
    \left(c_{f\bar{f}V}^{+-0}{}^{W^+}\right){}^{e_L}_{\overline{\nu}_L} &= -\frac{1-i}{\sqrt{2}}g\,, \\
    \left(c_{f\bar{f}V}^{+-0}{}^{Z}\right){}^{e_L}_{\overline{e}_L} &= -\sqrt{2}(\frac{g}{2}\cos\theta_W - g'\sin\theta_W)\,, \\
    \left(c_{f\bar{f}V}^{+-0}{}^{Z}\right){}^{\nu_L}_{\overline{\nu}_L} &= \sqrt{2}(\frac{g}{2}\cos\theta_W + g'\sin\theta_W)\,, \\
    \left(c_{f\bar{f}V}^{+-0}{}^{A}\right){}^{e_L}_{\overline{e}_L} &= -\sqrt{2}(-\frac{g}{2}\sin\theta_W + g'\cos\theta_W)\,, \\
    \left(c_{f\bar{f}V}^{+-0}{}^{A}\right){}^{\nu_L}_{\overline{\nu}_L} &= -\sqrt{2}(\frac{g}{2}\sin\theta_W + g'\cos\theta_W)\,,\\
    \left(c_{f\bar{f}V}^{+-0}{}^{Z}\right){}^{e_L}_{\overline{e}_R} &= i\frac{Y}{2}\,.
\end{align}

\paragraph{\boldsymbol{$f^2V$}}

\begin{align}
    \left(c_{f^2V}^{---}\,^{A}\right){}^{\nu_L\nu_L} &= \frac{v^2}{\Lambda^3}\left(\cos\theta_W C_{L^2B_LH^2} + \sin\theta_W C_{L^2W_LH^2}\right)\,,\\
    \left(c_{f^2V}^{---}\,^{Z}\right){}^{\nu_L\nu_L} &= \frac{v^2}{\Lambda^3}\left(-\sin\theta_W C_{L^2B_LH^2} + \cos\theta_W C_{L^2W_LH^2}\right)\,,\\
    \left(c_{f^2V}^{---}\,^{W^+}\right){}^{\nu_L e_L} &= \frac{v^2}{\sqrt{2}\Lambda^3} C_{L^2W_LH^2}\,.
\end{align}

\paragraph{$\boldsymbol{V^3}$}

\begin{align}
    \left(c_{V^3}^{---}\right){}^{W^+W^-Z} &= \frac{C_{W_L^3}}{2\Lambda^2}\cos\theta_W\,,\\
    \left(c_{V^3}^{---}\right){}^{W^+W^-A} &= \frac{C_{W_L^3}}{2\Lambda^2}\sin\theta_W\,,
\end{align}
\begin{align}
    \left(c_{V^3}^{--0}\right){}^{AAZ} &= -\frac{v}{\sqrt{2}\Lambda^2}\left(\cos^2\theta_W C_{B_L^2HH^\dagger} - \sin\theta_W\cos\theta_W C_{B_LW_LHH^\dagger}\right)\,,\\
    \left(c_{V^3}^{--0}\right){}^{ZZZ} &= -\frac{v}{\sqrt{2}\Lambda^2}\left(\sin^2\theta_W C_{B_L^2HH^\dagger} + \sin\theta_W\cos\theta_W C_{B_LW_LHH^\dagger}\right)\,,\\
    \left(c_{V^3}^{--0}\right){}^{AZZ} &= -\frac{v}{\sqrt{2}\Lambda^2}\left(-\sin\theta_W\cos\theta_W C_{B_L^2HH^\dagger} - \cos^2\theta_W C_{B_LW_LHH^\dagger} + \sin\theta_W\cos\theta_W C_{W_L^2HH^\dagger}\right)\,,\\
    \left(c_{V^3}^{--0}\right){}^{W^+W^-Z} &= -\frac{v}{\sqrt{2}\Lambda^2}C_{W_L^2HH^\dagger}\,,\\
    \left(c_{V^3}^{--0}\right){}^{AW^-W^+} &= -\frac{\sqrt{2}v}{\Lambda^2}\cos\theta_W C_{B_LW_LHH^\dagger}\,,\\
    \left(c_{V^3}^{--0}\right){}^{ZW^-W^+} &= \frac{\sqrt{2}v}{\Lambda^2}\sin\theta_W C_{B_LW_LHH^\dagger}\,.
\end{align}

\paragraph{$\boldsymbol{V^2h}$}
\begin{align}
     \left(c_{V^2h}^{--0}\,{}^{AA}\right){}^{h} &= \frac{v}{2\Lambda^2}\left(C_{B_L^2HH^\dagger}\cos^2\theta_W +C_{B_LW_LHH^\dagger}\cos\theta_W \sin\theta_W + C_{W_L^2HH^\dagger} \sin\theta_W^2  \right)\,,\\
     \left(c_{V^2h}^{--0}\,{}^{ZZ}\right){}^{h} &= \frac{v}{2\Lambda^2}\left(C_{B_L^2HH^\dagger}\sin^2\theta_W -C_{B_LW_LHH^\dagger}\cos\theta_W \sin\theta_W + C_{W_L^2HH^\dagger} \cos\theta_W^2  \right)\,,\\
     \left(c_{V^2h}^{--0}\,{}^{W^+W^-}\right){}^{h} &= \frac{v}{2\Lambda^2}C_{W_L^2HH^\dagger}\,.
\end{align}

\paragraph{$\boldsymbol{h^3}$}

\begin{align}
    \left(c_{h^3,1}^{000}\right){}^{hhh} &= \frac{v}{4\Lambda^2}\left(C_{D^2H^2H^\dagger{}^2,1} + C_{D^2H^2H^\dagger{}^2,2}\right)\,,\\
    \left(c_{h^3,2}^{000}\right){}^{hhh} &= \frac{v^3}{8\Lambda^2}C_{H^3H^\dagger{}^3}\,.
\end{align}

\subsection{4-point Matching Result}

\paragraph{\boldsymbol{$f^2\bar{f}{}^2$}}
\begin{align}
    \left(\mathcal{M}_{f^2\bar{f}{}^2}\right){}^{\hi{1}\hi{2}}_{\hi{3}\hi{4}} \left(\mathbf{\frac{1}{2},\frac{1}{2},\frac{1}{2},\frac{1}{2}}\right) &= \left(c_{f^2\bar{f}{}^2,1}^{--++}\right){}^{\hi{1}\hi{2}}_{\hi{3}\hi{4}}\lra{\mathbf{12}}[\mathbf{34}] + \left(c_{f^2\bar{f}{}^2,1}^{++--}\right){}^{\hi{1}\hi{2}}_{\hi{3}\hi{4}}[\mathbf{12}]\lra{\mathbf{34}} +  \left(c_{f^2\bar{f}{}^2,1}^{-+-+}\right){}^{\hi{1}\hi{2}}_{\hi{3}\hi{4}}\lra{\mathbf{13}}[\mathbf{24}]\notag \\
    & + \left(c_{f^2\bar{f}{}^2,2}^{--++}\right){}^{\hi{1}\hi{2}}_{\hi{3}\hi{4}}\lra{\mathbf{12}}[\mathbf{34}]s_{34} + \left(c_{f^2\bar{f}{}^2,3}^{--++}\right){}^{\hi{1}\hi{2}}_{\hi{3}\hi{4}}\lra{\mathbf{1|p_3p_4|2}}[\mathbf{34}]\notag \\
    & + \left(c_{f^2\bar{f}{}^2,2}^{++--}\right){}^{\hi{1}\hi{2}}_{\hi{3}\hi{4}}[\mathbf{12}]\lra{\mathbf{34}}s_{34} + \left(c_{f^2\bar{f}{}^2,3}^{++--}\right){}^{\hi{1}\hi{2}}_{\hi{3}\hi{4}}[\mathbf{1|p_3p_4|2}]\lra{\mathbf{34}}\notag \\
    & + \left(c_{f^2\bar{f}{}^2,2}^{-+-+}\right){}^{\hi{1}\hi{2}}_{\hi{3}\hi{4}}\lra{\mathbf{13}}[\mathbf{24}]s_{34} + \left(c_{f^2\bar{f}{}^2,3}^{-+-+}\right){}^{\hi{1}\hi{2}}_{\hi{3}\hi{4}}\lra{\mathbf{1|p_3p_4|3}}[\mathbf{24}]\notag \\
    & + \left(c_{f^2\bar{f}{}^2,1}^{-+--}\right){}_{\hi{3}\hi{4}}^{\hi{1}\hi{2}} \lra{\mathbf{14}}[\mathbf{2|p_4|3}\rangle + \left(c_{f^2\bar{f}{}^2,2}^{--+-}\right){}_{\hi{3}\hi{4}}^{\hi{1}\hi{2}} \lra{\mathbf{14}}[\mathbf{3|p_4|2}\rangle \notag \\
    & + \left(c_{f^2\bar{f}{}^2,1}^{----}\right){}^{\hi{1}\hi{2}}_{\hi{3}\hi{4}} \lra{\mathbf{12}}\lra{\mathbf{34}} + \left(c_{f^2\bar{f}{}^2,2}^{----}\right){}^{\hi{1}\hi{2}}_{\hi{3}\hi{4}} \lra{\mathbf{13}}\lra{\mathbf{24}} + \text{h.c.}\,,\label{eq:cls_f2bar2}
\end{align}

\begin{align}
    \left(c_{f^2\bar{f}{}^2,1}^{--++}\right){}^{\hi{1}\hi{2}}_{\hi{3}\hi{4}} &= \left(\frac{C_{L^2L^\dagger{}^2}\delta^{\hat{i}_1}_{\hat{i}_3}\delta^{\hat{i}_2}_{\hat{i}_4}}{\Lambda^2}\right)\Omega_{\hat{i}_1}^{\hi{1}}\Omega_{\hat{i}_2}^{\hi{2}}\Omega^{\hat{i}_3}_{\hi{3}}\Omega^{\hat{i}_4}_{\hi{4}}\,,\\
    \left(c_{f^2\bar{f}{}^2,1}^{++--}\right){}^{\hi{1}\hi{2}}_{\hi{3}\hi{4}} &= \left(\frac{C_{e_\mathbbm{C}^2e_\mathbbm{C}^\dagger{}^2}\delta^{\hat{i}_1}_4\delta^{\hat{i}_2}_4\delta_{\hat{i}_3}^4\delta_{\hat{i}_4}^4}{\Lambda^2} \right)\Omega_{\hat{i}_1}^{\hi{1}}\Omega_{\hat{i}_2}^{\hi{2}}\Omega^{\hat{i}_3}_{\hi{3}}\Omega^{\hat{i}_4}_{\hi{4}}\,,\\
    \left(c_{f^2\bar{f}{}^2,1}^{-+-+}\right){}^{\hi{1}\hi{2}}_{\hi{3}\hi{4}} &= \left(\frac{C_{Le_\mathbbm{C}^\dagger e_\mathbbm{C} L^\dagger}\delta^{\hat{i}_1}_{\hat{i}_4}\delta^{\hat{i}_2}_4\delta_{\hat{i}_3}^4}{\Lambda^2}\right)\Omega_{\hat{i}_1}^{\hi{1}}\Omega_{\hat{i}_2}^{\hi{2}}\Omega^{\hat{i}_3}_{\hi{3}}\Omega^{\hat{i}_4}_{\hi{4}}\,,\\
    \left(c_{f^2\bar{f}{}^2,2}^{--++}\right){}^{\hi{1}\hi{2}}_{\hi{3}\hi{4}} &= \left(\frac{C_{D^2L^2L^\dagger{}^2,1}\delta^{\hat{i}_1}_{\hat{i}_3}\delta^{\hat{i}_2}_{\hat{i}_4}}{\Lambda^4}\right)\Omega_{\hat{i}_1}^{\hi{1}}\Omega_{\hat{i}_2}^{\hi{2}}\Omega^{\hat{i}_3}_{\hi{3}}\Omega^{\hat{i}_4}_{\hi{4}}\,,\\
    \left(c_{f^2\bar{f}{}^2,2}^{++--}\right){}^{\hi{1}\hi{2}}_{\hi{3}\hi{4}} &= \left(\frac{C_{D^2e_\mathbbm{C}^2e_\mathbbm{C}^\dagger{}^2,1}\delta^{\hat{i}_1}_4\delta^{\hat{i}_2}_4\delta_{\hat{i}_3}^4\delta_{\hat{i}_4}^4}{\Lambda^4} \right)\Omega_{\hat{i}_1}^{\hi{1}}\Omega_{\hat{i}_2}^{\hi{2}}\Omega^{\hat{i}_3}_{\hi{3}}\Omega^{\hat{i}_4}_{\hi{4}}\,,\\
    \left(c_{f^2\bar{f}{}^2,2}^{-+-+}\right){}^{\hi{1}\hi{2}}_{\hi{3}\hi{4}} &= \left(\frac{C_{D^2Le_\mathbbm{C}^\dagger e_\mathbbm{C} L^\dagger,1}\delta^{\hat{i}_1}_{\hat{i}_4}\delta^{\hat{i}_2}_4\delta_{\hat{i}_3}^4}{\Lambda^4}\right)\Omega_{\hat{i}_1}^{\hi{1}}\Omega_{\hat{i}_2}^{\hi{2}}\Omega^{\hat{i}_3}_{\hi{3}}\Omega^{\hat{i}_4}_{\hi{4}}\,,\\
    \left(c_{f^2\bar{f}{}^2,3}^{--++}\right){}^{\hi{1}\hi{2}}_{\hi{3}\hi{4}} &= \left(\frac{C_{D^2L^2L^\dagger{}^2,2}\delta^{\hat{i}_1}_{\hat{i}_3}\delta^{\hat{i}_2}_{\hat{i}_4}}{\Lambda^4}\right)\Omega_{\hat{i}_1}^{\hi{1}}\Omega_{\hat{i}_2}^{\hi{2}}\Omega^{\hat{i}_3}_{\hi{3}}\Omega^{\hat{i}_4}_{\hi{4}}\,,\\
    \left(c_{f^2\bar{f}{}^2,3}^{++--}\right){}^{\hi{1}\hi{2}}_{\hi{3}\hi{4}} &= \left(\frac{C_{D^2e_\mathbbm{C}^2e_\mathbbm{C}^\dagger{}^2,2}\delta^{\hat{i}_1}_4\delta^{\hat{i}_2}_4\delta_{\hat{i}_3}^4\delta_{\hat{i}_4}^4}{\Lambda^4} \right)\Omega_{\hat{i}_1}^{\hi{1}}\Omega_{\hat{i}_2}^{\hi{2}}\Omega^{\hat{i}_3}_{\hi{3}}\Omega^{\hat{i}_4}_{\hi{4}}\,,\\
    \left(c_{f^2\bar{f}{}^2,3}^{-+-+}\right){}^{\hi{1}\hi{2}}_{\hi{3}\hi{4}} &= \left(\frac{C_{D^2Le_\mathbbm{C}^\dagger e_\mathbbm{C} L^\dagger,2}\delta^{\hat{i}_1}_{\hat{i}_4}\delta^{\hat{i}_2}_4\delta_{\hat{i}_3}^4}{\Lambda^4}\right)\Omega_{\hat{i}_1}^{\hi{1}}\Omega_{\hat{i}_2}^{\hi{2}}\Omega^{\hat{i}_3}_{\hi{3}}\Omega^{\hat{i}_4}_{\hi{4}}\,,\\
    \left(c_{f^2\bar{f}{}^2,1}^{-+--}\right){}_{\hi{3}\hi{4}}^{\hi{1}\hi{2}} &= \left(\frac{C_{DLe_\mathbbm{C}^\dagger e_\mathbbm{C}^2H^\dagger,2}\delta^{\hat{i}_1}_{i_5}\delta^{\hat{i}_2}_4\delta_{\hat{i}_3}^4\delta_{\hat{i}_4}^4}{\Lambda^4}\right)\Omega_{\hat{i}_1}^{\hi{1}}\Omega_{\hat{i}_2}^{\hi{2}} \Omega^{\hat{i}_3}_{\hi{3}}\Omega^{\hat{i}_4}_{\hi{4}}\mathcal{U}^{i_5h}v\,,\\
    \left(c_{f^2\bar{f}{}^2,2}^{-+--}\right){}_{\hi{3}\hi{4}}^{\hi{1}\hi{2}} &= \left(\frac{C_{DL^2L^\dagger e_\mathbbm{C} H^\dagger,2}\delta^{\hat{i}_1}_{\hat{i}_3}\delta^{\hat{i}_2}_{i_5}\delta_{\hat{i}_4}^4} {\Lambda^4}\right)\Omega_{\hat{i}_1}^{\hi{1}}\Omega_{\hat{i}_2}^{\hi{2}} \Omega^{\hat{i}_3}_{\hi{3}}\Omega^{\hat{i}_4}_{\hi{4}}\mathcal{U}^{i_5h}v\,,\\
    \left(c_{f^2\bar{f}{}^2,1}^{----}\right){}^{\hi{1}\hi{2}}_{\hi{3}\hi{4}} &= \left(\frac{C_{L^2e_\mathbbm{C}^2H^\dagger{}^2,1}\delta^{\hat{i}_1}_{i_5}\delta^{\hat{i}_2}_{i_6}\delta^4_{\hat{i}_3}\delta^4_{\hat{i}_4}}{\Lambda^4}\right)\Omega_{\hat{i}_1}^{\hi{1}}\Omega_{\hat{i}_2}^{\hi{2}}\Omega_{\hi{3}}^{\hat{i}_3}\Omega_{\hi{4}}^{\hat{i}_4} \mathcal{U}^{i_5 h}\mathcal{U}^{i_6 h}v^2\,,\\
    \left(c_{f^2\bar{f}{}^2,2}^{----}\right){}^{\hi{1}\hi{2}}_{\hi{3}\hi{4}} &= \left(\frac{C_{L^2e_\mathbbm{C}^2H^\dagger{}^2,2}\delta^{\hat{i}_1}_{i_5}\delta^{\hat{i}_2}_{i_6}\delta^4_{\hat{i}_3}\delta^4_{\hat{i}_4}}{\Lambda^4}\right)\Omega_{\hat{i}_1}^{\hi{1}}\Omega_{\hat{i}_2}^{\hi{2}}\Omega_{\hi{3}}^{\hat{i}_3}\Omega_{\hi{4}}^{\hat{i}_4} \mathcal{U}^{i_5 h}\mathcal{U}^{i_6 h}v^2\,.
\end{align}

\paragraph{\boldsymbol{$f^3\bar{f}$}}
\begin{align}
    \left(\mathcal{M}_{f^3\bar{f}}\right){}_{\hi{4}}^{\hi{1}\hi{2}\hi{3}}\left(\mathbf{\frac{1}{2},\frac{1}{2},\frac{1}{2},\frac{1}{2}}\right) = 
    \left( c_{f^3\bar{f}}^{----}\right){}_{\hi{4}}^{\hi{1}\hi{2}\hi{3}} \lra{\mathbf{12}}\lra{\mathbf{34}} + \text{h.c.}\,, \label{eq:cls_f3bar}
\end{align}

\begin{align}
    \left( c_{f^3\bar{f}}^{----}\right){}_{\hi{4}}^{\hi{1}\hi{2}\hi{3}}&= 
    \left(\frac{C_{L^3e_\mathbbm{C}H}\epsilon^{\hat{i}_1 i_5}\epsilon^{\hat{i}_2 \hat{i}_3}\delta^{\hat{i}_4}_4}{\Lambda^3}\right)\Omega_{\hat{i}_1}^{\hi{1}}\Omega_{\hat{i}_2}^{\hi{2}}\Omega_{\hat{i}_3}^{\hi{3}}\Omega_{\hat{i}_4}^{\hi{4}}\mathcal{U}_{i_5}^h v\,.
\end{align}

\paragraph{$\boldsymbol{f^2h^2}$}
\begin{align}
    \mathcal{M}^{\hi{1}\hi{2}hh}_{f^2h^2}\left(\mathbf{\frac{1}{2},\frac{1}{2},0,0}\right) &= 
    \left(c^{--00}_{f^2h^2,1}\right){}^{\hi{1}\hi{2}hh}\lra{\mathbf{12}} 
    + \left(c^{--00}_{f^2h^2,2}\right){}^{\hi{1}\hi{2}hh}\lra{\mathbf{12}} s_{34} 
    & + \left(c^{--00}_{f^2h^2,3}\right){}^{\hi{1}\hi{2}hh}\lra{\mathbf{1|p_3p_4|2}}  \notag \\
    & + \left(c^{-+00}_{f^2h^2,1}\right){}^{\hi{1}\hi{2}hh} \langle\mathbf{1|p_3|2}]+ \text{h.c.}\,,\label{eq:cls_f2h2}
\end{align}

\begin{align}
    \left(c^{--00}_{f^2h^2,1}\right){}^{\hi{1}\hi{2}hh} &= \left(\frac{C_{L^2H^2}\epsilon^{\hat{i}_1 i_3}\epsilon^{\hat{i}_2 i_4}}{\Lambda}\right)\Omega_{\hat{i}_1}^{\hi{1}}\Omega_{\hat{i}_2}^{\hi{2}}\mathcal{U}^{h}_{i_3} \mathcal{U}^{h}_{i_4}\,,\\
    \left(c^{--00}_{f^2h^2,2}\right){}^{\hi{1}\hi{2}hh} &= \left(\frac{C_{D^2L^2H^2,1}\epsilon^{\hat{i}_1 i_3}\epsilon^{\hat{i}_2 i_4}}{\Lambda^3}\right)\Omega_{\hat{i}_1}^{\hi{1}}\Omega_{\hat{i}_2}^{\hi{2}}\mathcal{U}^{h}_{i_3} \mathcal{U}^{h}_{i_4}\,,\\
    \left(c^{--00}_{f^2h^2,3}\right){}^{\hi{1}\hi{2}hh} &= \left(\frac{C_{D^2L^2H^2,2}\epsilon^{\hat{i}_1 i_3}\epsilon^{\hat{i}_2 i_4}}{\Lambda^3}\right)\Omega_{\hat{i}_1}^{\hi{1}}\Omega_{\hat{i}_2}^{\hi{2}}\mathcal{U}^{h}_{i_3} \mathcal{U}^{h}_{i_4}\,,\\
    \left(c^{-+00}_{f^2h^2,1}\right){}^{\hi{1}\hi{2}hh} &= \left(\frac{C_{DLe_\mathbbm{C}^\dagger H^3}\epsilon^{\hat{i}_1 i_5}\epsilon^{i_3 i_4}\delta^{\hat{i}_2}_4}{\Lambda^3}\right)\Omega_{\hat{i}_1}^{\hi{1}}\Omega_{\hat{i}_2}^{\hi{2}}\mathcal{U}^{h}_{i_3} \mathcal{U}^{h}_{i_4}\mathcal{U}^{h}_{i_5} v\,.
\end{align}

\paragraph{$\boldsymbol{f\bar{f}h^2}$}
\begin{align}
    \left(\mathcal{M}_{f\bar{f}h^2}\right){}^{\hi{1}hh}_{\hi{2}}\left(\mathbf{\frac{1}{2},\frac{1}{2},0,0}\right) &=  \left(c_{f\bar{f}h^2,1}^{+-00}\right){}_{\hi{2}}^{\hi{1}hh} [\mathbf{1|p_3|2}\rangle + \left(c_{f\bar{f}h^2,1}^{-+00}\right){}_{\hi{2}}^{\hi{1}hh} [\mathbf{2|p_3|1}\rangle + \left(c_{f^2h^2,1}^{--00}\right){}_{\hi{2}}^{\hi{1}hh}\lra{\mathbf{12}} \notag \\
    & 
    + \left(c^{-+00}_{f\bar{f}h^2,2}\right){}^{\hi{1}hh}_{\hi{2}}\langle\mathbf{1|p_4|2}]s_{34} + \left(c^{-+00}_{f\bar{f}h^2,3}\right){}^{\hi{1}hh}_{\hi{2}}\langle\mathbf{1|p_4|2}]s_{24} \notag \\
    & + \left(c^{+-00}_{f\bar{f}h^2,2}\right){}^{\hi{1}hh}_{\hi{2}}\langle\mathbf{2|p_4|1}]s_{34} + \left(c^{+-00}_{f\bar{f}h^2,3}\right){}^{\hi{1}hh}_{\hi{2}}\langle\mathbf{2|p_4|1}]s_{24}\notag \\
    & + \left(c_{f\bar{f}h^2,2}^{--00}\right){}_{\hi{2}}^{\hi{1}hh}\lra{\mathbf{12}}s_{34} + \left(c_{f\bar{f}h^2,3}^{--00}\right){}_{\hi{2}}^{\hi{1}hh}\lra{\mathbf{1|p_3p_4|2}}
+ \text{h.c.}\,, \label{eq:cls_fbarh2}
\end{align}

\begin{align}
    \left(c_{f\bar{f}h^2,1}^{+-00}\right){}_{\hi{2}}^{\hi{1}hh} &= \left(\frac{C_{De_\mathbbm{C}^\dagger e_\mathbbm{C} HH^\dagger}\delta^{\hat{i}_1}_4\delta_{\hat{i}_2}^4 \delta^{i_3}_{i_4}}{\Lambda^2}\right) \Omega_{\hat{i}_1}^{\hi{1}}\Omega_{\hi{2}}^{\hat{i}_2}\mathcal{U}_{i_3}^{h} \mathcal{U}^{i_4h}\,,\\
    \left(c_{f\bar{f}h^2,1}^{+-00}\right){}_{\hi{2}}^{\hi{1}hh} &= \left(\frac{C_{DL^\dagger L HH^\dagger,1}\delta^{\hat{i}_1}_{\hat{i}_2}\delta^{i_3}_{i_4}}{\Lambda^2} + \frac{C_{DL^\dagger L HH^\dagger,2}\delta^{\hat{i}_1}_{i_4}\delta_{\hat{i}_2}^{i_3}}{\Lambda^2}\right) \Omega_{\hat{i}_1}^{\hi{1}}\Omega_{\hi{2}}^{\hat{i}_2}\mathcal{U}_{i_3}^{h} \mathcal{U}^{i_4h}\,,\\
    \left(c_{f^2h^2,1}^{--00}\right){}_{\hi{2}}^{\hi{1}hh} &=  \left(\frac{C_{Le_\mathbbm{C}HH^\dagger{}^2}\delta^{\hat{i}_1}_{i_4}\delta^{i_3}_{i_5}\delta_{\hat{i}_2}^4}{\Lambda^2}\right)\Omega_{\hat{i}_1}^{\hi{1}}\Omega_{\hi{2}}^{\hat{i}_2}\mathcal{U}_{i_3}^{h} \mathcal{U}^{i_4h}\mathcal{U}^{i_5h} v\,, \\
    \left(c^{-+00}_{f\bar{f}h^2,2}\right){}^{\hi{1}hh}_{\hi{2}} &= \left(\frac{C_{D^3LL^\dagger HH^\dagger,1}\delta^{\hat{i}_1}_{\hat{i}_2}\delta^{i_3}_{i_4}}{\Lambda^4} + \frac{C_{D^3LL^\dagger HH^\dagger,2}\delta^{\hat{i}_1}_{i_4}\delta^{i_3}_{\hat{i}_2}}{\Lambda^4} \right)\Omega_{\hat{i}_1}^{\hi{1}}\Omega^{\hat{i}_2}_{\hi{2}}\mathcal{U}_{i_3}^h\mathcal{U}^{i_4h}\,,\\
    \left(c^{-+00}_{f\bar{f}h^2,3}\right){}^{\hi{1}hh}_{\hi{2}} &= \left(\frac{C_{D^3LL^\dagger HH^\dagger,3}\delta^{\hat{i}_1}_{\hat{i}_2}\delta^{i_3}_{i_4}}{\Lambda^4} + \frac{C_{D^3LL^\dagger HH^\dagger,4}\delta^{\hat{i}_1}_{i_4}\delta^{i_3}_{\hat{i}_2}}{\Lambda^4} \right)\Omega_{\hat{i}_1}^{\hi{1}}\Omega^{\hat{i}_2}_{\hi{2}}\mathcal{U}_{i_3}^h\mathcal{U}^{i_4h}\,,\\
    \left(c^{+-00}_{f\bar{f}h^2,2}\right){}^{\hi{1}hh}_{\hi{2}} &= \left(\frac{C_{D^3e_\mathbbm{C}^\dagger e_\mathbbm{C} HH^\dagger,1}\delta^{\hat{i}_1}_{\hat{i}_2}\delta^{i_3}_{i_4}}{\Lambda^4} \right)\Omega_{\hat{i}_1}^{\hi{1}}\Omega^{\hat{i}_2}_{\hi{2}}\mathcal{U}_{i_3}^h\mathcal{U}^{i_4h}\,,\\
    \left(c^{+-00}_{f\bar{f}h^2,3}\right){}^{\hi{1}hh}_{\hi{2}} &= \left(\frac{C_{D^3e_\mathbbm{C}^\dagger e_\mathbbm{C} HH^\dagger,2}\delta^{\hat{i}_1}_{i_4}\delta^{i_3}_{\hat{i}_2}}{\Lambda^4} \right)\Omega_{\hat{i}_1}^{\hi{1}}\Omega^{\hat{i}_2}_{\hi{2}}\mathcal{U}_{i_3}^h\mathcal{U}^{i_4h}\,,\\
    \left(c_{f^2h^2,2}^{--00}\right){}_{\hi{2}}^{\hi{1}hh} &= \left(\frac{C_{D^2Le_\mathbbm{C}HH^\dagger{}^2,3}\delta^{\hat{i}_1}_{i_4}\delta^{i_3}_{i_5}\delta_{\hat{i}_2}^4}{\Lambda^4}\right)\Omega_{\hat{i}_1}^{\hi{1}}\Omega_{\hi{2}}^{\hat{i}_2}\mathcal{U}_{i_3}^{h} \mathcal{U}^{i_4h} \mathcal{U}^{i_5h}v\,,\\
    \left(c_{f^2h^2,3}^{--00}\right){}_{\hi{2}}^{\hi{1}hh} &= \left(\frac{C_{D^2Le_\mathbbm{C}HH^\dagger{}^2,6}\delta^{\hat{i}_1}_{i_4}\delta^{i_3}_{i_5}\delta_{\hat{i}_2}^4}{\Lambda^4}\right)\Omega_{\hat{i}_1}^{\hi{1}}\Omega_{\hi{2}}^{\hat{i}_2}\mathcal{U}_{i_3}^{h} \mathcal{U}^{i_4h} \mathcal{U}^{i_5h}v\,.
\end{align}

\paragraph{$\boldsymbol{f\bar{f}Vh}$}
\begin{align}
    \left(\mathcal{M}_{f\bar{f}Vh}^{\mathbf{I}_3}\right){}_{\hi{2}}^{\hi{1}h}\left(\mathbf{\frac{1}{2},\frac{1}{2},1,0}\right) &= 
    \left(c_{f\bar{f}Vh,1}^{---0}\,{}^{\mathbf{I}_3}\right){}^{\hi{1}h}_{\hi{2}} \lra{\mathbf{1}\mathbb{3}}\lra{\mathbf{2}\mathbb{3}}
    +\left(c_{f\bar{f}Vh,2}^{---0}\,{}^{\mathbf{I}_3}\right){}^{\hi{1}h}_{\hi{2}} \lra{\mathbf{1}\mathbb{3}}\lra{\mathbf{2}\mathbb{3}}s_{24}  \notag \\
    & + \left(c_{f\bar{f}Vh,3}^{---0}\,{}^{\mathbf{I}_3}\right){}^{\hi{1}h}_{\hi{2}} \lra{\mathbf{1}\mathbb{3}}\lra{\mathbf{2}\mathbb{3}}s_{34}+ \left(c_{f\bar{f}Vh,1}^{+-00}\,{}^{\mathbf{I}_3}\right){}_{\hi{2}}^{\hi{1}h} [\mathbf{13}]\lra{\mathbf{23}} \notag \\
    & + \left(c_{f\bar{f}Vh,1}^{-+00}\,{}^{\mathbf{I}_3}\right){}_{\hi{2}}^{\hi{1}h} \lra{\mathbf{13}}[\mathbf{23}] + \left(c_{f\bar{f}Vh}^{--+0}\,{}^{\mathbf{I}_3}\right){}_{\hi{2}}^{\hi{1}h} \langle\mathbf{1|p_2|}\mathbb{3}]\langle\mathbf{2|p_4|}\mathbb{3}] \notag \\
    &
    + \left(c^{-+00}_{f\bar{f}Vh,2}\,{}^{\mathbf{I}_3}\right){}^{\hi{1}h}_{\hi{2}}\langle\mathbf{1|p_4|2}]\langle\mathbf{3|p_4|3 }]+ \left(c^{-+00}_{f\bar{f}Vh,3}\,{}^{\mathbf{I}_3}\right){}^{\hi{1}h}_{\hi{2}}\langle\mathbf{1|p_4|2}]\langle\mathbf{3|p_1|3 }] \notag \\
    & + \left(c^{+-00}_{f\bar{f}Vh,2}\,{}^{\mathbf{I}_3}\right){}^{\hi{1}h}_{\hi{2}}\langle\mathbf{2|p_4|1}]\langle\mathbf{3|p_4|3 }]+ \left(c^{+-00}_{f\bar{f}Vh,3}\,{}^{\mathbf{I}_3}\right){}^{\hi{1}h}_{\hi{2}}\langle\mathbf{2|p_4|1}]\langle\mathbf{3|p_1|3 }]\notag \\
    & +\left(c_{f\bar{f}Vh,1}^{+--0}\,^{\mathbf{I}_3}\right){}^{\hi{1}h}_{\hi{2}}[\mathbf{1|p_1|}\mathbb{3}\rangle \lra{\mathbf{2}\mathbb{3}} +\left(c_{f\bar{f}Vh,2}^{+--0}\,^{\mathbf{I}_3}\right){}^{\hi{1}h}_{\hi{2}}[\mathbf{1|p_4|}\mathbb{3}\rangle \lra{\mathbf{2}\mathbb{3}}\notag \\
    & +\left(c_{f\bar{f}Vh,1}^{-+-0}\,^{\mathbf{I}_3}\right){}^{\hi{1}h}_{\hi{2}}\lra{\mathbf{1}\mathbb{3}}[\mathbf{2|p_1|}\mathbb{3}\rangle +\left(c_{f\bar{f}Vh,2}^{-+-0}\,^{\mathbf{I}_3}\right){}^{\hi{1}h}_{\hi{2}}\lra{\mathbf{1}\mathbb{3}}[\mathbf{2|p_4|}\mathbb{3}\rangle \notag \\
    & + \left(c_{f\bar{f}Vh,1}^{--00}\,{}^{\mathbf{I}_3}\right){}_{\hi{2}}^{\hi{1}h}\lra{\mathbf{12}}\langle\mathbf{3|p_4|3}] + \left(c_{f\bar{f}Vh,2}^{--00}\,{}^{\mathbf{I}_3}\right){}_{\hi{2}}^{\hi{1}h}\lra{\mathbf{13}}\langle\mathbf{2|p_4|3}]+ \text{h.c.}\,, \label{eq:cls_fbarVh}
\end{align}

\begin{align}
    \left(c_{f\bar{f}Vh,1}^{---0}\,{}^{\mathbf{I}_3}\right){}^{\hi{1}h}_{\hi{2}} &= \left(\frac{C_{Le_\mathbbm{C}B_L H^\dagger}\delta^{I_34} \delta^{\hat{i}_1}_{i_4}\delta^4_{\hat{i}_2}}{\Lambda^2} + \frac{C_{Le_\mathbbm{C}W_LH^\dagger}(\tau^{I_3})^{\hat{i}_1}_{i_4}\delta_{\hat{i}_2}^4}{\Lambda^2}\right)O_{I_3}^{\mathbf{I}_3} \Omega_{\hat{i}_1}^{\hi{1}}\Omega^{\hat{i}_2}_{\hi{2}}\mathcal{U}^{i_4h}\,,\\
    \left(c_{f\bar{f}Vh,2}^{---0}\,{}^{\mathbf{I}_3}\right){}^{\hi{1}h}_{\hi{2}} &= \left(\frac{C_{D^2Le_\mathbbm{C}B_LH^\dagger,1}\delta^{\hat{i}_1}_{i_4}\delta^4_{\hat{i}_2}\delta^{I_34}}{\Lambda^4} + \frac{C_{D^2Le_\mathbbm{C}W_LH^\dagger,1}(\tau^{I_3}){}^{\hat{i}_1}_{i_4}\delta^4_{\hat{i}_2}}{\Lambda^4}\right)O^{\mathbf{I}_3}_{I_3}\Omega_{\hat{i}_1}^{\hi{1}}\Omega^{\hat{i}_2}_{\hi{2}}\mathcal{U}^{i_4h}\,,\\
    \left(c_{f\bar{f}Vh,3}^{---0}\,{}^{\mathbf{I}_3}\right){}^{\hi{1}h}_{\hi{2}} &= \left(\frac{C_{D^2Le_\mathbbm{C}B_LH^\dagger,2}\delta^{\hat{i}_1}_{i_4}\delta^4_{\hat{i}_2}\delta^{I_34}}{\Lambda^4} + \frac{C_{D^2Le_\mathbbm{C}W_LH^\dagger,2}(\tau^{I_3}){}^{\hat{i}_1}_{i_4}\delta^4_{\hat{i}_2}}{\Lambda^4}\right)O^{\mathbf{I}_3}_{I_3}\Omega_{\hat{i}_1}^{\hi{1}}\Omega^{\hat{i}_2}_{\hi{2}}\mathcal{U}^{i_4h}\,,\\
    \left(c_{f\bar{f}Vh,1}^{+-00}\,{}^{\mathbf{I}_3}\right){}_{\hi{2}}^{\hi{1}h} &= \left(\frac{C_{De_\mathbbm{C}^\dagger e_\mathbbm{C} HH^\dagger}\delta^{\hat{i}_1}_3\delta_{\hat{i}_2}^4 \delta^{i_3}_{i_4}}{\Lambda^2} \right)\Omega_{\hat{i}_1}^{\hi{1}}\Omega_{\hi{2}}^{\hat{i}_2}(\mathcal{U}^{\mathbf{I}_3}_{i_3} \mathcal{U}^{hi_4} + \mathcal{U}^{\mathbf{I}_3i_4} \mathcal{U}^{h}_{i_3}) \,,\\
    \left(c_{f\bar{f}Vh,1}^{-+00}\,{}^{\mathbf{I}_3}\right){}_{\hi{2}}^{\hi{1}h}&= \left(\frac{C_{DLL^\dagger  HH^\dagger,1}\delta^{\hat{i}_1}_{\hat{i}_2}\delta^{i_3}_{i_4}}{\Lambda^2} + \frac{C_{DLL^\dagger  HH^\dagger,2}\delta^{\hat{i}_1}_{i_4}\delta_{\hat{i}_2}^{i_3}}{\Lambda^2}\right)\Omega_{\hat{i}_1}^{\hi{1}}\Omega_{\hi{2}}^{\hat{i}_2}(\mathcal{U}^{\mathbf{I}_3}_{i_3} \mathcal{U}^{hi_4} + \mathcal{U}^{\mathbf{I}_3i_4} \mathcal{U}^{h}_{i_3}) \,,\\
    \left(c_{f\bar{f}Vh}^{--+0}\,{}^{\mathbf{I}_3}\right){}_{\hi{2}}^{\hi{1}h} &= \left(\frac{C_{D^2Le_{\mathbbm{C}}B_RH^\dagger}\delta^{\hat{i}_1}_{i_4}\delta^4_{\hat{i}_2}\delta^{I_34}}{\Lambda^4} + \frac{C_{D^2Le_{\mathbbm{C}}W_RH^\dagger}(\tau^{I_3}){}^{\hat{i}_1}_{i_4}\delta^4_{\hat{i}_2}}{\Lambda^4}\right)O_{I_3}^{\mathbf{I}_3} \Omega_{\hat{i}_1}^{\hi{1}}\Omega^{\hat{i}_2}_{\hi{2}}\mathcal{U}^{i_4h}\,,\\
    \left(c^{-+00}_{f\bar{f}Vh,2}\,{}^{\mathbf{I}_3}\right){}^{\hi{1}h}_{\hi{2}} &= \left(\frac{C_{D^3LL^\dagger HH^\dagger,1}\delta^{\hat{i}_1}_{\hat{i}_2}\delta^{i_3}_{i_4}}{\Lambda^4} + \frac{C_{D^3LL^\dagger HH^\dagger,2}\delta^{\hat{i}_1}_{i_4}\delta^{i_3}_{\hat{i}_2}}{\Lambda^4} \right)\Omega_{\hat{i}_1}^{\hi{1}}\Omega^{\hat{i}_2}_{\hi{2}}(\mathcal{U}^{\mathbf{I}_3}_{i_3} \mathcal{U}^{hi_4} + \mathcal{U}^{\mathbf{I}_3i_4} \mathcal{U}^{h}_{i_3})\,,\\
    \left(c^{-+00}_{f\bar{f}Vh,3}\,{}^{\mathbf{I}_3}\right){}^{\hi{1}h}_{\hi{2}} &= \left(\frac{C_{D^3LL^\dagger HH^\dagger,3}\delta^{\hat{i}_1}_{\hat{i}_2}\delta^{i_3}_{i_4}}{\Lambda^4} + \frac{C_{D^3LL^\dagger HH^\dagger,4}\delta^{\hat{i}_1}_{i_4}\delta^{i_3}_{\hat{i}_2}}{\Lambda^4} \right)\Omega_{\hat{i}_1}^{\hi{1}}\Omega^{\hat{i}_2}_{\hi{2}}(\mathcal{U}^{\mathbf{I}_3}_{i_3} \mathcal{U}^{hi_4} + \mathcal{U}^{\mathbf{I}_3i_4} \mathcal{U}^{h}_{i_3})\,, \\
    \left(c^{+-00}_{f\bar{f}Vh,1}\,{}^{\mathbf{I}_3}\right){}^{\hi{1}h}_{\hi{2}} &= \left(\frac{C_{D^3e_\mathbbm{C}^\dagger e_\mathbbm{C} HH^\dagger,2}\delta^{\hat{i}_1}_{\hat{i}_2}\delta^{i_3}_{i_4}}{\Lambda^4}\right)\Omega_{\hat{i}_1}^{\hi{1}}\Omega^{\hat{i}_2}_{\hi{2}}(\mathcal{U}^{\mathbf{I}_3}_{i_3} \mathcal{U}^{hi_4} + \mathcal{U}^{\mathbf{I}_3i_4} \mathcal{U}^{h}_{i_3})\,,\\
    \left(c^{+-00}_{f\bar{f}Vh,2}\,{}^{\mathbf{I}_3}\right){}^{\hi{1}h}_{\hi{2}} &= \left(\frac{C_{D^3e_\mathbbm{C}^\dagger e_\mathbbm{C} HH^\dagger,3}\delta^{\hat{i}_1}_{i_4}\delta^{i_3}_{\hat{i}_2}}{\Lambda^4} \right)\Omega_{\hat{i}_1}^{\hi{1}}\Omega^{\hat{i}_2}_{\hi{2}}(\mathcal{U}^{\mathbf{I}_3}_{i_3} \mathcal{U}^{hi_4} + \mathcal{U}^{\mathbf{I}_3i_4} \mathcal{U}^{h}_{i_3})\,,
\end{align}

\begin{align}
    \left(c_{f\bar{f}Vh,1}^{+--0}\,^{\mathbf{I}_3}\right){}^{\hi{1}h}_{\hi{2}} &= \left(\frac{C_{De_\mathbbm{C}^\dagger e_{\mathbbm{C}}B_LHH^\dagger,1}\delta^{\hat{i}_1}_4\delta^4_{\hat{i}_2}\delta^{I_34}\delta^{i_4}_{i_5}}{\Lambda^4} + \frac{C_{De_\mathbbm{C}^\dagger e_{\mathbbm{C}}W_LHH^\dagger,1}\delta^{\hat{i}_1}_4\delta^4_{\hat{i}_2}(\tau^{I_3}){}^{i_4}_{i_5}}{\Lambda^4}\right)\notag \\
    & \times \Omega_{\hat{i}_1}^{\hi{1}}\Omega^{\hat{i}_2}_{\hi{2}}O^{\mathbf{I}_3}_{I_3}\mathcal{U}^{h}_{i_4}\mathcal{U}^{i_5h} v\\
    \left(c_{f\bar{f}Vh,2}^{+--0}\,^{\mathbf{I}_3}\right){}^{\hi{1}h}_{\hi{2}} &= \left(\frac{C_{De_\mathbbm{C}^\dagger e_{\mathbbm{C}}B_LHH^\dagger,2}\delta^{\hat{i}_1}_4\delta^4_{\hat{i}_2}\delta^{I_34}\delta^{i_4}_{i_5}}{\Lambda^4} + \frac{C_{De_\mathbbm{C}^\dagger e_{\mathbbm{C}}W_LHH^\dagger,2}\delta^{\hat{i}_1}_4\delta^4_{\hat{i}_2}(\tau^{I_3}){}^{i_4}_{i_5}}{\Lambda^4}\right)\notag \\
    & \times \Omega_{\hat{i}_1}^{\hi{1}}\Omega^{\hat{i}_2}_{\hi{2}}O^{\mathbf{I}_3}_{I_3}\mathcal{U}^{h}_{i_4}\mathcal{U}^{i_5h} v\\
    \left(c_{f\bar{f}Vh,1}^{-+-0}\,^{\mathbf{I}_3}\right){}^{\hi{1}h}_{\hi{2}} &= \left(\frac{C_{DLL^\dagger B_LHH^\dagger,1} \delta^{\hat{i}_1}_{\hat{i}_2}\delta^{i_4}_{i_5}\delta^{I_34}}{\Lambda^4} + \frac{C_{DLL^\dagger B_LHH^\dagger,2} \delta^{\hat{i}_1}_{i_5}\delta^{i_4}_{\hat{i}_2}\delta^{I_34}}{\Lambda^4} + \frac{C_{DLL^\dagger W_LHH^\dagger,1} (\tau^{I_3}){}^{\hat{i}_1}_{\hat{i}_2}\delta^{i_4}_{i_5}}{\Lambda^4}\right.\notag \\
    &\left.+ \frac{C_{DLL^\dagger W_LHH^\dagger,2} (\tau^{I_3}){}^{\hat{i}_1}_{i_5}\delta^{i_4}_{\hat{i}_2}}{\Lambda^4} + \frac{C_{DLL^\dagger W_LHH^\dagger,3} (\tau^{I_3}){}^{i_4}_{i_5}\delta^{\hat{i}_1}_{\hat{i}_2}}{\Lambda^4}\right)\Omega_{\hat{i}_1}^{\hi{1}}\Omega^{\hat{i}_2}_{\hi{2}}O^{\mathbf{I}_3}_{I_3}\mathcal{U}^{h}_{i_4}\mathcal{U}^{i_5h} v\,,\\
    \left(c_{f\bar{f}Vh,2}^{-+-0}\,^{\mathbf{I}_3}\right){}^{\hi{1}h}_{\hi{2}} &= \left(\frac{C_{DLL^\dagger B_LHH^\dagger,3} \delta^{\hat{i}_1}_{\hat{i}_2}\delta^{i_4}_{i_5}\delta^{I_34}}{\Lambda^4} + \frac{C_{DLL^\dagger B_LHH^\dagger,4} \delta^{\hat{i}_1}_{i_5}\delta^{i_4}_{\hat{i}_2}\delta^{I_34}}{\Lambda^4} + \frac{C_{DLL^\dagger W_LHH^\dagger,4} (\tau^{I_3}){}^{\hat{i}_1}_{\hat{i}_2}\delta^{i_4}_{i_5}}{\Lambda^4}\right.\notag \\
    &\left.+ \frac{C_{DLL^\dagger W_LHH^\dagger,5} (\tau^{I_3}){}^{\hat{i}_1}_{i_5}\delta^{i_4}_{\hat{i}_2}}{\Lambda^4} + \frac{C_{DLL^\dagger W_LHH^\dagger,6} (\tau^{I_3}){}^{i_4}_{i_5}\delta^{\hat{i}_1}_{\hat{i}_2}}{\Lambda^4}\right)\Omega_{\hat{i}_1}^{\hi{1}}\Omega^{\hat{i}_2}_{\hi{2}}O^{\mathbf{I}_3}_{I_3}\mathcal{U}^{h}_{i_4}\mathcal{U}^{i_5h} v\,,\\
    \left(c_{f\bar{f}Vh,1}^{--00}\,{}^{\mathbf{I}_3}\right){}_{\hi{2}}^{\hi{1}h} &= \left(\frac{C_{D^2Le_\mathbbm{C}HH^\dagger{}^2,3}\delta^{\hat{i}_1}_{i_4}\delta^{i_3}_{i_5}\delta_{\hat{i}_2}^4}{\Lambda^4}\right)\Omega_{\hat{i}_1}^{\hi{1}}\Omega_{\hi{2}}^{\hat{i}_2}\mathcal{U}^{i_5h}\left(\mathcal{U}_{i_3}^{\mathbf{I}_3} \mathcal{U}^{i_4h} +\mathcal{U}^{\mathbf{I}_3i_4} \mathcal{U}^{h}_{i_3}\right)v\,,\\
    \left(c_{f\bar{f}Vh,2}^{--00}\,{}^{\mathbf{I}_3}\right){}_{\hi{2}}^{\hi{1}h} &= \left(\frac{C_{D^2Le_\mathbbm{C}HH^\dagger{}^2,6}\delta^{\hat{i}_1}_{i_4}\delta^{i_3}_{i_5}\delta_{\hat{i}_2}^4}{\Lambda^4}\right)\Omega_{\hat{i}_1}^{\hi{1}}\Omega_{\hi{2}}^{\hat{i}_2}\mathcal{U}^{i_5h}\left(\mathcal{U}_{i_3}^{\mathbf{I}_3} \mathcal{U}^{i_4h} +\mathcal{U}^{\mathbf{I}_3i_4} \mathcal{U}^{h}_{i_3}\right)v\,.
\end{align}

\paragraph{$\boldsymbol{f^2Vh}$}
\begin{align}
    \left(\mathcal{M}_{f^2Vh}^{\mathbf{I}_3}\right){}^{\hi{1}\hi{2}h}\left(\mathbf{\frac{1}{2},\frac{1}{2},1,0}\right) = \left(c_{f^2Vh}^{---}\,^{\mathbf{I}_3}\right){}^{\hi{1}\hi{2}h}\lra{\mathbf{1}\mathbb{3}}\lra{\mathbf{2}\mathbb{3}} + \text{h.c.}\,, \label{eq:cls_f2Vh}
\end{align}

\begin{align}
    \left(c_{f^2Vh}^{---}\,^{\mathbf{I}_3}\right){}^{\hi{1}\hi{2}h} &= \left(\frac{C_{L^2B_LH^2}\epsilon^{\hat{i}_1 i_4}\epsilon^{\hat{i}_2 i_5}\delta^{I_43}}{\Lambda^3} + \frac{C_{L^2W_LH^2}\epsilon^{\hat{i}_1 i_4}(\tau^{I_3})^{\hat{i}_2 i_5}}{\Lambda^3}\right)\Omega_{\hat{i}_1}^{\hi{1}}\Omega_{\hat{i}_2}^{\hi{2}}O^{\mathbf{I}_3}_{I_3} \mathcal{U}_{i_4}^h\mathcal{U}_{i_5}^h v\,.
\end{align}

\paragraph{$\boldsymbol{f\bar{f}V^2}$}
\begin{align}
    \left(\mathcal{M}_{f\bar{f}V^2}^{\mathbf{I}_3\mathbf{I}_4}\right){}^{\hi{1}}_{\hi{2}}\left(\mathbf{\frac{1}{2},\frac{1}{2},1,1}\right) &= \left(c_{f\bar{f}V^2}^{-+--}\,{}^{\mathbf{I}_3\mathbf{I}_4}\right){}^{\hi{1}}_{\hi{2}} \langle\mathbf{1|\mathbb{p_4}|2}]\lra{\mathbb{34}}^2 +\left(c_{f\bar{f}V^2,1}^{---0}\,{}^{\mathbf{I}_3\mathbf{I}_4}\right){}^{\hi{1}}_{\hi{2}} \lra{\mathbf{1}\mathbb{3}}\lra{\mathbf{2}\mathbb{3}}\langle\mathbf{4|p_2|4}]  \notag \\
    & + \left(c_{f\bar{f}V^2,2}^{---0}\,{}^{\mathbf{I}_3\mathbf{I}_4}\right){}^{\hi{1}}_{\hi{2}} \lra{\mathbf{1}\mathbb{3}}\lra{\mathbf{2}\mathbb{3}}\langle\mathbf{4|p_3|4}] + \left(c_{f\bar{f}V^2}^{-+-+}\,{}^{\mathbf{I}_3\mathbf{I}_4}\right){}^{\hi{1}}_{\hi{2}} \lra{\mathbf{1}\mathbb{3}}[\mathbf{2}\mathbb{4}]\langle\mathbf{\mathbb{3}|p_2|\mathbb{4}}] \notag \\
    & + \left(c_{f\bar{f}V^2}^{+--+}\,{}^{\mathbf{I}_3\mathbf{I}_4}\right){}^{\hi{1}}_{\hi{2}} [\mathbf{1}\mathbb{4}]\lra{\mathbf{2}\mathbb{3}}\langle\mathbf{\mathbb{3}|p_2|\mathbb{4}}] + \left(c_{f\bar{f}V^2}^{--+0}\,{}^{\mathbf{I}_3\mathbf{I}_4}\right){}_{\hi{2}}^{\hi{1}} \langle\mathbf{1|p_2|}\mathbb{3}]\langle\mathbf{24}\rangle [\mathbb{3}\mathbf{4}]\notag \\
    &
    + \left(c^{-+00}_{f\bar{f}V^2}\,{}^{\mathbf{I}_3\mathbf{I}_4}\right){}^{\hi{1}}_{\hi{2}}\langle\mathbf{1|p_4|2}]\lra{\mathbf{34}}[\mathbf{34}] + \left(c^{+-00}_{f\bar{f}V^2}\,{}^{\mathbf{I}_3\mathbf{I}_4}\right){}^{\hi{1}}_{\hi{2}}\langle\mathbf{2|p_4|1}]\lra{\mathbf{34}}[\mathbf{34}]\notag \\
    & +\left(c_{f\bar{f}V^2,1}^{----}\,^{\mathbf{I}_3\mathbf{I}_4}\right){}^{\hi{1}}_{\hi{2}} \lra{\mathbf{12}}\lra{\mathbb{34}}^2 +  \left(c_{f\bar{f}V^2,2}^{----}\,^{\mathbf{I}_3\mathbf{I}_4}\right){}^{\hi{1}}_{\hi{2}} \lra{\mathbf{1}\mathbb{3}}\lra{\mathbf{2}\mathbb{4}}\lra{\mathbb{34}} \notag \\
    &+\left(c_{f\bar{f}V^2}^{+--0}\,^{\mathbf{I}_3\mathbf{I}_4}\right){}^{\hi{1}}_{\hi{2}}[\mathbf{14}] \langle\mathbb{3}\mathbf{4}\rangle \lra{\mathbf{2}\mathbb{3}} +\left(c_{f\bar{f}V^2}^{-+-0}\,^{\mathbf{I}_3\mathbf{I}_4}\right){}^{\hi{1}}_{\hi{2}}\lra{\mathbf{1}\mathbb{3}}[\mathbf{24}] \langle\mathbb{3}\mathbf{4}\rangle \notag \\
     & + \left(c_{f\bar{f}V^2,1}^{--00}\,{}^{\mathbf{I}_3\mathbf{I}_4}\right){}_{\hi{2}}^{\hi{1}}\lra{\mathbf{12}}\lra{\mathbf{34}}[\mathbf{34}] + \left(c_{f\bar{f}V^2,2}^{--00}\,{}^{\mathbf{I}_3\mathbf{I}_4}\right){}_{\hi{2}}^{\hi{1}}\lra{\mathbf{13}}\lra{\mathbf{24}}[\mathbf{34}] \notag \\
     &+ \left(c_{f\bar{f}V^2,1}^{--++}\,^{\mathbf{I}_3\mathbf{I}_4}\right){}^{\hi{1}}_{\hi{2}} \lra{\mathbf{12}}[\mathbb{34}]^2 + \text{h.c.}\,,\label{eq:cls_fbarV2}
\end{align}

\begin{align}
    \left(c_{f\bar{f}V^2}^{-+--}\,{}^{\mathbf{I}_3\mathbf{I}_4}\right){}^{\hi{1}}_{\hi{2}} &= \left(\frac{C_{DLL^\dagger B_LW_L}\delta^{I_34} (\tau^{I_4}){}^{\hat{i}_1}_{\hat{i}_2}}{\Lambda^4} + \frac{C_{DLL^\dagger W_L^2} \epsilon^{I_3I_4K}(\tau^{K}){}^{\hat{i}_1}_{\hat{i}_2}}{\Lambda^4}\right) O_{I_3}^{\mathbf{I}_3} O_{I_4}^{\mathbf{I}_4}\Omega_{\hat{i}_1}^{\hi{i}}\Omega_{\hi{2}}^{\hat{i}_2}\,,\\
    \left(c_{f\bar{f}V^2,1}^{---0}\,{}^{\mathbf{I}_3\mathbf{I}_4}\right){}^{\hi{1}}_{\hi{2}} &= \left(\frac{C_{D^2Le_\mathbbm{C}B_LH^\dagger,1}\delta^{\hat{i}_1}_{i_4}\delta^4_{\hat{i}_2}\delta^{I_34}}{\Lambda^4} + \frac{C_{D^2Le_\mathbbm{C}W_LH^\dagger,1}(\tau^{I_3}){}^{\hat{i}_1}_{i_4}\delta^4_{\hat{i}_2}}{\Lambda^4}\right)O^{\mathbf{I}_3}_{I_3}\Omega_{\hat{i}_1}^{\hi{1}}\Omega^{\hat{i}_2}_{\hi{2}}\mathcal{U}^{i_4\mathbf{I}_4}\,,\\
    \left(c_{f\bar{f}V^2,1}^{---0}\,{}^{\mathbf{I}_3\mathbf{I}_4}\right){}^{\hi{1}}_{\hi{2}} &= \left(\frac{C_{D^2Le_\mathbbm{C}B_LH^\dagger,2}\delta^{\hat{i}_1}_{i_4}\delta^4_{\hat{i}_2}\delta^{I_34}}{\Lambda^4} + \frac{C_{D^2Le_\mathbbm{C}W_LH^\dagger,2}(\tau^{I_3}){}^{\hat{i}_1}_{i_4}\delta^4_{\hat{i}_2}}{\Lambda^4}\right)O^{\mathbf{I}_3}_{I_3}\Omega_{\hat{i}_1}^{\hi{1}}\Omega^{\hat{i}_2}_{\hi{2}}\mathcal{U}^{i_4\mathbf{I}_4}\,,\\
    \left(c_{f\bar{f}V^2}^{-+-+}\,{}^{\mathbf{I}_3\mathbf{I}_4}\right){}^{\hi{1}}_{\hi{2}} &= \left(\frac{C_{DLL^\dagger B_LB_R}\delta^{\hat{i}_1}_{\hat{i}_2}\delta^{I_34}\delta^{I_44}}{\Lambda^4} + \frac{C_{DLL^\dagger B_LW_R}(\tau^{I_4}){}^{\hat{i}_1}_{\hat{i}_2}\delta^{I_34}}{\Lambda^4} + \frac{C_{DLL^\dagger W_LW_R}\epsilon^{I_3I_4K}(\tau^K){}^{\hat{i}_1}_{\hat{i}_2}}{\Lambda^4} \right.\notag \\
    &\left.+ \frac{C_{DLL^\dagger W_LW_R}\delta^{I_3I_4}\delta^{\hat{i}_1}_{\hat{i}_2}}{\Lambda^4}\right)O_{I_3}^{\mathbf{I}_3} O_{I_4}^{\mathbf{I}_4}\Omega_{\hat{i}_1}^{\hi{i}}\Omega_{\hi{2}}^{\hat{i}_2}\,,\\
    \left(c_{f\bar{f}V^2}^{+--+}\,{}^{\mathbf{I}_3\mathbf{I}_4}\right){}^{\hi{1}}_{\hi{2}} &= \left(\frac{C_{De_\mathbbm{C}^\dagger e_\mathbbm{C}B_LB_R}\delta^{\hat{i}_1}_4\delta_{\hat{i}_2}^4\delta^{I_34}\delta^{I_44}}{\Lambda^4} + \frac{C_{De_\mathbbm{C}^\dagger e_\mathbbm{C}W_LW_R}\delta^{\hat{i}_1}_4\delta_{\hat{i}_2}^4\delta^{I_3I_4}}{\Lambda^4}\right)O_{I_3}^{\mathbf{I}_3} O_{I_4}^{\mathbf{I}_4}\Omega_{\hat{i}_1}^{\hi{i}}\Omega_{\hi{2}}^{\hat{i}_2}\,,\\
    \left(c_{f\bar{f}V^2}^{--+0}\,{}^{\mathbf{I}_3\mathbf{I}_4}\right){}_{\hi{2}}^{\hi{1}} &= \left(\frac{C_{D^2Le_{\mathbbm{C}}B_RH^\dagger}\delta^{\hat{i}_1}_{i_4}\delta^4_{\hat{i}_2}\delta^{I_34}}{\Lambda^4} + \frac{C_{D^2Le_{\mathbbm{C}}W_RH^\dagger}(\tau^{I_3}){}^{\hat{i}_1}_{i_4}\delta^4_{\hat{i}_2}}{\Lambda^4}\right)O_{I_3}^{\mathbf{I}_3} \Omega_{\hat{i}_1}^{\hi{1}}\Omega^{\hat{i}_2}_{\hi{2}}\mathcal{U}^{i_4\mathbf{I}_4}\,,\\
    \left(c^{-+00}_{f\bar{f}V^2}\,{}^{\mathbf{I}_3\mathbf{I}_4}\right){}^{\hi{1}}_{\hi{2}} &=  \left(\frac{C_{D^3LL^\dagger HH^\dagger,1}\delta^{\hat{i}_1}_{\hat{i}_2}\delta^{i_3}_{i_4}}{\Lambda^4} + \frac{C_{D^3LL^\dagger HH^\dagger,2}\delta^{\hat{i}_1}_{i_4}\delta^{i_3}_{\hat{i}_2}}{\Lambda^4} \right)\Omega_{\hat{i}_1}^{\hi{1}}\Omega^{\hat{i}_2}_{\hi{2}}\mathcal{U}_{i_3}^{\mathbf{I}_3}\mathcal{U}^{i_4\mathbf{I}_4}\,,\\
    \left(c^{+-00}_{f\bar{f}V^2}\,{}^{\mathbf{I}_3\mathbf{I}_4}\right){}^{\hi{1}}_{\hi{2}} &=  \left(\frac{C_{D^3e_{\mathbbm{C}}^\dagger e_{\mathbbm{C}} HH^\dagger,1}\delta^{\hat{i}_1}_{\hat{i}_2}\delta^{i_3}_{i_4}}{\Lambda^4}\right)\Omega_{\hat{i}_1}^{\hi{1}}\Omega^{\hat{i}_2}_{\hi{2}}\mathcal{U}_{i_3}^{\mathbf{I}_3}\mathcal{U}^{i_4\mathbf{I}_4}\,,\\
    \left(c_{f\bar{f}V^2,1}^{----}\,^{\mathbf{I}_3\mathbf{I}_4}\right){}^{\hi{1}}_{\hi{2}} &= \left(\frac{C_{Le_{\mathbbm{C}}B_L^2H^\dagger}\delta^{\hat{i}_1}_{i_5}\delta_{\hat{i}_2}^4\delta^{I_3 4}\delta^{I_44}}{\Lambda^4} + \frac{C_{Le_{\mathbbm{C}}W_LB_LH^\dagger,1}(\tau^{I_3}){}^{\hat{i}_1}_{i_5}\delta_{\hat{i}_2}^4\delta^{I_44}}{\Lambda^4} + \frac{C_{Le_{\mathbbm{C}}W_L^2H^\dagger,1}\delta^{\hat{i}_1}_{i_5}\delta_{\hat{i}_2}^4\delta^{I_3I_4}}{\Lambda^4}\right)\notag \\
    &\times \Omega_{\hat{i}_1}^{\hi{1}}\Omega^{\hat{i}_2}_{\hi{2}}O_{I_3}^{\mathbf{I}_3}O_{I_4}^{\mathbf{I}_4}\mathcal{U}^{i_5 h}v\,,\\
    \left(c_{f\bar{f}V^2,2}^{----}\,^{\mathbf{I}_3\mathbf{I}_4}\right){}^{\hi{1}}_{\hi{2}} &= \left(\frac{C_{Le_{\mathbbm{C}}W_LB_LH^\dagger,2}(\tau^{I_3}){}^{\hat{i}_1}_{i_5}\delta_{\hat{i}_2}^4\delta^{I_44}}{\Lambda^4} + \frac{C_{Le_{\mathbbm{C}}W_L^2H^\dagger,2}(\tau^K){}^{\hat{i}_1}_{i_5}\delta_{\hat{i}_2}^4\epsilon^{I_3I_4K}}{\Lambda^4}\right)\notag \\
    &\times\Omega_{\hat{i}_1}^{\hi{1}}\Omega^{\hat{i}_2}_{\hi{2}}O_{I_3}^{\mathbf{I}_3}O_{I_4}^{\mathbf{I}_4}\mathcal{U}^{i_5 h}v\,,\\
    \left(c_{f\bar{f}V^2,1}^{--++}\,^{\mathbf{I}_3\mathbf{I}_4}\right){}^{\hi{1}}_{\hi{2}} &= \left(\frac{C_{Le_{\mathbbm{C}}B_R^2H^\dagger}\delta^{\hat{i}_1}_{i_5}\delta_{\hat{i}_2}^4\delta^{I_3 4}\delta^{I_44}}{\Lambda^4} + \frac{C_{Le_{\mathbbm{C}}W_RB_RH^\dagger}(\tau^{I_3}){}^{\hat{i}_1}_{i_5}\delta_{\hat{i}_2}^4\delta^{I_44}}{\Lambda^4} + \frac{C_{Le_{\mathbbm{C}}W_R^2H^\dagger}\delta^{\hat{i}_1}_{i_5}\delta_{\hat{i}_2}^4\delta^{I_3I_4}}{\Lambda^4}\right)\notag \\
    &\times \Omega_{\hat{i}_1}^{\hi{1}}\Omega^{\hat{i}_2}_{\hi{2}}O_{I_3}^{\mathbf{I}_3}O_{I_4}^{\mathbf{I}_4}\mathcal{U}^{i_5 h}v\,,
\end{align}

\begin{align}
    \left(c_{f\bar{f}V^2}^{+--0}\,^{\mathbf{I}_3\mathbf{I}_4}\right){}^{\hi{1}}_{\hi{2}}&= \left(\frac{C_{De_\mathbbm{C}^\dagger e_{\mathbbm{C}}B_LHH^\dagger,1}\delta^{\hat{i}_1}_4\delta^4_{\hat{i}_2}\delta^{I_34}\delta^{i_4}_{i_5}}{\Lambda^4} + \frac{C_{De_\mathbbm{C}^\dagger e_{\mathbbm{C}}W_LHH^\dagger,1}\delta^{\hat{i}_1}_4\delta^4_{\hat{i}_2}(\tau^{I_3}){}^{i_4}_{i_5}}{\Lambda^4}\right)\notag \\
    &\times \Omega_{\hat{i}_1}^{\hi{1}}\Omega^{\hat{i}_2}_{\hi{2}}O^{\mathbf{I}_3}_{I_3}\left(\mathcal{U}^{\mathbf{I}_4}_{i_4}\mathcal{U}^{i_5h}+\mathcal{U}^{\mathbf{I}_4i_5}\mathcal{U}_{i_4}^{h}\right)v\,, \\
    \left(c_{f\bar{f}V^2}^{-+-0}\,^{\mathbf{I}_3\mathbf{I}_4}\right){}^{\hi{1}}_{\hi{2}} &= \left(\frac{C_{DLL^\dagger B_LHH^\dagger,1} \delta^{\hat{i}_1}_{\hat{i}_2}\delta^{i_4}_{i_5}\delta^{I_34}}{\Lambda^4} + \frac{C_{DLL^\dagger B_LHH^\dagger,2} \delta^{\hat{i}_1}_{i_5}\delta^{i_4}_{\hat{i}_2}\delta^{I_34}}{\Lambda^4} + \frac{C_{DLL^\dagger W_LHH^\dagger,1} (\tau^{I_3}){}^{\hat{i}_1}_{\hat{i}_2}\delta^{i_4}_{i_5}}{\Lambda^4}\right.\notag \\
    &\left.+ \frac{C_{DLL^\dagger W_LHH^\dagger,2} (\tau^{I_3}){}^{\hat{i}_1}_{i_5}\delta^{i_4}_{\hat{i}_2}}{\Lambda^4} + \frac{C_{DLL^\dagger W_LHH^\dagger,3} (\tau^{I_3}){}^{i_4}_{i_5}\delta^{\hat{i}_1}_{\hat{i}_2}}{\Lambda^4} \right)\notag \\
    &\times \Omega_{\hat{i}_1}^{\hi{1}}\Omega^{\hat{i}_2}_{\hi{2}}O^{\mathbf{I}_3}_{I_3}\left(\mathcal{U}^{\mathbf{I}_4}_{i_4}\mathcal{U}^{i_5h}+\mathcal{U}^{\mathbf{I}_4i_5}\mathcal{U}_{i_4}^{h}\right)v\,,
\end{align}

\begin{align}
    \left(c_{f\bar{f}V^2,1}^{--00}\,{}^{\mathbf{I}_3\mathbf{I}_4}\right){}_{\hi{2}}^{\hi{1}} &= \left(\frac{C_{D^2Le_\mathbbm{C}HH^\dagger{}^2,3}\delta^{\hat{i}_1}_{i_4}\delta^{i_3}_{i_5}\delta_{\hat{i}_2}^4}{\Lambda^4}\right)\Omega_{\hat{i}_1}^{\hi{1}}\Omega_{\hi{2}}^{\hat{i}_2}\mathcal{U}^{i_5h}\left(\mathcal{U}_{i_3}^{\mathbf{I}_3} \mathcal{U}^{i_4\mathbf{I}_4} + \mathcal{U}^{i_4\mathbf{I}_3} \mathcal{U}_{i_3}^{\mathbf{I}_4}\right)v\,,\\
    \left(c_{f\bar{f}V^2,2}^{--00}\,{}^{\mathbf{I}_3\mathbf{I}_4}\right){}_{\hi{2}}^{\hi{1}} &= \left(\frac{C_{D^2Le_\mathbbm{C}HH^\dagger{}^2,6}\delta^{\hat{i}_1}_{i_4}\delta^{i_3}_{i_5}\delta_{\hat{i}_2}^4}{\Lambda^4}\right)\Omega_{\hat{i}_1}^{\hi{1}}\Omega_{\hi{2}}^{\hat{i}_2}\mathcal{U}^{i_5h}\left(\mathcal{U}_{i_3}^{\mathbf{I}_3} \mathcal{U}^{i_4\mathbf{I}_4} + \mathcal{U}^{i_4\mathbf{I}_3} \mathcal{U}_{i_3}^{\mathbf{I}_4}\right)v\,.
\end{align}

\paragraph{\boldsymbol{$V^4$}}

\begin{align}
    \left(\mathcal{M}_{V^4}^{\mathbf{I}_1\mathbf{I}_2\mathbf{I}_3\mathbf{I}_4}\right)\left(\mathbf{1,1,1,1}\right) &= 
    \left(c_{V^4,1}^{----}\,{}^{\mathbf{I}_1\mathbf{I}_2\mathbf{I}_3\mathbf{I}_4}\right) \lra{\mathbb{12}}\lra{\mathbb{13}}\lra{\mathbb{24}}\lra{\mathbb{34}} + \left(c_{V^4,2}^{----}\,{}^{\mathbf{I}_1\mathbf{I}_2\mathbf{I}_3\mathbf{I}_4}\right) \lra{\mathbb{12}}^2\lra{\mathbb{34}}^2 \notag \\
    & + \left(c_{V^4,1}^{--00}\,{}^{\mathbf{I}_1\mathbf{I}_2\mathbf{I}_3\mathbf{I}_4}\right)\lra{\mathbb{12}}^2\lra{\mathbf{34}}[\mathbf{34}] + \left(c_{V^4,2}^{--00}\,{}^{\mathbf{I}_1\mathbf{I}_2\mathbf{I}_3\mathbf{I}_4}\right)\lra{\mathbb{12}}\lra{\mathbb{1}\mathbf{3}}\lra{\mathbb{2}\mathbf{4}}[\mathbf{34}] \notag \\
    & + \left(c_{V^4}^{++--}\,{}^{\mathbf{I}_1\mathbf{I}_2\mathbf{I}_3\mathbf{I}_4}\right)\lra{\mathbb{12}}^2[\mathbb{34}]^2 + \left(c_{V^4}^{-+00}\,{}^{\mathbf{I}_1\mathbf{I}_2 \mathbf{I}_3\mathbf{I}_4}\right) \lra{\mathbb{1}\mathbf{3}}[\mathbb{2}\mathbf{3}]\langle\mathbb{1}\mathbf{4}\rangle[\mathbb{2}\mathbf{4}]\notag \\
    & + \left(c_{V^4,1}^{0000}\,{}^{\mathbf{I}_1\mathbf{I}_2\mathbf{I}_3\mathbf{I}_4}\right) \lra{\mathbf{12}}[\mathbf{12}]\lra{\mathbf{34}}[\mathbf{34}] + \left(c_{V^4,2}^{0000}\,{}^{\mathbf{I}_1\mathbf{I}_2\mathbf{I}_3\mathbf{I}_4}\right)\lra{\mathbf{13}}[\mathbf{13}]\lra{\mathbf{24}}[\mathbf{24}] \notag \\
    & + \left(c_{V^4,3}^{0000}\,{}^{\mathbf{I}_1\mathbf{I}_2\mathbf{I}_3\mathbf{I}_4}\right) \lra{\mathbf{14}}[\mathbf{14}]\lra{\mathbf{23}}[\mathbf{23}] + \text{h.c.}\,, \label{eq:cls_V4}
\end{align}

\begin{align}
    \left(c_{V^4,1}^{----}\,{}^{\mathbf{I}_1\mathbf{I}_2\mathbf{I}_3\mathbf{I}_4}\right) &= \left(\frac{C_{B_L^4}\delta^{I_14}\delta^{I_24}\delta^{I_34}\delta^{I_44}}{\Lambda^4} + \frac{C_{B_L^2W_L^2,1}\delta^{I_14}\delta^{I_24}\delta^{I_3I_4}}{\Lambda^4} + \frac{C_{W_L^4,1}\delta^{I_1I_2}\delta^{I_3I_4}}{\Lambda^4} \right) \notag \\
    & \times O_{I_1}^{\mathbf{I}_1}O_{I_2}^{\mathbf{I}_2}O_{I_3}^{\mathbf{I}_3}O_{I_4}^{\mathbf{I}_4}\,, \\
    \left(c_{V^4,2}^{----}\,{}^{\mathbf{I}_1\mathbf{I}_2\mathbf{I}_3\mathbf{I}_4}\right) &= \left(\frac{C_{B_L^2W_L^2,2}\delta^{I_14}\delta^{I_24}\delta^{I_3I_4}}{\Lambda^4} + \frac{C_{W_L^4,2}\delta^{I_1I_2}\delta^{I_3I_4}}{\Lambda^4} \right) O_{I_1}^{\mathbf{I}_1}O_{I_2}^{\mathbf{I}_2}O_{I_3}^{\mathbf{I}_3}O_{I_4}^{\mathbf{I}_4}\,,\\
    \left(c_{V^4,1}^{--00}\,{}^{\mathbf{I}_1\mathbf{I}_2\mathbf{I}_3\mathbf{I}_4}\right) &= \left(\frac{C_{D^2B_L^2HH^\dagger}\delta^{I_14}\delta^{I_24}\delta^{i_3}_{i_4}}{\Lambda^4} + \frac{C_{D^2B_LW_LHH^\dagger,1}\delta^{I_14}(\tau^{I_2}){}^{i_3}_{i_4}}{\Lambda^4} + \frac{C_{D^2W_L^2HH^\dagger,1}\delta^{I_1I_2}\delta^{i_3}_{i_4}}{\Lambda^4}\right)\notag \\
    & \times O^{\mathbf{I}_1}_{I_1} O^{\mathbf{I}_2}_{I_2}\left(\mathcal{U}^{\mathbf{I}_3}_{i_3}\mathcal{U}^{\mathbf{I}_4 i_4} + \mathcal{U}^{\mathbf{I}_3i_4}\mathcal{U}_{i_3}^{\mathbf{I}_4}\right)\,, \\
    \left(c_{V^4,2}^{--00}\,{}^{\mathbf{I}_1\mathbf{I}_2\mathbf{I}_3\mathbf{I}_4}\right) &= \left(\frac{C_{D^2B_LW_LHH^\dagger,2}\delta^{I_14}(\tau^{I_2}){}^{i_3}_{i_4}}{\Lambda^4} + \frac{C_{D^2W_L^2HH^\dagger,2}\epsilon^{I_1i_2K}(\tau^K){}^{i_3}_{i_4}}{\Lambda^4}\right)\notag \\
    & \times O^{\mathbf{I}_1}_{I_1} O^{\mathbf{I}_2}_{I_2}\left(\mathcal{U}^{\mathbf{I}_3}_{i_3}\mathcal{U}^{\mathbf{I}_4 i_4} + \mathcal{U}^{\mathbf{I}_3i_4}\mathcal{U}_{i_3}^{\mathbf{I}_4}\right)\,, \\
    \left(c_{V^4}^{--++}\,{}^{\mathbf{I}_1\mathbf{I}_2\mathbf{I}_3\mathbf{I}_4}\right) &= \left(\frac{C_{B_L^2B_R^2}\delta^{I_14}\delta^{I_24}\delta^{I_34}\delta^{I_44}}{\Lambda^4} + \frac{C_{B_L^2W_R^2}\delta^{I_14}\delta^{I_24}\delta^{I_3I_4}}{\Lambda^4} + \frac{C_{B_LW_LB_RW_R}\delta^{I_14}\delta^{I_34}\delta^{I_2I_4}}{\Lambda^4} + \right. \notag \\
    & \left.\frac{C_{W_L^2W_R^2,1}\delta^{I_1I_2}\delta^{I_3I_4}}{\Lambda^4} + \frac{C_{W_L^2W_R^2,1}\delta^{I_1I_3}\delta^{I_2I_4}}{\Lambda^4}\right)O_{I_1}^{\mathbf{I}_1}O_{I_2}^{\mathbf{I}_2}O_{I_3}^{\mathbf{I}_3}O_{I_4}^{\mathbf{I}_4}\,, \\
    \left(c_{V^4}^{-+00}\,{}^{\mathbf{I}_1\mathbf{I}_2 \mathbf{I}_3\mathbf{I}_4}\right) &= \left(\frac{C_{D^2B_LB_RHH^\dagger}\delta^{I_14}\delta^{I_24}\delta^{i_3}_{i_4}}{\Lambda^4} + \frac{C_{D^2B_LW_RHH^\dagger}\delta^{I_14}(\tau^{I_2}){}^{i_3}_{i_4}}{\Lambda^4} + \frac{C_{D^2W_LW_RHH^\dagger,1}\delta^{I_1I_2}\delta^{i_3}_{i_4}}{\Lambda^4} \right.\notag \\
    & \left.+ \frac{C_{D^2W_LW_RHH^\dagger,2}\epsilon^{I_1I_2K}(\tau^K){}^{i_3}_{i_4}}{\Lambda^4}\right)O^{\mathbf{I}_1}_{I_1} O^{\mathbf{I}_2}_{I_2}O^{\mathbf{I}_1}_{I_1} O^{\mathbf{I}_2}_{I_2}\mathcal{U}^{\mathbf{I}_3}_{i_3}\mathcal{U}^{\mathbf{I}_4 i_4}\,,\\
    \left(c_{V^4,1}^{0000}\,{}^{\mathbf{I}_1\mathbf{I}_2\mathbf{I}_3\mathbf{I}_4}\right) &= \left(\frac{C_{D^4H^2H^\dagger{}^2,1}}{\Lambda^4} \delta^{i_1}_{i_3}\delta^{i_2}_{i_4}\right)\left(\calu^{\mathbf{I}_1}_{i_1} \calu^{\mathbf{I}_2}_{i_2}\calu^{i_3\mathbf{I}_3} \calu^{i_4\mathbf{I}_4}\right)\,,  \\
    \left(c_{V^4,2}^{0000}\,{}^{\mathbf{I}_1\mathbf{I}_2\mathbf{I}_3\mathbf{I}_4}\right) &= \left(\frac{C_{D^4H^2H^\dagger{}^2,2}}{\Lambda^4} \delta^{i_1}_{i_3}\delta^{i_2}_{i_4}\right)\left(\calu^{\mathbf{I}_1}_{i_1} \calu^{\mathbf{I}_2}_{i_2}\calu^{i_3\mathbf{I}^3} \calu^{i_4\mathbf{I}_4}\right)\,,  \\
    \left(c_{V^4,3}^{0000}\,{}^{\mathbf{I}_1\mathbf{I}_2\mathbf{I}_3\mathbf{I}_4}\right) &= \left(\frac{C_{D^4H^2H^\dagger{}^2,3}}{\Lambda^4} \delta^{i_1}_{i_3}\delta^{i_2}_{i_4}\right)\left(\calu^{\mathbf{I}_1}_{i_1} \calu^{\mathbf{I}_2}_{i_2}\calu^{i_3\mathbf{I}^3} \calu^{i_4\mathbf{I}_4}\right)\,,  
\end{align}

\paragraph{$\boldsymbol{V^3h}$}
\begin{align}
    \mathcal{M}^{\mathbf{I}_1\mathbf{I}_2\mathbf{I}_3h}_{V^3h}\left(\mathbf{1,1,1,0}\right) &= \left(c_{V^3h,1}^{--00}\,{}^{\mathbf{I}_1\mathbf{I}_2\mathbf{I}_3}\right){}^h\lra{\mathbb{12}}^2\langle\mathbf{3|p_4|3}] + \left(c_{V^3h,2}^{--00}\,{}^{\mathbf{I}_1\mathbf{I}_2\mathbf{I}_3}\right){}^h\lra{\mathbb{12}}\lra{\mathbb{1}\mathbf{3}} \langle\mathbb{2}\mathbf{p_4|3}] \notag \\
    & + \left(c_{V^3h}^{-+00}\,{}^{\mathbf{I}_1\mathbf{I}_2\mathbf{I}_3}\right){}^{h} \lra{\mathbb{1}\mathbf{3}}[\mathbb{2}\mathbf{3}]\langle\mathbb{1}\mathbf{|p_4}|\mathbb{2}] + \left(c_{V^3h,1}^{0000}\,{}^{\mathbf{I}_1\mathbf{I}_2\mathbf{I}_3}\right){}^{h} \lra{\mathbf{12}}[\mathbf{12}]\langle\mathbf{3|p_4|3}]\notag \\
    &+ \left(c_{V^3h,2}^{0000}\,{}^{\mathbf{I}_1\mathbf{I}_2\mathbf{I}_3}\right){}^{h}\lra{\mathbf{13}}[\mathbf{13}]\langle\mathbf{2|p_4|2}] + \left(c_{V^3h,3}^{0000}\,{}^{\mathbf{I}_1\mathbf{I}_2\mathbf{I}_3}\right){}^{h} \langle\mathbf{1|p_4|1}]\lra{\mathbf{23}}[\mathbf{23}] \notag \\
    & + \left(c_{V^3h}^{---0}\,{}^{\mathbf{I}_1\mathbf{I}_2\mathbf{I}_3}\right){}^{h}\lra{\mathbb{12}}\lra{\mathbb{13}}\lra{\mathbb{23}} + \left(c_{V^3h}^{-000}\,{}^{\mathbf{I}_1\mathbf{I}_2\mathbf{I}_3}\right){}^{h}\langle\mathbb{1}\mathbf{2}\rangle \langle\mathbb{1}\mathbf{3}\rangle[\mathbf{23}] +\text{h.c.}\,,\label{eq:cls_V3h}
\end{align}

\begin{align}
    \left(c_{V^3h,1}^{--00}\,{}^{\mathbf{I}_1\mathbf{I}_2\mathbf{I}_3}\right){}^h &= \left(\frac{C_{D^2B_L^2HH^\dagger}\delta^{I_14}\delta^{I_24}\delta^{i_3}_{i_4}}{\Lambda^4} + \frac{C_{D^2B_LW_LHH^\dagger,1}\delta^{I_14}(\tau^{I_2}){}^{i_3}_{i_4}}{\Lambda^4} + \frac{C_{D^2W_L^2HH^\dagger,1}\delta^{I_1I_2}\delta^{i_3}_{i_4}}{\Lambda^4}\right) \notag \\
    & \times O^{\mathbf{I}_1}_{I_1} O^{\mathbf{I}_2}_{I_2}\left(\calu^{\mathbf{I}_3}_{i_3} \calu^{i_4h} + \calu^{\mathbf{I}_3i_4} \calu_{i_4}^h\right)\,, \\
    \left(c_{V^3h,2}^{--00}\,{}^{\mathbf{I}_1\mathbf{I}_2\mathbf{I}_3}\right){}^h &= \left(\frac{C_{D^2B_LW_LHH^\dagger,2}\delta^{I_14}(\tau^{I_2}){}^{i_3}_{i_4}}{\Lambda^4} + \frac{C_{D^2W_L^2HH^\dagger,2}\epsilon^{I_1i_2K}(\tau^K){}^{i_3}_{i_4}}{\Lambda^4}\right)\notag \\
    & \times O^{\mathbf{I}_1}_{I_1} O^{\mathbf{I}_2}_{I_2}\left(\calu^{\mathbf{I}_3}_{i_3} \calu^{i_4h} + \calu^{\mathbf{I}_3i_4} \calu_{i_4}^h\right)\,,\\
    \left(c_{V^3h}^{-+00}\,{}^{\mathbf{I}_1\mathbf{I}_2\mathbf{I}_3}\right){}^{h}  &= \left(\frac{C_{D^2B_LB_RHH^\dagger}\delta^{I_14}\delta^{I_24}\delta^{i_3}_{i_4}}{\Lambda^4} + \frac{C_{D^2B_LW_RHH^\dagger}\delta^{I_14}(\tau^{I_2}){}^{i_3}_{i_4}}{\Lambda^4} + \frac{C_{D^2W_LW_RHH^\dagger,1}\delta^{I_1I_2}\delta^{i_3}_{i_4}}{\Lambda^4} \right.\notag \\
    & \left.+ \frac{C_{D^2W_LW_RHH^\dagger,2}\epsilon^{I_1I_2K}(\tau^K){}^{i_3}_{i_4}}{\Lambda^4}\right)O^{\mathbf{I}_1}_{I_1} O^{\mathbf{I}_2}_{I_2}\left(\calu^{\mathbf{I}_3}_{i_3} \calu^{i_4h} + \calu^{\mathbf{I}_3i_4} \calu_{i_4}^h\right)\,,\\
    \left(c_{V^3h,1}^{0000}\,{}^{\mathbf{I}_1\mathbf{I}_2\mathbf{I}_3}\right){}^{h} &= \left(\frac{C_{D^4H^2H^\dagger{}^2,1}}{\Lambda^4} \delta^{i_1}_{i_3}\delta^{i_2}_{i_4}\right)\left(\calu^{\mathbf{I}_1}_{i_1} \calu^{\mathbf{I}_2}_{i_2}\calu^{i_3\mathbf{I}_3} \calu^{i_4h} + perms\right)\,,  \\
    \left(c_{V^3h,2}^{0000}\,{}^{\mathbf{I}_1\mathbf{I}_2\mathbf{I}_3}\right){}^{h} &= \left(\frac{C_{D^4H^2H^\dagger{}^2,2}}{\Lambda^4} \delta^{i_1}_{i_3}\delta^{i_2}_{i_4}\right)\left(\calu^{\mathbf{I}_1}_{i_1} \calu^{\mathbf{I}_2}_{i_2}\calu^{i_3\mathbf{I}^3} \calu^{i_4h} + perms\right)\,,  \\
    \left(c_{V^3h,3}^{0000}\,{}^{\mathbf{I}_1\mathbf{I}_2\mathbf{I}_3}\right){}^{h} &= \left(\frac{C_{D^4H^2H^\dagger{}^2,3}}{\Lambda^4} \delta^{i_1}_{i_3}\delta^{i_2}_{i_4}\right)\left(\calu^{\mathbf{I}_1}_{i_1} \calu^{\mathbf{I}_2}_{i_2}\calu^{i_3\mathbf{I}^3} \calu^{i_4h} + perms\right)\,,  \\
    \left(c_{V^3h}^{---0}\,{}^{\mathbf{I}_1\mathbf{I}_2\mathbf{I}_3}\right){}^{h} &= \left(\frac{C_{B_LW_L^2HH^\dagger}\delta^{I_14}\epsilon^{I_2I_3K}(\tau^K)^{i_4}_{i_5}}{\Lambda^4} + \frac{C_{W_L^3HH^\dagger}\epsilon^{I_1I_2I_3}\delta^{i_4}_{i_5}}{\Lambda^4} \right)O_{I_1}^{\mathbf{I}_1}O_{I_2}^{\mathbf{I}_2} O_{I_3}^{\mathbf{I}_3}\calu_{i_4}^h\calu^{i_5 h} v\,,\\
    \left(c_{V^3h}^{-000}\,{}^{\mathbf{I}_1\mathbf{I}_2\mathbf{I}_3}\right){}^{h} &= \left(\frac{C_{D^2B_LH^2H^\dagger{}^2}\delta^{I_14}\delta^{i_2}_{i_3}\delta^{i_4}_{i_5}}{\Lambda^4} + \frac{C_{D^2W_LH^2H^\dagger{}^2,1}(\tau^{I_1}){}^{i_4}_{i_5}\delta^{i_2}_{i_3}}{\Lambda^4} + \frac{C_{D^2W_LH^2H^\dagger{}^2,2}(\tau^{I_1}){}^{i_2}_{i_3}\delta^{i_4}_{i_5}}{\Lambda^4} \right)\notag \\
    &\times O_{I_1}^{\mathbf{I}_1} \calu^{i_3\mathbf{I}_3} \calu_{i_2}^{\mathbf{I}_2}\calu_{i_4}^h\calu^{i_5h}v\,,
\end{align}

\paragraph{$\boldsymbol{V^2h^2}$}
\begin{align}
    \mathcal{M}^{\mathbf{I}_1\mathbf{I}_2 hh}_{V^2h^2}\left(\mathbf{1,1,0,0}\right) &= 
    \left(c_{V^2h^2,1}^{--00}\right){}^{\mathbf{I}_1\mathbf{I}_2 hh}\lra{\mathbb{12}}^2  
    + \left(c_{V^2h^2,2}^{--00}\right){}^{\mathbf{I}_1\mathbf{I}_2 hh}\lra{\mathbb{12}}^2 s_{34} \notag \\
    &+ \left(c_{V^2h^2,3}^{--00}\right){}^{\mathbf{I}_1\mathbf{I}_2 hh}\lra{\mathbb{12}}^2 \lra{\mathbb{1}\mathbf{|p_3p_4|}\mathbb{2}}+ \left(c_{V^2h^2}^{-+00}\right){}^{\mathbf{I}_1\mathbf{I}_2 hh} \langle\mathbb{1}\mathbf{|p_3}|\mathbb{2}]\langle\mathbb{1}\mathbf{|p_4}|\mathbb{2}]\notag \\
    & + \left(c_{V2h^2,1}^{0000}\,{}^{\mathbf{I}_1\mathbf{I}_2}\right){}^{hh} \lra{\mathbf{12}}[\mathbf{12}]s_{34} + \left(c_{V^2h^2,2}^{0000}\,{}^{\mathbf{I}_1\mathbf{I}_2}\right){}^{hh}\langle\mathbf{1|p_3|1}]\langle\mathbf{2|p_4|2}] \notag \\
    & + \left(c_{V^2h^2,3}^{0000}\,{}^{\mathbf{I}_1\mathbf{I}_2}\right){}^{hh} \langle\mathbf{1|p_4|1}]\langle\mathbf{2|p_3|2}] + \left(c_{V^2h^2}^{-000}\,{}^{\mathbf{I}_1\mathbf{I}_2}\right){}^{hh}\langle\mathbb{1}\mathbf{2}\rangle [\mathbf{2|p_3|}\mathbb{1}\rangle +\text{h.c.}\,, \label{eq:cls_V2h2}
\end{align}
\begin{align}
    \left(c_{V^2h^2,1}^{--0}\right){}^{\mathbf{I}_1\mathbf{I}_2hh} &= \left(\frac{C_{B_L^2HH^\dagger}\delta^{I_1 4}\delta^{I_2 4}\delta^{i_3}_{i_4}}{\Lambda^2} + \frac{C_{B_LW_LHH^\dagger}\delta^{I_1}_4(\tau^{I_2})^{i_3}_{i_4}}{\Lambda^2} + \frac{C_{W_L^2HH^\dagger}\delta^{I_1I_2}\delta^{i_3}_{i_4}}{\Lambda^2}\right) \notag \\
    & \times O^{\mathbf{I}_1}_{I_1} O^{\mathbf{I}_2}_{I_2}\calu^h_{i_3} \calu^{i_4h}\,, \\
    \left(c_{V^2h^2,2}^{--00}\right){}^{\mathbf{I}_1\mathbf{I}_2 hh} & =\left(\frac{C_{D^2B_L^2HH^\dagger}\delta^{I_14}\delta^{I_24}\delta^{i_3}_{i_4}}{\Lambda^4} + \frac{C_{D^2B_LW_LHH^\dagger,1}\delta^{I_14}(\tau^{I_2}){}^{i_3}_{i_4}}{\Lambda^4} + \frac{C_{D^2W_L^2HH^\dagger,1}\delta^{I_1I_2}\delta^{i_3}_{i_4}}{\Lambda^4}\right)\notag \\
    & \times O^{\mathbf{I}_1}_{I_1} O^{\mathbf{I}_2}_{I_2}\calu^h_{i_3} \calu^{i_4h}\,, \\
    \left(c_{V^2h^2,3}^{--00}\right){}^{\mathbf{I}_1\mathbf{I}_2 hh} &= \left(\frac{C_{D^2B_LW_LHH^\dagger,2}\delta^{I_14}(\tau^{I_2}){}^{i_3}_{i_4}}{\Lambda^4} + \frac{C_{D^2W_L^2HH^\dagger,2}\epsilon^{I_1i_2K}(\tau^K){}^{i_3}_{i_4}}{\Lambda^4}\right)O^{\mathbf{I}_1}_{I_1} O^{\mathbf{I}_2}_{I_2}\calu^h_{i_3} \calu^{i_4h}\,,\\
    \left(c_{V^2h^2}^{-+00}\right){}^{\mathbf{I}_1\mathbf{I}_2 hh} &= \left(\frac{C_{D^2B_LB_RHH^\dagger}\delta^{I_14}\delta^{I_24}\delta^{i_3}_{i_4}}{\Lambda^4} + \frac{C_{D^2B_LW_RHH^\dagger}\delta^{I_14}(\tau^{I_2}){}^{i_3}_{i_4}}{\Lambda^4} + \frac{C_{D^2W_LW_RHH^\dagger,1}\delta^{I_1I_2}\delta^{i_3}_{i_4}}{\Lambda^4} \right.\notag \\
    & \left.+ \frac{C_{D^2W_LW_RHH^\dagger,2}\epsilon^{I_1I_2K}(\tau^K){}^{i_3}_{i_4}}{\Lambda^4}\right)O^{\mathbf{I}_1}_{I_1} O^{\mathbf{I}_2}_{I_2}\calu^h_{i_3} \calu^{i_4h}\,,\\
    \left(c_{V^2h^2,1}^{0000}\,{}^{\mathbf{I}_1\mathbf{I}_2}\right){}^{hh} &= \frac{C_{D^2H^3H^\dagger{}^3,1}}{\Lambda^4} \delta^{i_1}_{i_3}\delta^{i_2}_{i_4}\left(\calu^{\mathbf{I}_1}_{i_1}\calu^{i_3h} + \calu^{i_3\mathbf{I}_1}\calu_{i_1}^{h}\right)\left(\calu^{\mathbf{I}_2}_{i_2} \calu^{i_4h} + \calu^{i_4\mathbf{I}_2} \calu_{i_2}^{h}\right)\,,  \\
    \left(c_{V^2h^2,2}^{0000}\,{}^{\mathbf{I}_1\mathbf{I}_2}\right){}^{hh} &= \frac{C_{D^2H^3H^\dagger{}^3,2}}{\Lambda^4} \delta^{i_1}_{i_3}\delta^{i_2}_{i_4}\left(\calu^{\mathbf{I}_1}_{i_1}\calu^{i_3h} + \calu^{i_3\mathbf{I}_1}\calu_{i_1}^{h}\right)\left(\calu^{\mathbf{I}_2}_{i_2} \calu^{i_4h} + \calu^{i_4\mathbf{I}_2} \calu_{i_2}^{h}\right)\,,  \\
    \left(c_{V^2h^2,3}^{0000}\,{}^{\mathbf{I}_1\mathbf{I}_2}\right){}^{hh} &= \frac{C_{D^2H^3H^\dagger{}^3,3}}{\Lambda^4} \delta^{i_1}_{i_3}\delta^{i_2}_{i_4}\left(\calu^{\mathbf{I}_1}_{i_1}\calu^{i_3h} + \calu^{i_3\mathbf{I}_1}\calu_{i_1}^{h}\right)\left(\calu^{\mathbf{I}_2}_{i_2} \calu^{i_4h} + \calu^{i_4\mathbf{I}_2} \calu_{i_2}^{h}\right)\,,  \\
    \left(c_{V^2h^2}^{-000}\,{}^{\mathbf{I}_1\mathbf{I}_2}\right){}^{hh} &= \left(\frac{C_{D^2B_LH^2H^\dagger{}^2}\delta^{I_14}\delta^{i_2}_{i_3}\delta^{i_4}_{i_5}}{\Lambda^4} + \frac{C_{D^2W_LH^2H^\dagger{}^2,1}(\tau^{I_1}){}^{i_4}_{i_5}\delta^{i_2}_{i_3}}{\Lambda^4} + \frac{C_{D^2W_LH^2H^\dagger{}^2,2}(\tau^{I_1}){}^{i_2}_{i_3}\delta^{i_4}_{i_5}}{\Lambda^4} \right)\notag \\
    &\times O_{I_1}^{\mathbf{I}_1} \left(\calu_{i_2}^{\mathbf{I}_2}\calu^{i_3h} + \calu^{i_3\mathbf{I}_2}\calu_{i_2}^{h}\right)\left(\calu_{i_4}^h\calu^{i_5h} + \calu^{i_5h}\calu_{i_4}^{h} \right)v\,,
\end{align}

\paragraph{$\boldsymbol{Vh^3}$}
\begin{align}
    \mathcal{M}^{\mathbf{I}_1hhh}_{Vh^3} \left(\mathbf{1,0,0,0}\right) &= 
    \left(c_{Vh^3,1}^{0000}\,{}^{\mathbf{I}_1}\right){}^{hhh} \langle\mathbf{1|p_2|1}]s_{34} + \left(c_{Vh^3,2}^{0000}\,{}^{\mathbf{I}_1}\right){}^{hhh}\langle\mathbf{1|p_3|1}]s_{24} \notag \\
    & + \left(c_{Vh^3,3}^{0000}\,{}^{\mathbf{I}_1}\right){}^{hhh} \langle\mathbf{1|p_4|1}]s_{23}  + \left(c_{Vh^3}^{-000}\,{}^{\mathbf{I}_1}\right){}^{hhh}\langle\mathbb{1}\mathbf{|p_2p_3|}\mathbb{1}\rangle + \text{h.c.}\,, \label{eq:cls_Vh3}
\end{align}

\begin{align}
    \left(c_{Vh^3,1}^{0000}\,{}^{\mathbf{I}_1}\right){}^{hhh} &= \left(\frac{C_{D^2H^3H^\dagger{}^3,1}}{\Lambda^4} \delta^{i_1}_{i_3}\delta^{i_2}_{i_4} \right)\calu^{i_4h}\calu^{h}_{i_2}\left(\calu^{\mathbf{I}_1}_{i_1} \calu^{i_3h} + \calu^{\mathbf{I}_1i_3} \calu_{i_1}^{h}\right)\,,  \\
    \left(c_{Vh^3,2}^{0000}\,{}^{\mathbf{I}_1}\right){}^{hhh} &= \left(\frac{C_{D^2H^3H^\dagger{}^3,2}}{\Lambda^4} \delta^{i_1}_{i_3}\delta^{i_2}_{i_4}\right) \calu^{i_4h}\calu^{h}_{i_2}\left(\calu^{\mathbf{I}_1}_{i_1} \calu^{i_3h} + \calu^{\mathbf{I}_1i_3} \calu_{i_1}^{h}\right)\,,  \\
    \left(c_{Vh^3,3}^{0000}\,{}^{\mathbf{I}_1}\right){}^{hhh} &= \left(\frac{C_{D^2H^3H^\dagger{}^3,3}}{\Lambda^4} \delta^{i_1}_{i_3}\delta^{i_2}_{i_4} \right)\calu^{i_4h}\calu^{h}_{i_2}\left(\calu^{\mathbf{I}_1}_{i_1} \calu^{i_3h} + \calu^{\mathbf{I}_1i_3} \calu_{i_1}^{h}\right)\,,  \\
    \left(c_{Vh^3}^{-000}\,{}^{\mathbf{I}_1}\right){}^{hhh} &=  \left(\frac{C_{D^2B_LH^2H^\dagger{}^2}\delta^{I_14}\delta^{i_2}_{i_3}\delta^{i_4}_{i_5}}{\Lambda^4} + \frac{C_{D^2W_LH^2H^\dagger{}^2,1}(\tau^{I_1}){}^{i_4}_{i_5}\delta^{i_2}_{i_3}}{\Lambda^4} + \frac{C_{D^2W_LH^2H^\dagger{}^2,2}(\tau^{I_1}){}^{i_2}_{i_3}\delta^{i_4}_{i_5}}{\Lambda^4} \right)\notag \\
    &\times O_{I_1}^{\mathbf{I}_1} \calu_{i_2}^h\calu_{i_4}^h\calu^{i_3h}\calu^{i_5h}v\,,
\end{align}

\paragraph{$\boldsymbol{h^4}$}
\begin{align}
    \mathcal{M}^{hhhh}_{h^4} \left(\mathbf{0,0,0,0}\right) &= 
    \left(c_{h^4,1}^{0000}\right){}^{hhhh} s_{12} + \left(c_{h^4,2}^{0000}\right){}^{hhhh}
    + \left(c_{h^4,3}^{0000}\right){}^{hhhh} s_{12}s_{34} \notag \\
    & + \left(c_{h^4,4}^{0000}\right){}^{hhhh} s_{13}s_{24} + \left(c_{h^4,5}^{0000}\right){}^{hhhh} s_{14}s_{23}   \,, \label{eq:cls_h4}
\end{align}

\begin{align}
    & \left(c_{h^4,1}^{0000}\right){}^{hhhh} = \left(\frac{C_{D^2H^2H^\dagger{}^2,1}\epsilon^{i_1i_2}\epsilon^{i_3i_4}}{\Lambda^2} + \frac{C_{D^2H^2H^\dagger{}^2,2}\epsilon^{i_1i_3}\epsilon^{i_2i_4}}{\Lambda^2}\right)\calu^{h}_{i_1}\calu^{h}_{i_2}\calu^{h}_{i_3} \calu^h_{i_4}\,,\\
    & \left(c_{h^4,2}^{0000}\right){}^{hhhh} = \left(\frac{C_{H^3H^\dagger{}^3}}{\Lambda^2} \epsilon^{i_1i_4}\epsilon^{i_2i_5}\epsilon^{i_3i_6}\right)\calu^h_{i_1}\calu^h_{i_2}\calu^h_{i_3}\calu^h_{i_4}\calu^h_{i_5}\calu^h_{i_6}v^2  + 4\lambda \delta^{i_1}_{i_3}\delta^{i_2}_{i_4} \mathcal{U}_{i_1}^h \mathcal{U}_{i_2}^h \mathcal{U}^{i_3 h} \mathcal{U}^{i_4h}\,,\\
    & \left(c_{h^4,3}^{0000}\right){}^{hhhh} = \left(\frac{C_{D^2H^3H^\dagger{}^3,1}}{\Lambda^4} \delta^{i_1}_{i_3}\delta^{i_2}_{i_4}\right)\calu^{h}_{i_1} \calu^{h}_{i_2}\calu^{i_3h} \calu^{i_4h}\,,  \\
    & \left(c_{h^4,4}^{0000}\right){}^{hhhh} = \left(\frac{C_{D^2H^3H^\dagger{}^3,2}}{\Lambda^4} \delta^{i_1}_{i_3}\delta^{i_2}_{i_4}\right)\calu^{h}_{i_1} \calu^{h}_{i_2}\calu^{i_3h} \calu^{i_4h}\,,  \\
    & \left(c_{h^4,5}^{0000}\right){}^{hhhh} = \left(\frac{C_{D^2H^3H^\dagger{}^3,3}}{\Lambda^4} \delta^{i_1}_{i_3}\delta^{i_2}_{i_4}\right)\calu^{h}_{i_1} \calu^{h}_{i_2}\calu^{i_3h} \calu^{i_4h}\,,  
\end{align}

For the 4-point amplitudes, new matching from the derivatives of the Higgs fields to the vector fields emerges, for example, the 4-point class $V^4$. One of its Wilson coefficients is 
\begin{equation}
    \begin{aligned}
        \left(c_{V^4,2}^{--00}\,{}^{\mathbf{I}_1\mathbf{I}_2\mathbf{I}_3\mathbf{I}_4}\right) &= \left(\frac{C_{D^2B_LW_LHH^\dagger,2}\delta^{I_14}(\tau^{I_2}){}^{i_3}_{i_4}}{\Lambda^4} + \frac{C_{D^2W_L^2HH^\dagger,2}\epsilon^{I_1 I_2 K}(\tau^K){}^{i_3}_{i_4}}{\Lambda^4}\right)\notag \\
    & \times O^{\mathbf{I}_1}_{I_1} O^{\mathbf{I}_2}_{I_2}\left(\mathcal{U}^{\mathbf{I}_3}_{i_3}\mathcal{U}^{\mathbf{I}_4 i_4} + \mathcal{U}^{\mathbf{I}_3i_4}\mathcal{U}_{i_3}^{\mathbf{I}_4}\right)\,, 
    \end{aligned}
\end{equation}
where the first two vector bosons are matched from the gauge bosons, while the last two are matched from the Higgs fields. Considering a specific class $ZZW^+W^-$, the Wilson coefficient can be obtained as
\begin{align}
    c_{ZZW^+W^-,2}^{--00} & \equiv \left(c_{V^4,2}^{--00}\,{}^{ZZW^+W^-}\right) = \frac{1}{\Lambda^4} C_{D^2B_LW_LHH^\dagger,2} (\tau^3)_1^1 \times  \calu^{W^+}_{W^1} \calu^{W^- W^1} O^{Z}_{B} O^Z_{W^3} \times 2\notag \\
    &= \frac{2}{\Lambda^4}\sin\theta_W\cos\theta_W C_{D^2B_LW_LHH^\dagger,2}\,,
\end{align}
where the contribution from $C_{D^2W_L^2HH^\dagger,2}$ vanishes due to the antisymmetric tensor. On the other hand, the massive class $W^+W^-ZZ$ is matched from the other one solely, 
\begin{align}
    c_{W^+W^-ZZ,2}^{--00} & \equiv \left(c_{V^4,2}^{--00}\,{}^{W^+W^-ZZ}\right) = \frac{1}{\Lambda^4}C_{D^2W_L^2HH^\dagger,2} \epsilon^{123}(\tau^3)_2^2 \times O^{W^+}_{W^1} O^{W^-}_{W^2}  \calu^{Z}_{H^2} \calu^{Z H_2^\dagger}\times 2 \notag \\
    &= \frac{1}{2\Lambda^4}C_{D^2W_L^2HH^\dagger,2}\,.
\end{align}
Since other coefficients can be obtained similarly, we do not list them all for the other amplitudes.

\subsection{5-point Matching Result}

\paragraph{\boldsymbol{$f^2\bar{f}{}^2h$}}
\begin{align}
    \left(\mathcal{M}_{f^2\bar{f}{}^2h}\right){}_{\hi{3}\hi{4}}^{\hi{1}\hi{2}h}\left(\mathbf{\frac{1}{2},\frac{1}{2},\frac{1}{2},\frac{1}{2},0}\right) &= \left(c_{f^2\bar{f}{}^2h,1}^{-+--0}\right){}_{\hi{3}\hi{4}}^{\hi{1}\hi{2}h} \lra{\mathbf{13}}[\mathbf{2|p_5|4}\rangle + \left(c_{f^2\bar{f}{}^2h,2}^{-+--0}\right){}_{\hi{3}\hi{4}}^{\hi{1}\hi{2}h} \lra{\mathbf{14}}[\mathbf{2|p_4|3}\rangle \notag \\
    &+ \left(c_{f^2\bar{f}{}^2h,3}^{-+--0}\right){}_{\hi{3}\hi{4}}^{\hi{1}\hi{2}h} \lra{\mathbf{14}}[\mathbf{2|p_5|3}\rangle + \left(c_{f^2\bar{f}{}^2h,1}^{--+-0}\right){}_{\hi{3}\hi{4}}^{\hi{1}\hi{2}h} \lra{\mathbf{12}}[\mathbf{3|p_5|4}\rangle \notag \\
    & + \left(c_{f^2\bar{f}{}^2h,2}^{--+-0}\right){}_{\hi{3}\hi{4}}^{\hi{1}\hi{2}h} \lra{\mathbf{14}}[\mathbf{3|p_4|2}\rangle + \left(c_{f^2\bar{f}{}^2h,3}^{--+-0}\right){}_{\hi{3}\hi{4}}^{\hi{1}\hi{2}h} \lra{\mathbf{14}}[\mathbf{3|p_5|2}\rangle \notag \\
    &+ \left(c_{f^2\bar{f}{}^2h,1}^{----0}\right){}^{\hi{1}\hi{2}h}_{\hi{3}\hi{4}} \lra{\mathbf{12}}\lra{\mathbf{34}} + \left(c_{f^2\bar{f}{}^2h,2}^{----0}\right){}^{\hi{1}\hi{2}h}_{\hi{3}\hi{4}} \lra{\mathbf{13}}\lra{\mathbf{24}} \notag \\
    & + \left(c_{f^2\bar{f}{}^2h}^{++--0}\right){}^{\hi{1}\hi{2}h}_{\hi{3}\hi{4}}[\mathbf{12}]\lra{\mathbf{34}} + \left(c_{f^2\bar{f}{}^2h}^{-+-+0}\right){}^{\hi{1}\hi{2}h}_{\hi{3}\hi{4}}\lra{\mathbf{13}}[\mathbf{24}] \notag \\
    & + \left(c_{f^2\bar{f}{}^2h}^{--++0}\right){}^{\hi{1}\hi{2}h}_{\hi{3}\hi{4}}\lra{\mathbf{12}}[\mathbf{34}] + \text{h.c.} \,,\label{eq:cls_f2bar2h}
\end{align}

\begin{align}
    \left(c_{f^2\bar{f}{}^2h,1}^{-+--0}\right){}_{\hi{3}\hi{4}}^{\hi{1}\hi{2}h} &= \left(\frac{C_{DLe_\mathbbm{C}^\dagger e_\mathbbm{C}^2H^\dagger,1}\delta^{\hat{i}_1}_{i_5}\delta^{\hat{i}_2}_4\delta_{\hat{i}_3}^4\delta_{\hat{i}_4}^4}{\Lambda^4}\right)\Omega_{\hat{i}_1}^{\hi{1}}\Omega_{\hat{i}_2}^{\hi{2}} \Omega^{\hat{i}_3}_{\hi{3}}\Omega^{\hat{i}_4}_{\hi{4}}\mathcal{U}^{i_5h}\,,\\
    \left(c_{f^2\bar{f}{}^2h,2}^{-+--0}\right){}_{\hi{3}\hi{4}}^{\hi{1}\hi{2}h} &= \left(\frac{C_{DLe_\mathbbm{C}^\dagger e_\mathbbm{C}^2H^\dagger,2}\delta^{\hat{i}_1}_{i_5}\delta^{\hat{i}_2}_4\delta_{\hat{i}_3}^4\delta_{\hat{i}_4}^4}{\Lambda^4}\right)\Omega_{\hat{i}_1}^{\hi{1}}\Omega_{\hat{i}_2}^{\hi{2}} \Omega^{\hat{i}_3}_{\hi{3}}\Omega^{\hat{i}_4}_{\hi{4}}\mathcal{U}^{i_5h}\,,\\
    \left(c_{f^2\bar{f}{}^2h,3}^{-+--0}\right){}_{\hi{3}\hi{4}}^{\hi{1}\hi{2}h} &= \left(\frac{C_{DLe_\mathbbm{C}^\dagger e_\mathbbm{C}^2H^\dagger,3}\delta^{\hat{i}_1}_{i_5}\delta^{\hat{i}_2}_4\delta_{\hat{i}_3}^4\delta_{\hat{i}_4}^4}{\Lambda^4}\right)\Omega_{\hat{i}_1}^{\hi{1}}\Omega_{\hat{i}_2}^{\hi{2}} \Omega^{\hat{i}_3}_{\hi{3}}\Omega^{\hat{i}_4}_{\hi{4}}\mathcal{U}^{i_5h}\,,\\
    \left(c_{f^2\bar{f}{}^2h,1}^{--+-0}\right){}_{\hi{3}\hi{4}}^{\hi{1}\hi{2}h} &= \left(\frac{C_{DL^2L^\dagger e_\mathbbm{C} H^\dagger,1}\delta^{\hat{i}_1}_{\hat{i}_3}\delta^{\hat{i}_2}_{i_5}\delta_{\hat{i}_4}^4} {\Lambda^4}\right)\Omega_{\hat{i}_1}^{\hi{1}}\Omega_{\hat{i}_2}^{\hi{2}} \Omega^{\hat{i}_3}_{\hi{3}}\Omega^{\hat{i}_4}_{\hi{4}}\mathcal{U}^{i_5h}\,,\\
    \left(c_{f^2\bar{f}{}^2h,2}^{--+-0}\right){}_{\hi{3}\hi{4}}^{\hi{1}\hi{2}h} &= \left(\frac{C_{DL^2L^\dagger e_\mathbbm{C} H^\dagger,2}\delta^{\hat{i}_1}_{\hat{i}_3}\delta^{\hat{i}_2}_{i_5}\delta_{\hat{i}_4}^4} {\Lambda^4}\right)\Omega_{\hat{i}_1}^{\hi{1}}\Omega_{\hat{i}_2}^{\hi{2}} \Omega^{\hat{i}_3}_{\hi{3}}\Omega^{\hat{i}_4}_{\hi{4}}\mathcal{U}^{i_5h}\,,\\
    \left(c_{f^2\bar{f}{}^2h,3}^{--+-0}\right){}_{\hi{3}\hi{4}}^{\hi{1}\hi{2}h} &= \left(\frac{C_{DL^2L^\dagger e_\mathbbm{C} H^\dagger,3}\delta^{\hat{i}_1}_{\hat{i}_3}\delta^{\hat{i}_2}_{i_5}\delta_{\hat{i}_4}^4} {\Lambda^4}\right)\Omega_{\hat{i}_1}^{\hi{1}}\Omega_{\hat{i}_2}^{\hi{2}} \Omega^{\hat{i}_3}_{\hi{3}}\Omega^{\hat{i}_4}_{\hi{4}}\mathcal{U}^{i_5h}\,,\\
    \left(c_{f^2\bar{f}{}^2h,1}^{----0}\right){}^{\hi{1}\hi{2}h}_{\hi{3}\hi{4}} &= \left(\frac{C_{L^2e_\mathbbm{C}^2H^\dagger{}^2,1}\delta^{\hat{i}_1}_{i_5}\delta^{\hat{i}_2}_{i_6}\delta^4_{\hat{i}_3}\delta^4_{\hat{i}_4}}{\Lambda^4}\right)\Omega_{\hat{i}_1}^{\hi{1}}\Omega_{\hat{i}_2}^{\hi{2}}\Omega_{\hi{3}}^{\hat{i}_3}\Omega_{\hi{4}}^{\hat{i}_4} \mathcal{U}^{i_5 h}\mathcal{U}^{i_6 h}v\,,\\
    \left(c_{f^2\bar{f}{}^2h,2}^{----0}\right){}^{\hi{1}\hi{2}h}_{\hi{3}\hi{4}} &= \left(\frac{C_{L^2e_\mathbbm{C}^2H^\dagger{}^2,2}\delta^{\hat{i}_1}_{i_5}\delta^{\hat{i}_2}_{i_6}\delta^4_{\hat{i}_3}\delta^4_{\hat{i}_4}}{\Lambda^4}\right)\Omega_{\hat{i}_1}^{\hi{1}}\Omega_{\hat{i}_2}^{\hi{2}}\Omega_{\hi{3}}^{\hat{i}_3}\Omega_{\hi{4}}^{\hat{i}_4} \mathcal{U}^{i_5 h}\mathcal{U}^{i_6 h}v\,,\\
    \left(c_{f^2\bar{f}{}^2h}^{++--0}\right){}^{\hi{1}\hi{2}h}_{\hi{3}\hi{4}} &= \left(\frac{C_{e_\mathbbm{C}^\dagger{}^2e_\mathbbm{C}^2HH^\dagger}\delta^{\hat{i}_1}_4\delta^{\hat{i}_2}_4\delta_{\hat{i}_3}^4\delta_{\hat{i}_4}^4\delta^{i_5}_{i_6}}{\Lambda^4}\right)\Omega_{\hat{i}_1}^{\hi{1}}\Omega_{\hat{i}_2}^{\hi{2}}\Omega_{\hi{3}}^{\hat{i}_3}\Omega_{\hi{4}}^{\hat{i}_4} \mathcal{U}_{i_5}^h\mathcal{U}^{i_6 h}v \,,\\
    \left(c_{f^2\bar{f}{}^2h}^{-+-+0}\right){}^{\hi{1}\hi{2}h}_{\hi{3}\hi{4}} &= \left(\frac{C_{Le_\mathbbm{C}^\dagger e_\mathbbm{C} L^\dagger HH^\dagger,1}\delta^{\hat{i}_1}_{i_6}\delta^{\hat{i}_2}_4\delta_{\hat{i}_3}^4\delta_{\hat{i}_4}^{i_5}}{\Lambda^4} + \frac{C_{Le_\mathbbm{C}^\dagger e_\mathbbm{C} L^\dagger HH^\dagger,2}\delta^{\hat{i}_1}_{i_4}\delta^{\hat{i}_2}_4\delta_{\hat{i}_3}^4\delta_{i_5}^{i_6}}{\Lambda^4}\right)\notag \\
    & \times \Omega_{\hat{i}_1}^{\hi{1}}\Omega_{\hat{i}_2}^{\hi{2}}\Omega_{\hi{3}}^{\hat{i}_3}\Omega_{\hi{4}}^{\hat{i}_4} \mathcal{U}_{i_5}^h\mathcal{U}^{i_6 h}v\,,\\
    \left(c_{f^2\bar{f}{}^2h}^{--++0}\right){}^{\hi{1}\hi{2}h}_{\hi{3}\hi{4}} &= \left(\frac{C_{L^2L^\dagger{}^2 HH^\dagger,1}\delta^{\hat{i}_1}_{\hat{i}_3}\delta^{\hat{i}_2}_{i_6}\delta_{\hat{i}_4}^{i_5}}{\Lambda^4} + \frac{C_{L^2L^\dagger{}^2 HH^\dagger,2}\delta^{\hat{i}_1}_{\hat{i}_3}\delta^{\hat{i}_2}_{i_4}\delta_{\hat{i}_6}^{i_5}}{\Lambda^4}\right)\notag \\
    & \times \Omega_{\hat{i}_1}^{\hi{1}}\Omega_{\hat{i}_2}^{\hi{2}}\Omega_{\hi{3}}^{\hat{i}_3}\Omega_{\hi{4}}^{\hat{i}_4} \mathcal{U}_{i_5}^h\mathcal{U}^{i_6 h}v\,.
\end{align}

\paragraph{\boldsymbol{$f^3\bar{f}h$}}
\begin{align}
    \left(\mathcal{M}_{f^3\bar{f}h}\right){}_{\hi{4}}^{\hi{1}\hi{2}\hi{3}h}\left(\mathbf{\frac{1}{2},\frac{1}{2},\frac{1}{2},\frac{1}{2},0}\right) = \left( c_{f^3\bar{f}h}^{----0}\right){}_{\hi{4}}^{\hi{1}\hi{2}\hi{3}h} \lra{\mathbf{12}}\lra{\mathbf{34}} + \text{h.c.}\,, \label{eq:cls_f3barh}
\end{align}

\begin{align}
    \left( c_{f^3\bar{f}h}^{----0}\right){}_{\hi{4}}^{\hi{1}\hi{2}\hi{3}h}&= \left(\frac{C_{L^3e_\mathbbm{C}H}\epsilon^{\hat{i}_1 i_5}\epsilon^{\hat{i}_2 \hat{i}_3}\delta^{\hat{i}_4}_4}{\Lambda^3}\right)\Omega_{\hat{i}_1}^{\hi{1}}\Omega_{\hat{i}_2}^{\hi{2}}\Omega_{\hat{i}_3}^{\hi{3}}\Omega_{\hat{i}_4}^{\hi{4}}\mathcal{U}_{i_5}^h\,.
\end{align}

\paragraph{\boldsymbol{$f^2\bar{f}{}^2V$}}
\begin{align}
    \left(\mathcal{M}_{f^2\bar{f}{}^2V}\,{}^{\mathbf{I}_5}\right){}_{\hi{3}\hi{4}}^{\hi{1}\hi{2}}\left(\mathbf{\frac{1}{2},\frac{1}{2},\frac{1}{2},\frac{1}{2},1}\right) &= \left(c_{f^2\bar{f}{}^2V}^{++---}\,{}^{\mathbf{I}_5}\right){}_{\hi{3}\hi{4}}^{\hi{1}\hi{2}} [\mathbf{12}]\lra{\mathbf{35}}\lra{\mathbf{45}} +\left(c_{f^2\bar{f}{}^2V}^{--++-}\,{}^{\mathbf{I}_5}\right){}_{\hi{3}\hi{4}}^{\hi{1}\hi{2}} [\mathbf{34}]\lra{\mathbf{15}}\lra{\mathbf{25}}\notag \\
    & + \left(c_{f^2\bar{f}{}^2V}^{-+-+-}\,{}^{\mathbf{I}_5}\right){}_{\hi{3}\hi{4}}^{\hi{1}\hi{2}} [\mathbf{24}]\lra{\mathbf{15}}\lra{\mathbf{35}}+ \left(c_{f^2\bar{f}{}^2h,1}^{-+--0}\,{}^{\mathbf{I}_5}\right){}_{\hi{3}\hi{4}}^{\hi{1}\hi{2}} \lra{\mathbf{13}}[\mathbf{25}]\lra{\mathbf{45}}\notag \\
    & + \left(c_{f^2\bar{f}{}^2h,2}^{-+--0},{}^{\mathbf{I}_5}\right){}_{\hi{3}\hi{4}}^{\hi{1}\hi{2}} \lra{\mathbf{14}}[\mathbf{25}]\lra{\mathbf{35}} + \left(c_{f^2\bar{f}{}^2h,1}^{--+-0},{}^{\mathbf{I}_5}\right){}_{\hi{3}\hi{4}}^{\hi{1}\hi{2}} \lra{\mathbf{12}}[\mathbf{35}]\lra{\mathbf{45}}\notag \\
    & + \left(c_{f^2\bar{f}{}^2h,2}^{--+-0},{}^{\mathbf{I}_5}\right){}_{\hi{3}\hi{4}}^{\hi{1}\hi{2}} \lra{\mathbf{14}}[\mathbf{35}]\lra{\mathbf{25}} + \textbf{h.c.}\,, \label{eq:cls_f2bar2V}
\end{align}

\begin{align}
    \left(c_{f^2\bar{f}{}^2V}^{++---}\,{}^{\mathbf{I}_5}\right){}_{\hi{3}\hi{4}}^{\hi{1}\hi{2}} &= \left(\frac{C_{e_\mathbbm{C}^\dagger{}^2 e_\mathbbm{C}^2B_L}\delta^{\hat{i}_1}_4\delta^{\hat{i}_2}_4\delta^4_{\hat{i}_3}\delta^4_{\hat{i}_4}\delta^{I_54}}{\Lambda^4} \right)\Omega_{\hat{i}_1}^{\hi{1}}\Omega_{\hat{i}_2}^{\hi{2}}\Omega^{\hat{i}_3}_{\hi{3}}\Omega^{\hat{i}_4}_{\hi{4}}O^{\mathbf{I}_5}_{I_5}\,,\\
    \left(c_{f^2\bar{f}{}^2V}^{--++-}\,{}^{\mathbf{I}_5}\right){}_{\hi{3}\hi{4}}^{\hi{1}\hi{2}} &= \left(\frac{C_{L^2L^\dagger{}^2B_L}\delta^{\hat{i}_1}_{\hat{i}_3}\delta^{\hat{i}_2}_{\hat{i}_4}\delta^{I_54}}{\Lambda^4} + \frac{C_{L^2L^\dagger{}^2W_L}(\tau^{I_5}){}^{\hat{i}_1}_{\hat{i}_3}\delta^{\hat{i}_2}_{\hat{i}_4}}{\Lambda^4} \right)\Omega_{\hat{i}_1}^{\hi{1}}\Omega_{\hat{i}_2}^{\hi{2}}\Omega^{\hat{i}_3}_{\hi{3}}\Omega^{\hat{i}_4}_{\hi{4}}O^{\mathbf{I}_5}_{I_5}\,,\\
    \left(c_{f^2\bar{f}{}^2V}^{-+-+-}\,{}^{\mathbf{I}_5}\right){}_{\hi{3}\hi{4}}^{\hi{1}\hi{2}} &= \left(\frac{C_{Le_\mathbbm{C}^\dagger e_\mathbbm{C}L^\dagger B_L}\delta^{\hat{i}_1}_{\hat{i}_4}\delta^{\hat{i}_2}_4\delta_{\hat{i}_3}^4\delta^{I_54}}{\Lambda^4} + \frac{C_{Le_\mathbbm{C}^\dagger e_\mathbbm{C}L^\dagger W_L}(\tau^{I_5}){}^{\hat{i}_1}_{\hat{i}_4}\delta^{\hat{i}_2}_4\delta_{\hat{i}_3}^4}{\Lambda^4}\right)\notag \\
    & \times \Omega_{\hat{i}_1}^{\hi{1}}\Omega_{\hat{i}_2}^{\hi{2}}\Omega^{\hat{i}_3}_{\hi{3}}\Omega^{\hat{i}_4}_{\hi{4}}O^{\mathbf{I}_5}_{I_5}\,,\\
    \left(c_{f^2\bar{f}{}^2V,1}^{-+--0}\,{}^{\mathbf{I}_5}\right){}_{\hi{3}\hi{4}}^{\hi{1}\hi{2}} &= \left(\frac{C_{DLe_\mathbbm{C}^\dagger e_\mathbbm{C}^2H^\dagger,1}\delta^{\hat{i}_1}_{i_5}\delta^{\hat{i}_2}_4\delta_{\hat{i}_3}^4\delta_{\hat{i}_4}^4}{\Lambda^4}\right)\Omega_{\hat{i}_1}^{\hi{1}}\Omega_{\hat{i}_2}^{\hi{2}} \Omega^{\hat{i}_3}_{\hi{3}}\Omega^{\hat{i}_4}_{\hi{4}}\mathcal{U}^{i_5\mathbf{I}_5}\,,\\
    \left(c_{f^2\bar{f}{}^2V,2}^{-+--0}\,{}^{\mathbf{I}_5}\right){}_{\hi{3}\hi{4}}^{\hi{1}\hi{2}} &= \left(\frac{C_{DLe_\mathbbm{C}^\dagger e_\mathbbm{C}^2H^\dagger,3}\delta^{\hat{i}_1}_{i_5}\delta^{\hat{i}_2}_4\delta_{\hat{i}_3}^4\delta_{\hat{i}_4}^4}{\Lambda^4}\right)\Omega_{\hat{i}_1}^{\hi{1}}\Omega_{\hat{i}_2}^{\hi{2}} \Omega^{\hat{i}_3}_{\hi{3}}\Omega^{\hat{i}_4}_{\hi{4}}\mathcal{U}^{i_5\mathbf{I}_5}\,,\\
    \left(c_{f^2\bar{f}{}^2V,1}^{--+-0}\,{}^{\mathbf{I}_5}\right){}_{\hi{3}\hi{4}}^{\hi{1}\hi{2}} &= \left(\frac{C_{DL^2L^\dagger e_\mathbbm{C} H^\dagger,1}\delta^{\hat{i}_1}_{\hat{i}_3}\delta^{\hat{i}_2}_{i_5}\delta_{\hat{i}_4}^4} {\Lambda^4}\right)\Omega_{\hat{i}_1}^{\hi{1}}\Omega_{\hat{i}_2}^{\hi{2}} \Omega^{\hat{i}_3}_{\hi{3}}\Omega^{\hat{i}_4}_{\hi{4}}\mathcal{U}^{i_5\mathbf{I}_5}\,,\\
    \left(c_{f^2\bar{f}{}^2V,2}^{--+-0}\,{}^{\mathbf{I}_5}\right){}_{\hi{3}\hi{4}}^{\hi{1}\hi{2}} &= \left(\frac{C_{DL^2L^\dagger e_\mathbbm{C} H^\dagger,3}\delta^{\hat{i}_1}_{\hat{i}_3}\delta^{\hat{i}_2}_{i_5}\delta_{\hat{i}_4}^4} {\Lambda^4}\right)\Omega_{\hat{i}_1}^{\hi{1}}\Omega_{\hat{i}_2}^{\hi{2}} \Omega^{\hat{i}_3}_{\hi{3}}\Omega^{\hat{i}_4}_{\hi{4}}\mathcal{U}^{i_5\mathbf{I}_5}\,,
\end{align}

\paragraph{\boldsymbol{$f\bar{f}h^3$}}

\begin{align}
    \left(\mathcal{M}_{f\bar{f}h^3}\right){}_{\hi{2}}^{\hi{1}hhh} &= \left(c_{f\bar{f}h^3,1}^{--000}\right){}_{\hi{2}}^{\hi{1}hhh}\lra{\mathbf{12}} +  \left(c_{f\bar{f}h^3,2}^{--000}\right){}_{\hi{2}}^{\hi{1}hhh}\lra{\mathbf{12}}s_{45} + \left(c_{f\bar{f}h^3,3}^{--000}\right){}_{\hi{2}}^{\hi{1}hhh}\lra{\mathbf{12}}s_{35}\notag \\
    & + \left(c_{f\bar{f}h^3,4}^{--000}\right){}_{\hi{2}}^{\hi{1}hhh}\lra{\mathbf{12}}s_{34}+ \left(c_{f\bar{f}h^3,5}^{--000}\right){}_{\hi{2}}^{\hi{1}hhh}\lra{\mathbf{1|p_4p_5|2}}\notag \\
    & + \left(c_{f\bar{f}h^3,6}^{--000}\right){}_{\hi{2}}^{\hi{1}hhh}\lra{\mathbf{1|p_3p_5|2}} + \left(c_{f\bar{f}h^3,7}^{--000}\right){}_{\hi{2}}^{\hi{1}hhh}\lra{\mathbf{1|p_3p_4|2}}\notag \\
    & +\left(c_{f\bar{f}h^3}^{+-000}\right){}^{\hi{1}hhh}_{\hi{2}}[\mathbf{1|p_3|2}\rangle + \left(c_{f\bar{f}h^3,1}^{-+000}\right){}^{\hi{1}hhh}_{\hi{2}}[\mathbf{2|p_3|1}\rangle \notag \\
    & + \left(c_{f\bar{f}h^3,2}^{-+000}\right){}^{\hi{1}hhh}_{\hi{2}}[\mathbf{2|p_4|1}\rangle + \left(c_{f\bar{f}h^3,3}^{-+000}\right){}^{\hi{1}hhh}_{\hi{2}}[\mathbf{2|p_5|1}\rangle + \text{h.c.}\,, \label{eq:cls_fbarh3}
\end{align}

\begin{align}
    \left(c_{f\bar{f}h^3,1}^{--000}\right){}_{\hi{2}}^{\hi{1}hhh} &= \left(\frac{C_{Le_\mathbbm{C}HH^\dagger{}^2}\delta^{\hat{i}_1}_{i_4}\delta^{i_3}_{i_5}\delta_{\hat{i}_2}^4}{\Lambda^2} \right) \Omega_{\hat{i}_1}^{\hi{1}}\Omega_{\hi{2}}^{\hat{i}_2}\mathcal{U}_{i_3}^{h} \mathcal{U}^{i_4h} \mathcal{U}^{i_5h}\,,\\
    \left(c_{f\bar{f}h^3,2}^{--000}\right){}_{\hi{2}}^{\hi{1}hhh} &= \left(\frac{C_{D^2Le_\mathbbm{C}HH^\dagger{}^2,1}\delta^{\hat{i}_1}_{i_4}\delta^{i_3}_{i_5}\delta_{\hat{i}_2}^4}{\Lambda^4}\right)\Omega_{\hat{i}_1}^{\hi{1}}\Omega_{\hi{2}}^{\hat{i}_2}\mathcal{U}_{i_3}^{h} \mathcal{U}^{i_4h} \mathcal{U}^{i_5h}\,,\\
    \left(c_{f\bar{f}h^3,3}^{--000}\right){}_{\hi{2}}^{\hi{1}hhh} &= \left(\frac{C_{D^2Le_\mathbbm{C}HH^\dagger{}^2,2}\delta^{\hat{i}_1}_{i_4}\delta^{i_3}_{i_5}\delta_{\hat{i}_2}^4}{\Lambda^4}\right)\Omega_{\hat{i}_1}^{\hi{1}}\Omega_{\hi{2}}^{\hat{i}_2}\mathcal{U}_{i_3}^{h} \mathcal{U}^{i_4h} \mathcal{U}^{i_5h}\,,\\
    \left(c_{f\bar{f}h^3,4}^{--000}\right){}_{\hi{2}}^{\hi{1}hhh} &= \left(\frac{C_{D^2Le_\mathbbm{C}HH^\dagger{}^2,3}\delta^{\hat{i}_1}_{i_4}\delta^{i_3}_{i_5}\delta_{\hat{i}_2}^4}{\Lambda^4}\right)\Omega_{\hat{i}_1}^{\hi{1}}\Omega_{\hi{2}}^{\hat{i}_2}\mathcal{U}_{i_3}^{h} \mathcal{U}^{i_4h} \mathcal{U}^{i_5h}\,,\\
    \left(c_{f\bar{f}h^3,5}^{--000}\right){}_{\hi{2}}^{\hi{1}hhh} &= \left(\frac{C_{D^2Le_\mathbbm{C}HH^\dagger{}^2,4}\delta^{\hat{i}_1}_{i_4}\delta^{i_3}_{i_5}\delta_{\hat{i}_2}^4}{\Lambda^4}\right)\Omega_{\hat{i}_1}^{\hi{1}}\Omega_{\hi{2}}^{\hat{i}_2}\mathcal{U}_{i_3}^{h} \mathcal{U}^{i_4h} \mathcal{U}^{i_5h}\,,\\
    \left(c_{f\bar{f}h^3,6}^{--000}\right){}_{\hi{2}}^{\hi{1}hhh} &= \left(\frac{C_{D^2Le_\mathbbm{C}HH^\dagger{}^2,5}\delta^{\hat{i}_1}_{i_4}\delta^{i_3}_{i_5}\delta_{\hat{i}_2}^4}{\Lambda^4}\right)\Omega_{\hat{i}_1}^{\hi{1}}\Omega_{\hi{2}}^{\hat{i}_2}\mathcal{U}_{i_3}^{h} \mathcal{U}^{i_4h} \mathcal{U}^{i_5h}\,,\\
    \left(c_{f\bar{f}h^3,7}^{--000}\right){}_{\hi{2}}^{\hi{1}hhh} &= \left(\frac{C_{D^2Le_\mathbbm{C}HH^\dagger{}^2,6}\delta^{\hat{i}_1}_{i_4}\delta^{i_3}_{i_5}\delta_{\hat{i}_2}^4}{\Lambda^4}\right)\Omega_{\hat{i}_1}^{\hi{1}}\Omega_{\hi{2}}^{\hat{i}_2}\mathcal{U}_{i_3}^{h} \mathcal{U}^{i_4h} \mathcal{U}^{i_5h}\,,\\
    \left(c_{f\bar{f}h^3}^{+-000}\right){}^{\hi{1}hhh}_{\hi{2}} &= \left(\frac{C_{De_\mathbbm{C}^\dagger e_\mathbbm{C}H^2H^\dagger{}^2}\delta^{\hat{i}_1}_4\delta_{\hat{i}_2}^4\delta^{i_3}_{i_5}\delta^{i_4}_{i_6}}{\Lambda^4}\right)\Omega_{\hat{i}_1}^{\hi{1}}\Omega_{\hi{2}}^{\hat{i}_2}\mathcal{U}_{i_3}^h\mathcal{U}_{i_4}^h \mathcal{U}^{i_5 h}\mathcal{U}^{i_6h}v\,,\\
    \left(c_{f\bar{f}h^3,1}^{-+000}\right){}^{\hi{1}hhh}_{\hi{2}} &= \left(\frac{C_{DLL^\dagger H^2H^\dagger{}^2,1}\delta^{\hat{i}_1}_{\hat{i}_2}\delta^{i_3}_{i_5}\delta^{i_4}_{i_6}}{\Lambda^4} + \frac{C_{DLL^\dagger H^2H^\dagger{}^2,2}\delta^{i_3}_{\hat{i}_2}\delta^{\hat{i}_1}_{i_5}\delta^{i_4}_{i_6}}{\Lambda^4}\right) \notag \\
    & \times \Omega_{\hat{i}_1}^{\hi{1}}\Omega_{\hi{2}}^{\hat{i}_2}\mathcal{U}_{i_3}^h\mathcal{U}_{i_4}^h \mathcal{U}^{i_5 h}\mathcal{U}^{i_6h}v\,,\\
    \left(c_{f\bar{f}h^3,2}^{-+000}\right){}^{\hi{1}hhh}_{\hi{2}} &= \left(\frac{C_{DLL^\dagger H^2H^\dagger{}^2,3}\delta^{i_4}_{\hat{i}_2}\delta^{\hat{i}_1}_{i_5}\delta^{i_3}_{i_6}}{\Lambda^4}\right) \Omega_{\hat{i}_1}^{\hi{1}}\Omega_{\hi{2}}^{\hat{i}_2}\mathcal{U}_{i_3}^h\mathcal{U}_{i_4}^h \mathcal{U}^{i_5 h}\mathcal{U}^{i_6h}v\,,\\
    \left(c_{f\bar{f}h^3,3}^{-+000}\right){}^{\hi{1}hhh}_{\hi{2}} &= \left(\frac{C_{DLL^\dagger H^2H^\dagger{}^2,4}\delta^{i_3}_{\hat{i}_2}\delta^{\hat{i}_1}_{i_5}\delta^{i_4}_{i_6}}{\Lambda^4}\right) \Omega_{\hat{i}_1}^{\hi{1}}\Omega_{\hi{2}}^{\hat{i}_2}\mathcal{U}_{i_3}^h\mathcal{U}_{i_4}^h \mathcal{U}^{i_5 h}\mathcal{U}^{i_6h}v\,.
\end{align}


\paragraph{\boldsymbol{$f^2h^3$}}

\begin{align}
    & \mathcal{M}_{f^2h^3}^{\hi{1}\hi{1}hhh}\left(\mathbf{\frac{1}{2},\frac{1}{2},0,0,0}\right) = 
    \left(c_{f^2h^3}^{-+000}\right){}^{\hi{1}\hi{1}hhh} \langle\mathbf{1|p_3|2}] 
    + \left(c_{f^2h^3}^{--000}\right){}^{\hi{1}\hi{1}hhh} \lra{\mathbf{12}} 
    + \text{h.c.}\,, \label{eq:cls_f2h3}
\end{align}

\begin{align}
    \left(c_{f^2h^3}^{-+000}\right){}^{\hi{1}\hi{1}hhh} &= \left(\frac{C_{DLe_\mathbbm{C}^\dagger H^3}\epsilon^{\hat{i}_1 i_5}\epsilon^{i_3 i_4}\delta^{\hat{i}_2}_4}{\Lambda^3}\right)\Omega_{\hat{i}_1}^{\hi{1}}\Omega_{\hat{i}_2}^{\hi{2}}\mathcal{U}^{h}_{i_3} \mathcal{U}^{h}_{i_4}\mathcal{U}^{h}_{i_5}\,,\\
    \left(c_{f^2h^3}^{--000}\right){}^{\hi{1}\hi{1}hhh} &= \left(\frac{C_{L^2H^3H^\dagger}\epsilon^{\hat{i}_1 i_3}\epsilon^{\hat{i}_2 i_4}\delta^{i_5}_{i_6}}{\Lambda^3}\right) \Omega_{\hat{i}_1}^{\hi{1}} \Omega_{\hat{i}_2}^{\hi{2}} \mathcal{U}_{i_3}^h \mathcal{U}_{i_4}^h \mathcal{U}_{i_5}^h \mathcal{U}^{i_6h} v\,.
\end{align}

\paragraph{\boldsymbol{$f^2Vh^2$}}
\begin{align}
    \left(\mathcal{M}_{f^2Vh^2}^{\mathbf{I}_3}\right){}^{\hi{1}\hi{2}hh}\left(\mathbf{\frac{1}{2},\frac{1}{2},1,0,0}\right) = \left(c_{f^2Vh^2}^{---00}\,^{\mathbf{I}_3}\right){}^{\hi{1}\hi{2}hh}\lra{\mathbf{1}\mathbb{3}}\lra{\mathbf{2}\mathbb{3}} \,, \label{eq:cls_f2Vh2}
\end{align}

\begin{align}
    \left(c_{f^2Vh^2}^{---00}\,^{\mathbf{I}_3}\right){}^{\hi{1}\hi{2}hh} &= \left(\frac{C_{L^2B_LH^2}\epsilon^{\hat{i}_1 i_4}\epsilon^{\hat{i}_2 i_5}\delta^{I_43}}{\Lambda^3} + \frac{C_{L^2W_LH^2}\epsilon^{\hat{i}_1 i_4}(\tau^{I_3})^{\hat{i}_2 i_5}}{\Lambda^3}\right)\Omega_{\hat{i}_1}^{\hi{1}}\Omega_{\hat{i}_2}^{\hi{2}}O^{\mathbf{I}_3}_{I_3} \mathcal{U}_{i_4}^h\mathcal{U}_{i_5}^h\,,
\end{align}

\paragraph{\boldsymbol{$f\bar{f}Vh^2$}}
\begin{align}
    \left(\mathcal{M}_{f\bar{f}Vh^2}^{\mathbf{I}_3}\right){}^{\hi{1}hh}_{\hi{2}}\left(\mathbf{\frac{1}{2},\frac{1}{2},1,0,0}\right) &= \left(c_{f\bar{f}Vh^2,1}^{+--00}\,^{\mathbf{I}_3}\right){}^{\hi{1}hh}_{\hi{2}}[\mathbf{1|p_5|}\mathbb{3}\rangle \lra{\mathbf{2}\mathbb{3}} +\left(c_{f\bar{f}Vh^2,2}^{+--00}\,^{\mathbf{I}_3}\right){}^{\hi{1}hh}_{\hi{2}}[\mathbf{1|p_4|}\mathbb{3}\rangle \lra{\mathbf{2}\mathbb{3}}\notag \\
    & +\left(c_{f\bar{f}Vh^2,1}^{-+-00}\,^{\mathbf{I}_3}\right){}^{\hi{1}hh}_{\hi{2}}\lra{\mathbf{1}\mathbb{3}}[\mathbf{2|p_5|}\mathbb{3}\rangle +\left(c_{f\bar{f}Vh^2,2}^{-+-00}\,^{\mathbf{I}_3}\right){}^{\hi{1}hh}_{\hi{2}}\lra{\mathbf{1}\mathbb{3}}[\mathbf{2|p_4|}\mathbb{3}\rangle\notag \\
    & + \left(c_{f\bar{f}Vh^2,1}^{--000}\,{}^{\mathbf{I}_3}\right){}_{\hi{2}}^{\hi{1}hh}\lra{\mathbf{12}}\langle\mathbf{3|p_5|3}] + \left(c_{f\bar{f}Vh^2,2}^{--000}\,{}^{\mathbf{I}_3}\right){}_{\hi{2}}^{\hi{1}hh}\lra{\mathbf{12}}\langle\mathbf{3|p_4|3}]\notag \\
    & + \left(c_{f\bar{f}Vh^2,3}^{--000}\,{}^{\mathbf{I}_3}\right){}_{\hi{2}}^{\hi{1}hh}\lra{\mathbf{13}}[\mathbf{3|p_5|2}\rangle + \left(c_{f\bar{f}Vh^2,4}^{--000}\,{}^{\mathbf{I}_3}\right){}_{\hi{2}}^{\hi{1}hh}\lra{\mathbf{13}}[\mathbf{3|p_4|2}\rangle \notag \\
    & + \left(c^{---00}_{f\bar{f}Vh^2}\,{}^{\mathbf{I}_3}\right){}_{\hi{2}}^{\hi{1}hh} \lra{\mathbf{1}\mathbb{3}}\lra{\mathbf{2}\mathbb{3}} +\left(c_{f\bar{f}Vh^2}^{+-000}\,{}^{\mathbf{I}_3}\right){}^{\hi{1}hh}_{\hi{2}}[\mathbf{1|p_3|2}\rangle\notag \\
    & + \left(c_{f\bar{f}Vh^2,1}^{-+000}\,{}^{\mathbf{I}_3}\right){}^{\hi{1}hh}_{\hi{2}}[\mathbf{2|p_3|1}\rangle + \left(c_{f\bar{f}Vh^2,2}^{-+000}\,{}^{\mathbf{I}_3}\right){}^{\hi{1}hh}_{\hi{2}}[\mathbf{2|p_4|1}\rangle \notag \\
    & + \left(c_{f\bar{f}Vh^2,3}^{-+000}\,{}^{\mathbf{I}_3}\right){}^{\hi{1}hh}_{\hi{2}}[\mathbf{2|p_5|1}\rangle + \text{h.c.} \,,\label{eq:cls_fbarVh2}
\end{align}

\begin{align}
    \left(c_{f\bar{f}Vh^2,1}^{+--00}\,^{\mathbf{I}_3}\right){}^{\hi{1}hh}_{\hi{2}} &= \left(\frac{C_{De_\mathbbm{C}^\dagger e_{\mathbbm{C}}B_LHH^\dagger,1}\delta^{\hat{i}_1}_4\delta^4_{\hat{i}_2}\delta^{I_34}\delta^{i_4}_{i_5}}{\Lambda^4} + \frac{C_{De_\mathbbm{C}^\dagger e_{\mathbbm{C}}W_LHH^\dagger,1}\delta^{\hat{i}_1}_4\delta^4_{\hat{i}_2}(\tau^{I_3}){}^{i_4}_{i_5}}{\Lambda^4}\right)\notag \\
    & \times \Omega_{\hat{i}_1}^{\hi{1}}\Omega^{\hat{i}_2}_{\hi{2}}O^{\mathbf{I}_3}_{I_3}\mathcal{U}^{h}_{i_4}\mathcal{U}^{i_5h}\\
    \left(c_{f\bar{f}Vh^2,2}^{+--00}\,^{\mathbf{I}_3}\right){}^{\hi{1}hh}_{\hi{2}} &= \left(\frac{C_{De_\mathbbm{C}^\dagger e_{\mathbbm{C}}B_LHH^\dagger,2}\delta^{\hat{i}_1}_4\delta^4_{\hat{i}_2}\delta^{I_34}\delta^{i_4}_{i_5}}{\Lambda^4} + \frac{C_{De_\mathbbm{C}^\dagger e_{\mathbbm{C}}W_LHH^\dagger,2}\delta^{\hat{i}_1}_4\delta^4_{\hat{i}_2}(\tau^{I_3}){}^{i_4}_{i_5}}{\Lambda^4}\right)\notag \\
    & \times \Omega_{\hat{i}_1}^{\hi{1}}\Omega^{\hat{i}_2}_{\hi{2}}O^{\mathbf{I}_3}_{I_3}\mathcal{U}^{h}_{i_4}\mathcal{U}^{i_5h}\\
    \left(c_{f\bar{f}Vh^2,1}^{-+-00}\,^{\mathbf{I}_3}\right){}^{\hi{1}hh}_{\hi{2}} &= \left(\frac{C_{DLL^\dagger B_LHH^\dagger,1} \delta^{\hat{i}_1}_{\hat{i}_2}\delta^{i_4}_{i_5}\delta^{I_34}}{\Lambda^4} + \frac{C_{DLL^\dagger B_LHH^\dagger,2} \delta^{\hat{i}_1}_{i_5}\delta^{i_4}_{\hat{i}_2}\delta^{I_34}}{\Lambda^4} + \frac{C_{DLL^\dagger W_LHH^\dagger,1} (\tau^{I_3}){}^{\hat{i}_1}_{\hat{i}_2}\delta^{i_4}_{i_5}}{\Lambda^4}\right.\notag \\
    &\left.+ \frac{C_{DLL^\dagger W_LHH^\dagger,2} (\tau^{I_3}){}^{\hat{i}_1}_{i_5}\delta^{i_4}_{\hat{i}_2}}{\Lambda^4} + \frac{C_{DLL^\dagger W_LHH^\dagger,3} (\tau^{I_3}){}^{i_4}_{i_5}\delta^{\hat{i}_1}_{\hat{i}_2}}{\Lambda^4}\right)\Omega_{\hat{i}_1}^{\hi{1}}\Omega^{\hat{i}_2}_{\hi{2}}O^{\mathbf{I}_3}_{I_3}\mathcal{U}^{h}_{i_4}\mathcal{U}^{i_5h}\,,\\
    \left(c_{f\bar{f}Vh^2,2}^{-+-00}\,^{\mathbf{I}_3}\right){}^{\hi{1}hh}_{\hi{2}} &= \left(\frac{C_{DLL^\dagger B_LHH^\dagger,3} \delta^{\hat{i}_1}_{\hat{i}_2}\delta^{i_4}_{i_5}\delta^{I_34}}{\Lambda^4} + \frac{C_{DLL^\dagger B_LHH^\dagger,4} \delta^{\hat{i}_1}_{i_5}\delta^{i_4}_{\hat{i}_2}\delta^{I_34}}{\Lambda^4} + \frac{C_{DLL^\dagger W_LHH^\dagger,4} (\tau^{I_3}){}^{\hat{i}_1}_{\hat{i}_2}\delta^{i_4}_{i_5}}{\Lambda^4}\right.\notag \\
    &\left.+ \frac{C_{DLL^\dagger W_LHH^\dagger,5} (\tau^{I_3}){}^{\hat{i}_1}_{i_5}\delta^{i_4}_{\hat{i}_2}}{\Lambda^4} + \frac{C_{DLL^\dagger W_LHH^\dagger,6} (\tau^{I_3}){}^{i_4}_{i_5}\delta^{\hat{i}_1}_{\hat{i}_2}}{\Lambda^4}\right)\Omega_{\hat{i}_1}^{\hi{1}}\Omega^{\hat{i}_2}_{\hi{2}}O^{\mathbf{I}_3}_{I_3}\mathcal{U}^{h}_{i_4}\mathcal{U}^{i_5h}\,,\\
     \left(c_{f\bar{f}Vh^2,1}^{--000}\,{}^{\mathbf{I}_3}\right){}_{\hi{2}}^{\hi{1}hh} &= \left(\frac{C_{D^2Le_\mathbbm{C}HH^\dagger{}^2,2}\delta^{\hat{i}_1}_{i_4}\delta^{i_3}_{i_5}\delta_{\hat{i}_2}^4}{\Lambda^4}\right)\Omega_{\hat{i}_1}^{\hi{1}}\Omega_{\hi{2}}^{\hat{i}_2}\mathcal{U}^{i_4h} \left(\mathcal{U}_{i_3}^{\mathbf{I}_3} \mathcal{U}^{i_5h}+\mathcal{U}^{i_5\mathbf{I}_3} \mathcal{U}_{i_3}^{h}\right)\,,\\
     \left(c_{f\bar{f}Vh^2,2}^{--000}\,{}^{\mathbf{I}_3}\right){}_{\hi{2}}^{\hi{1}hh} &= \left(\frac{C_{D^2Le_\mathbbm{C}HH^\dagger{}^2,3}\delta^{\hat{i}_1}_{i_4}\delta^{i_3}_{i_5}\delta_{\hat{i}_2}^4}{\Lambda^4}\right)\Omega_{\hat{i}_1}^{\hi{1}}\Omega_{\hi{2}}^{\hat{i}_2}\mathcal{U}^{i_4h} \left(\mathcal{U}_{i_3}^{\mathbf{I}_3} \mathcal{U}^{i_5h}+\mathcal{U}^{i_5\mathbf{I}_3} \mathcal{U}_{i_3}^{h}\right)\,,\\
     \left(c_{f\bar{f}Vh^2,3}^{--000}\,{}^{\mathbf{I}_3}\right){}_{\hi{2}}^{\hi{1}hh} &= \left(\frac{C_{D^2Le_\mathbbm{C}HH^\dagger{}^2,5}\delta^{\hat{i}_1}_{i_4}\delta^{i_3}_{i_5}\delta_{\hat{i}_2}^4}{\Lambda^4}\right)\Omega_{\hat{i}_1}^{\hi{1}}\Omega_{\hi{2}}^{\hat{i}_2}\mathcal{U}^{i_4h} \left(\mathcal{U}_{i_3}^{\mathbf{I}_3} \mathcal{U}^{i_5h}+\mathcal{U}^{i_5\mathbf{I}_3} \mathcal{U}_{i_3}^{h}\right)\,,\\
     \left(c_{f\bar{f}Vh^2,4}^{--000}\,{}^{\mathbf{I}_3}\right){}_{\hi{2}}^{\hi{1}hh} &= \left(\frac{C_{D^2Le_\mathbbm{C}HH^\dagger{}^2,6}\delta^{\hat{i}_1}_{i_4}\delta^{i_3}_{i_5}\delta_{\hat{i}_2}^4}{\Lambda^4}\right)\Omega_{\hat{i}_1}^{\hi{1}}\Omega_{\hi{2}}^{\hat{i}_2}\mathcal{U}^{i_4h} \left(\mathcal{U}_{i_3}^{\mathbf{I}_3} \mathcal{U}^{i_5h}+\mathcal{U}^{i_5\mathbf{I}_3} \mathcal{U}_{i_3}^{h}\right)\,,\\
     \left(c^{---00}_{f\bar{f}Vh^2}\,{}^{\mathbf{I}_3}\right){}_{\hi{2}}^{\hi{1}hh}  &= \left(\frac{C_{Le_{\mathbbm{C}}B_LHH^\dagger{}^2}\delta^{\hat{i}_1}_{i_5}\delta^{i_4}_{i_6}\delta_{\hat{i}_2}^4\delta^{I_34}}{\Lambda^4} + \frac{C_{Le_{\mathbbm{C}}W_LHH^\dagger{}^2,1}\delta^{\hat{i}_1}_{i_5}(\tau^{I_3}){}^{i_4}_{i_6}\delta_{\hat{i}_2}^4}{\Lambda^4}\right.\notag \\
    & \left.+ \frac{C_{Le_{\mathbbm{C}}W_LHH^\dagger{}^2,2}\delta^{i_4}_{i_6}(\tau^{I_3}){}^{\hat{i}_1}_{i_5}\delta_{\hat{i}_2}^4}{\Lambda^4}\right) \Omega_{\hat{i}_1}^{\hi{1}}\Omega^{\hat{i}_2}_{\hi{2}}\mathcal{U}^{\mathbf{I}_3}_{I_3}\mathcal{U}_{i_4}^h\mathcal{U}^{i_5h}\mathcal{U}^{i_6h}v\,,\\
    \left(c_{f\bar{f}Vh^2}^{+-000}\,{}^{\mathbf{I}_3}\right){}^{\hi{1}hh}_{\hi{2}} &= \left(\frac{C_{De_\mathbbm{C}^\dagger e_\mathbbm{C}H^2H^\dagger{}^2}\delta^{\hat{i}_1}_4\delta_{\hat{i}_2}^4\delta^{i_3}_{i_5}\delta^{i_4}_{i_6}}{\Lambda^4}\right)\Omega_{\hat{i}_1}^{\hi{1}}\Omega_{\hi{2}}^{\hat{i}_2}\mathcal{U}_{i_3}^h\mathcal{U}_{i_4}^h \mathcal{U}^{i_5 h}\mathcal{U}^{i_6h}v\,,\\
    \left(c_{f\bar{f}Vh^2,1}^{-+000}\,{}^{\mathbf{I}_3}\right){}^{\hi{1}hh}_{\hi{2}} &= \left(\frac{C_{DLL^\dagger H^2H^\dagger{}^2,1}\delta^{\hat{i}_1}_{\hat{i}_2}\delta^{i_3}_{i_5}\delta^{i_4}_{i_6}}{\Lambda^4} + \frac{C_{DLL^\dagger H^2H^\dagger{}^2,2}\delta^{i_3}_{\hat{i}_2}\delta^{\hat{i}_1}_{i_5}\delta^{i_4}_{i_6}}{\Lambda^4}\right)\notag \\
    & \times \Omega_{\hat{i}_1}^{\hi{1}}\Omega_{\hi{2}}^{\hat{i}_2}\mathcal{U}_{i_3}^h\mathcal{U}_{i_4}^h \mathcal{U}^{i_5 h}\mathcal{U}^{i_6h}v\,,\\
    \left(c_{f\bar{f}Vh^2,2}^{-+000}\,{}^{\mathbf{I}_3}\right){}^{\hi{1}hh}_{\hi{2}} &= \left(\frac{C_{DLL^\dagger H^2H^\dagger{}^2,3}\delta^{i_4}_{\hat{i}_2}\delta^{\hat{i}_1}_{i_5}\delta^{i_3}_{i_6}}{\Lambda^4}\right) \Omega_{\hat{i}_1}^{\hi{1}}\Omega_{\hi{2}}^{\hat{i}_2}\mathcal{U}_{i_3}^h\mathcal{U}_{i_4}^h \mathcal{U}^{i_5 h}\mathcal{U}^{i_6h}v\,,\\
    \left(c_{f\bar{f}Vh^2,3}^{-+000}\,{}^{\mathbf{I}_3}\right){}^{\hi{1}hh}_{\hi{2}} &= \left(\frac{C_{DLL^\dagger H^2H^\dagger{}^2,4}\delta^{i_3}_{\hat{i}_2}\delta^{\hat{i}_1}_{i_5}\delta^{i_4}_{i_6}}{\Lambda^4}\right) \Omega_{\hat{i}_1}^{\hi{1}}\Omega_{\hi{2}}^{\hat{i}_2}\mathcal{U}_{i_3}^h\mathcal{U}_{i_4}^h \mathcal{U}^{i_5 h}\mathcal{U}^{i_6h}v\,,\\
\end{align}

\paragraph{\boldsymbol{$f\bar{f}V^2h$}}
\begin{align}
    \left(\mathcal{M}_{f\bar{f}V^2h}^{\mathbf{I}_3\mathbf{I}_4}\right){}^{\hi{1}h}_{\hi{2}}\left(\mathbf{\frac{1}{2},\frac{1}{2},1,1,0}\right) &= \left(c_{f\bar{f}V^2h,1}^{----0}\,^{\mathbf{I}_3\mathbf{I}_4}\right){}^{\hi{1}h}_{\hi{2}} \lra{\mathbf{12}}\lra{\mathbb{34}}^2 +  \left(c_{f\bar{f}V^2h,2}^{----0}\,^{\mathbf{I}_3\mathbf{I}_4}\right){}^{\hi{1}h}_{\hi{2}} \lra{\mathbf{1}\mathbb{3}}\lra{\mathbf{2}\mathbb{4}}\lra{\mathbb{34}}\notag \\
    & + \left(c_{f\bar{f}V^2h,1}^{--++0}\,^{\mathbf{I}_3\mathbf{I}_4}\right){}^{\hi{1}h}_{\hi{2}} \lra{\mathbf{12}}[\mathbb{34}]^2
    +\left(c_{f\bar{f}V^2h}^{+--00}\,^{\mathbf{I}_3\mathbf{I}_4}\right){}^{\hi{1}h}_{\hi{2}}[\mathbf{14}] \langle\mathbb{3}\mathbf{4}\rangle \lra{\mathbf{2}\mathbb{3}} \notag \\
    & +\left(c_{f\bar{f}V^2h}^{-+-00}\,^{\mathbf{I}_3\mathbf{I}_4}\right){}^{\hi{1}h}_{\hi{2}}\lra{\mathbf{1}\mathbb{3}}[\mathbf{24}] \langle\mathbb{3}\mathbf{4}\rangle + \left(c_{f\bar{f}V^2h,1}^{--000}\,{}^{\mathbf{I}_3\mathbf{I}_4}\right){}_{\hi{2}}^{\hi{1}h}\lra{\mathbf{12}}\langle\mathbf{34}\rangle[\mathbf{34}] \notag \\
    & + \left(c_{f\bar{f}V^2h,2}^{--000}\,{}^{\mathbf{I}_3\mathbf{I}_4}\right){}_{\hi{2}}^{\hi{1}h}\lra{\mathbf{13}}\langle\mathbf{24}\rangle[\mathbf{34}] + \text{h.c.}\,, \label{eq:cls_fbarV2h}
\end{align}

\begin{align}
    \left(c_{f\bar{f}V^2h,1}^{----0}\,^{\mathbf{I}_3\mathbf{I}_4}\right){}^{\hi{1}h}_{\hi{2}} &= \left(\frac{C_{Le_{\mathbbm{C}}B_L^2H^\dagger}\delta^{\hat{i}_1}_{i_5}\delta_{\hat{i}_2}^4\delta^{I_3 4}\delta^{I_44}}{\Lambda^4} + \frac{C_{Le_{\mathbbm{C}}W_LB_LH^\dagger,1}(\tau^{I_3}){}^{\hat{i}_1}_{i_5}\delta_{\hat{i}_2}^4\delta^{I_44}}{\Lambda^4} + \frac{C_{Le_{\mathbbm{C}}W_L^2H^\dagger,1}\delta^{\hat{i}_1}_{i_5}\delta_{\hat{i}_2}^4\delta^{I_3I_4}}{\Lambda^4}\right)\notag \\
    & \times \Omega_{\hat{i}_1}^{\hi{1}}\Omega^{\hat{i}_2}_{\hi{2}}O_{I_3}^{\mathbf{I}_3}O_{I_4}^{\mathbf{I}_4}\mathcal{U}^{i_5 h}\,,\\
    \left(c_{f\bar{f}V^2h,2}^{----0}\,^{\mathbf{I}_3\mathbf{I}_4}\right){}^{\hi{1}h}_{\hi{2}} &= \left(\frac{C_{Le_{\mathbbm{C}}W_LB_LH^\dagger,2}(\tau^{I_3}){}^{\hat{i}_1}_{i_5}\delta_{\hat{i}_2}^4\delta^{I_44}}{\Lambda^4} + \frac{C_{Le_{\mathbbm{C}}W_L^2H^\dagger,2}(\tau^K){}^{\hat{i}_1}_{i_5}\delta_{\hat{i}_2}^4\epsilon^{I_3I_4K}}{\Lambda^4}\right)\notag \\
    & \times \Omega_{\hat{i}_1}^{\hi{1}}\Omega^{\hat{i}_2}_{\hi{2}}O_{I_3}^{\mathbf{I}_3}O_{I_4}^{\mathbf{I}_4}\mathcal{U}^{i_5 h}\,,\\
    \left(c_{f\bar{f}V^2h,1}^{--++0}\,^{\mathbf{I}_3\mathbf{I}_4}\right){}^{\hi{1}h}_{\hi{2}} &= \left(\frac{C_{Le_{\mathbbm{C}}B_R^2H^\dagger}\delta^{\hat{i}_1}_{i_5}\delta_{\hat{i}_2}^4\delta^{I_3 4}\delta^{I_44}}{\Lambda^4} + \frac{C_{Le_{\mathbbm{C}}W_RB_RH^\dagger}(\tau^{I_3}){}^{\hat{i}_1}_{i_5}\delta_{\hat{i}_2}^4\delta^{I_44}}{\Lambda^4} + \frac{C_{Le_{\mathbbm{C}}W_R^2H^\dagger}\delta^{\hat{i}_1}_{i_5}\delta_{\hat{i}_2}^4\delta^{I_3I_4}}{\Lambda^4}\right)\notag \\
    & \times \Omega_{\hat{i}_1}^{\hi{1}}\Omega^{\hat{i}_2}_{\hi{2}}O_{I_3}^{\mathbf{I}_3}O_{I_4}^{\mathbf{I}_4}\mathcal{U}^{i_5 h}\,,\\
    \left(c_{f\bar{f}V^2h}^{+--00}\,^{\mathbf{I}_3\mathbf{I}_4}\right){}^{\hi{1}h}_{\hi{2}}&= \left(\frac{C_{De_\mathbbm{C}^\dagger e_{\mathbbm{C}}B_LHH^\dagger,1}\delta^{\hat{i}_1}_4\delta^4_{\hat{i}_2}\delta^{I_34}\delta^{i_4}_{i_5}}{\Lambda^4} + \frac{C_{De_\mathbbm{C}^\dagger e_{\mathbbm{C}}W_LHH^\dagger,1}\delta^{\hat{i}_1}_4\delta^4_{\hat{i}_2}(\tau^{I_3}){}^{i_4}_{i_5}}{\Lambda^4} + \frac{C_{De_\mathbbm{C}^\dagger e_{\mathbbm{C}}B_LHH^\dagger,2}\delta^{\hat{i}_1}_4\delta^4_{\hat{i}_2}\delta^{I_34}\delta^{i_4}_{i_5}}{\Lambda^4} \right.\notag \\
    &\left.+ \frac{C_{De_\mathbbm{C}^\dagger e_{\mathbbm{C}}W_LHH^\dagger,2}\delta^{\hat{i}_1}_4\delta^4_{\hat{i}_2}(\tau^{I_3}){}^{i_4}_{i_5}}{\Lambda^4}\right)\Omega_{\hat{i}_1}^{\hi{1}}\Omega^{\hat{i}_2}_{\hi{2}}O^{\mathbf{I}_3}_{I_3}\left(\mathcal{U}^{\mathbf{I}_4}_{i_4}\mathcal{U}^{i_5h}+\mathcal{U}^{i_5\mathbf{I}_4}\mathcal{U}_{i_4}^{h}\right) \\
    \left(c_{f\bar{f}V^2h}^{-+-00}\,^{\mathbf{I}_3\mathbf{I}_4}\right){}^{\hi{1}h}_{\hi{2}} &= \left(\frac{C_{DLL^\dagger B_LHH^\dagger,1} \delta^{\hat{i}_1}_{\hat{i}_2}\delta^{i_4}_{i_5}\delta^{I_34}}{\Lambda^4} + \frac{C_{DLL^\dagger B_LHH^\dagger,2} \delta^{\hat{i}_1}_{i_5}\delta^{i_4}_{\hat{i}_2}\delta^{I_34}}{\Lambda^4} + \frac{C_{DLL^\dagger W_LHH^\dagger,1} (\tau^{I_3}){}^{\hat{i}_1}_{\hat{i}_2}\delta^{i_4}_{i_5}}{\Lambda^4}\right.\notag \\
    &\left.+ \frac{C_{DLL^\dagger W_LHH^\dagger,2} (\tau^{I_3}){}^{\hat{i}_1}_{i_5}\delta^{i_4}_{\hat{i}_2}}{\Lambda^4} + \frac{C_{DLL^\dagger W_LHH^\dagger,3} (\tau^{I_3}){}^{i_4}_{i_5}\delta^{\hat{i}_1}_{\hat{i}_2}}{\Lambda^4} + \frac{C_{DLL^\dagger B_LHH^\dagger,3} \delta^{\hat{i}_1}_{\hat{i}_2}\delta^{i_4}_{i_5}\delta^{I_34}}{\Lambda^4} \right.\\
    &\left.+ \frac{C_{DLL^\dagger B_LHH^\dagger,4} \delta^{\hat{i}_1}_{i_5}\delta^{i_4}_{\hat{i}_2}\delta^{I_34}}{\Lambda^4} + \frac{C_{DLL^\dagger W_LHH^\dagger,4} (\tau^{I_3}){}^{\hat{i}_1}_{\hat{i}_2}\delta^{i_4}_{i_5}}{\Lambda^4}+  \frac{C_{DLL^\dagger W_LHH^\dagger,5} (\tau^{I_3}){}^{\hat{i}_1}_{i_5}\delta^{i_4}_{\hat{i}_2}}{\Lambda^4} \right.\notag \\
    & \left.+ \frac{C_{DLL^\dagger W_LHH^\dagger,6} (\tau^{I_3}){}^{i_4}_{i_5}\delta^{\hat{i}_1}_{\hat{i}_2}}{\Lambda^4}\right)\Omega_{\hat{i}_1}^{\hi{1}}\Omega^{\hat{i}_2}_{\hi{2}}O^{\mathbf{I}_3}_{I_3}\left(\mathcal{U}^{\mathbf{I}_4}_{i_4}\mathcal{U}^{i_5h}+\mathcal{U}^{i_5\mathbf{I}_4}\mathcal{U}_{i_4}^{h}\right)\,,\\
    \left(c_{f\bar{f}V^2h,1}^{--000}\,{}^{\mathbf{I}_3\mathbf{I}_4}\right){}_{\hi{2}}^{\hi{1}h} &= \left(\frac{C_{D^2Le_\mathbbm{C}HH^\dagger{}^2,3}\delta^{\hat{i}_1}_{i_4}\delta^{i_3}_{i_5}\delta_{\hat{i}_2}^4}{\Lambda^4}\right)\Omega_{\hat{i}_1}^{\hi{1}}\Omega_{\hi{2}}^{\hat{i}_2} \mathcal{U}^{i_4\mathbf{I}_4}\left(\mathcal{U}_{i_3}^{\mathbf{I}_3} \mathcal{U}^{i_5h}+\mathcal{U}^{i_5\mathbf{I}_3} \mathcal{U}_{i_3}^{h}\right)\,,\\
    \left(c_{f\bar{f}V^2h,2}^{--000}\,{}^{\mathbf{I}_3\mathbf{I}_4}\right){}_{\hi{2}}^{\hi{1}h} &= \left(\frac{C_{D^2Le_\mathbbm{C}HH^\dagger{}^2,6}\delta^{\hat{i}_1}_{i_4}\delta^{i_3}_{i_5}\delta_{\hat{i}_2}^4}{\Lambda^4}\right)\Omega_{\hat{i}_1}^{\hi{1}}\Omega_{\hi{2}}^{\hat{i}_2}\mathcal{U}^{i_4\mathbf{I}_4}\left(\mathcal{U}_{i_3}^{\mathbf{I}_3} \mathcal{U}^{i_5h}+\mathcal{U}^{i_5\mathbf{I}_3} \mathcal{U}_{i_3}^{h}\right)\,.
\end{align}

\paragraph{\boldsymbol{$V^3h^2$}}
\begin{align}
    \mathcal{M}^{\mathbf{I}_1\mathbf{I}_2\mathbf{I}_3hh}_{V^3h^2} \left(\mathbf{1,1,1,0,0}\right) &=\left(c_{V^3h^2}^{---00}\,{}^{\mathbf{I}_1\mathbf{I}_2\mathbf{I}_3}\right){}^{hh}\lra{\mathbb{12}}\lra{\mathbb{13}}\lra{\mathbb{23}} \notag \\
    & + \left(c_{V^3h^2}^{-0000}\,{}^{\mathbf{I}_1\mathbf{I}_2\mathbf{I}_3}\right){}^{hh}\langle\mathbb{1}\mathbf{2}\rangle \lra{\mathbb{1}\mathbf{3}}[\mathbf{23}] +\text{h.c.} \,, \label{eq:cls_V3h2}
\end{align}
    
\begin{align}
    \left(c_{V^3h^2}^{---00}\,{}^{\mathbf{I}_1\mathbf{I}_2\mathbf{I}_3}\right){}^{hh} &= \left(\frac{C_{B_LW_L^2HH^\dagger}\delta^{I_14}\epsilon^{I_2I_3K}(\tau^K)^{i_4}_{i_5}}{\Lambda^4} + \frac{C_{W_L^3HH^\dagger}\epsilon^{I_1I_2I_3}\delta^{i_4}_{i_5}}{\Lambda^4} \right)O_{I_1}^{\mathbf{I}_1}O_{I_2}^{\mathbf{I}_2} O_{I_3}^{\mathbf{I}_3}\calu_{i_4}^h\calu^{i_5 h}\,,\\
    \left(c_{V^3h^2}^{-0000}\,{}^{\mathbf{I}_1\mathbf{I}_2\mathbf{I}_3}\right){}^{hh} &= \left(\frac{C_{D^2B_LH^2H^\dagger{}^2}\delta^{I_14}\delta^{i_2}_{i_4}\delta^{i_3}_{i_5}}{\Lambda^4} + \frac{C_{D^2W_LH^2H^\dagger{}^2,1}(\tau^{I_1}){}^{i_3}_{i_5}\delta^{i_2}_{i_4}}{\Lambda^4} + \frac{C_{D^2W_LH^2H^\dagger{}^2,2}(\tau^{I_1}){}^{i_2}_{i_4}\delta^{i_3}_{i_5}}{\Lambda^4} \right)\notag \\
     & \times O_{I_1}^{\mathbf{I}_1} \left(\calu_{i_2}^{\mathbf{I}_2}\calu^{i_4\mathbf{I}_3} + \calu^{i_4\mathbf{I}_2}\calu_{i_2}^{\mathbf{I}_3}\right)\left(\calu_{i_3}^h\calu^{i_5h}+\calu^{i_5h}\calu_{i_3}^{h}\right)\,,
\end{align}

\paragraph{\boldsymbol{$V^2h^3$}}
\begin{align}
    \mathcal{M}^{\mathbf{I}_1\mathbf{I}_2hhh}_{V^2h^3} \left(\mathbf{1,1,0,0,0}\right) &= \left(c_{V^2h^3}^{-0000}\,{}^{\mathbf{I}_1\mathbf{I}_2}\right){}^{hhh}\langle\mathbb{1}\mathbf{2}\rangle \langle\mathbb{1}\mathbf{|p_4|2}] + \left(c_{V^2h^3}^{--0000}\,{}^{\mathbf{I}_1\mathbf{I}_2}\right){}^{hhh} \lra{\mathbb{12}}^2  \notag \\
    & + \left(c_{V^2h^3}^{00000}\,{}^{\mathbf{I}_1\mathbf{I}_2}\right){}^{hhh} \lra{\mathbf{12}}[\mathbf{12}]+\text{h.c.}\,, \label{eq:cls_V2h3}
\end{align}

\begin{align}
     \left(c_{V^2h^3}^{-0000}\,{}^{\mathbf{I}_1\mathbf{I}_2}\right){}^{hhh} &= \left(\frac{C_{D^2B_LH^2H^\dagger{}^2}\delta^{I_14}\delta^{i_2}_{i_4}\delta^{i_3}_{i_5}}{\Lambda^4} + \frac{C_{D^2W_LH^2H^\dagger{}^2,1}(\tau^{I_1}){}^{i_3}_{i_5}\delta^{i_2}_{i_4}}{\Lambda^4} + \frac{C_{D^2W_LH^2H^\dagger{}^2,2}(\tau^{I_1}){}^{i_2}_{i_4}\delta^{i_3}_{i_5}}{\Lambda^4} \right)\notag \\
     & \times O_{I_1}^{\mathbf{I}_1} \left(\calu_{i_2}^{\mathbf{I}_2}\calu^{i_4\mathbf{I}_3} + \calu^{i_4\mathbf{I}_2}\calu_{i_2}^{\mathbf{I}_3}\right)\left(\calu_{i_3}^h\calu^{i_5h}+\calu^{i_5h}\calu_{i_3}^{h}\right)\,,\\
     \left(c_{V^2h^3}^{--0000}\,{}^{\mathbf{I}_1\mathbf{I}_2}\right){}^{hhh} &= \left(\frac{C_{B_L^2H^2H^\dagger{}^2}\delta^{I_14}\delta^{I_24} \delta^{i_3}_{i_4}\delta^{i_5}_{i_6}}{\Lambda^4} + \frac{C_{B_LW_LH^2H^\dagger{}^2}\delta^{I_14} (\tau^{I_2}){}^{i_3}_{i_4}\delta^{i_5}_{i_6}}{\Lambda^4} + \frac{C_{W_L^2H^2H^\dagger{}^2,1}\delta^{I_1I_2} \delta^{i_3}_{i_4}\delta^{i_5}_{i_6}}{\Lambda^4} \right.\notag \\
    &\left.+ \frac{C_{W_L^2H^2H^\dagger{}^2,2}(\tau^{I_1}){}^{i_3}_{i_4}(\tau^{I_2}){}^{i_5}_{i_6}}{\Lambda^4}\right)\times O^{\mathbf{I}_1}_{I_1}O^{\mathbf{I}_2}_{I_2} \calu^h_{i_3}\calu_{i_5}^h\calu^{i_4h}\calu^{i_5h}v\,,\\
    \left(c_{V^2h^3}^{00000}\,{}^{\mathbf{I}_1\mathbf{I}_2}\right){}^{hhh} &= \frac{C_{D^2H^3H^\dagger{}^3,1}\delta^{i_1}_{i_2}\delta^{i_3}_{i_4}\delta^{i_5}_{i_6}}{\Lambda^4} \calu^{\mathbf{I}_1}_{i_1}\calu^{\mathbf{I}_2i_2} \calu^h_{i_3}\calu^h_{i_5} \calu^{i_4 h}\calu^{i_6 h}v\notag \\
    & + \frac{C_{D^2H^3H^\dagger{}^3,2}\delta^{i_1}_{i_2}\delta^{i_3}_{i_4}\delta^{i_5}_{i_6}}{\Lambda^4}\left(\calu^{\mathbf{I}_1}_{i_1}\calu^{hi_2} + \calu^{i_2\mathbf{I}_1}\calu_{i_1}^{h}\right)\notag \\
    & \times \left(\calu^{\mathbf{I}_2}_{i_3}\calu^{hi_4} + \calu^{i_4\mathbf{I}_2}\calu_{i_3}^{h} \right) \calu^h_{i_5}\calu^{hi_6}v\,,
\end{align}

\paragraph{\boldsymbol{$Vh^4$}}
\begin{align}
    \mathcal{M}^{\mathbf{I}_1hhhh}_{Vh^4} \left(\mathbf{1,0,0,0,0}\right) &= \left(c_{Vh^4}^{-0000}\,{}^{\mathbf{I}_1}\right){}^{hhhh}\langle\mathbb{1}\mathbf{|p_2p_4|}\mathbb{1}\rangle + \left(c_{Vh^4,1}^{000000}\,{}^{\mathbf{I}_1}\right){}^{hhhh} \langle \mathbf{1|p_2|1}]\notag \\
    & + \left(c_{Vh^4,2}^{00000}\,{}^{\mathbf{I}_1}\right){}^{hhhh} \langle \mathbf{1|p_3|1}] + \text{h.c.} \,, \label{eq:cls_Vh4}
\end{align}

\begin{align}
    \left(c_{Vh^4}^{-0000}\,{}^{\mathbf{I}_1}\right){}^{hhhh} &= \left(\frac{C_{D^2B_LH^2H^\dagger{}^2}\delta^{I_14}\delta^{i_2}_{i_4}\delta^{i_3}_{i_5}}{\Lambda^4} + \frac{C_{D^2W_LH^2H^\dagger{}^2,1}(\tau^{I_1}){}^{i_3}_{i_5}\delta^{i_2}_{i_4}}{\Lambda^4} + \frac{C_{D^2W_LH^2H^\dagger{}^2,2}(\tau^{I_1}){}^{i_2}_{i_4}\delta^{i_3}_{i_5}}{\Lambda^4} \right)\notag \\
    &\times O_{I_1}^{\mathbf{I}_1} \calu_{i_2}^h\calu_{i_3}^h\calu^{i_4h}\calu^{i_5h}\,,\\
    \left(c_{Vh^4,1}^{00000}\,{}^{\mathbf{I}_1}\right){}^{hhhh} &= \frac{C_{D^2H^3H^\dagger{}^3,1}\delta^{i_1}_{i_2}\delta^{i_3}_{i_4}\delta^{i_5}_{i_6}}{\Lambda^4}\left(\calu_{i_1}^\mathbf{I_1}\calu^{i_2h}+\calu^{i_2\mathbf{I_1}}\calu_{i_1}^{h}\right)\calu_{i_3}^h\calu^{i_4h}\calu_{i_5}^h\calu^{i_6h}\,,\\
    \left(c_{Vh^4,2}^{00000}\,{}^{\mathbf{I}_1}\right){}^{hhhh} &=  \frac{C_{D^2H^3H^\dagger{}^3,2}\delta^{i_1}_{i_2}\delta^{i_3}_{i_4}\delta^{i_5}_{i_6}}{\Lambda^4}\left(\calu_{i_1}^\mathbf{I_1}\calu^{i_2h}+\calu^{i_2\mathbf{I_1}}\calu_{i_1}^{h}\right)\calu_{i_3}^h\calu^{i_4h}\calu_{i_5}^h\calu^{i_6h}v\,.
\end{align}

\paragraph{\boldsymbol{$h^5$}}
\begin{align}
    \mathcal{M}^{hhhhh}_{h^5,1} \left(\mathbf{0,0,0,0,0}\right) =\left(c_{h^5}^{00000}\right){}^{hhhhh} + \left(c_{h^5,2}^{000000}\right){}^{hhhhhh}s_{12} +  \left(c_{h^5,3}^{000000}\right){}^{hhhhhh}s_{13} + \text{h.c.}  \,, \label{eq:cls_h5}
\end{align}

\begin{align}
    & \left(c_{h^5,1}^{00000}\right){}^{hhhhh} = \left(\frac{C_{H^3H^\dagger{}^3}}{\Lambda^2} \epsilon^{i_1i_4}\epsilon^{i_2i_5}\epsilon^{i_3i_6}\right)\calu^h_{i_1}\calu^h_{i_2}\calu^h_{i_3}\calu^h_{i_4}\calu^h_{i_5}\calu^h_{i_6} v\,,\\
    & \left(c_{h^5,2}^{000000}\right){}^{hhhhhh} = \left(\frac{C_{D^2H^3H^\dagger{}^3,1}\delta^{i_1}_{i_2}\delta^{i_3}_{i_4}\delta^{i_5}_{i_6}}{\Lambda^4}\right)\calu_{i_1}^h\calu_{i_3}^h\calu_{i_5}^h\calu^{i_2h}\calu^{i_4h}\calu^{i_6h}v\,,\\
    & \left(c_{h^5,3}^{000000}\right){}^{hhhhhh} = \left(\frac{C_{D^2H^3H^\dagger{}^3,2}\delta^{i_1}_{i_2}\delta^{i_3}_{i_4}\delta^{i_5}_{i_6}}{\Lambda^4}\right)\calu_{i_1}^h\calu_{i_3}^h\calu_{i_5}^h\calu^{i_2h}\calu^{i_4h}\calu^{i_6h}v\,,
\end{align}

\subsection{6, 7, and 8-point Matching Results}

\paragraph{\boldsymbol{$f^2\bar{f}{}^2h^2$}}
\begin{align}
    \left(\mathcal{M}_{f^2\bar{f}{}^2h^2}\right){}^{\hi{1}\hi{2}hh}_{\hi{3}\hi{4}}\left(\mathbf{\frac{1}{2},\frac{1}{2},\frac{1}{2},\frac{1}{2},0,0}\right) &= \left(c_{f^2\bar{f}{}^2h^2,1}^{----00}\right){}^{\hi{1}\hi{2}hh}_{\hi{3}\hi{4}} \lra{\mathbf{12}}\lra{\mathbf{34}} + \left(c_{f^2\bar{f}{}^2h^2,2}^{----00}\right){}^{\hi{1}\hi{2}hh}_{\hi{3}\hi{4}} \lra{\mathbf{13}}\lra{\mathbf{24}}\notag \\
    &+ \left(c_{f^2\bar{f}{}^2h^2}^{++--00}\right){}^{\hi{1}\hi{2}hh}_{\hi{3}\hi{4}}[\mathbf{12}]\lra{\mathbf{34}} + \left(c_{f^2\bar{f}{}^2h^2}^{-+-+00}\right){}^{\hi{1}\hi{2}hh}_{\hi{3}\hi{4}}\lra{\mathbf{13}}[\mathbf{24}] \notag \\
    &+ \left(c_{f^2\bar{f}{}^2h^2}^{--++00}\right){}^{\hi{1}\hi{2}hh}_{\hi{3}\hi{4}}\lra{\mathbf{12}}[\mathbf{34}] + \text{h.c.}\,,\label{eq:cls_f2bar2h2}
\end{align}

\begin{align}
    \left(c_{f^2\bar{f}{}^2h^2,1}^{----00}\right){}^{\hi{1}\hi{2}hh}_{\hi{3}\hi{4}} &= \left(\frac{C_{L^2e_\mathbbm{C}^2H^\dagger{}^2,1}\delta^{\hat{i}_1}_{i_5}\delta^{\hat{i}_2}_{i_6}\delta^4_{\hat{i}_3}\delta^4_{\hat{i}_4}}{\Lambda^4}\right)\Omega_{\hat{i}_1}^{\hi{1}}\Omega_{\hat{i}_2}^{\hi{2}}\Omega_{\hi{3}}^{\hat{i}_3}\Omega_{\hi{4}}^{\hat{i}_4} \mathcal{U}^{i_5 h}\mathcal{U}^{i_6 h}\,,\\
    \left(c_{f^2\bar{f}{}^2h^2,2}^{----00}\right){}^{\hi{1}\hi{2}hh}_{\hi{3}\hi{4}} &= \left(\frac{C_{L^2e_\mathbbm{C}^2H^\dagger{}^2,2}\delta^{\hat{i}_1}_{i_5}\delta^{\hat{i}_2}_{i_6}\delta^4_{\hat{i}_3}\delta^4_{\hat{i}_4}}{\Lambda^4}\right)\Omega_{\hat{i}_1}^{\hi{1}}\Omega_{\hat{i}_2}^{\hi{2}}\Omega_{\hi{3}}^{\hat{i}_3}\Omega_{\hi{4}}^{\hat{i}_4} \mathcal{U}^{i_5 h}\mathcal{U}^{i_6 h}\,,\\
    \left(c_{f^2\bar{f}{}^2h^2}^{++--00}\right){}^{\hi{1}\hi{2}hh}_{\hi{3}\hi{4}} &= \left(\frac{C_{e_\mathbbm{C}^\dagger{}^2e_\mathbbm{C}^2HH^\dagger}\delta^{\hat{i}_1}_4\delta^{\hat{i}_2}_4\delta_{\hat{i}_3}^4\delta_{\hat{i}_4}^4\delta^{i_5}_{i_6}}{\Lambda^4}\right)\Omega_{\hat{i}_1}^{\hi{1}}\Omega_{\hat{i}_2}^{\hi{2}}\Omega_{\hi{3}}^{\hat{i}_3}\Omega_{\hi{4}}^{\hat{i}_4} \mathcal{U}_{i_5}^h\mathcal{U}^{i_6 h} \,,\\
    \left(c_{f^2\bar{f}{}^2h^2}^{-+-+00}\right){}^{\hi{1}\hi{2}hh}_{\hi{3}\hi{4}} &= \left(\frac{C_{Le_\mathbbm{C}^\dagger e_\mathbbm{C} L^\dagger HH^\dagger,1}\delta^{\hat{i}_1}_{i_6}\delta^{\hat{i}_2}_4\delta_{\hat{i}_3}^4\delta_{\hat{i}_4}^{i_5}}{\Lambda^4} + \frac{C_{Le_\mathbbm{C}^\dagger e_\mathbbm{C} L^\dagger HH^\dagger,2}\delta^{\hat{i}_1}_{i_4}\delta^{\hat{i}_2}_4\delta_{\hat{i}_3}^4\delta_{i_5}^{i_6}}{\Lambda^4}\right)\notag \\
    & \times \Omega_{\hat{i}_1}^{\hi{1}}\Omega_{\hat{i}_2}^{\hi{2}}\Omega_{\hi{3}}^{\hat{i}_3}\Omega_{\hi{4}}^{\hat{i}_4} \mathcal{U}_{i_5}^h\mathcal{U}^{i_6 h}\,,\\
    \left(c_{f^2\bar{f}{}^2h^2}^{--++00}\right){}^{\hi{1}\hi{2}hh}_{\hi{3}\hi{4}} &= \left(\frac{C_{L^2L^\dagger{}^2 HH^\dagger,1}\delta^{\hat{i}_1}_{\hat{i}_3}\delta^{\hat{i}_2}_{i_6}\delta_{\hat{i}_4}^{i_5}}{\Lambda^4} + \frac{C_{L^2L^\dagger{}^2 HH^\dagger,2}\delta^{\hat{i}_1}_{\hat{i}_3}\delta^{\hat{i}_2}_{i_4}\delta_{\hat{i}_6}^{i_5}}{\Lambda^4}\right)\Omega_{\hat{i}_1}^{\hi{1}}\Omega_{\hat{i}_2}^{\hi{2}}\Omega_{\hi{3}}^{\hat{i}_3}\Omega_{\hi{4}}^{\hat{i}_4} \mathcal{U}_{i_5}^h\mathcal{U}^{i_6 h}\,.
\end{align}

\paragraph{\boldsymbol{$f^2h^4$}}

\begin{align}
    & \mathcal{M}_{f^2h^6}^{\hi{1}\hi{2}hhhh}\left(\mathbf{\frac{1}{2},\frac{1}{2},0,0,0,0}\right) = 
    \left(c_{f^2h^3}^{--0000}\right){}^{\hi{1}\hi{2}hhhh} \lra{\mathbf{12}} \,,\label{eq:cls_f2h4}
\end{align}

\begin{align}
    \left(c_{f^2h^3}^{--0000}\right){}^{\hi{1}\hi{2}hhhh} &= \left(\frac{C_{L^2H^3H^\dagger}\epsilon^{\hat{i}_1 i_3}\epsilon^{\hat{i}_2 i_4}\delta^{i_5}_{i_6}}{\Lambda^3}\right) \Omega_{\hat{i}_1}^{\hi{1}} \Omega_{\hat{i}_2}^{\hi{2}} \mathcal{U}_{i_3}^h \mathcal{U}_{i_4}^h \mathcal{U}_{i_5}^h \mathcal{U}^{i_6h} + \text{h.c.}\,,
\end{align}

\paragraph{\boldsymbol{$f\bar{f}h^4$}}
\begin{align}
    \mathcal{M}_{f\bar{f}h^4}^{\hi{1}\hi{2}hhhh}\left(\mathbf{\frac{1}{2},\frac{1}{2},0,0,0,0}\right) &= \left(c_{f\bar{f}h^4}^{+-0000}\right){}^{\hi{1}hhhh}_{\hi{2}}[\mathbf{1|p_3|2}\rangle + \left(c_{f\bar{f}h^4,1}^{-+0000}\right){}^{\hi{1}hhhh}_{\hi{2}}[\mathbf{2|p_3|1}\rangle \notag \\
    & + \left(c_{f\bar{f}h^4,2}^{-+0000}\right){}^{\hi{1}hhhh}_{\hi{2}}[\mathbf{2|p_4|1}\rangle + \left(c_{f\bar{f}h^4,3}^{-+0000}\right){}^{\hi{1}hhhh}_{\hi{2}}[\mathbf{2|p_5|1}\rangle  \notag \\
    & + \left(c_{f\bar{f}h^4}^{--0000}\right){}^{\hi{1}hhhh}_{\hi{2}} \lra{\mathbf{12}}
    + \text{h.c.}\,,\label{eq:cls_fbarh4}
\end{align}

\begin{align}
    \left(c_{f\bar{f}h^4}^{+-0000}\right){}^{\hi{1}hhhh}_{\hi{2}} &= \left(\frac{C_{De_\mathbbm{C}^\dagger e_\mathbbm{C}H^2H^\dagger{}^2}\delta^{\hat{i}_1}_4\delta_{\hat{i}_2}^4\delta^{i_3}_{i_5}\delta^{i_4}_{i_6}}{\Lambda^4}\right)\Omega_{\hat{i}_1}^{\hi{1}}\Omega_{\hi{2}}^{\hat{i}_2}\mathcal{U}_{i_3}^h\mathcal{U}_{i_4}^h \mathcal{U}^{i_5 h}\mathcal{U}^{i_6h}\,,\\
    \left(c_{f\bar{f}h^4,1}^{-+0000}\right){}^{\hi{1}hhhh}_{\hi{2}} &= \left(\frac{C_{DLL^\dagger H^2H^\dagger{}^2,1}\delta^{\hat{i}_1}_{\hat{i}_2}\delta^{i_3}_{i_5}\delta^{i_4}_{i_6}}{\Lambda^4} + \frac{C_{DLL^\dagger H^2H^\dagger{}^2,2}\delta^{i_3}_{\hat{i}_2}\delta^{\hat{i}_1}_{i_5}\delta^{i_4}_{i_6}}{\Lambda^4}\right) \notag \\
    & \times \Omega_{\hat{i}_1}^{\hi{1}}\Omega_{\hi{2}}^{\hat{i}_2}\mathcal{U}_{i_3}^h\mathcal{U}_{i_4}^h \mathcal{U}^{i_5 h}\mathcal{U}^{i_6h}\,,\\
    \left(c_{f\bar{f}h^4,2}^{-+0000}\right){}^{\hi{1}hhhh}_{\hi{2}} &= \left(\frac{C_{DLL^\dagger H^2H^\dagger{}^2,3}\delta^{i_4}_{\hat{i}_2}\delta^{\hat{i}_1}_{i_5}\delta^{i_3}_{i_6}}{\Lambda^4}\right) \Omega_{\hat{i}_1}^{\hi{1}}\Omega_{\hi{2}}^{\hat{i}_2}\mathcal{U}_{i_3}^h\mathcal{U}_{i_4}^h \mathcal{U}^{i_5 h}\mathcal{U}^{i_6h}\,,\\
    \left(c_{f\bar{f}h^4,3}^{-+0000}\right){}^{\hi{1}hhhh}_{\hi{2}} &= \left(\frac{C_{DLL^\dagger H^2H^\dagger{}^2,4}\delta^{i_3}_{\hat{i}_2}\delta^{\hat{i}_1}_{i_5}\delta^{i_4}_{i_6}}{\Lambda^4}\right) \Omega_{\hat{i}_1}^{\hi{1}}\Omega_{\hi{2}}^{\hat{i}_2}\mathcal{U}_{i_3}^h\mathcal{U}_{i_4}^h \mathcal{U}^{i_5 h}\mathcal{U}^{i_6h}\,,\\
    \left(c_{f\bar{f}h^4}^{--0000}\right){}^{\hi{1}hhhh}_{\hi{2}} &= \left(\frac{C_{Le_\mathbbm{C}H^2H^\dagger{}^3}\delta^{\hat{i}_1}_{i_5}\delta_{\hat{i}_2}^4\delta^{i_3}_{i_6}\delta^{i_4}_{i_7}}{\Lambda^4}\right)\Omega_{\hat{i}_1}^{\hi{1}}\Omega^{\hat{i}_2}_{\hi{2}}\mathcal{U}_{i_3}^h\mathcal{U}_{i_4}^h\mathcal{U}^{i_5 h}\mathcal{U}^{i_6 h}\mathcal{U}^{i_7 h}v\,.
\end{align}

\paragraph{\boldsymbol{$f\bar{f}Vh^3$}}
\begin{align}
    \left(\mathcal{M}^{\mathbf{I}_3}_{f\bar{f}Vh^3}\right){}_{\hi{2}}^{\hi{1}hhh}\left(\mathbf{\frac{1}{2},\frac{1}{2},1,0,0,0}\right) &= \left(c^{---000}_{f\bar{f}Vh^3}\,{}^{\mathbf{I}_3}\right){}_{\hi{2}}^{\hi{1}hhh} \lra{\mathbf{1}\mathbb{3}}\lra{\mathbf{2}\mathbb{3}} + \left(c^{+-0000}_{f\bar{f}Vh^3}\,{}^{\mathbf{I}_3}\right){}_{\hi{2}}^{\hi{1}hhh}[\mathbf{13}]\lra{\mathbf{23}} \notag \\
    & + \left(c^{-+0000}_{f\bar{f}Vh^3}\,{}^{\mathbf{I}_3}\right){}_{\hi{2}}^{\hi{1}hhh}\lra{\mathbf{13}}[\mathbf{23}] + \text{h.c.}\,, \label{eq:cls_fbarVh3}
\end{align}

\begin{align}
    \left(c^{---000}_{f\bar{f}Vh^3}\,{}^{\mathbf{I}_3}\right){}_{\hi{2}}^{\hi{1}hhh} &= \left(\frac{C_{Le_{\mathbbm{C}}B_LHH^\dagger{}^2}\delta^{\hat{i}_1}_{i_5}\delta^{i_4}_{i_6}\delta_{\hat{i}_2}^4\delta^{I_34}}{\Lambda^4} + \frac{C_{Le_{\mathbbm{C}}W_LHH^\dagger{}^2,1}\delta^{\hat{i}_1}_{i_5}(\tau^{I_3}){}^{i_4}_{i_6}\delta_{\hat{i}_2}^4}{\Lambda^4}\right.\notag \\
    & \left.+ \frac{C_{Le_{\mathbbm{C}}W_LHH^\dagger{}^2,2}\delta^{i_4}_{i_6}(\tau^{I_3}){}^{\hat{i}_1}_{i_5}\delta_{\hat{i}_2}^4}{\Lambda^4}\right) \Omega_{\hat{i}_1}^{\hi{1}}\Omega^{\hat{i}_2}_{\hi{2}}\mathcal{U}^{\mathbf{I}_3}_{I_3}\mathcal{U}_{i_4}^h\mathcal{U}^{i_5h}\mathcal{U}^{i_6h}\,,\\
    \left(c^{+-0000}_{f\bar{f}Vh^3}\,{}^{\mathbf{I}_3}\right){}_{\hi{2}}^{\hi{1}hhh} &= \left(\frac{C_{De_\mathbbm{C}^\dagger e_\mathbbm{C}H^2H^\dagger{}^2}\delta^{\hat{i}_1}_4\delta_{\hat{i}_2}^4\delta^{i_3}_{i_5}\delta^{i_4}_{i_6}}{\Lambda^4}\right)\Omega_{\hat{i}_1}^{\hi{1}}\Omega_{\hi{2}}^{\hat{i}_2}\mathcal{U}_{i_4}^h \mathcal{U}^{i_6h}\left(\mathcal{U}_{i_3}^{\mathbf{I}_3}\mathcal{U}^{i_5 h}+\mathcal{U}^{i_5\mathbf{I}_3}\mathcal{U}_{i_3}^{ h}\right)\,,\\
    \left(c^{-+0000}_{f\bar{f}Vh^3}\,{}^{\mathbf{I}_3}\right){}_{\hi{2}}^{\hi{1}hhh} &= \frac{C_{DLL^\dagger H^2H^\dagger{}^2,1}\delta^{\hat{i}_1}_{\hat{i}_2}\delta^{i_3}_{i_5}\delta^{i_4}_{i_6}}{\Lambda^4}\Omega_{\hat{i}_1}^{\hi{1}}\Omega_{\hi{2}}^{\hat{i}_2}\mathcal{U}_{i_4}^h \mathcal{U}^{i_6h}\left(\mathcal{U}_{i_3}^{\mathbf{I}_3}\mathcal{U}^{i_5 h}+\mathcal{U}^{i_5\mathbf{I}_3}\mathcal{U}_{i_3}^{ h}\right)\notag \\
    & + \frac{C_{DLL^\dagger H^2H^\dagger{}^2,2}\delta^{i_3}_{\hat{i}_2}\delta^{\hat{i}_1}_{i_5}\delta^{i_4}_{i_6}}{\Lambda^4} \Omega_{\hat{i}_1}^{\hi{1}}\Omega_{\hi{2}}^{\hat{i}_2}\mathcal{U}_{i_3}^{\mathbf{I}_3}\mathcal{U}_{i_4}^h \mathcal{U}^{i_5 h}\mathcal{U}^{i_6h}\,,\\
\end{align}

\paragraph{\boldsymbol{$V^2h^4$}}
\begin{align}
    & \mathcal{M}_{V^2h^4}^{\mathbf{I}_1\mathbf{I}_2hhhh}\left(\mathbf{1,1,0,0,0,0}\right) = \left(c_{V^2h^4}^{--0000}\,{}^{\mathbf{I}_1\mathbf{I}_2}\right){}^{hhhh} \lra{\mathbb{12}}^2 + \left(c_{V^2h^4}^{000000}\,{}^{\mathbf{I}_1\mathbf{I}_2}\right){}^{hhhh} \lra{\mathbf{12}}[\mathbf{12}] + \text{h.c.} \,,\label{eq:cls_V2h4}
\end{align}

\begin{align}
    \left(c_{V^2h^4}^{--0000}\,{}^{\mathbf{I}_1\mathbf{I}_2h}\right){}^{hhh} &= \left(\frac{C_{B_L^2H^2H^\dagger{}^2}\delta^{I_14}\delta^{I_24} \delta^{i_3}_{i_4}\delta^{i_5}_{i_6}}{\Lambda^4} + \frac{C_{B_LW_LH^2H^\dagger{}^2}\delta^{I_14} (\tau^{I_2}){}^{i_3}_{i_4}\delta^{i_5}_{i_6}}{\Lambda^4} + \frac{C_{W_L^2H^2H^\dagger{}^2,1}\delta^{I_1I_2} \delta^{i_3}_{i_4}\delta^{i_5}_{i_6}}{\Lambda^4} \right.\notag \\
    &\left.+ \frac{C_{W_L^2H^2H^\dagger{}^2,2}(\tau^{I_1}){}^{i_3}_{i_4}(\tau^{I_2}){}^{i_5}_{i_6}}{\Lambda^4}\right)\times O^{\mathbf{I}_1}_{I_1}O^{\mathbf{I}_2}_{I_2} \calu^h_{i_3}\calu_{i_5}^h\calu^{i_4h}\calu^{i_5h}\,,\\
    \left(c_{V^2h^4}^{000000}\,{}^{\mathbf{I}_1\mathbf{I}_2}\right){}^{hhhh} &= \frac{C_{D^2H^3H^\dagger{}^3,1}\delta^{i_1}_{i_2}\delta^{i_3}_{i_4}\delta^{i_5}_{i_6}}{\Lambda^4} \calu^{\mathbf{I}_1}_{i_1}\calu^{\mathbf{I}_2i_2} \calu^h_{i_3}\calu^h_{i_5} \calu^{i_4 h}\calu^{i_6 h}\notag \\
    & + \frac{C_{D^2H^3H^\dagger{}^3,2}\delta^{i_1}_{i_2}\delta^{i_3}_{i_4}\delta^{i_5}_{i_6}}{\Lambda^4}\left(\calu^{\mathbf{I}_1}_{i_1}\calu^{hi_2} + \calu^{i_2\mathbf{I}_1}\calu_{i_1}^{h}\right)\notag \\
    &\times \left(\calu^{\mathbf{I}_2}_{i_3}\calu^{hi_4} + \calu^{i_4\mathbf{I}_2}\calu_{i_3}^{h}\right) \calu^h_{i_5}\calu^{hi_6}\,,
\end{align}

\paragraph{\boldsymbol{$Vh^5$}}
\begin{align}
    & \mathcal{M}_{Vh^5}^{\mathbf{I}_1hhhhh}\left(\mathbf{1,0,0,0,0,0}\right) = \left(c_{Vh^5,1}^{000000}\,{}^{\mathbf{I}_1}\right){}^{hhhhh} \langle \mathbf{1|p_2|1}] + \left(c_{Vh^5,2}^{000000}\,{}^{\mathbf{I}_1}\right){}^{hhhhh} \langle \mathbf{1|p_3|1}] + \text{h.c.} \,,\label{eq:cls_Vh5}
\end{align}

\begin{align}
    \left(c_{Vh^5,1}^{000000}\,{}^{\mathbf{I}_1}\right){}^{hhhhh} &= \frac{C_{D^2H^3H^\dagger{}^3,1}\delta^{i_1}_{i_2}\delta^{i_3}_{i_4}\delta^{i_5}_{i_6}}{\Lambda^4}\left(\calu_{i_1}^\mathbf{I_1}\calu^{i_2h}+\calu^{i_2\mathbf{I_1}}\calu_{i_1}^{h}\right)\calu_{i_3}^h\calu^{i_4h}\calu_{i_5}^h\calu^{i_6h}\,,\\
    \left(c_{Vh^5,2}^{000000}\,{}^{\mathbf{I}_1}\right){}^{hhhhh} &=  \frac{C_{D^2H^3H^\dagger{}^3,2}\delta^{i_1}_{i_2}\delta^{i_3}_{i_4}\delta^{i_5}_{i_6}}{\Lambda^4}\left(\calu_{i_1}^\mathbf{I_1}\calu^{i_2h}+\calu^{i_2\mathbf{I_1}}\calu_{i_1}^{h}\right)\calu_{i_3}^h\calu^{i_4h}\calu_{i_5}^h\calu^{i_6h}\,,
\end{align}

\paragraph{\boldsymbol{$h^6$}}
\begin{align}
    \mathcal{M}^{hhhhhh}_{h^6} \left(\mathbf{0,0,0,0,0,0}\right) &=  \left(c_{h^6,1}^{000000}\right){}^{hhhhhh} + \left(c_{h^6,2}^{000000}\right){}^{hhhhhh}s_{12} \notag \\
    & + \left(c_{h^6,3}^{000000}\right){}^{hhhhhh}s_{13} + \text{h.c.}\,,\label{eq:cls_h6}
\end{align}

\begin{align}
    & \left(c_{h^6,1}^{000000}\right){}^{hhhhhh} = \frac{C_{H^3H^\dagger{}^2}}{\Lambda^2} \epsilon^{i_1i_4}\epsilon^{i_2i_5}\epsilon^{i_3i_6}\calu^h_{i_1}\calu^h_{i_2}\calu^h_{i_3}\calu^h_{i_4}\calu^h_{i_5}\calu^h_{i_6}\,,\\
    & \left(c_{h^6,2}^{000000}\right){}^{hhhhhh} = \frac{C_{D^2H^3H^\dagger{}^3,1}\delta^{i_1}_{i_2}\delta^{i_3}_{i_4}\delta^{i_5}_{i_6}}{\Lambda^4}\calu_{i_1}^h\calu_{i_3}^h\calu_{i_5}^h\calu^{i_2h}\calu^{i_4h}\calu^{i_6h}\,,\\
    & \left(c_{h^6,3}^{000000}\right){}^{hhhhhh} = \frac{C_{D^2H^3H^\dagger{}^3,2}\delta^{i_1}_{i_2}\delta^{i_3}_{i_4}\delta^{i_5}_{i_6}}{\Lambda^4}\calu_{i_1}^h\calu_{i_3}^h\calu_{i_5}^h\calu^{i_2h}\calu^{i_4h}\calu^{i_6h}\,,
\end{align}

\paragraph{\boldsymbol{$f\bar{f}h^5$}}
\begin{align}
     \left(\mathcal{M}_{f\bar{f}h^5}\right){}^{\hi{1}hhhhh}_{\hi{2}} \left(\mathbf{\frac{1}{2},\frac{1}{2},0,0,0,0,0}\right) &= \left(c_{f\bar{f}h^5}^{--00000}\right){}^{\hi{1}hhhhh}_{\hi{2}} \lra{\mathbf{12}} + \text{h.c.}\,,\label{eq:cls_fbarh5}
\end{align}

\begin{align}
    \left(c_{f\bar{f}h^5}^{--00000}\right){}^{\hi{1}hhhhh}_{\hi{2}}  &= \left(\frac{C_{Le_\mathbbm{C}H^2H^\dagger{}^3}\delta^{i_1}_{i_5}\delta_{i_2}^4\delta^{i_3}_{i_6}\delta^{i_4}_{i_7}}{\Lambda^4}\right)\Omega_{i_1}^{\hi{1}}\Omega^{i_2}_{\hi{2}}\calu_{i_3}^h\calu_{i_4}^h\calu^{i_5 h}\calu^{i_6 h}\calu^{i_7 h}\,.
\end{align}

\paragraph{\boldsymbol{$h^7$}}
\begin{align}
    &\mathcal{M}^{hhhhhhh}_{h^7} \left(\mathbf{0,0,0,0,0,0,0}\right) =  \left(c_{h^7}^{0000000}\right){}^{hhhhhhh} \,,\label{eq:cls_h7}
\end{align}

\begin{align}
    \left(c_{h^7}^{0000000}\right){}^{hhhhhhh} &= \frac{C_{H^4H^\dagger{}^4}\delta^{i_1}_{i_2}\delta^{i_3}_{i_4}\delta^{i_5}_{i_6}\delta^{i_7}_{i_8}}{\Lambda^4} \calu_{i_1}^h\calu_{i_3}^h\calu_{i_5}^h\calu^{i_2h}\calu^{i_4h}\calu^{i_6h}v\,,
\end{align}

\paragraph{\boldsymbol{$h^8$}}
\begin{align}
    &\mathcal{M}^{hhhhhhhh}_{h^8} \left(\mathbf{0,0,0,0,0,0,0,0}\right) =  \left(c_{h^8}^{00000000}\right){}^{hhhhhhhh} \,, \label{eq:cls_h8}
\end{align}

\begin{align}
    \left(c_{h^8}^{00000000}\right){}^{hhhhhhhh} &= \frac{C_{H^4H^\dagger{}^4}\delta^{i_1}_{i_2}\delta^{i_3}_{i_4}\delta^{i_5}_{i_6}\delta^{i_7}_{i_8}}{\Lambda^4} \calu_{i_1}^h\calu_{i_3}^h\calu_{i_5}^h\calu^{i_2h}\calu^{i_4h}\calu^{i_6h}\,.
\end{align}

%% file: sec6-con.tex
\section{Summary}\label{sec:con}

We have developed a systematic correspondence between massless contact amplitudes in the unbroken phase and massive contact amplitudes in theories with spontaneous symmetry breaking. Our construction uses the spin-transversality (ST) basis, in which a massive particle carries both its $SU(2)$ little-group spin and a transversality quantum number. Resolving an ST amplitude into minimal-helicity-chirality (MHC) components organizes its high-energy behavior in powers of $\mathbf m/E$ and makes its connection to massless spinor-helicity amplitudes explicit. In particular, the $U(2)=SU(2)\times U(1)_t$ symmetry for single massive particle state allows the SSYT construction of massless Lorentz structures to be extended directly to the massive case.

The matching is governed by the first non-vanishing MHC component. When this component occurs at leading order, a massive ST structure is obtained by direct matching to a massless contact amplitude of the same UV origin. In matching a Goldstone mode to a longitudinal vector, the corresponding massless amplitude must additionally obey the Adler-zero condition. This direct branch applies to general multiplicity and supplies the lowest-dimension three- and four-point ST bases from the associated massless bases.

We also identified the obstruction to this simple picture when the first non-vanishing MHC component is at sub-leading order. Some massive structures have a vanishing leading high-energy limit even though they cannot be written with an overall explicit mass factor. They originate from conserved currents and require exceptional matching: their first non-zero descendant is fixed by the covariant-derivative interaction $A_\mu J^\mu$, supplemented where appropriate by operator contributions related by the Goldstone equivalence theorem. In a generic massive theory, the exceptional classes are $VVS$, $VSS$, $ffV$, $VVV$, and $VVVV$. For the SM particle content, only $VVS$, $ffV$, and $VVV$ occur; the $VSS$ and $VVVV$ classes require particle species absent from the SM spectrum.

We applied this framework to electroweak symmetry breaking, $SU(2)_L\times U(1)_Y\to U(1)_{\mathrm{em}}$, in the one-flavor electroweak sector of the SMEFT. Using the field-transformation projectors, we expressed broken-phase amplitude coefficients in terms of unbroken-phase Wilson coefficients and, for exceptional structures, renormalizable SM couplings. We presented explicit matching results through dimension eight for amplitudes with three to eight external particles. Quark and gluon fields can be retained through the generic field labels, but explicit $SU(3)_C$ color tensors were not listed; because color is unbroken, they factorize from the electroweak matching and can be attached separately.

The ST basis is designed to retain UV information rather than to be minimal at every infrared mass dimension. Consequently, it can be overcomplete before higher-dimensional UV origins are specified. This is the trade-off that makes the high-energy expansion and the matching to unbroken-phase operator coefficients transparent. Once the allowed UV operators are included, the resulting relations recover the redundancies expected in more minimal massive bases.

Several extensions are immediate. The explicit SMEFT results can be enlarged to include quark and gluon sectors, additional flavor structure, and operators of higher mass dimension. More generally, the construction depends only on the particle content, gauge symmetry, and pattern of spontaneous symmetry breaking, and can therefore be applied to other effective theories and to models with higher-spin massive states. The ST--MHC correspondence thus provides a unified on-shell framework for tracing local interactions from an unbroken UV description to massive amplitudes in the infrared.

%% file: sec7-app.tex
\section{Dual Group vs Little Group}
Within the spinor description, the global linear representation of the Lorentz group automatically generates little-group transformations. In the massless case, the little group acts merely via global $\mathrm{U}(1)$ phase rotations
\begin{eqnarray}
\text{Little group } U(1):\quad \Lambda_\alpha^{\;\beta}\lambda_\beta(p)=\lambda_\alpha(\Lambda p) e^{-\frac{i}{2}\theta(\Lambda,p)}, \quad \Lambda_{\dot\alpha}^{*\dot\beta}\tilde{\lambda}_{\dot\beta}(p)={\tilde\lambda}_{\dot\alpha}(\Lambda p) e^{\frac{i}{2}\theta(\Lambda,p)}.
\end{eqnarray}
This phase nature of the little group imposes stringent constraints on the structure of scattering amplitudes. We now introduce a dual group transformation bearing formal similarities yet distinct from the foregoing little-group transformation:
\begin{eqnarray}
    \text{Dual group } U(1):\quad\lambda_\alpha(p)\to  \lambda_\alpha(p)e^{-i\varphi},\quad {\tilde\lambda}_{\dot\alpha}(p)\to  {\tilde\lambda}_{\dot\alpha}(p)e^{i\varphi}.
\end{eqnarray}
We refer to this $\mathrm{U}(1)$ group as the dual group, since its action commutes with that of the Poincar\'e group. In the case of a massless particle, the action of the dual group is nearly indistinguishable from that of the little group; in effect, it may be interpreted as a phase redefinition of the particle state. Given the perfect overlap between little-group transformations and dual-group actions, the dual group fully governs the helicity (spin) properties of the particle. Furthermore, since the dual group commutes with momentum generators, it leaves the four-momentum of the particle unchanged, which accounts for its striking similarity to the little group. As we shall demonstrate, an analogous structure persists in the massive case.

For massive particles, the standard reference momentum is conventionally taken to be the rest-frame four-momentum, and the corresponding little group is the rotation group $\mathrm{SU}(2)$. Under Lorentz transformations, the spinor representation induces the following little-group transformation:
\begin{eqnarray}
\text{Little group } SU(2):\quad \Lambda_\alpha^{\;\beta}\lambda_\beta^I(p)=\lambda_\alpha^J(\Lambda p) W_J^I(\Lambda,p), \quad \Lambda_{\dot\alpha}^{*\dot\beta}\tilde{\lambda}^I_{\dot\beta}(p)={\tilde\lambda}^J_{\dot\alpha}(\Lambda p) W_J^I(\Lambda,p).
\end{eqnarray}
For a single spinor, $W_J^I$ furnishes the spin-$1/2$ representation. General spin-$s$ representations can be constructed upon this fundamental building block.
We emphasize that $W$ is a little-group element induced from $\Lambda$, whose action leaves the momentum $k$ invariant. Analogous to the massless case, we may define the following dual group transformation in $\mathrm{U}(2)$ that commutes with the Poincar\'e group:
\begin{eqnarray}\label{eq:DualgroupSU2}
\text{Dual group } U(2):\quad \lambda_\alpha^I(p)\to\lambda_\alpha^J(p) w_J^Ie^{-i\varphi}, \quad \tilde{\lambda}^I_{\dot\alpha}(p)\to{\tilde\lambda}^J_{\dot\alpha}( p) w_J^Ie^{i\varphi}.
\end{eqnarray}
This constitutes the dual group associated with a single massive particle. Analogous to the massless scenario, the dual group may be regarded as a redefinition of spinors. Its $\mathrm{SU}(2)$ component $w^{\,I}_{J}$ corresponds to a reorientation of the particle’s spin axis, while the $\mathrm{U}(1)$ factor $e^{i\varphi}$ implements a phase redefinition. Once the spin axis of the particle is fixed, $w^{\,I}_{J}$ is uniquely determined and the redundant $\mathrm{SU}(2)$ freedom is eliminated. Imposing the constraint $\det(\lambda_\alpha^I)=\det(\tilde{\lambda}_{\dot\alpha I})=m$ further removes the residual $\mathrm{U}(1)$ ambiguity. Fixing a representative $\lambda_{\alpha}^I=\lambda_{\alpha}^I(\mathbf{p})$, the $\mathrm{SL}(2,\mathbb{C})$ Lorentz action on $\lambda(p)$ induces the little-group transformation in $\Lambda |\mathbf{p}^I\rangle=|\Lambda \mathbf{p}^J\rangle W_J^I$ with Wigner rotation $W$, motivating the conventional name "little-group index" for $I$.

Nevertheless, the little group (the momentum-stabilizing Lorentz subgroup) acts on $\lambda$ from the left, while $W$ is merely the induced $\mathrm{SU}(2)$ transformation on the index $I$ and is not the little-group element itself. The transformation $\lambda\to\lambda w$ defines the dual group of the Poincar\'e group, whose action commutes with Poincar\'e transformations and differs fundamentally from the little group. To illustrate this distinction, we consider the rest momentum $k$, whose spinor takes the parameterized form
\begin{eqnarray}
    \lambda_\alpha^I=\sqrt{m}\begin{pmatrix}
        \cos \frac{2\alpha+\pi}{4} e^{i\frac{\beta+\gamma}{4}} & \sin \frac{2\alpha+\pi}{4} ie^{i\frac{\gamma-\beta}{4}}\\
        \sin\frac{2\alpha+\pi}{4} ie^{i\frac{\beta-\gamma}{4}}& \cos\frac{2\alpha+\pi}{4} e^{-i\frac{\beta+\gamma}{4}}
    \end{pmatrix}.
\end{eqnarray}
This spinor describes a stationary particle with spin axis $(\cos\alpha\sin\frac{\gamma}{2} ,\cos\alpha\cos\frac{\gamma}{2},-\sin\alpha)$. The little group and dual group act naturally upon it:
\begin{eqnarray}
    \exp({i\boldsymbol{\theta}\cdot \boldsymbol{J}})\lambda_\alpha^I=R_\alpha^\beta(\boldsymbol{\theta}) \lambda_\beta^I,\quad \exp({i\boldsymbol{\theta}'\cdot \boldsymbol{\mathcal{K}}})\lambda_\alpha^I=\lambda_\alpha^J  R_J^I(\boldsymbol{\theta}'),
\end{eqnarray}
where $\boldsymbol{\theta}\cdot \boldsymbol{J}=\sum_{i=1}^3\theta_i\, J_i$, and similarly for $\boldsymbol{\theta}'\cdot \boldsymbol{\mathcal{K}}$. The little-group rotation $\exp({i\boldsymbol{\theta}\cdot \boldsymbol{J}})$ leaves the rest momentum $k$ invariant and rotates the particle spin components. Meanwhile, the dual group $\exp({i\boldsymbol{\theta}'\cdot \boldsymbol{\mathcal{K}}})$ acts to rotate the spin axis, and may alternatively be viewed as a redefinition of the particle state. The little-group generators $J$ of this spinor can be extracted from the Lorentz generators 
\begin{equation}
J_3  = 2i\partial_\gamma,\quad
J_- = -i e^{\frac{i\gamma}{2}} \left(\partial_\alpha - 2i\tan\alpha\,\partial_\gamma - 2i\sec\alpha\,\partial_\beta\right),
\end{equation}
with $J_-=J_1-i J_2$. The dual-group generators $\mathcal{K}^I_J$ can be derived using standard group-theoretic techniques:
\begin{equation}
\mathcal{K}_0  = 2i\partial_\beta,\quad
\mathcal{K}_-  = -i e^{\frac{i\beta}{2}} \left(\partial_\alpha - 2i\tan\alpha\,\partial_\beta - 2i\sec\alpha\,\partial_\gamma\right).
\label{eq:A.7}
\end{equation}
The commutator of the little-group and dual-group generators satisfies $[J_i,\mathcal{K}_j]=0$. Taking $J_3$ and $K_-$ as illustrative examples, evaluating their corresponding matrix elements in the vicinity of $\beta = \gamma = 2\alpha+\pi = 0$ yields
\begin{eqnarray}
    J_3=\frac{1}{2}\begin{pmatrix}
        -1&0\\0&1
    \end{pmatrix}\otimes \mathbf{1}_{2\times2},\quad 
    \mathcal{K_-}=\mathbf{1}_{2\times2}\otimes\begin{pmatrix}
        0 & 1\\
        0&0
    \end{pmatrix},
\end{eqnarray}
where $\mathbf{1}_{2\times2}$ denotes the identity matrix. The Kronecker product $\otimes$ arises because $J_I$ acts on the first index $\alpha$ of $\lambda_{\alpha}^{I}$, while $\mathcal{K}_i$ acts on the second index $I$~\footnote{If one fails to distinguish the respective indices on which these generators act and naively computes their representation matrices, inconsistencies $\Big[\frac{1}{2}\begin{pmatrix}
    -1&0\\0&1    
\end{pmatrix},
\begin{pmatrix}
        0 & 1\\
        0&0
\end{pmatrix}\Big]\neq0,$ will emerge.}.

We observe that the spinor for a massive particle decomposes into two massless spinors, which leads to a dual group of structure $\mathrm{U}(2)$. Generalising this construction, a system containing $N$ massless particles is equipped with $N$ independent massless spinors, and the corresponding dual group becomes $\mathrm{U}(N)$
\begin{eqnarray}
    \text{Dual group }U(N):\quad
    \begin{cases}
    (|1\rangle,|2\rangle,\dots,|N\rangle)&\to (|i\rangle U_{i1},|i\rangle U_{i2},\dots,|i\rangle U_{iN}),\\
    ([1|,[2|,\dots,[N|)&\to (U_{i1}^* [i|,U_{i2}^* [i|,\dots,U_{iN}^* [i|)
    \end{cases}.
\end{eqnarray}
Treating the particle labels as indices $i$ of $N$ massless spinors, we may rewrite the above expression in a more abstract form:
\begin{eqnarray}
    \lambda_\alpha^i\to \lambda_\alpha^j U_{ji},\quad\tilde{\lambda}_{\dot\alpha i}=(\lambda_\alpha^i)\to \tilde{\lambda}_{\dot{\alpha}j}U^*_{ji},\quad i,j=1,2,\dots,N,
\end{eqnarray}
which matches the transformation Eq.~\eqref{eq:DualgroupSU2} for the massive dual group exactly. The helicity of a single massless particle is fixed equivalently by the $\mathrm{U}(1)$ factor of either its little group or the dual group, meaning the particle transforms in a definite $\mathrm{U}(1)$ representation. For scattering amplitudes involving $N$ massless particles, this yields a $\mathrm{U}(1)^N$ representation. The remaining generators of the full $\mathrm{U}(N)$ group encode intrinsic internal information among these massless particles. This setup corresponds to inducing representations of the subgroup $\mathrm{U}(1)^N$ from representations of $\mathrm{U}(N)$; the number of inequivalent induced representations directly dictates the number of distinct independent amplitude structures.

Going further, suppose the system consists of $N$ massless particles and $N'$ massive particles; there are altogether $N + 2N'$ massless spinors, and the total dual group is therefore $\mathrm{U}(N+2N')$. Consistent with the purely massless case, the individual dual group of each particle determines its intrinsic spin property and transversality condition, which is equivalent to specifying a representation of $\mathrm{U}(2)$. For an $(N+N')$-point scattering process, once the single-particle quantum info encoded in representations of $\mathrm{U}(1)^N\otimes\mathrm{U}(2)^{N'}$ are fixed, combining these with the global internal data carried by $\mathrm{U}(N+2N')$ yields inequivalent induced representations of $\mathrm{U}(1)^N\otimes\mathrm{U}(2)^{N'}$, each corresponding to an independent amplitude structure. Furthermore, taking the high-energy limit for each massive particle, the high-energy components dominate each spinor at the leading order; the amplitude structure is then approximately governed by $N+N'$ effective spinors and characterised by the group $\mathrm{U}(N+N')$.


\section{Brief Review of Selected Massive Bases}
\label{app:basis}

The basis of massive amplitudes of certain dimensions is not unique. Different strategies of reduction lead to different bases.
Refs.~\cite{Arkani-Hamed:2017jhn,Dong:2021vxo} use a chiral basis, in which all the external fields are of the left-handed spinor representation. Its shortcoming is that there could be many momentum insertions, making the dimension of the corresponding operators implicit. On the contrary, Refs.~\cite{Durieux:2019eor,Durieux:2020gip} choose another strategy to extract the momentum contractions from spinor bilinears. Consequently, a general amplitude takes the form
\begin{equation}
\label{eq:sctexpansion}
    \mathcal{M} = \sum_n \mathcal{S}_n^{\{I\}} L_n\left[s_{ij}, \epsilon(\mathbf{p}_i,\mathbf{p}_j,\mathbf{p}_k,\mathbf{p}_l)\right]\,.
\end{equation}
$L_n$'s are polynomials of the Lorentz invariants $s_{ij}, \epsilon(\mathbf{p}_i,\mathbf{p}_j,\mathbf{p}_k,\mathbf{p}_l)$, and $ \mathcal{S}_n^{\{I\}}$ are the structures composed of spinors of the external fields, carrying the little-group indices $\{I\}$. By reductions, we can always obtain a finite set of $ \mathcal{S}_n^{\{I\}}$, in which the momentum insertions are minimal. Essentially, we consider only the polynomials $L_n$ and structures $ \mathcal{S}_n^{\{I\}}$ with non-negative powers of the Lorentz invariants. Consequently, for an amplitude of a specific dimension, its expansion such as Eq.~\eqref{eq:sctexpansion} is determined by the structures $ \mathcal{S}_n^{\{I\}}$. In this point of view, the finite set of $ \mathcal{S}_n^{\{I\}}$ is referred to as the stripped contact terms (SCTs). The SCTs form a basis called the SCT basis, which has specific dimensions. Compared to the ST basis, the SCT basis eliminates not only the EOM and IBP, but also all the scalars. Mathematically, the amplitudes are regarded as a module over the ring of the dynamic invariants, $s_{ij}, \epsilon(\mathbf{p}_i,\mathbf{p}_j,\mathbf{p}_k,\mathbf{p}_l)$. The SCTs are actually the generators of the module.

Furthermore, if we remove the restriction that all the functions $L_n$ are of non-negative powers, which means they could be fractional functions such as $s_{ij}/s_{kl}$.
The amplitudes are regarded as modules over not the ring but a field composed of $s_{ij}, \epsilon(\mathbf{p}_i,\mathbf{p}_j,\mathbf{p}_k,\mathbf{p}_l)$. Consequently, these scalars can be divided, and the SCT basis can be reduced further.
The resultant basis, which exploits all the independent relations among the SCTs, is called the spinor-structure basis. Any amplitudes can be expanded in terms of the spinor-structure basis as in Eq.~\eqref{eq:sctexpansion}, but the powers of the prefactors $F_n$ are not necessarily non-negative.

For both the SCT basis and the spinor-structure basis, the constructions rely heavily on the H.E. limit and the massless amplitudes.
A typical example is the 3-point class $\boldsymbol{VVV}$ of $(\mathbf1^1,\mathbf2^1,\mathbf3^1)$, the following six
dimension-3 structures occur for
$t=(\pm1,0,0),(0,\pm1,0),(0,0,\pm1)$:
\begin{equation} \begin{aligned} \label{eq:VVV_six}
&[\mathbf{12}][\mathbf{23}]\langle\mathbf{13}\rangle,\quad
[\mathbf{13}][\mathbf{23}]\langle\mathbf{12}\rangle,\quad
[\mathbf{21}][\mathbf{13}]\langle\mathbf{23}\rangle,\\
&[\mathbf{12}]\langle\mathbf{23}\rangle\langle\mathbf{13}\rangle,\quad
[\mathbf{23}]\langle\mathbf{21}\rangle\langle\mathbf{13}\rangle,\quad
[\mathbf{13}]\langle\mathbf{23}\rangle\langle\mathbf{12}\rangle.
\end{aligned} \end{equation}
For $t=(1,1,1)$ and $t=(-1,-1,-1)$, the two additional amplitudes are
\begin{equation}
[\mathbf{13}][\mathbf{23}][\mathbf{12}],\quad
\langle\mathbf{13}\rangle\langle\mathbf{23}\rangle\langle\mathbf{12}\rangle .
\end{equation}
Thus, these 8 amplitudes form the SCT basis, all of which can be obtained by matching the massless amplitudes. In particular, the LO matching of the amplitudes in Eq.~\eqref{eq:VVV_six} vanishes, and they are contributed only from NLO and higher-order matching. For example, the dimension-8 type $F_R^2D^2\phi^2$ and $D^4\phi^4$ match to them. For the case of $D^4\phi^4$, if we require the 4 scalar fields to be soft, we have the basic amplitudes that
\begin{align}
a^{\{0,0,0,0\}}_{D^4\phi^4,1} &= \langle 12\rangle \langle 34\rangle [12][34]\,,\\
a^{\{0,0,0,0\}}_{D^4\phi^4,2} &= \langle 13\rangle \langle 24\rangle [13][24]\,,\\
a^{\{0,0,0,0\}}_{D^4\phi^4,3} &= \langle 14\rangle \langle 23\rangle [14][23]\,.
\end{align}
The Fierz identity gives that
\begin{equation}
    a^{\{0,0,0,0\}}_{D^4\phi^4,1} = \langle 12\rangle \langle 34\rangle [12][34]= \langle13\rangle\langle 24\rangle [12][34] -\langle 14\rangle\langle23\rangle[12][34] = \langle12\rangle\langle34\rangle[13][24]-\langle12\rangle\langle34\rangle[14][23]\,.
\end{equation}
If the first field is flipped, we have 
\begin{equation}
    a^{\{0,0,0,0\}}_{D^4\phi^4,1} \rightarrow -\langle\mathbf{12}\rangle[\mathbf{12}]\langle \mathbf{3}\mathbf{p}_1\mathbf{3}] = -\langle\mathbf{13}\rangle[\mathbf{12}]\langle \mathbf{2}\mathbf{p}_1\mathbf{3}] + m_1 [\mathbf{12}][\mathbf{13}]\langle\mathbf{23}\rangle = -\langle \mathbf{12}\rangle[\mathbf{13}] \langle \mathbf{3} \mathbf{p}_1\mathbf{2}] + \tilde{m}_1\langle\mathbf{12}\rangle\langle\mathbf{13}\rangle [\mathbf{23}]\,.
\end{equation}
Using massive momentum conservation, the second equation gives
\begin{eqnarray}
    m_1\langle\mathbf{12}\rangle\langle\mathbf{13}\rangle[\mathbf{23}]+m_2\langle\mathbf{12}\rangle\langle\mathbf{23}\rangle[\mathbf{13}]+m_3\langle\mathbf{23}\rangle\langle\mathbf{13}\rangle[\mathbf{12}] \notag\\
    =m_1[\mathbf{12}][\mathbf{13}]\langle\mathbf{23}\rangle+m_1[\mathbf{12}][\mathbf{23}]\langle\mathbf{13}\rangle+m_1[\mathbf{13}][\mathbf{23}]\langle\mathbf{12}\rangle\,,\label{eq:sctrelation}
\end{eqnarray}
where we have identified $m=\tilde{m}$.
This relation is symmetric under the permutations of the three fields, so the same analysis of the other two amplitudes $a^{\{0,0,0,0\}}_{D^4\phi^4,2,3}$ gives no more independent relations.
Once the division of the masses is allowed, we can solve any one of them using the other five independent structures.  
Thus, there are 7 amplitudes in the spinor-structure basis, as summarized in Table~\ref{tab:3ptmassive}.
An alternative way to eliminate the redundancy is via a numerical check, as proposed in Ref.~\cite{AccettulliHuber:2021uoa,DeAngelis:2022qco}.

In summary, not only the SCT basis but also the relations leading to the spinor-structure basis can be obtained by the massless-massive correspondence. However, the relations usually appear beyond LO matching and merge amplitudes across different transversalities, making the construction of the spinor-structure basis difficult.
On the one hand, since the relations such as Eq.~\eqref{eq:sctrelation} usually appear at high dimension, reflecting UV constraints of the corresponding massless amplitudes, the spinor-structure basis is not adequate for the LO matching.
On the other hand, since the operators in the effective Lagrangian are all local, the SCT basis corresponds directly to the effective operators. While for the spinor-structure basis, they are the independent ones by quotienting out all the linear relations. 
Thus, they are also inadequate when matching to effective operators. For example, although the relation in Eq.~\eqref{eq:sctrelation} eliminates one amplitude of dimension 3, it is valid only at dimension 4. Thus, if we are concerned only with the dimension 3 amplitudes and their corresponding operators, the spinor-structure basis is not enough.


\begin{table}
\centering
\begin{tabular}{c|c|c}
\mbox{external particles} & \mbox{SM structures} & \mbox{EFT structures} \\
\hline
$(\mathbf{1}^{1/2}, \mathbf{2}^{1/2}, \mathbf{3}^{1})$ &  $\langle\mathbf{23}\rangle [\mathbf{31}], [\mathbf{23}] \langle\mathbf{31}\rangle$ & $[\mathbf{23}][\mathbf{31}],\langle\mathbf{23}\rangle \langle\mathbf{31}\rangle $\\
\hline
$(\mathbf{1}^{1/2}, \mathbf{2}^{1/2}, \mathbf{3}^{0})$ & $[\mathbf{12}], \langle\mathbf{12}\rangle$ &  \\
\hline
$(\mathbf{1}^{1}, \mathbf{2}^{1}, \mathbf{3}^{1})$ &$ \makecell{ [\mathbf{12}] [\mathbf{23}] \langle\mathbf{31}\rangle, [\mathbf{12}] \langle\mathbf{23}\rangle [\mathbf{31}], \langle\mathbf{12}\rangle [\mathbf{23}] [\mathbf{31}],$  $\\$  $\langle\mathbf{12}\rangle [\mathbf{23}] \langle\mathbf{31}\rangle,[\mathbf{12}]\langle\mathbf{23}\rangle \langle\mathbf{31}\rangle,\color{blue}{\langle\mathbf{12}\rangle \langle\mathbf{23}\rangle [\mathbf{31}]} }$
& $\makecell{[\mathbf{12}] [\mathbf{23}] [\mathbf{31}],\\\langle\mathbf{12}\rangle \langle\mathbf{23}\rangle \langle\mathbf{31}\rangle}$ \\
\hline
$(\mathbf{1}^{1}, \mathbf{2}^{1}, \mathbf{3}^{0})$ &  $\langle\mathbf{12}\rangle [\mathbf{12}]$ & $\langle\mathbf{12}\rangle^2,[\mathbf{12}]^2$ \\
 \hline
 $(\mathbf{1}^{1}, \mathbf{2}^{0}, \mathbf{3}^{0})$ & $\langle\mathbf{1}|\mathbf{p}_2|\mathbf{1}]$ \\
\hline
$(\mathbf{1}^{0}, \mathbf{2}^{0}, \mathbf{3}^{0})$ & $1 $&
\end{tabular}
\caption{This table gives all the 3-point structures for spin $\le 1$ particles. The blue term is the redundant structure and is not included in the SCT basis.}
\label{tab:3ptmassive}
\end{table}

\section{Detailed Diagrammatic Construction of MHC Amplitudes}
\label{app:detail_MHC}

In this appendix, we provide a more detailed diagrammatic representation and current decomposition discussion for five types of MHC amplitudes: $ffS$, $ffV$, $VVS$, $VVV$ and $VVVV$. Both direct and exceptional matching are covered. As we will see, when multiple vector bosons are present, the amplitude admits more than one current decomposition, which will simplify the subsequent discussion.

\subsection{\texorpdfstring{$ffS$, $ffV$, and $VVS$}{ffS, ffV, and VVS}}

\paragraph{1. $ffS$}

The primary MHC amplitudes for the $ffS$ class are not derived from $[\mathcal J]_0\cdot [\mathbf{A}]_0$,
\eqs{
(-\frac{1}{2},-\frac{1}{2},0): \ c_{ffS}^{--0}\langle12\rangle = \Ampthree{1^-}{2^-}{3^0}{\fer{red}{i1}{v1}}{\antfer{cyan}{i2}{v1}}{\sca{i3}} \xrightarrow{H.E.}  \Ampthree{1^-}{2^-}{3^0}{\propag[fer] (i1) to (v1)}{\propag[antfer] (i2) to (v1)}{\propag[sca] (i3) to (v1); \vertex[dot] (v) at (0,0)} =Y\langle12\rangle, \\
(+\frac{1}{2},+\frac{1}{2},0): \ c_{ffS}^{++0} [12] = \Ampthree{1^+}{2^+}{3^0}{\fer{cyan}{i1}{v1}}{\antfer{red}{i2}{v1}}{\sca{i3}} \xrightarrow{H.E.}  \Ampthree{1^+}{2^+}{3^0}{\propag[fer] (i1) to (v1)}{\propag[antfer] (i2) to (v1)}{\propag[sca] (i3) to (v1); \vertex[dot] (v) at (0,0) }
=Y^* [12].
} 
Thus, in the H.E. limit, these amplitudes reduce to the Yukawa interaction, and the coefficients are given by
\eq{
c_{ffS}^{--0} = Y, \quad c_{ffS}^{++0} = Y^*.
}

\paragraph{2. $VVS$}

There are two primary MHC amplitudes in the $VVS$ class that are not generated by $[\mathcal J]_0\cdot [\mathbf{A}]_0$,
\eqs{
(-1,-1,0): \ c^{--0}_{VVS}\langle12\rangle^2 = \Ampthree{1^-}{2^-}{3^0}{\bos{i1}{red}}{\bos{i2}{red}}{\sca{i3}} \xrightarrow{H.E.} \Ampthree{1^-}{2^-}{3^0}{\propag[bos] (i1) to (v1)}{\propag[bos] (i2) to (v1)}{\propag[sca] (i3) to (v1); \vertex[dot] (v) at (0,0){}}
=\frac{1}{\Lambda} C_{F_L^2 \phi} \langle12\rangle^2, \\
(+1,+1,0): \ c^{++0}_{VVS}{[12]}^2 = \Ampthree{1^+}{2^+}{3^0}{\bos{i1}{cyan}}{\bos{i2}{cyan}}{\sca{i3}} \xrightarrow{H.E.} \Ampthree{1^+}{2^+}{3^0}{\propag[bos] (i1) to (v1)}{\propag[bos] (i2) to (v1)}{\propag[sca] (i3) to (v1); \vertex[dot] (v) at (0,0){}}
=\frac{1}{\Lambda} C_{F_R^2 \phi} {[12]}^2.
}
They correspond to the massless EFT operators $F_L^2\phi$ and $F_R^2\phi$ respectively. Therefore, these two primary MHC amplitudes are identified with massive EFT operators, with the following Wilson coefficients
\eq{
c^{--0}_{VVS} = \frac{1}{\Lambda} C_{F_L^2 \phi}, \quad
c^{++0}_{VVS} = \frac{1}{\Lambda} C_{F_R^2 \phi}.
}

The other primary amplitude is generated by $[\mathcal J]_0\cdot [\mathbf{A}]_0$,
\eq{
(0,0,0): \ \langle12\rangle [21] 
=& [\mathcal{J}(2^{0},3^{0})]_0\cdot[\mathbf A(1^{0})]_0 =[\mathcal{J}(1^{0},3^{0})]_0\cdot[\mathbf A(2^{0})]_0 \\
=&\Ampthree{1^0}{2^0}{3^0}{\bos{i1}{brown}}{\bos{i2}{brown}}{\sca{i3}} \xrightarrow{H.E.} 0.
}
This shows that we can choose either particle 1 or 2 as $[\mathbf{A}]_0$, so there are two equivalent ways to perform the current decomposition. The primary $VS$ current $[\mathcal J]_0$ is defined as
\begin{equation} \begin{aligned}
\relax
[\mathcal{J}(2^{0},3^{0})]_0&=|2]^{\dot\alpha}\langle2|^{\alpha}, \\
[\mathcal{J}(1^{0},3^{0})]_0&=|1]^{\dot\alpha}\langle1|^{\alpha}.
\end{aligned}\end{equation}

Similarly, the 1st descendant MHC amplitude also admits two current decompositions: $[\mathcal J]_0\cdot [\mathbf{A}]_1$ and $[\mathcal J]_1\cdot [\mathbf{A}]_0$. They are
\eq{
(-1,0,0): \ c^{-00}_{VVS} m_1 \langle12\rangle [2\eta_1] 
=& [\mathcal{J}(1^{-},3^{0})]_1 \cdot [\mathbf{A}(2^{0})]_0 = [\mathcal{J}(2^{0},3^{0})]_0 \cdot [\mathbf{A}(1^{-})]_{1} \\
=& \Ampthree{1^-}{2^0}{3^0}{\bosflip{1}{180}{brown}{red}}{\bos{i2}{brown}}{\sca{i3}}
\xrightarrow{H.E.} \Ampthree{1^-}{2^0}{3^0}{\propag[bos] (i1) to (v1)}{\propag[gho] (i2) to (v1)}{\propag[sca] (i3) to (v1)} = g\frac{\langle 12\rangle\langle 31\rangle}{\langle 23\rangle},
}
\eq{
(+1,0,0): \ -c^{+00}_{VVS} \tilde{m}_1 \langle\eta_12\rangle [21] 
=& [\mathcal{J}(1^{+},3^{0})]_1 \cdot [\mathbf{A}(2^{0})]_0 = [\mathcal{J}(2^{0},3^{0})]_0 \cdot [\mathbf{A}(1^{+})]_{1} \\
=& \Ampthree{1^+}{2^0}{3^0}{\bosflip{1}{180}{brown}{cyan}}{\bos{i2}{brown}}{\sca{i3}}  
\xrightarrow{H.E.} \Ampthree{1^+}{2^0}{3^0}{\propag[bos] (i1) to (v1)}{\propag[gho] (i2) to (v1)}{\propag[sca] (i3) to (v1)} = g\frac{[12][31]}{[23]},
}
\eq{
(0,-1,0): \ c^{0-0}_{VVS} m_2 \langle12\rangle [\eta_21] 
=& [\mathcal{J}(2^{-},3^{0})]_1 \cdot [\mathbf{A}(1^{0})]_0 = [\mathcal{J}(1^{0},3^{0})]_0 \cdot [\mathbf{A}(2^{-})]_{1} \\
=& \Ampthree{1^0}{2^-}{3^0}{\bos{i1}{brown}}{\bosflip{1}{55}{brown}{red}}{\sca{i3}}  
\xrightarrow{H.E.} \Ampthree{1^0}{2^-}{3^0}{\propag[bos] (i2) to (v1)}{\propag[gho] (i1) to (v1)}{\propag[sca] (i3) to (v1)} = g\frac{\langle 23\rangle\langle 12\rangle}{\langle 31\rangle},
}
\eq{
(0,+1,0): \ -c^{0+0}_{VVS} \tilde{m}_2 \langle1\eta_2\rangle [21] 
=& [\mathcal{J}(2^{-},3^{0})]_1 \cdot [\mathbf{A}(1^{0})]_0 = [\mathcal{J}(1^{0},3^{0})]_0 \cdot [\mathbf{A}(2^{-})]_{1} \\
=& \Ampthree{1^0}{2^+}{3^0}{\bos{i1}{brown}}{\bosflip{1}{55}{brown}{cyan}}{\sca{i3}} 
\xrightarrow{H.E.} \Ampthree{1^0}{2^+}{3^0}{\propag[bos] (i2) to (v1)}{\propag[gho] (i1) to (v1)}{\propag[sca] (i3) to (v1)} = g\frac{[23][12]}{[31]}.
}
For the first decomposition $[\mathcal J]_0\cdot [\mathbf{A}]_1$, as proposed in Ref.~\cite{Ni:2026wiz}, it corresponds to massless SM contact terms involving a gauge boson. The H.E limit of these MHC amplitudes is $igA_{\mu}(\phi^*\partial^{\mu}\phi-\phi\partial^{\mu}\phi^*)$, which is a subleading component of the kinetic term $D_{\mu}\phi (D^{\mu}\phi)^*$,
\eq{
\begin{array}{ccccc}
& & \mbox{leading} & \mbox{subleading} & \mbox{subsubleading} \\
D_{\mu}\phi (D^{\mu}\phi)^* & = & \partial_{\mu} \phi \partial^{\mu}\phi^* & +igA_{\mu}(\phi^*\partial^{\mu}\phi-\phi\partial^{\mu}\phi^*) & +g^2 A_{\mu} A^{\mu} \phi \phi^*.
\end{array}
}
The subleading component is not a gauge-invariant operator but merely a contact term, so the above MHC amplitudes do not originate from an operator. Thus, their coefficients are determined by the gauge coupling:
\begin{equation}
c^{-00}_{VVS} \mathbf{m}_1^2 =
c^{+00}_{VVS} \mathbf{m}_1^2=
c^{0-0}_{VVS} \mathbf{m}_2^2=
c^{0+0}_{VVS} \mathbf{m}_2^2=g.
\end{equation}

The 2nd descendant $VVS$ amplitude is given by
\eq{
(0,0,0): \quad & [\mathcal{J}(1^{0},3^{0})]_2 \cdot [\mathbf{A}(2^0)]_0 +[\mathcal{J}(2^{0},3^{0})]_2 \cdot [\mathbf{A}(1^0)]_0 \\ =& -c_{VVS,1}^{000} m_1 \tilde{m}_1 \langle\eta_12\rangle [2\eta_1] -c_{VVS,2}^{000} m_2 \tilde{m}_2 \langle\eta_21\rangle [1\eta_2] \\
=& \Ampthree{1^0}{2^0}{3^0}{\bosflipflip{1}{180}{brown}{red}{brown}}{\bos{i2}{brown}}{\sca{i3}} +\Ampthree{1^0}{2^0}{3^0}{\bos{i1}{brown}}{\bosflipflip{1}{55}{brown}{red}{brown}}{\sca{i3}}.
}
The spin group $SU(2)$ constrains the coefficients to satisfy $c_1 \mathbf{m}_1^2= c_2 \mathbf{m}_2^2$, so we need not consider the case $c_1 \neq c_2$. Therefore, its H.E. limit is the $\phi^3$ operator,
\eq{
\Ampthree{1^0}{2^0}{3^0}{\bosflipflip{1}{180}{brown}{red}{brown}}{\bos{i2}{brown}}{\sca{i3}} +\Ampthree{1^0}{2^0}{3^0}{\bos{i1}{brown}}{\bosflipflip{1}{55}{brown}{red}{brown}}{\sca{i3}}
\xrightarrow{H.E.} \Ampthree{1^0}{2^0}{3^0}{\propag[gho] (i1) to (v1)}{\propag[gho] (i2) to (v1)}{\propag[sca] (i3) to (v1)} = \lambda_3,
}
requiring the coefficients to satisfy
\eq{
c_{VVS,1}^{000} \mathbf{m}_1^2 = c_{VVS,2}^{000} \mathbf{m}_2^2 = \frac{\lambda_3}{(\mathbf{m}_3^2-\mathbf{m}_1^2-\mathbf{m}_2^2)} .
}

\paragraph{3. $ffV$}

There are also two primary MHC amplitudes in the $ffV$ class that are not generated by $[\mathcal J]_0\cdot [\mathbf{A}]_0$,
\begin{align}
(-\frac{1}{2},-\frac{1}{2},-1): \ c_{ffV}^{---}\langle13\rangle \langle23\rangle &= 
\Ampthree{1^-}{2^-}{3^-}{\fer{red}{i1}{v1}}{\antfer{cyan}{i2}{v1}}{\bos{i3}{red}} 
\xrightarrow{H.E.} \Ampthree{1^-}{2^-}{3^-}{\propag[fer] (i1) to (v1)}{\propag[antfer] (i2) to (v1)}{\propag[bos] (i3) to (v1); \vertex[dot] (v) at (0,0) {}} 
= \frac{1}{\Lambda} C_{\psi^2 F_L} \langle13\rangle \langle23\rangle
, \\
(+\frac{1}{2},+\frac{1}{2},+1): \ c_{ffV}^{+++} {[13]} {[23]} &= 
\Ampthree{1^+}{2^+}{3^+}{\fer{cyan}{i1}{v1}}{\antfer{red}{i2}{v1}}{\bos{i3}{cyan}} 
\xrightarrow{H.E.} \Ampthree{1^+}{2^+}{3^+}{\propag[fer] (i1) to (v1)}{\propag[antfer] (i2) to (v1)}{\propag[bos] (i3) to (v1); \vertex[dot] (v) at (0,0) {}}
= \frac{1}{\Lambda} C_{\bar{\psi}^2 F_R} {[13]} {[23]}.
\end{align}
They correspond to the massless EFT operators $\psi^2 F_L$ and $\bar{\psi}^2 F_R$ respectively, with the Wilson coefficients
\eq{
c_{ffV}^{---} = \frac{1}{\Lambda} C_{\psi^2 F_L}, \quad
c_{ffV}^{+++} = \frac{1}{\Lambda} C_{\bar{\psi}^2 F_R}.
}

The other primary MHC amplitudes contain the $ff$ current
\begin{equation} \begin{aligned}
[\mathcal{J}(1^+,2^-)]_0 &= |1]^{\dot{\alpha}} \langle2|^{\alpha}, \\
[\mathcal{J}(1^-,2^+)]_0 &= |2]^{\dot{\alpha}} \langle1|^{\alpha}.
\end{aligned}
\end{equation}
Therefore, the H.E. limits of the following primary MHC amplitudes vanish,
\eqs{
(+\frac{1}{2},-\frac{1}{2},0): \ \langle23\rangle [31] = [\mathcal{J}(1^+,2^-)]_0 \cdot [\mathbf{A}(3^0)]_0 = \Ampthree{1^+}{2^-}{3^0}{\fer{cyan}{i1}{v1}}{\antfer{cyan}{i2}{v1}}{\bos{i3}{brown}} \xrightarrow{H.E.} 0, \\
(-\frac{1}{2},+\frac{1}{2},0): \ \langle13\rangle [32] = [\mathcal{J}(1^-,2^+)]_0 \cdot [\mathbf{A}(3^0)]_0 = \Ampthree{1^-}{2^+}{3^0}{\fer{red}{i1}{v1}}{\antfer{red}{i2}{v1}}{\bos{i3}{brown}} \xrightarrow{H.E.} 0. 
}

We have shown in Ref.~\cite{Ni:2026wiz} that some of the 1st descendant amplitudes are given by $[\mathcal J]_0\cdot [\mathbf{A}]_1$,
\eq{
(+\frac{1}{2},-\frac{1}{2},-1): \ c_{ffV}^{+--} m_3 \langle23\rangle [\eta_31] 
=& [\mathcal{J}(1^+,2^-)]_0 \cdot [\mathbf{A}(3^-)]_1 \\ 
=& \Ampthree{1^+}{2^-}{3^-}{\fer{cyan}{i1}{v1}}{\antfer{cyan}{i2}{v1}}{\bosflip{1}{-55}{brown}{red}} 
\xrightarrow{H.E.} \Ampthree{1^+}{2^-}{3^-}{\propag[fer] (i1) to (v1)}{\propag[antfer] (i2) to (v1)}{\propag[bos] (i3) to (v1)}=g\frac{\langle 23\rangle^2}{\langle 12 \rangle}, }
\eq{
(+\frac{1}{2},-\frac{1}{2},+1): \ -c_{ffV}^{+-+} \tilde{m}_3 \langle2\eta_3\rangle [31] 
=& [\mathcal{J}(1^+,2^-)]_0 \cdot [\mathbf{A}(3^+)]_1 \\
=& \Ampthree{1^+}{2^-}{3^+}{\fer{cyan}{i1}{v1}}{\antfer{cyan}{i2}{v1}}{\bosflip{1}{-55}{brown}{cyan}}
\xrightarrow{H.E.} \Ampthree{1^+}{2^-}{3^+}{\propag[fer] (i1) to (v1)}{\propag[antfer] (i2) to (v1)}{\propag[bos] (i3) to (v1)}=g\frac{[13]^2}{[12]}, }
\eq{
(-\frac{1}{2},+\frac{1}{2},-1): \ c_{ffV}^{-+-} m_3 \langle13\rangle [\eta_32] 
=& [\mathcal{J}(1^-,2^+)]_0 \cdot [\mathbf{A}(3^-)]_1 \\
=& \Ampthree{1^-}{2^+}{3^-}{\fer{red}{i1}{v1}}{\antfer{red}{i2}{v1}}{\bosflip{1}{-55}{brown}{red}}
\xrightarrow{H.E.} \Ampthree{1^-}{2^+}{3^-}{\propag[fer] (i1) to (v1)}{\propag[antfer] (i2) to (v1)}{\propag[bos] (i3) to (v1)}=g\frac{\langle 13\rangle^2}{\langle 12 \rangle},, \label{eq:ffV-ffA1}}
\eq{
(-\frac{1}{2},+\frac{1}{2},+1): \ -c_{ffV}^{-++} \tilde{m}_3 \langle1\eta_3\rangle [32] 
=& [\mathcal{J}(1^-,2^+)]_0 \cdot [\mathbf{A}(3^+)]_1 \\
=& \Ampthree{1^-}{2^+}{3^+}{\fer{red}{i1}{v1}}{\antfer{red}{i2}{v1}}{\bosflip{1}{-55}{brown}{cyan}}
\xrightarrow{H.E.} \Ampthree{1^-}{2^+}{3^+}{\propag[fer] (i1) to (v1)}{\propag[antfer] (i2) to (v1)}{\propag[bos] (i3) to (v1)}=g\frac{[23]^2}{[12]}. \label{eq:ffV-ffA2}
}
Their H.E. limit correspond to contact term $\bar{\psi} A_\mu \gamma^\mu \psi$, which is a subleading component of the kinetic term $\bar{\psi} D_\mu \gamma^\mu \psi$. Thus the coefficient is given by the gauge coupling
\begin{equation}
c_{ffV}^{+--} \mathbf{m}_3^2=
c_{ffV}^{+-+} \mathbf{m}_3^2=
c_{ffV}^{-+-} \mathbf{m}_3^2=
c_{ffV}^{-++} \mathbf{m}_3^2=g.
\end{equation}

Meanwhile, the other 1st descendant amplitudes are given by $[\mathcal J]_1\cdot[\mathbf{A}]_0$, corresponding to the Yukawa coupling
\eqs{
(+\frac{1}{2},+\frac{1}{2},0):&& [\mathcal{J}(1^+,2^+)]_1 \cdot [\mathbf{A}(3^0)]_0 
= -c_{ffV,1}^{++0} \tilde{m}_2 \langle\eta_23\rangle [31] -c_{ffV,2}^{++0} \tilde{m}_1 \langle\eta_13\rangle [32]  \nonumber \\
&=& \Ampthree{1^+}{2^+}{3^0}{\fer{cyan}{i1}{v1}}{\antferflip{1}{55}{red}{cyan}}{\bos{i3}{brown}} +\Ampthree{1^+}{2^+}{3^0}{\ferflip{1}{180}{cyan}{red}}{\antfer{red}{i2}{v1}}{\bos{i3}{brown}} \xrightarrow{H.E.} \Ampthree{+\frac{1}{2}}{+\frac{1}{2}}{0}{\propag[fer] (i1) to (v1)}{\propag[antfer] (i2) to (v1)}{\propag[gho] (i3) to (v1)}, \label{eq:ffV-Yukawa1} \\
(-\frac{1}{2},-\frac{1}{2},0):&& [\mathcal{J}(1^-,2^-)]_1 \cdot [\mathbf{A}(3^0)]_0 
= c_{ffV,1}^{--0} m_1 \langle23\rangle [3\eta_1] +c_{ffV,2}^{--0} m_2 \langle13\rangle [3\eta_2] \nonumber \\
&=& \Ampthree{1^-}{2^-}{3^0}{\ferflip{1}{180}{red}{cyan}}{\antfer{cyan}{i2}{v1}}{\bos{i3}{brown}} +\Ampthree{1^-}{2^-}{3^0}{\fer{red}{i1}{v1}}{\antferflip{1}{55}{cyan}{red}}{\bos{i3}{brown}} \xrightarrow{H.E.} \Ampthree{+\frac{1}{2}}{+\frac{1}{2}}{0}{\propag[fer] (i1) to (v1)}{\propag[antfer] (i2) to (v1)}{\propag[gho] (i3) to (v1)}. \label{eq:ffV-Yukawa3}
}
There remains some freedom in the coefficients:
\eq{
c_{ffV,1}^{++0} \mathbf{m}_2^2 -c_{ffV,2}^{++0} \mathbf{m}_1^2 = Y^*, \quad
-c_{ffV,1}^{--0} \mathbf{m}_1^2 +c_{ffV,2}^{--0} \mathbf{m}_2^2 = Y.
}
They would be settled after introducing the gauge group.

\subsection{\texorpdfstring{$VVV$}{VVV}}

There are eight independent primary MHC amplitudes in the $VVV$ class, two of which do not vanish in the H.E. limit:
\eq{
(-1,-1,-1): \ c_{V^3}^{---}\langle12\rangle \langle23\rangle \langle31\rangle = \Ampthree{-1}{-1}{-1}{\bos{i1}{red}}{\bos{i2}{red}}{\bos{i3}{red}} \xrightarrow{H.E.} \Ampthree{-1}{-1}{-1}{\propag[bos] (i1) to (v1)}{\propag[bos] (i2) to (v1)}{\propag[bos] (i3) to (v1); \vertex[dot] (v) at (0,0) {}}
=\frac{C_{F_L^3}}{\Lambda^2}\langle12\rangle \langle23\rangle \langle31\rangle, \\
(+1,+1,+1): \ c_{V^3}^{+++}[12] [23] [31] = \Ampthree{+1}{+1}{+1}{\bos{i1}{cyan}}{\bos{i2}{cyan}}{\bos{i3}{cyan}} \xrightarrow{H.E.} \Ampthree{+1}{+1}{+1}{\propag[bos] (i1) to (v1)}{\propag[bos] (i2) to (v1)}{\propag[bos] (i3) to (v1); \vertex[dot] (v) at (0,0) {}}
=\frac{C_{F_R^3}}{\Lambda^2} [12] [23] [31].
}
The Wilson coefficients of the MHC amplitudes above are given by the $F_L^3$ and $F_R^3$ massless operators,
\eq{
c_{V^3}^{---} = \frac{C_{F_L^3}}{\Lambda^2}, \quad
c_{V^3}^{+++} = \frac{C_{F_R^3}}{\Lambda^2}. 
}

The other six primary MHC amplitudes vanish in the H.E. limit,
\eqs{
(-1,0,0): \ \langle12\rangle [23] \langle31\rangle 
= [\mathcal{J}(1^-,2^0)]_0 \cdot [\mathbf{A}(3^0)]_0 = [\mathcal{J}(1^-,3^0)]_0 \cdot [\mathbf{A}(2^0)]_0
= \Ampthree{1^-}{2^0}{3^0}{\bos{i1}{red}}{\bos{i2}{brown}}{\bos{i3}{brown}} \xrightarrow{H.E.} 0, \\
(0,-1,0): \ \langle12\rangle \langle23\rangle [31] 
= [\mathcal{J}(1^0,2^-)]_0 \cdot [\mathbf{A}(3^0)]_0 = [\mathcal{J}(2^-,3^0)]_0 \cdot [\mathbf{A}(1^0)]_0
= \Ampthree{1^0}{2^-}{3^0}{\bos{i1}{brown}}{\bos{i2}{red}}{\bos{i3}{brown}} \xrightarrow{H.E.} 0, \\
(0,0,-1): \ [12] \langle23\rangle \langle31\rangle 
= [\mathcal{J}(1^0,3^-)]_0 \cdot [\mathbf{A}(2^0)]_0 = [\mathcal{J}(2^0,3^-)]_0 \cdot [\mathbf{A}(1^0)]_0
= \Ampthree{1^0}{2^0}{3^-}{\bos{i1}{brown}}{\bos{i2}{brown}}{\bos{i3}{red}} \xrightarrow{H.E.} 0, \\
(+1,0,0): \ [12] \langle23\rangle [31] 
= [\mathcal{J}(1^+,2^0)]_0 \cdot [\mathbf{A}(3^0)]_0 = [\mathcal{J}(1^+,3^0)]_0 \cdot [\mathbf{A}(2^0)]_0
= \Ampthree{1^+}{2^0}{3^0}{\bos{i1}{cyan}}{\bos{i2}{brown}}{\bos{i3}{brown}} \xrightarrow{H.E.} 0, \\
(0,+1,0): \ [12] [23] \langle31\rangle 
= [\mathcal{J}(1^0,2^+)]_0 \cdot [\mathbf{A}(3^0)]_0 = [\mathcal{J}(2^+,3^0)]_0 \cdot [\mathbf{A}(1^0)]_0
= \Ampthree{1^0}{2^+}{3^0}{\bos{i1}{brown}}{\bos{i2}{cyan}}{\bos{i3}{brown}} \xrightarrow{H.E.} 0, \\
(0,0,+1): \ \langle12\rangle [23] [31] 
= [\mathcal{J}(1^0,3^+)]_0 \cdot [\mathbf{A}(2^0)]_0 = [\mathcal{J}(2^0,3^+)]_0 \cdot [\mathbf{A}(1^0)]_0
= \Ampthree{1^0}{2^0}{3^+}{\bos{i1}{brown}}{\bos{i2}{brown}}{\bos{i3}{cyan}} \xrightarrow{H.E.} 0. 
}
The primary MHC currents are
\begin{align}
{[\mathcal{J}(1^-,2^0)]}_0 &= -\langle12\rangle |2]^{\dot\alpha}\langle1|^{\alpha},&
{[\mathcal{J}(1^-,3^0)]}_0 &= -\langle31\rangle |3]^{\dot{\alpha}} \langle1|^{\alpha},& \\
{[\mathcal{J}(1^0,2^-)]}_0 &= -\langle12\rangle |1]^{\dot\alpha}\langle2|^{\alpha},&
{[\mathcal{J}(2^-,3^0)]}_0 &= -\langle23\rangle |3]^{\dot{\alpha}} \langle2|^{\alpha},& \\
{[\mathcal{J}(1^0,3^-)]}_0 &= -\langle31\rangle |1]^{\dot{\alpha}} \langle1|^{\alpha},&
{[\mathcal{J}(2^0,3^-)]}_0 &= -\langle23\rangle |2]^{\dot{\alpha}} \langle2|^{\alpha},& \\
{[\mathcal{J}(1^+,2^0)]}_0 &= -[12] |1]^{\dot\alpha} \langle2|^{\alpha},&
{[\mathcal{J}(1^+,3^0)]}_0 &= -[31] |1]^{\dot{\alpha}} \langle3|^{\alpha},& \\
{[\mathcal{J}(1^0,2^+)]}_0 &= -[12] |2]^{\dot\alpha} \langle1|^{\alpha},&
{[\mathcal{J}(2^+,3^0)]}_0 &= -[23] |2]^{\dot{\alpha}} \langle3|^{\alpha},& \\
{[\mathcal{J}(1^0,3^+)]}_0 &= -[31] |3]^{\dot{\alpha}} \langle1|^{\alpha},&
{[\mathcal{J}(2^0,3^+)]}_0 &= -[23] |3]^{\dot{\alpha}} \langle2|^{\alpha}.&
\end{align}
Their massless counterpart is the conserved current $J_{\mu} = F_{\mu\nu} \partial^{\nu} \phi$.
It can be easily proven that their coupling with $\partial \phi$ vanishes, $ [\mathcal J]_0\cdot [\mathbf{A}]_0 \xrightarrow{H.E.} J \cdot \partial \phi =0$, consistent with the fact that the $h=\pm1$ primary amplitudes vanish in the H.E. limit. 

The 1st descendant amplitudes are either $[\mathcal J]_1\cdot [\mathbf{A}]_0$ or $[\mathcal J]_0\cdot[\mathbf{A}]_1$. The latter can always be converted to the former in the $(\pm\pm0)$ or $(\pm\mp0)$ helicity categories, as well as their permutations. Therefore, we present the $[\mathcal J]_1$ for the two kinds of helicity categories:
\begin{align}
[\mathcal{J}(1^-,2^-)]_1 &= -c^{--}_1 m_2 \langle12\rangle |\eta_2]^{\dot{\alpha}} \langle1|^{\alpha} -c^{--}_2 m_1 \langle12\rangle |\eta_1]^{\dot{\alpha}} \langle2|^{\alpha}, \\
[\mathcal{J}(1^-,3^-)]_1 &= -c^{--}_3 m_3 \langle13\rangle |\eta_3]^{\dot{\alpha}} \langle1|^{\alpha} -c^{--}_4 m_1 \langle13\rangle |\eta_1]^{\dot{\alpha}} \langle3|^{\alpha}, \\
[\mathcal{J}(2^-,3^-)]_1 &= -c^{--}_5 m_2 \langle32\rangle |\eta_2]^{\dot{\alpha}} \langle3|^{\alpha} -c^{--}_6 m_3 \langle32\rangle |\eta_3]^{\dot{\alpha}} \langle2|^{\alpha}, \\
[\mathcal{J}(1^+,2^+)]_1 &= c^{++}_1 \tilde{m}_2 [12] |1]^{\dot{\alpha}} \langle\eta_2|^{\alpha} + c^{++}_2 \tilde{m}_1 [12] |2]^{\dot{\alpha}} \langle\eta_1|^{\alpha}, \\
[\mathcal{J}(1^+,3^+)]_1 &= c^{++}_3 \tilde{m}_3 [13] |1]^{\dot{\alpha}} \langle\eta_3|^{\alpha} + c^{++}_4 \tilde{m}_1 [13] |3]^{\dot{\alpha}} \langle\eta_1|^{\alpha}, \\
[\mathcal{J}(2^+,3^+)]_1 &= c^{++}_5 \tilde{m}_2 [32] |3]^{\dot{\alpha}} \langle\eta_2|^{\alpha} + c^{++}_6 \tilde{m}_3 [32] |2]^{\dot{\alpha}} \langle\eta_3|^{\alpha}, \\
[\mathcal{J}(1^-,2^+)]_1 &= c^{-+}_1 \tilde{m}_2 \langle1\eta_2\rangle |2]^{\dot{\alpha}} \langle1|^{\alpha} - c^{-+}_2 m_1 [\eta_12] |2]^{\dot{\alpha}} \langle1|^{\alpha}, \label{eq:J+-1} \\
[\mathcal{J}(1^-,3^+)]_1 &= c^{-+}_3 \tilde{m}_3 \langle1\eta_3\rangle |3]^{\dot{\alpha}} \langle1|^{\alpha} - c^{-+}_4 m_1 [\eta_13] |3]^{\dot{\alpha}} \langle1|^{\alpha}, \\
[\mathcal{J}(2^-,3^+)]_1 &= c^{-+}_5 \tilde{m}_3 \langle2\eta_3\rangle |3]^{\dot{\alpha}} \langle2|^{\alpha} - c^{-+}_6 m_2 [\eta_23] |3]^{\dot{\alpha}} \langle2|^{\alpha}, \\
[\mathcal{J}(1^+,2^-)]_1 &= c^{+-}_1 \tilde{m}_1 \langle2\eta_1\rangle |1]^{\dot{\alpha}} \langle2|^{\alpha} - c^{+-}_2 m_2 [\eta_21] |1]^{\dot{\alpha}} \langle2|^{\alpha}, \\
[\mathcal{J}(1^+,3^-)]_1 &= c^{+-}_3 \tilde{m}_1 \langle3\eta_1\rangle |1]^{\dot{\alpha}} \langle3|^{\alpha} - c^{+-}_4 m_3 [\eta_31] |1]^{\dot{\alpha}} \langle3|^{\alpha}, \\
[\mathcal{J}(2^+,3^-)]_1 &= c^{+-}_5 \tilde{m}_2 \langle3\eta_2\rangle |2]^{\dot{\alpha}} \langle3|^{\alpha} - c^{+-}_6 m_3 [\eta_32] |2]^{\dot{\alpha}} \langle3|^{\alpha}. \label{eq:J+-6}
\end{align}

The helicity-$(\pm\mp)$ MHC currents, Eqs.~(\ref{eq:J+-1}-\ref{eq:J+-6}), are all conserved in the H.E. limit. The helicity-$(\pm\pm)$ MHC currents, in contrast, lead to either the conserved current $[\mathcal{J}_c]_1$ or non-conserved ones $[\mathcal{J}_n]_1$ in the H.E. limit, depending on the choice of coefficients. The former are
\begin{align}
[\mathcal{J}_c(1^-,2^-)]_1 &= -\frac{1}{\tilde{m}_2} \langle12\rangle |\eta_2]^{\dot{\alpha}} \langle1|^{\alpha} -\frac{1}{\tilde{m}_1} \langle12\rangle |\eta_1]^{\dot{\alpha}} \langle2|^{\alpha}, \\
[\mathcal{J}_c(1^-,3^-)]_1 &= -\frac{1}{\tilde{m}_3} \langle13\rangle |\eta_3]^{\dot{\alpha}} \langle1|^{\alpha} -\frac{1}{\tilde{m}_1} \langle13\rangle |\eta_1]^{\dot{\alpha}} \langle3|^{\alpha}, \\
[\mathcal{J}_c(2^-,3^-)]_1 &= -\frac{1}{\tilde{m}_2} \langle32\rangle |\eta_2]^{\dot{\alpha}} \langle3|^{\alpha} -\frac{1}{\tilde{m}_3} \langle32\rangle |\eta_3]^{\dot{\alpha}} \langle2|^{\alpha}, \\
[\mathcal{J}_c(1^+,2^+)]_1 &= \frac{1}{m_2} [12] |1]^{\dot{\alpha}} \langle\eta_2|^{\alpha} +\frac{1}{m_1} [12] |2]^{\dot{\alpha}} \langle\eta_1|^{\alpha}, \\
[\mathcal{J}_c(1^+,3^+)]_1 &= \frac{1}{m_3} [13] |1]^{\dot{\alpha}} \langle\eta_3|^{\alpha} +\frac{1}{m_1} [13] |3]^{\dot{\alpha}} \langle\eta_1|^{\alpha}, \\
[\mathcal{J}_c(2^+,3^+)]_1 &= \frac{1}{m_2} [32] |3]^{\dot{\alpha}} \langle\eta_2|^{\alpha} +\frac{1}{m_3} [32] |2]^{\dot{\alpha}} \langle\eta_3|^{\alpha}.
\end{align}
Take $[\mathcal{J}_c(1^-,2^-)]_1 $ as an example. Its coefficients satisfy $c_1^{--} \mathbf{m}_2^2 = c_2^{--} \mathbf{m}_1^2$. The two particles in a current are anti-symmetric, requiring a non-abelian gauge symmetry. One would expect an $f^{IJK}$ gauge structure attached to the currents above, so that the gauge bosons are symmetric under exchange. This leads to the Noether current $F_{\mu\nu} A^{\nu}$ for the Yang-Mills field. 

If the coefficients use another combination $-c_1^{--} \mathbf{m}_2^2 = c_2^{--} \mathbf{m}_1^2=c_{VVV}^{--0}$, the H.E. limit of $[\mathcal J']_1\cdot[\mathbf{A}]_0$ would be an EFT operator,
\eq{
(-1,-1,0): \ [\mathcal{J}_n(1^-,2^-)]_1 \cdot [\mathbf{A}(3^0)]_0 =& c_{V^3}^{--0} \left(\frac{1}{2\tilde{m}_2} \langle12\rangle [\eta_23] \langle31\rangle -\frac{1}{2\tilde{m}_1} \langle12\rangle \langle23\rangle [3\eta_1]^{\dot{\alpha}} \right) \\
=& \Ampthree{1^-}{2^-}{3^0}{\bos{i1}{red}}{\bosflip{1}{55}{brown}{red}}{\bos{i3}{brown}} +\Ampthree{1^-}{2^-}{3^0}{\bosflip{1}{180}{brown}{red}}{\bos{i2}{red}}{\bos{i3}{brown}} \\
\xrightarrow{H.E.}& \Ampthree{1^-}{2^-}{3^0}{\propag[bos] (i1) to (v1)}{ \propag[bos] (i2) to (v1)}{\propag[gho] (i3) to (v1); \vertex[dot] (v) at (0,0) {}} = \frac{C_{F_L^2 \phi} v}{\Lambda^2} \langle12\rangle^2,
}
where $c_{V^3}^{--0} = \frac{C_{F_L^2 \phi} v}{\Lambda^2}$. Exchange symmetry does not require a non-abelian gauge group. If a non-abelian gauge group were present, the gauge structure would be represented as $\delta^{IJ}$. This is quite different from the conserved currents $[\mathcal J_c]_1$, which correspond to the SM massless currents.
Similarly, the other five $[\mathcal J_n]_1\cdot[\mathbf{A}]_0$ amplitudes are also contributed by the $F_L^2\phi$ or $F_R^2\phi$ operator:
\eq{
(-1,0,-1): \ [\mathcal{J}_n(1^-,3^-)]_1 \cdot [\mathbf{A}(2^0)]_0 =& c_{V^3}^{-0-} \left(\frac{1}{2\tilde{m}_3} \langle13\rangle [\eta_32] \langle21\rangle -\frac{1}{2\tilde{m}_1} \langle13\rangle \langle32\rangle [2\eta_1] \right) \\
=& \Ampthree{1^-}{2^0}{3^-}{\bos{i1}{red}}{\bos{i3}{brown}}{\bosflip{1}{55}{brown}{red}} +\Ampthree{1^-}{2^0}{3^-}{\bosflip{1}{180}{brown}{red}}{\bos{i2}{brown}}{\bos{i3}{red}} \\
\xrightarrow{H.E.}& \Ampthree{1^-}{2^0}{3^-}{\propag[bos] (i1) to (v1)}{ \propag[gho] (i2) to (v1)}{\propag[bos] (i3) to (v1); \vertex[dot] (v) at (0,0) {}} = \frac{C_{F_L^2 \phi} v}{\Lambda^2} \langle13\rangle^2,
}
\eq{
(0,-1,-1): \ [\mathcal{J}_n(2^-,3^-)]_1 \cdot [\mathbf{A}(1^0)]_0 =& c_{V^3}^{0--} \left(\frac{1}{2\tilde{m}_2} \langle32\rangle [\eta_21] \langle13\rangle -\frac{1}{2\tilde{m}_3} \langle32\rangle \langle21\rangle [1\eta_3] \right) \\
=& \Ampthree{1^0}{2^-}{3^-}{\bos{i1}{brown}}{\bosflip{1}{55}{brown}{red}}{\bos{i3}{red}} +\Ampthree{1^0}{2^-}{3^-}{\bos{i1}{brown}}{\bos{i2}{red}}{\bosflip{1}{-55}{brown}{red}} \\
\xrightarrow{H.E.}& \Ampthree{1^0}{2^-}{3^-}{\propag[gho] (i1) to (v1)}{ \propag[bos] (i2) to (v1)}{\propag[bos] (i3) to (v1); \vertex[dot] (v) at (0,0) {}} = \frac{C_{F_L^2 \phi} v}{\Lambda^2} \langle23\rangle^2,
}
\eq{
(+1,+1,0): \ [\mathcal{J}_n(1^+,2^+)]_1 \cdot [\mathbf{A}(3^0)]_0 =& c_{V^3}^{++0} \left(-\frac{1}{2{m}_2} [12] \langle\eta_23\rangle [31] +\frac{1}{2{m}_1} [12] [23] \langle3\eta_1\rangle \right) \\
=& \Ampthree{1^+}{2^+}{3^0}{\bos{i1}{cyan}}{\bosflip{1}{55}{brown}{cyan}}{\bos{i3}{brown}} +\Ampthree{1^+}{2^+}{3^0}{\bosflip{1}{180}{brown}{cyan}}{\bos{i2}{cyan}}{\bos{i3}{brown}} \\
\xrightarrow{H.E.}& \Ampthree{1^+}{2^+}{3^0}{\propag[bos] (i1) to (v1)}{ \propag[bos] (i2) to (v1)}{\propag[gho] (i3) to (v1); \vertex[dot] (v) at (0,0) {}} = \frac{C_{F_R^2 \phi} v}{\Lambda^2} [12]^2,
}
\eq{
(+1,0,+1): \ [\mathcal{J}_n(1^+,3^+)]_1 \cdot [\mathbf{A}(2^0)]_0 =& c_{V^3}^{+0+} \left(-\frac{1}{2{m}_3} [13] \langle\eta_32\rangle [21] +\frac{1}{2{m}_1} [13] [32] \langle2\eta_1\rangle \right) \\
=& \Ampthree{1^+}{2^0}{3^+}{\bos{i1}{cyan}}{\bos{i2}{brown}}{\bosflip{1}{-55}{brown}{cyan}} +\Ampthree{1^+}{2^0}{3^+}{\bosflip{1}{180}{brown}{cyan}}{\bos{i2}{brown}}{\bos{i3}{cyan}} \\
\xrightarrow{H.E.}& \Ampthree{1^+}{2^0}{3^+}{\propag[bos] (i1) to (v1)}{ \propag[gho] (i2) to (v1)}{\propag[bos] (i3) to (v1); \vertex[dot] (v) at (0,0) {}} = \frac{C_{F_R^2 \phi} v}{\Lambda^2} [13]^2,
}
\eq{
(0,+1,+1): \ [\mathcal{J}_n(2^+,3^+)]_1 \cdot [\mathbf{A}(1^0)]_0 =& c_{V^3}^{0++} \left(-\frac{1}{2{m}_2} [32] \langle\eta_21\rangle [13] +\frac{1}{2{m}_3} [32] [21] \langle1\eta_3\rangle \right) \\
=& \Ampthree{1^0}{2^+}{3^+}{\bos{i1}{brown}}{\bosflip{1}{55}{brown}{cyan}}{\bos{i3}{cyan}} +\Ampthree{1^0}{2^+}{3^+}{\bos{i1}{brown}}{\bos{i2}{cyan}}{\bosflip{1}{-55}{brown}{cyan}} \\
\xrightarrow{H.E.}& \Ampthree{1^0}{2^+}{3^+}{\propag[gho] (i1) to (v1)}{ \propag[bos] (i2) to (v1)}{\propag[bos] (i3) to (v1); \vertex[dot] (v) at (0,0) {}} = \frac{C_{F_R^2 \phi} v}{\Lambda^2} [23]^2.
}
These are the six independent MHC amplitudes that correspond to the massless leading operators. So far, we have listed all eight massless leading operators that contribute to the $VVV$ MHC amplitudes. It still satisfies the proposition that the number of independent primary MHC amplitudes equals the number of massless leading operators that contribute to the MHC class. 

The rest of the 1st descendant MHC currents are the helicity-$(00)$ ones:
\eq{
[\mathcal{J}(1^0,2^0)]_1 = c_{11} \tilde{m}_1 (\langle\eta_12\rangle |2]^{\dot{\alpha}} \langle1|^{\alpha} +\langle12\rangle |2]^{\dot{\alpha}} \langle\eta_1|^{\alpha}) +c_{12} \tilde{m}_2 (\langle1\eta_2\rangle |1]^{\dot{\alpha}} \langle2|^{\alpha} +\langle12\rangle |1]^{\dot{\alpha}} \langle\eta_2|^{\alpha}) \\ -c_{13} m_1 ([\eta_12] |1]^{\dot{\alpha}} \langle2|^{\alpha} +[12] |\eta_1]^{\dot{\alpha}} \langle2|^{\alpha}) -c_{14} m_2 ([1\eta_2] |2]^{\dot{\alpha}} \langle1|^{\alpha} +[12] |\eta_2]^{\dot{\alpha}} \langle1|^{\alpha}), 
}
\eq{
[\mathcal{J}(1^0,3^0)]_1 = c_{21} \tilde{m}_1 (\langle\eta_13\rangle |3]^{\dot{\alpha}} \langle1|^{\alpha} +\langle13\rangle |3]^{\dot{\alpha}} \langle\eta_1|^{\alpha}) +c_{22} \tilde{m}_3 (\langle1\eta_3\rangle |1]^{\dot{\alpha}} \langle3|^{\alpha} +\langle13\rangle |1]^{\dot{\alpha}} \langle\eta_3|^{\alpha}) \\ -c_{23} m_1 ([\eta_13] |1]^{\dot{\alpha}} \langle3|^{\alpha} +[13] |\eta_1]^{\dot{\alpha}} \langle3|^{\alpha}) -c_{24} m_3 ([1\eta_3] |3]^{\dot{\alpha}} \langle1|^{\alpha} +[13] |\eta_3]^{\dot{\alpha}} \langle1|^{\alpha}), 
}
\eq{
[\mathcal{J}(2^0,3^0)]_1 = c_{31} \tilde{m}_3 (\langle\eta_32\rangle |2]^{\dot{\alpha}} \langle3|^{\alpha} +\langle32\rangle |2]^{\dot{\alpha}} \langle\eta_3|^{\alpha}) +c_{32} \tilde{m}_2 (\langle3\eta_2\rangle |3]^{\dot{\alpha}} \langle2|^{\alpha} +\langle32\rangle |3]^{\dot{\alpha}} \langle\eta_2|^{\alpha}) \\ -c_{33} m_3 ([\eta_32] |3]^{\dot{\alpha}} \langle2|^{\alpha} +[32] |\eta_3]^{\dot{\alpha}} \langle2|^{\alpha}) -c_{34} m_2 ([3\eta_2] |2]^{\dot{\alpha}} \langle3|^{\alpha} +[32] |\eta_2]^{\dot{\alpha}} \langle3|^{\alpha}). 
}
There are four independent choices of coefficients for each current. We study $[\mathcal{J}(1^0,2^0)]_1$ without loss of generality. The H.E. limit of the following current is conserved $[\mathcal{J}_c]_1$,
\eq{
[\mathcal{J}_c(1^0,2^0)]_1 = \frac{c}{m_1} (\langle\eta_12\rangle |2]^{\dot{\alpha}} \langle1|^{\alpha} +\langle12\rangle |2]^{\dot{\alpha}} \langle\eta_1|^{\alpha}) -\frac{c}{m_2} (\langle1\eta_2\rangle |1]^{\dot{\alpha}} \langle2|^{\alpha} +\langle12\rangle |1]^{\dot{\alpha}} \langle\eta_2|^{\alpha}) \\ -\frac{c'}{\tilde{m}_1} ([\eta_12] |1]^{\dot{\alpha}} \langle2|^{\alpha} +[12] |\eta_1]^{\dot{\alpha}} \langle2|^{\alpha}) +\frac{c'}{\tilde{m}_2} ([1\eta_2] |2]^{\dot{\alpha}} \langle1|^{\alpha} +[12] |\eta_2]^{\dot{\alpha}} \langle1|^{\alpha}),
}
including two sets of choice of coefficients. Although the other two choices correspond to non-conserved current $[\mathcal{J}_n]_1$, the MHC amplitude $[\mathcal{J}_n(1^0,2^0)]_1\cdot [\mathbf{A}(3^0)]_0$ still vanishes in the H.E. limit. This is because we can identify particle 1 or 2 as the vector boson $[\mathbf{A}(1^0)]_0$ or $[\mathbf{A}(2^0)]_0$, and the corresponding current becomes a conserved current 
\eq{
[\mathcal{J}_n(1^0,2^0)]_1 \cdot [\mathbf{A}(3^0)]_0 \to [\mathcal{J}_c(1^0,3^0)]_1 \cdot [\mathbf{A}(2^0)]_0 +[\mathcal{J}_c(2^0,3^0)]_1 \cdot [\mathbf{A}(1^0)]_0, 
}
Therefore, the helicity-$(000)$ 1st descendant MHC amplitudes constructed by $[\mathcal J]_1\cdot[\mathbf{A}]_0$ all vanish in the H.E. limit.

Then, we move on to the 2nd descendant MHC amplitudes. They are constructed by $[\mathcal J]_2\cdot[\mathbf{A}]_0$, $[\mathcal J]_1\cdot [\mathbf{A}]_1$ or $[\mathcal J]_0\cdot[\mathbf{A}]_2$. Since we can move between different current decompositions, $[\mathcal J]_0\cdot[\mathbf{A}]_2$ can be converted to the other two structures, so we do not need to consider it.

We first consider helicity-$(\pm\pm\mp)$ case, which only exists in the $[\mathcal J]_1\cdot [\mathbf{A}]_1$ case. Without loss of generality, we study the $(-1,-1,+1)$ amplitude, constructed by $[\mathcal{J}(1^-,2^-)]_1\cdot[\mathbf{A}(3^+)]_1$. For conserved curent $[\mathcal J_c]_1$, we have
\eq{
[\mathcal{J}_c(1^-,2^-)]_1\cdot[\mathbf{A}(3^+)]_1 =& -c_{V^3,1}^{--+} \left(\frac{1}{\tilde{m}_2 m_3} \langle12\rangle [3\eta_2] \langle1\eta_3\rangle +\frac{1}{\tilde{m}_1 m_3} \langle12\rangle [3\eta_1] \langle2\eta_3\rangle\right) \\
=& \Ampthree{1^-}{2^-}{3^+}{\bos{i1}{red}}{\bosflip{1}{55}{brown}{red}}{\bosflip{1}{-55}{brown}{cyan}} +\Ampthree{1^-}{2^-}{3^+}{\bosflip{1}{180}{brown}{red}}{\bos{i2}{red}}{\bosflip{1}{-55}{brown}{cyan}} \\
\xrightarrow{H.E.}& \Ampthree{1^-}{2^-}{3^+}{\propag[bos] (i1) to (v1)}{\propag[bos] (i2) to (v1)}{\propag[bos] (i3) to (v1)}
= g \frac{\langle12\rangle^3}{\langle23\rangle \langle31\rangle},
}
where $c_{V^3,1}^{--+} = -g$. Its H.E. limit is $g^{\nu\rho} A_{[\mu}A_{\nu}\partial^{\mu}A_{\rho]}$, the subleading term of the kinematic term $F_{\mu\nu} F^{\mu\nu}$. 

\begin{figure}[htbp]
\centering
\subfloat[\hspace*{3em}]{ \label{fig:VV_flip1}
\includegraphics[scale=1.5,valign=c]{image/VV_flip1.pdf}}
\hspace{2em}
\subfloat[\hspace*{3em}]{ \label{fig:VV_flip2}
\includegraphics[scale=1.5,valign=c]{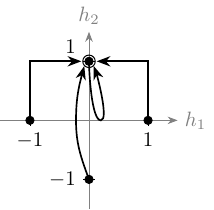}}
\caption{Illustration of the ladder operators $mJ^-$ and $\tilde{m}J^+$ acting on the $VV$ current. Black dots represent the primary current $[\mathcal J]_0$. Diagram (a) shows the flip from $[\mathcal J]_0$ to the 1st descendant current $[\mathcal J]_1$, indicated by red and cyan dots. Diagram (b) shows the flip from $[\mathcal J]_0$ to the 2nd descendant current $[\mathcal J]_2$, which occupies the same position as one of the $[\mathcal J]_0$ currents, represented by a black dot with a circle.}
\end{figure}

We then study the helicity-$(-1,0,0)$ amplitude, which involve two contributions ${[\mathcal J]}_1 \cdot [\mathbf{A}]_1$ and ${[\mathcal J]}_2 \cdot [\mathbf{A}]_0$. We first consider the ${[\mathcal J]}_1 \cdot [\mathbf{A}]_1$ contirbution, which is given by
\eq{\label{eq:00-_1}
[\mathcal{J}(2^0,3^0)]_1 \cdot [\mathbf{A}(1^-)]_1
=& -c_{31} m_1 \tilde{m}_3 (\langle\eta_32\rangle [\eta_12] \langle31\rangle +\langle32\rangle [\eta_12] \langle\eta_31\rangle) \\
&-c_{32} m_1 \tilde{m}_2 (\langle3\eta_2\rangle [\eta_13] \langle21\rangle +\langle32\rangle [\eta_13] \langle\eta_21\rangle) \\ 
&+c_{33} m_1 m_3 ([\eta_32] [\eta_13] \langle21\rangle +[32] [\eta_1\eta_3] \langle21\rangle) \\
&+c_{34} m_1 m_2 ([3\eta_2] [\eta_12] \langle31\rangle +[32] [\eta_1\eta_2] \langle31\rangle) \\
\xrightarrow{H.E.}& +2 \mathbf{m}_1^2 (c_{31} \tilde{m}_3 \langle31\rangle \langle\eta_31\rangle +c_{32} \tilde{m}_2 \langle12\rangle \langle\eta_21\rangle) \\ 
&-\mathbf{m}_1^2 ((c_{31} +c_{33}) \mathbf{m}_3^2 +(c_{32} +c_{34}) \mathbf{m}_2^2) \frac{\langle12\rangle \langle31\rangle}{\langle23\rangle}. 
}
Then we consider $[\mathcal J]_2\cdot[\mathbf{A}]_0$. This 2rd descendant current can be derived from the primary current, as shown by the self-pointed arrow in figure~\ref{fig:VV_flip2}. Thus, it gives
\eq{\label{eq:00-_2}
&[\mathcal{J}(1^-,2^0)]_2\cdot [\mathbf{A}(3^0)]_0
+[\mathcal{J}(1^-,3^0)]_2\cdot [\mathbf{A}(2^0)]_0 \\
=& -c_1 \langle1\eta_2\rangle [\eta_23] \langle31\rangle
-c_2 m_1 \tilde{m}_2 (\langle1\eta_2\rangle \langle23\rangle [3\eta_1] +\langle12\rangle \langle\eta_23\rangle [3\eta_1]) \\
&+c_3 m_1^2 [\eta_12] \langle23\rangle [3\eta_1] 
+c_4 m_1 m_2 ([\eta_1\eta_2] [23] \langle31\rangle +[\eta_12] [\eta_23] \langle31\rangle) \\
&-c_5 \langle1\eta_3\rangle [\eta_32] \langle21\rangle 
-c_6 m_1 \tilde{m}_3 (\langle1\eta_3\rangle \langle32\rangle [2\eta_1] +\langle13\rangle \langle\eta_32\rangle [2\eta_1]) \\
&+c_7 m_1^2 [\eta_13] \langle32\rangle [2\eta_1] 
+c_8 m_1 m_3 ([\eta_1\eta_3] [32] \langle21\rangle +[\eta_13] [\eta_32] \langle21\rangle) \\
\xrightarrow{H.E.}& -2 \mathbf{m}_1^2 \tilde{m}_2 (c_2 -\frac{c_1}{2 \mathbf{m}_1^2}) \langle12\rangle \langle\eta_21\rangle 
+2 \mathbf{m}_1^2 \tilde{m}_3 (c_6 -\frac{c_5}{2 \mathbf{m}_1^2}) \langle31\rangle \langle\eta_31\rangle \\
&+\mathbf{m}_1^2 ( (c_2+c_4) \mathbf{m}_2^2 -c_6 \mathbf{m}_3^2 +(c_7-c_3) \mathbf{m}_1^2 -c_8 \mathbf{m}_3^2) \frac{\langle12\rangle \langle31\rangle}{\langle23\rangle}.
}
Combining Eqs.~\eqref{eq:00-_1} and \eqref{eq:00-_2}, if their coefficients satisfy $c_{32} = c_2 -\frac{c_1}{2 \mathbf{m}_1^2}=0$ and $-c_{31} = c_6 -\frac{c_5}{2 \mathbf{m}_1^2}=0$, the H.E. limit would be 
\begin{equation}
\Ampthree{1^-}{2^0}{3^0}{\propag[bos] (i1) to (v1)}{\propag[gho] (i2) to (v1)}{\propag[gho] (i3) to (v1)} = g\frac{\langle12\rangle \langle31\rangle}{\langle23\rangle} \sim ig A_{\mu} (\phi^* \partial^{\mu} \phi -\phi \partial^{\mu} \phi^*).
\end{equation}

Finally, let us consider the 3rd descendant amplitude in helicity $(0,0,0)$. It can be written in the form $[\mathcal{J}(1^0,2^0)]_3 \cdot [\mathbf{A}(3^0)]_0$
\eq{
&[\mathcal{J}(1^0,2^0)]_3 \cdot [\mathbf{A}(3^0)]_0 +[\mathcal{J}(1^0,3^0)]_3 \cdot [\mathbf{A}(2^0)]_0 +[\mathcal{J}(2^0,3^0)]_3 \cdot [\mathbf{A}(1^0)]_0 \\
=& c_{11} \tilde{m}_1 (\langle\eta_1\eta_2\rangle [3\eta_2] \langle13\rangle +\langle1\eta_2\rangle [3\eta_2] \langle\eta_13\rangle) 
+c_{12} \tilde{m}_2 (\langle\eta_1\eta_2\rangle [3\eta_1] \langle23\rangle +\langle\eta_12\rangle [3\eta_1] \langle\eta_23\rangle) \\
&-c_{13} m_1 ([\eta_1\eta_2] [31] \langle\eta_23\rangle +[1\eta_2] [3\eta_1] \langle\eta_23\rangle) 
-c_{14} m_2 ([\eta_1\eta_2] [32] \langle\eta_13\rangle +[\eta_12] [3\eta_2] \langle\eta_13\rangle) +perm. \\
\xrightarrow{H.E.}& 2 (c_{11} -c_{12}) \tilde{m}_1 \tilde{m}_2 \langle\eta_1\eta_2\rangle \langle12\rangle  
+2 (c_{13} -c_{14}) m_1 m_2 [\eta_1\eta_2] [12]
+2 (c_{21} -c_{22}) \tilde{m}_1 \tilde{m}_3 \langle\eta_1\eta_3\rangle \langle13\rangle \\
&+2 (c_{23} -c_{24}) m_1 m_3 [\eta_1\eta_3] [13] 
+2 (c_{31} -c_{32}) \tilde{m}_3 \tilde{m}_2 \langle\eta_3\eta_2\rangle \langle32\rangle 
+2 (c_{33} -c_{34}) m_3 m_2 [\eta_3\eta_2] [32] \\ 
&+(c_{11} +c_{13}) \mathbf{m}_1^2 (2p_3\cdot\eta_2) 
-(c_{12} +c_{14}) \mathbf{m}_2^2 (2p_3\cdot\eta_1) 
+(c_{21} +c_{23}) \mathbf{m}_1^2 (2p_2\cdot\eta_3) \\
&-(c_{22} +c_{24}) \mathbf{m}_3^2 (2p_2\cdot\eta_1) 
+(c_{31} +c_{33}) \mathbf{m}_3^2 (2p_1\cdot\eta_2)
-(c_{32} +c_{34}) \mathbf{m}_2^2 (2p_1\cdot\eta_3).
}
No matter how we choose the coefficients, the H.E. limit cannot give the $\phi^3$ interaction. Thus, it can be inferred that the H.E. limit of $VVV$ type is not $\phi^3$, but the $F_L^2\phi$ and $F_R^2\phi$ operators with two Higgs insertions. It is consistent with the proposition.

\subsection{\texorpdfstring{$VVVV$}{VVVV}}

For the $VVVV$ class, the primary amplitude involves many helicity categories, as shown in Eq.~\eqref{eq:4V_massless}. In most helicity categories, the primary amplitude does not vanish, except for helicity $(0,0,0,0)$. In this subsection, we focus on this helicity category.

We first consider the structure with minimal mass dimensions. In this case, we can identify four independent current structures, among which one is conserved:
\eq{
[\mathcal{J}(1^0,2^0,3^0)]_0
=[\mathcal{J}_c(1^0,2^0,3^0)]_0+[\mathcal{J}_n(1^0,2^0,3^0)]_0,
}
where the conserved current $[\mathcal{J}_c]_0$ and non-conserved current $[\mathcal{J}_n]_0$ are defined as
\begin{equation} \begin{aligned}
[\mathcal{J}_c(1^0,2^0,3^0)]_0
&=c_1|1]\langle1|\langle23\rangle[23]+c_2|1]\langle2|
\langle13\rangle[23]+c_3|2] \langle2|\langle13\rangle[13], \\
[\mathcal{J}_n(1^0,2^0,3^0)]_0
&=c_4(|1]\langle2|[23]\langle13\rangle-|2]\langle1|[13]\langle23\rangle).
\end{aligned}
\end{equation}
See Eq.~\eqref{eq:VVVVconservedcurrent} for why we adopt this form. The $c_4$ term satisfies conservation in the H.E. limit, as verified by:
\eq{
c_4(|1]\langle2|[23]\langle13\rangle-|2]\langle1|[13]\langle23\rangle)\cdot (-|1]\langle1|-|2]\langle2|-|3]\langle3|)\xrightarrow{H.E.}0.
}

We first consider the non-conserved current $[\mathcal{J}_n ]_0$. The three terms yield three distinct massless amplitudes:
\begin{equation} \begin{aligned}
&[\mathcal{J}_n (1^0,2^0,3^0)]_0\cdot[\mathbf{A}(4^0)]_0\\
=&-c_1[41]\langle14\rangle\langle23\rangle[23]-c_2[41]\langle24\rangle
\langle13\rangle[23]-c_3[42] \langle24\rangle\langle13\rangle[13] \\
\xrightarrow{H.E.}&\frac{C_1}{\Lambda^4}[41]\langle14\rangle\langle23\rangle[23]+\frac{C_2}{\Lambda^4}[41]\langle24\rangle
\langle13\rangle[23]+\frac{C_3}{\Lambda^4}[42] \langle24\rangle\langle13\rangle[13]
=\begin{tikzpicture}[baseline=-0.1cm] \begin{feynhand}
\setlength{\feynhandarrowsize}{4pt}
\vertex [particle] (i1) at (-0.714,0.714) {$1^0$}; 
\vertex [particle] (i2) at (0.714,0.714) {$2^0$}; 
\vertex [particle] (i3) at (0.714,-0.714) {$3^0$}; 
\vertex [particle] (i4) at (-0.714,-0.714) {$4^0$}; 
\vertex[dot] (v1) at (0,0) {};
\propag[sca](i1) to (v1);
\propag[sca](i2) to (v1);
\propag[sca](i3) to (v1);
\propag[sca](i4) to (v1);
\end{feynhand} \end{tikzpicture},
\end{aligned}
\end{equation}
from which we derive the coefficient matching relations:
\eq{
c_1=-\frac{C_1}{\Lambda^4},\quad c_2=-\frac{C_2}{\Lambda^4},\quad c_3=-\frac{C_3}{\Lambda^4}.
}
This gives the leading-order matching result ($d=8$ or $n_\lambda=4$).

Then we consider the conserved current $[\mathcal{J}_c]_0$. The corresponding primary amplitude vanishes in the H.E. limit, 
\begin{equation} \begin{aligned}
[\mathcal{J}_c(1^0,2^0,3^0)]_0\cdot[\mathbf{A}(4^0)]_0
&=c_4 ([41]\langle24\rangle [23]\langle13\rangle - [42]\langle14\rangle [13]\langle14\rangle) 
=\Ampfour{1^0}{2^0}{3^0}{4^0}{\bos{i1}{brown}}{\bos{i2}{brown}}{\bos{i3}{brown}}{\bos{i4}{brown}}\xrightarrow{H.E.} 0.
\end{aligned} \end{equation}
Therefore, it does not contribute to the leading-order massless operator.

This current contributes to the massless amplitude at the 1st descendant level via angular momentum raising and lowering operators ($mJ^-,\Tilde{m}J^+$), taking the form:
\eq{
[\mathcal{J} (1^\pm,2^0,3^0)]_1\cdot[\mathbf{A}(4^0)]_0,\;\text{perm},
}
where:
\begin{eqnarray}
    [\mathcal{J} (1^+,2^0,3^0)]_1=\frac{c_1^+}{m_1}   \left(|1]\langle2|[23]\langle\eta_13\rangle-|2]\langle\eta_1|[13]\langle23\rangle\right),
\end{eqnarray}
\begin{eqnarray}
    [\mathcal{J} (1^-,2^0,3^0)]_1=\frac{c_1^-}{\Tilde{m}_1}\left(|\eta_1]\langle2|[23]\langle13\rangle-|2]\langle1|[\eta_13]\langle23\rangle\right),
\end{eqnarray}
with six additional structures obtained by permuting the above two expressions (totaling eight structures). Due to the triviality of permutations, we do not detail these further. These currents directly contribute to massless amplitudes as follows:
\eq{
[\mathcal{J}(1^+,2^0,3^0)]_1\cdot[\mathbf{A}_4]_0&=\frac{c_1^+}{m_1}\left([41]\langle24\rangle[23]\langle\eta_13\rangle-[42]\langle\eta_14\rangle[13]\langle23\rangle\right)\\
&=-\frac{c_1^+}{m_1}\left([31]\langle23\rangle[23]\langle\eta_13\rangle-[32]\langle\eta_13\rangle[13]\langle23\rangle-[12]\langle\eta_11\rangle[13]\langle23\rangle\right)\\
&=-c_1^+[12][13]\langle23\rangle=\Ampfour{1^+}{2^0}{3^0}{4^0}{\bosflip{1}{135}{brown}{red}}{\bos{i2}{brown}}{\bos{i3}{brown}}{\bos{i4}{brown}}\\
&\xrightarrow{H.E.}\frac{C_{F_{R} D^2 \phi^3}}{\Lambda^3}[12][13]\langle23\rangle=
\begin{tikzpicture}[baseline=-0.1cm] \begin{feynhand}
\setlength{\feynhandarrowsize}{4pt}
\vertex [particle] (i1) at (-0.714,0.714) {$1^+$}; 
\vertex [particle] (i2) at (0.714,0.714) {$2^0$}; 
\vertex [particle] (i3) at (0.714,-0.714) {$3^0$}; 
\vertex [particle] (i4) at (-0.714,-0.714) {$4^0$}; 
\vertex[dot] (v1) at (0,0) {};
\propag[bos](i1) to (v1);
\propag[sca](i2) to (v1);
\propag[sca](i3) to (v1);
\propag[sca](i4) to (v1);
\end{feynhand} \end{tikzpicture},
}
and its left-handed counterpart:
\eq{
[\mathcal{J}(1^-,2^0,3^0)]_1\cdot[\mathbf{A}(4^0)]_0&=\frac{c_1^-}{\tilde{m_1}}\left([4\eta_1]\langle24\rangle[23]\langle13\rangle-[42]\langle14\rangle[\eta_13]\langle23\rangle\right)\\
&=-\frac{c_1^-}{\tilde{m_1}}\left([1\eta_1]\langle21\rangle[23]\langle13\rangle+[3\eta_1]\langle23\rangle[23]\langle13\rangle-[32]\langle13\rangle[\eta_13]\langle23\rangle\right)\\
&=-c_1^-\langle21\rangle[23]\langle13\rangle=\Ampfour{1^-}{2^0}{3^0}{4^0}{\bosflip{1}{135}{brown}{cyan}}{\bos{i2}{brown}}{\bos{i3}{brown}}{\bos{i4}{brown}}\\
&\xrightarrow{H.E.}\frac{C_{F_{L} D^2 \phi^3}}{\Lambda^3}\langle21\rangle[23]\langle13\rangle=
\begin{tikzpicture}[baseline=-0.1cm] \begin{feynhand}
\setlength{\feynhandarrowsize}{4pt}
\vertex [particle] (i1) at (-0.714,0.714) {$1^-$}; 
\vertex [particle] (i2) at (0.714,0.714) {$2^0$}; 
\vertex [particle] (i3) at (0.714,-0.714) {$3^0$}; 
\vertex [particle] (i4) at (-0.714,-0.714) {$4^0$}; 
\vertex[dot] (v1) at (0,0) {};
\propag[bos](i1) to (v1);
\propag[sca](i2) to (v1);
\propag[sca](i3) to (v1);
\propag[sca](i4) to (v1);
\end{feynhand} \end{tikzpicture}.
}
Here, we have defined the massless operators $F_{L/R} D^2 \phi^3 =F_{L/R}^{\mu\nu}(D_\mu\phi_2 D_\nu\phi_3-D_\mu\phi_3 D_\nu\phi_2)\phi_4$. The above H.E. limits yield the final matching relations:
\eq{
c_1^+=-\frac{C_{F_{R} D^2 \phi^3}}{\Lambda^3},\quad 
c_1^-=-\frac{C_{F_{L} D^2 \phi^3}}{\Lambda^3}.
}
Notice that $\langle21\rangle[23]\langle13\rangle$ and $[12][13]\langle23\rangle$ do not satisfy Adler’s zero condition. In contrast, all operators that admit direct leading-order matching to massive operators necessarily satisfy Adler’s zero condition. Consequently, these structures (i.e., those violating Adler’s zero) do not possess leading-order massive amplitude matching and are therefore novel in type $VVVV$.

\section{Massive Structures with Conserved Current}
\label{app:exception}

In the main text, we noted that there are exceptional massive structures whose leading-order contributions vanish. These exceptional types are: $VVS,ffV,VSS,VVV,VVVV$. We now examine them in detail.

When matching massive amplitudes to massless ones, a massive momentum $\mathbf{p}$ is decomposed as
\begin{equation}
\mathbf{p}=p+\eta,
\end{equation}
where $p$ is the leading part and $\eta$ is the subleading part. At leading order, it suffices to replace each massive spinor $\boldsymbol{\lambda}$ with its massless counterpart $\lambda$. The massive amplitude $\mathbf{M}$ is then correspond to the massless amplitude $\mathcal{A}$:
\begin{equation}
\mathbf{M}(\boldsymbol{\lambda},\boldsymbol{\tilde{\lambda}})\quad \to\quad \mathcal{A}(\lambda,\tilde{\lambda}).
\end{equation}

When the massless counterpart vanishes, i.e. $\mathcal{A}=0$, the central question is whether the massive amplitude can be rewritten in the form
\begin{equation}
\mathbf{M}\quad \xRightarrow{?} \quad m\times\tilde{\mathbf{M}},
\end{equation}
where $\tilde{\mathbf{M}}$ is a massive structure of lower dimension than $\mathbf{M}$. There are two possibilities:

\begin{itemize}
\item In most cases, the massive amplitude can indeed be factorized in this way, so the vanishing result is explained by the explicit mass factor $m$. For example, the $ffS$ amplitude $\langle \mathbf{1}|\mathbf{2}|\mathbf{3}]=m_1 [\mathbf{13}]-\tilde{m}_3 \langle\mathbf{13}\rangle$ contains a mass factor in each term.

\item In a few exceptional cases, such factorization is not possible. For instance, the $ffV$ amplitude $\langle \mathbf{13}\rangle[\mathbf{23}]$ does not admit an overall mass factor.

\end{itemize}

In this appendix, we focus on the latter, exceptional cases.
The origin of this phenomenon lies in the fact that these exceptional massive amplitudes can be expressed as
\begin{equation}
\mathcal{M}=\mathcal{J}\cdot \mathbf{A},
\end{equation}
where $\mathcal{J}$ is a current and $\mathbf{A}$ is a massive vector field. In the high-energy limit, this reduces to current conservation, $A_\mu \mathcal{J}^\mu\to \partial\cdot \mathcal{J}=0$. As discussed more generally in Ref.~\cite{Ni:2025xkg}, the conserved currents in a renormalizable theory are finite in number. Consequently, the exceptional cases are also finite: for spins $s\le 1$, there are only 5 distinct cases which arise in types $VVS,ffV,VSS,VVV,VVVV$.

\paragraph{Exceptional structures}

\begin{enumerate}
\item{\textbf{VSS,ffV}} For these types, the current $\mathcal{J}$ does not contain a vector boson. Its general form is
\begin{eqnarray}
    [\mathcal{J}]_0=c_1(|1]\langle1|-|2]\langle2|)+c_2|1]\langle2|+c_3|2]\langle1|+c_4(|1]\langle1|+|2]\langle2|).
\end{eqnarray}
Current conservation imposes $c_4 = 0$, leaving three independent structures. Among these,  $c_2 $ and  $c_3$ correspond to the same structural form. Consequently, we ultimately obtain two exceptional massive bases:
\begin{eqnarray}
    A_{\alpha\dot\alpha}\left(\phi_1D^{\alpha\dot\alpha}\phi_2-\phi_2D^{\alpha\dot\alpha}\phi_1\right), \psi^\alpha A_{\alpha\dot\alpha} \psi^{\dagger\dot\alpha}.
    \end{eqnarray}
These correspond to the $VSS$ and $ffV$ structures, separately. Note that the $VSS$ structure requires two distinct massive scalar bosons; if $\phi_1$ and $\phi_2$ are identical, this structure vanishes.

\item {\textbf{VVS,VVV}}
For these types, the current $\mathcal{J}$ contains a vector. Therefore, we require that $|2]\langle2|$ appears at least once in each term as the counterpart of $V$. We write the general current structure as
\begin{eqnarray}
    [\mathcal{J}]_0=c_1|2]\langle2|+c_2|2]\langle1|\langle12\rangle+c_3|2]\langle1|[12]+c_4|1]\langle1| [12]\langle12\rangle.
\end{eqnarray}
The last term $|1]\langle1|[12]\langle12\rangle$ should not contribute to the exceptional structure, because $[31]\langle13\rangle[12]\langle12\rangle=-[32]\langle23\rangle^2$. This results in a massive contraction $\mathbf{p}_2\cdot \mathbf{p}_3=\frac{1}{2}(m_3^2-m_1^2-m_2^2)$. In the three-point case, this cannot be a primary amplitude, so we must set $c_4=0$.  Current conservation imposes no further constraints. We are left with three operators:
\begin{eqnarray}
    \phi_1A_2\cdot A_3, \;F_{1L}^{\alpha\beta}A_{2\alpha}^{\dot\alpha}A_{3\beta\dot\alpha},\;
F_{1R}^{\dot\alpha\dot\beta}A_{2\dot\alpha}^{\alpha}A_{3\alpha\dot\beta}.
\end{eqnarray}
The first corresponds to $VVS$, while the other two correspond to $VVV$.

    \item {\textbf{VVVV}} For this type, the current has the form $\mathcal{J}\sim A_1A_2A_3$, which is a dimension-3 polynomial. In the high-energy limit, the current $\mathcal{J}$ is constructed from $|1]\langle1|,|2]\langle2|,|3]\langle3|$, each with power 1. There are a total of 9 such Lorentz representations. Using the Schouten identities $|3][12]=|2][13]+|1][32]$ and $\langle3|\langle12\rangle=\langle2|\langle13\rangle+\langle1|\langle32\rangle$, the general form of $[\mathcal{J}]_0$ reduces to
    \begin{equation}
        [\mathcal{J}]_0=c_1 |1]\langle1| [23]\langle23\rangle+c_2|1]\langle2|[23]\langle13\rangle+c_3|2]\langle1|[13]\langle23\rangle+c_4|2]\langle2|[13]\langle13\rangle.
    \end{equation}
    Current conservation requires
    \begin{eqnarray}
        (\mathbf{p}_1+\mathbf{p}_2+\mathbf{p}_3)\cdot [\mathcal{J}]_0 &\sim& c_1 \left([31]\langle13\rangle+[21]\langle12\rangle\right)[23]\langle23\rangle+c_2[31]\langle23\rangle[23]\langle13\rangle+\\
        &&c_3[32]\langle13\rangle[13]\langle23\rangle+c_4\left([12]\langle21\rangle+[32]\langle23\rangle\right)[13]\langle13\rangle\nonumber\\
        &=&\left(c_1+c_2+c_3+c_4\right)[31]\langle23\rangle[23]\langle13\rangle+\left(c_1[23]\langle23\rangle+c_4[13]\langle13\rangle\right)[21]\langle12\rangle.\nonumber
    \end{eqnarray}
    Setting this to zero gives $c_1+c_2+c_3+c_4=0$ and $c_1=c_4=0$. Hence, there is only one independent structure
    \begin{eqnarray}\label{eq:VVVVconservedcurrent}
        A\cdot [\mathcal{J}]_0\sim [41]\langle24\rangle[23]\langle13\rangle-[42]\langle14\rangle[13]\langle23\rangle,
    \end{eqnarray}
    at the leading massless level. This corresponds to the unique $VVVV$ exceptional structure
    \begin{eqnarray}
        \epsilon^{\mu\nu\rho\sigma}A_{1\mu}A_{2\nu}A_{3\rho}A_{4\sigma},
    \end{eqnarray}
    which yields the massive amplitude  $[\mathbf{41}]\langle\mathbf{24}\rangle[\mathbf{23}]\langle\mathbf{13}\rangle-[\mathbf{42}]\langle\mathbf{14}\rangle[\mathbf{13}]\langle\mathbf{23}\rangle$. This structure requires four distinct massive vector bosons; otherwise, it vanishes.
\end{enumerate}

Up to now, we have obtained all exceptional structures, which correspond to five classes: $VVS$, $ffV$, $VSS$, $VVV$, $VVVV$. These represent the only possibilities based and further exceptions arise. A stricter proof will be given in Ref.~\cite{Ni:2026inpre}. Consequently, our massless-massive matching method can yield all primary amplitudes other than these exceptional structures.

\paragraph{Structures in higher-dimensions}

    The preceding discussion concerned exceptional structures that do not admit a mass-factor extraction. For structures with more derivatives, no such exceptions occur. By eliminating mass-factor terms, we can obtain a complete basis for the leading massive structures. 
    For 3-point structures, each derivative momentum reduces to a mass factor, so no additional independent leading structures exist. We therefore turn to four-point structures with more derivatives, i.e. Lorentz invariants of the form $V^4D^{2n}$ with transversality category $(0000)$.
    
    First, the power of derivatives must be even, the number of derivatives must be even, since a Lorentz scalar requires an equal number of left- and right-handed indices. Second, we consider primary operators, so mass-proportional terms may be neglected. Using the identities
    \begin{eqnarray}\label{eq:EoMofSpinor}
        |\mathbf{p}\rangle_{\beta} p_{\alpha\dot\alpha}-|\mathbf{p}\rangle_{\alpha} p_{\beta\dot\alpha}\sim\epsilon_{\alpha\beta }m[\mathbf{p}|_{\dot\alpha},\quad |\mathbf{p}]_{\dot\beta} p_{\alpha\dot\alpha}-|\mathbf{p}]_{\dot\alpha} p_{\alpha\dot\beta}\sim\epsilon_{\dot\alpha\dot\beta }\tilde{m}|\mathbf{p}\rangle_{\alpha},
    \end{eqnarray}
    the spinor indices of the wavefunctions and derivatives can be commuted. We have four wavefunctions $\epsilon$ and four derivatives contracted among themselves. General Lorentz scalars are built from dot products ($\epsilon_i\cdot \mathbf{p}_j,\;\mathbf{p}_i\cdot \mathbf{p}_j,\;\epsilon_i \cdot \epsilon_j$) and possibly the Levi-Civita tensor $\varepsilon(\cdot,\cdot,\cdot,\cdot)$. The vectors in $\varepsilon$ can be momenta or polarizations, and $\varepsilon$ can appear at most linearly. 
    
    We now perform a simplification for the structures involving $\varepsilon(\cdot,\cdot,\cdot,\cdot)$. Due to momentum conservation $\mathbf{p}_4=-\mathbf{p}_1-\mathbf{p}_2-\mathbf{p}_3$, the appearance of $\mathbf{p}_4$ is prohibited. In $\varepsilon(\cdot,\cdot,\cdot,\cdot)$, each particle’s momentum or polarization vector can appear only once; for example, the antisymmetric structure $\mathbf{p}_{1[\mu}\epsilon_{1\nu]}$ in $\varepsilon(\mathbf{p}_1,\epsilon_1,\cdot,\cdot)$ generates a mass term via Eq.~\eqref{eq:EoMofSpinor}, and thus vanishes. In $\varepsilon(\cdot,\cdot,\cdot,\cdot)$, each $\varepsilon_i$ appears exactly once. Otherwise, expressions like $\varepsilon(\mathbf{p}_1,\cdot,\cdot,\cdot)\,\epsilon_{1\mu}\cdots$. Using $\mathbf{p}_{1[\mu}\epsilon_{1\nu]}\sim 0$ iteratively, we may replace $p_1$ by $\epsilon_1$ within the Levi-Civita tensor. We therefore conclude that every $\varepsilon$ contribution reduces to $\varepsilon(\epsilon_1,\epsilon_2,\epsilon_3,\epsilon_4)$.

    We now show that the structures involving $\varepsilon(\epsilon_1,\epsilon_2,\epsilon_3,\epsilon_4)$ are further redundant when derivatives are present ($n\neq0$). As at least one derivative term is present, we can take $\mathbf{p}_4$ without loss of generality. Making use of $\mathbf{p}_{4[\mu}\epsilon_{4\nu]}\sim 0$, we replace $\epsilon_4$ by $\mathbf{p}_4$. Combining integration by parts $\mathbf{p}_4 = -\mathbf{p}_1 - \mathbf{p}_2 - \mathbf{p}_3$ and $\varepsilon(\mathbf{p}_i,\epsilon_i,\cdot,\cdot)\sim 0$, we conclude that all $\varepsilon$ terms can be dropped entirely. Thus, for the $V^4 D^2$ type with transversality category $(0000)$, only dot products $\epsilon_i\cdot \mathbf{p}_j$ survive. 
    
    Then we consider the simplification for the dot products $\epsilon_i\cdot \mathbf{p}_j$. We eliminate $\mathbf{p}_4$ via IBP. The wave function $\epsilon_4$ must contract with one of $\epsilon_1, \epsilon_2, \epsilon_3$; if it contracts with a momentum $\mathbf{p}_i$, we can convert the momentum back to an $\epsilon_i$ using the equivalence $[\mathbf{p}_i, \epsilon_i] \sim 0$. The amplitude then falls into three categories $(M_1, M_2, M_3)$ according to which $\epsilon_i$ contracts with $\epsilon_4$ contracts with. For $M_1$ (where $\epsilon_4$ contracts with $\epsilon_1$), the remaining contractions involve $\epsilon_2 \cdot \mathbf{p}_1$,$\epsilon_2 \cdot \mathbf{p}_3$, and $\epsilon_2 \cdot \epsilon_3$ with the latter two being equivalent. Thus,
    \eq{
    M_1=\epsilon_4\cdot\epsilon_1\times(\epsilon_2 \cdot \mathbf{p}_1\epsilon_3 \cdot \mathbf{p}_1 f_{11}(s_{ij}),\;\epsilon_2 \cdot \epsilon_3 f_{12}(s_{ij})),
    }
    \eq{
    M_2=\epsilon_4\cdot\epsilon_2\times(\epsilon_1 \cdot \mathbf{p}_2\epsilon_3 \cdot \mathbf{p}_2 f_{21}(s_{ij}),\;\epsilon_1 \cdot \epsilon_3 f_{22}(s_{ij})),
    }
    \eq{
    M_3=\epsilon_4\cdot\epsilon_3\times(\epsilon_2 \cdot \mathbf{p}_3\epsilon_1 \cdot \mathbf{p}_3 f_{31}(s_{ij}),\;\epsilon_2 \cdot \epsilon_1 f_{32}(s_{ij})),
    }
    where $s_{ij}=\mathbf{p}_i\cdot\mathbf{p}_j$ denotes the Mandelstam variable. Using the relation
\begin{eqnarray}
    -\epsilon_4\cdot\epsilon_3\mathbf{p}_3\cdot\epsilon_2\epsilon_1\cdot\mathbf{p}_3 
    &\sim& -\epsilon_4\cdot\mathbf{p}_3\epsilon_3\cdot\epsilon_2\epsilon_1\cdot\mathbf{p}_3\nonumber\\
    &\sim& (\epsilon_4\cdot\mathbf{p}_1+\epsilon_4\cdot\mathbf{p}_2)\epsilon_3\cdot\epsilon_2 \epsilon_1\cdot\mathbf{p}_3 \nonumber\\
    &\sim&\epsilon_4\cdot\epsilon_1\epsilon_3\cdot\epsilon_2\mathbf{p}_1\cdot\mathbf{p}_3+\epsilon_4\cdot\epsilon_2\epsilon_3\cdot\mathbf{p}_2 \epsilon_1\cdot\mathbf{p}_3 \nonumber\\
    &\sim&\epsilon_4\cdot\epsilon_1\epsilon_3\cdot\epsilon_2\mathbf{p}_1\cdot\mathbf{p}_3+\epsilon_4\cdot\epsilon_2\mathbf{p}_3\cdot\mathbf{p}_2 \epsilon_1\cdot\epsilon_3,
\end{eqnarray}
we can set $f_{11}=f_{21}=f_{31}=0$. This leaves
\begin{eqnarray}
        M_1 &=&\epsilon_4\cdot\epsilon_1\epsilon_2\cdot\epsilon_3(f_{12}),\\M_2 &=& \epsilon_4\cdot\epsilon_2\epsilon_3\cdot\epsilon_1(f_{22}),\\M_3 &=& \epsilon_4\cdot\epsilon_3\epsilon_2\cdot\epsilon_1(f_{32}).
\end{eqnarray}
Most terms in these sets are redundant. Since we are considering $V^4D^{2n}$ with $n\ge 1$, each $f$ contains at least one $s_{ij}$ factor. With the identity $s_{12} + s_{13} + s_{23} = 0$, there are only two independent $s_{ij}$  variables. Thus we may write
\begin{eqnarray}
        M_1 &=& \epsilon_4\cdot\epsilon_1\epsilon_2\cdot\epsilon_3(s_{12}^n,s_{12}^{n-1}s_{13},\dots,s_{13}^n),\\M_2 &=& \epsilon_4\cdot\epsilon_2\epsilon_3\cdot\epsilon_1(s_{12}^n,s_{12}^{n-1}s_{13},\dots,s_{13}^n),\\M_3 &=& \epsilon_4\cdot\epsilon_3\epsilon_2\cdot\epsilon_1(s_{12}^n,s_{12}^{n-1}s_{13},\dots,s_{13}^n).
\end{eqnarray}

In the next step, we can show that the terms in $M_1$ are redundant. Consider two relations
\begin{eqnarray}\label{eq:identityforV^4D^2eq1}
    -\epsilon_4\cdot\epsilon_1\epsilon_2\cdot\epsilon_3s_{12}\
    &\sim& -\epsilon_4\cdot\mathbf{p}_1\epsilon_2\cdot\epsilon_3\epsilon_1\cdot\mathbf{p}_2\nonumber\\
    &\sim&\epsilon_4\cdot\mathbf{p}_2\epsilon_2\cdot\epsilon_3\epsilon_1\cdot\mathbf{p}_2+\epsilon_4\cdot\mathbf{p}_3\epsilon_2\cdot\epsilon_3\epsilon_1\cdot\mathbf{p}_2\nonumber\\    &\sim&
    \epsilon_4\cdot\epsilon_2\mathbf{p}_2\cdot\epsilon_3\epsilon_1\cdot\mathbf{p}_2+\epsilon_4\cdot\epsilon_1\mathbf{p}_3\cdot\mathbf{p}_2\epsilon_1\cdot\epsilon_2,
\end{eqnarray}
and
\begin{eqnarray}\label{eq:identityforV^4D^2eq2}
    -\epsilon_4\cdot\epsilon_1\epsilon_2\cdot\epsilon_3s_{34}
    &\sim& -\epsilon_4\cdot\mathbf{p}_3\epsilon_2\cdot\epsilon_3\epsilon_1\cdot\mathbf{p}_4\nonumber\\
    &\sim&\epsilon_4\cdot\mathbf{p}_3\epsilon_2\cdot\epsilon_3\epsilon_1\cdot\mathbf{p}_2+\epsilon_4\cdot\mathbf{p}_3\epsilon_2\cdot\epsilon_3\epsilon_1\cdot\mathbf{p}_3\nonumber\\    &\sim&
    \epsilon_4\cdot\epsilon_3\epsilon_2\cdot\epsilon_3\mathbf{p}_3\cdot\mathbf{p}_2+\epsilon_4\cdot\epsilon_3\epsilon_2\cdot\mathbf{p}_3\epsilon_1\cdot\mathbf{p}_3.
\end{eqnarray}
Together with $s_{12}=s_{34}$, these imply
\begin{eqnarray}
    \epsilon_4\cdot\epsilon_3\epsilon_2\cdot\mathbf{p}_3\epsilon_1\cdot\mathbf{p}_3\sim\epsilon_4\cdot\epsilon_2\mathbf{p}_2\cdot\epsilon_3\epsilon_1\cdot\mathbf{p}_2.
\end{eqnarray}
More generally, we have the cyclic relation
\begin{eqnarray}
    \epsilon_3\cdot\epsilon_4\epsilon_2\cdot\mathbf{p}_4\epsilon_1\cdot\mathbf{p}_4\sim\epsilon_4\cdot\epsilon_3\epsilon_2\cdot\mathbf{p}_3\epsilon_1\cdot\mathbf{p}_3\sim\epsilon_4\cdot\epsilon_2\epsilon_3\cdot\mathbf{p}_2\epsilon_1\cdot\mathbf{p}_2\sim\epsilon_4\cdot\epsilon_1\epsilon_3\cdot\mathbf{p}_1\epsilon_2\cdot\mathbf{p}_1.
\end{eqnarray}
This implies that we can perform permutations on the indices $1,2,3,4$. Applying this property to Eq.~\eqref{eq:identityforV^4D^2eq1} and \eqref{eq:identityforV^4D^2eq2} yields the following chain of equalities:
\begin{eqnarray}
    \epsilon_4\cdot\epsilon_1\epsilon_2\cdot\epsilon_3 s_{12}+\epsilon_4\cdot\epsilon_3\epsilon_2\cdot\epsilon_1 s_{23}
    &\sim&\epsilon_4\cdot\epsilon_1\epsilon_2\cdot\epsilon_3 s_{13}+\epsilon_4\cdot\epsilon_2\epsilon_3\cdot\epsilon_1 s_{23}\nonumber\\
    &\sim&\epsilon_4\cdot\epsilon_2\epsilon_3\cdot\epsilon_1 s_{12}+\epsilon_4\cdot\epsilon_3\epsilon_2\cdot\epsilon_1 s_{13}.
    \end{eqnarray}
    This demonstrates that all terms in $M_1$ are redundant.
     
    For the remaining sets $M_2$ and $M_3$, there is still a redundancy among the basis elements:
    \begin{eqnarray}
    \epsilon_4\cdot\epsilon_3\epsilon_2\cdot\epsilon_1s_{13}^2+\epsilon_4\cdot\epsilon_2\epsilon_3\cdot\epsilon_1 s_{12}^2\sim0.
    \end{eqnarray}
    After eliminating these redundancies, the remaining operators are:
    \begin{eqnarray}\label{eq:VVVVSTamplitudes}
        M_2 &=&\epsilon_4\cdot\epsilon_2\epsilon_3\cdot\epsilon_1(s_{12}s_{13}^{n-1},s_{13}^n),\\
        M_3&=&\epsilon_4\cdot\epsilon_3\epsilon_2\cdot\epsilon_1(s_{12}^n,s_{12}^{n-1}s_{13},\dots,s_{13}^n).
\end{eqnarray}
These $n+3$ amplitudes have massless limits (for helicity category $(0000)$ ) given by 
\eq{\label{eq:VVVVmasslessamp}
(s_{12}^{n+2},s_{12}^{n+1}s_{13},\dots,s_{13}^{n+2}),
}
 which are all independent. Hence, the massive amplitudes themselves are independent and complete.

Thus, after eliminating all redundancies, we have obtained a complete and independent set of massive amplitudes for the $V^4 D^{2n}$ structures.

%% file: ref.bib
@article{Li:2020gnx,
    author = "Li, Hao-Lin and Ren, Zhe and Shu, Jing and Xiao, Ming-Lei and Yu, Jiang-Hao and Zheng, Yu-Hui",
    title = "{Complete set of dimension-eight operators in the standard model effective field theory}",
    eprint = "2005.00008",
    archivePrefix = "arXiv",
    primaryClass = "hep-ph",
    doi = "10.1103/PhysRevD.104.015026",
    journal = "Phys. Rev. D",
    volume = "104",
    number = "1",
    pages = "015026",
    year = "2021"
}

@article{Conde:2016izb,
    author = "Conde, Eduardo and Joung, Euihun and Mkrtchyan, Karapet",
    title = "{Spinor-Helicity Three-Point Amplitudes from Local Cubic Interactions}",
    eprint = "1605.07402",
    archivePrefix = "arXiv",
    primaryClass = "hep-th",
    doi = "10.1007/JHEP08(2016)040",
    journal = "JHEP",
    volume = "08",
    pages = "040",
    year = "2016"
}

@article{Li:2020xlh,
    author = "Li, Hao-Lin and Ren, Zhe and Xiao, Ming-Lei and Yu, Jiang-Hao and Zheng, Yu-Hui",
    title = "{Complete set of dimension-nine operators in the standard model effective field theory}",
    eprint = "2007.07899",
    archivePrefix = "arXiv",
    primaryClass = "hep-ph",
    doi = "10.1103/PhysRevD.104.015025",
    journal = "Phys. Rev. D",
    volume = "104",
    number = "1",
    pages = "015025",
    year = "2021"
}

@article{Li:2022tec,
    author = "Li, Hao-Lin and Ren, Zhe and Xiao, Ming-Lei and Yu, Jiang-Hao and Zheng, Yu-Hui",
    title = "{Operators for generic effective field theory at any dimension: on-shell amplitude basis construction}",
    eprint = "2201.04639",
    archivePrefix = "arXiv",
    primaryClass = "hep-ph",
    doi = "10.1007/JHEP04(2022)140",
    journal = "JHEP",
    volume = "04",
    pages = "140",
    year = "2022"
}

@article{Henning:2019enq,
    author = "Henning, Brian and Melia, Tom",
    title = "{Constructing effective field theories via their harmonics}",
    eprint = "1902.06754",
    archivePrefix = "arXiv",
    primaryClass = "hep-ph",
    doi = "10.1103/PhysRevD.100.016015",
    journal = "Phys. Rev. D",
    volume = "100",
    number = "1",
    pages = "016015",
    year = "2019"
}

@article{Ma:2019gtx,
    author = "Ma, Teng and Shu, Jing and Xiao, Ming-Lei",
    title = "{Standard model effective field theory from on-shell amplitudes*}",
    eprint = "1902.06752",
    archivePrefix = "arXiv",
    primaryClass = "hep-ph",
    doi = "10.1088/1674-1137/aca200",
    journal = "Chin. Phys. C",
    volume = "47",
    number = "2",
    pages = "023105",
    year = "2023"
}

@article{Dong:2024dce,
    author = "Dong, Ziyu and Ma, Teng and Yang, Chengjie and Zhou, Zizheng",
    title = "{Dark photons and high spin particles: complete EFT operator basis}",
    eprint = "2412.20096",
    archivePrefix = "arXiv",
    primaryClass = "hep-ph",
    doi = "10.1007/JHEP07(2025)104",
    journal = "JHEP",
    volume = "07",
    pages = "104",
    year = "2025"
}

@article{Arkani-Hamed:2017jhn,
    author = "Arkani-Hamed, Nima and Huang, Tzu-Chen and Huang, Yu-tin",
    title = "{Scattering amplitudes for all masses and spins}",
    eprint = "1709.04891",
    archivePrefix = "arXiv",
    primaryClass = "hep-th",
    reportNumber = "NCTS-TH/1714, NCTS-TH-1714",
    doi = "10.1007/JHEP11(2021)070",
    journal = "JHEP",
    volume = "11",
    pages = "070",
    year = "2021"
}

@article{Henning:2019mcv,
    author = "Henning, Brian and Melia, Tom",
    title = "{Conformal-helicity duality {\textbackslash}{\&} the Hilbert space of free CFTs}",
    eprint = "1902.06747",
    archivePrefix = "arXiv",
    primaryClass = "hep-th",
    month = "2",
    year = "2019"
}

@article{Shadmi:2018xan,
    author = "Shadmi, Yael and Weiss, Yaniv",
    title = "{Effective Field Theory Amplitudes the On-Shell Way: Scalar and Vector Couplings to Gluons}",
    eprint = "1809.09644",
    archivePrefix = "arXiv",
    primaryClass = "hep-ph",
    doi = "10.1007/JHEP02(2019)165",
    journal = "JHEP",
    volume = "02",
    pages = "165",
    year = "2019"
}

@article{Li:2021tsq,
    author = "Li, Hao-Lin and Ren, Zhe and Xiao, Ming-Lei and Yu, Jiang-Hao and Zheng, Yu-Hui",
    title = "{Operator bases in effective field theories with sterile neutrinos: d \ensuremath{\leq} 9}",
    eprint = "2105.09329",
    archivePrefix = "arXiv",
    primaryClass = "hep-ph",
    doi = "10.1007/JHEP11(2021)003",
    journal = "JHEP",
    volume = "11",
    pages = "003",
    year = "2021"
}

@article{Arkani-Hamed:2019ymq,
    author = "Arkani-Hamed, Nima and Huang, Yu-tin and O'Connell, Donal",
    title = "{Kerr black holes as elementary particles}",
    eprint = "1906.10100",
    archivePrefix = "arXiv",
    primaryClass = "hep-th",
    reportNumber = "NCTS-TH/1905",
    doi = "10.1007/JHEP01(2020)046",
    journal = "JHEP",
    volume = "01",
    pages = "046",
    year = "2020"
}

@article{Guevara:2019fsj,
    author = "Guevara, Alfredo and Ochirov, Alexander and Vines, Justin",
    title = "{Black-hole scattering with general spin directions from minimal-coupling amplitudes}",
    eprint = "1906.10071",
    archivePrefix = "arXiv",
    primaryClass = "hep-th",
    doi = "10.1103/PhysRevD.100.104024",
    journal = "Phys. Rev. D",
    volume = "100",
    number = "10",
    pages = "104024",
    year = "2019"
}

@article{Ren:2022tvi,
    author = "Ren, Zhe and Yu, Jiang-Hao",
    title = "{A complete set of the dimension-8 Green\textquoteright{}s basis operators in the Standard Model effective field theory}",
    eprint = "2211.01420",
    archivePrefix = "arXiv",
    primaryClass = "hep-ph",
    doi = "10.1007/JHEP02(2024)134",
    journal = "JHEP",
    volume = "02",
    pages = "134",
    year = "2024"
}

@article{Li:2020zfq,
    author = "Li, Hao-Lin and Shu, Jing and Xiao, Ming-Lei and Yu, Jiang-Hao",
    title = "{Depicting the Landscape of Generic Effective Field Theories}",
    eprint = "2012.11615",
    archivePrefix = "arXiv",
    primaryClass = "hep-ph",
    month = "12",
    year = "2020"
}

@article{Durieux:2019siw,
    author = "Durieux, Gauthier and Machado, Camila S.",
    title = "{Enumerating higher-dimensional operators with on-shell amplitudes}",
    eprint = "1912.08827",
    archivePrefix = "arXiv",
    primaryClass = "hep-ph",
    reportNumber = "MITP/19-090",
    doi = "10.1103/PhysRevD.101.095021",
    journal = "Phys. Rev. D",
    volume = "101",
    number = "9",
    pages = "095021",
    year = "2020"
}

@article{Li:2020tsi,
    author = "Li, Hao-Lin and Ren, Zhe and Xiao, Ming-Lei and Yu, Jiang-Hao and Zheng, Yu-Hui",
    title = "{Low energy effective field theory operator basis at d \ensuremath{\leq} 9}",
    eprint = "2012.09188",
    archivePrefix = "arXiv",
    primaryClass = "hep-ph",
    doi = "10.1007/JHEP06(2021)138",
    journal = "JHEP",
    volume = "06",
    pages = "138",
    year = "2021"
}

@article{Aoude:2019tzn,
    author = "Aoude, Rafael and Machado, Camila S.",
    title = "{The Rise of SMEFT On-shell Amplitudes}",
    eprint = "1905.11433",
    archivePrefix = "arXiv",
    primaryClass = "hep-ph",
    doi = "10.1007/JHEP12(2019)058",
    journal = "JHEP",
    volume = "12",
    pages = "058",
    year = "2019"
}

@article{Dong:2021vxo,
    author = "Dong, Zi-Yu and Ma, Teng and Shu, Jing",
    title = "{Constructing on-shell operator basis for all masses and spins}",
    eprint = "2103.15837",
    archivePrefix = "arXiv",
    primaryClass = "hep-ph",
    doi = "10.1103/PhysRevD.107.L111901",
    journal = "Phys. Rev. D",
    volume = "107",
    number = "11",
    pages = "L111901",
    year = "2023"
}

@article{Ballav:2021ahg,
    author = "Ballav, Sourav and Manna, Arkajyoti",
    title = "{Recursion relations for scattering amplitudes with massive particles II: Massive vector bosons}",
    eprint = "2109.06546",
    archivePrefix = "arXiv",
    primaryClass = "hep-th",
    doi = "10.1016/j.nuclphysb.2022.115935",
    journal = "Nucl. Phys. B",
    volume = "983",
    pages = "115935",
    year = "2022"
}

@article{Dong:2022mcv,
    author = "Dong, Zi-Yu and Ma, Teng and Shu, Jing and Zheng, Yu-Hui",
    title = "{Constructing generic effective field theory for all masses and spins}",
    eprint = "2202.08350",
    archivePrefix = "arXiv",
    primaryClass = "hep-ph",
    doi = "10.1103/PhysRevD.106.116010",
    journal = "Phys. Rev. D",
    volume = "106",
    number = "11",
    pages = "116010",
    year = "2022"
}

@article{Badger:2023eqz,
    author = "Badger, Simon and Henn, Johannes and Plefka, Jan Christoph and Zoia, Simone",
    title = "{Scattering Amplitudes in Quantum Field Theory}",
    eprint = "2306.05976",
    archivePrefix = "arXiv",
    primaryClass = "hep-th",
    doi = "10.1007/978-3-031-46987-9",
    journal = "Lect. Notes Phys.",
    volume = "1021",
    pages = "pp.",
    year = "2024"
}

@article{Goldberg:2024eot,
    author = "Goldberg, Jared M. and Liu, Hongkai and Shadmi, Yael",
    title = "{Dimension-8 SMEFT contact-terms for vector-pair production via on-shell Higgsing}",
    eprint = "2407.07945",
    archivePrefix = "arXiv",
    primaryClass = "hep-ph",
    doi = "10.1007/JHEP12(2024)057",
    journal = "JHEP",
    volume = "12",
    pages = "057",
    year = "2024"
}

@article{Bachu:2023fjn,
    author = "Bachu, Brad",
    title = "{Spontaneous symmetry breaking from an on-shell perspective}",
    eprint = "2305.02502",
    archivePrefix = "arXiv",
    primaryClass = "hep-th",
    doi = "10.1007/JHEP02(2024)098",
    journal = "JHEP",
    volume = "02",
    pages = "098",
    year = "2024"
}

@article{Liu:2023jbq,
    author = "Liu, Hongkai and Ma, Teng and Shadmi, Yael and Waterbury, Michael",
    title = "{An EFT hunter\textquoteright{}s guide to two-to-two scattering: HEFT and SMEFT on-shell amplitudes}",
    eprint = "2301.11349",
    archivePrefix = "arXiv",
    primaryClass = "hep-ph",
    doi = "10.1007/JHEP05(2023)241",
    journal = "JHEP",
    volume = "05",
    pages = "241",
    year = "2023"
}

@article{Wu:2021nmq,
    author = "Wu, Chao and Zhu, Shou-Hua",
    title = "{Massive on-shell recursion relations for n-point amplitudes}",
    eprint = "2112.12312",
    archivePrefix = "arXiv",
    primaryClass = "hep-th",
    doi = "10.1007/JHEP06(2022)117",
    journal = "JHEP",
    volume = "06",
    pages = "117",
    year = "2022"
}

@article{Ema:2024rss,
    author = "Ema, Yohei and Gao, Ting and Ke, Wenqi and Liu, Zhen and Lyu, Kun-Feng and Mahbub, Ishmam",
    title = "{Momentum shift and on-shell recursion relation for electroweak theory}",
    eprint = "2407.14587",
    archivePrefix = "arXiv",
    primaryClass = "hep-ph",
    reportNumber = "UMN-TH-4325/24, FTPI-MINN-24-16",
    doi = "10.1103/PhysRevD.110.105002",
    journal = "Phys. Rev. D",
    volume = "110",
    number = "10",
    pages = "105002",
    year = "2024"
}

@article{Ballav:2020ese,
    author = "Ballav, Sourav and Manna, Arkajyoti",
    title = "{Recursion relations for scattering amplitudes with massive particles}",
    eprint = "2010.14139",
    archivePrefix = "arXiv",
    primaryClass = "hep-th",
    doi = "10.1007/JHEP03(2021)295",
    journal = "JHEP",
    volume = "03",
    pages = "295",
    year = "2021"
}

@article{DeAngelis:2022qco,
    author = "De Angelis, Stefano",
    title = "{Amplitude bases in generic EFTs}",
    eprint = "2202.02681",
    archivePrefix = "arXiv",
    primaryClass = "hep-th",
    reportNumber = "QMUL-PH-22-05, SAGEX-22-18-E",
    doi = "10.1007/JHEP08(2022)299",
    journal = "JHEP",
    volume = "08",
    pages = "299",
    year = "2022"
}

@article{Franken:2019wqr,
    author = "Franken, Robert and Schwinn, Christian",
    title = "{On-shell constructibility of Born amplitudes in spontaneously broken gauge theories}",
    eprint = "1910.13407",
    archivePrefix = "arXiv",
    primaryClass = "hep-th",
    reportNumber = "TTK-19-43",
    doi = "10.1007/JHEP02(2020)073",
    journal = "JHEP",
    volume = "02",
    pages = "073",
    year = "2020"
}

@article{Wigner:1939cj,
    author = "Wigner, Eugene P.",
    editor = "Kim, Y. S. and Zachary, W. W.",
    title = "{On Unitary Representations of the Inhomogeneous Lorentz Group}",
    doi = "10.2307/1968551",
    journal = "Annals Math.",
    volume = "40",
    pages = "149--204",
    year = "1939"
}

@inproceedings{Cheung:2017pzi,
    author = "Cheung, Clifford",
    title = "{TASI lectures on scattering amplitudes.}",
    booktitle = "{TASI}: {Anticipating the Next Discoveries in Particle Physics}",
    eprint = "1708.03872",
    archivePrefix = "arXiv",
    primaryClass = "hep-ph",
    reportNumber = "CALT-TH-2017-041",
    doi = "10.1142/9789813233348_0008",
    pages = "571--623",
    year = "2018"
}

@article{Elvang:2013cua,
    author = "Elvang, Henriette and Huang, Yu-tin",
    title = "{Scattering Amplitudes}",
    eprint = "1308.1697",
    archivePrefix = "arXiv",
    primaryClass = "hep-th",
    month = "8",
    year = "2013"
}

@article{Durieux:2020gip,
    author = "Durieux, Gauthier and Kitahara, Teppei and Machado, Camila S. and Shadmi, Yael and Weiss, Yaniv",
    title = "{Constructing massive on-shell contact terms}",
    eprint = "2008.09652",
    archivePrefix = "arXiv",
    primaryClass = "hep-ph",
    reportNumber = "MITP/20-046, MITP/20-046",
    doi = "10.1007/JHEP12(2020)175",
    journal = "JHEP",
    volume = "12",
    pages = "175",
    year = "2020"
}

@inproceedings{Dixon:1996wi,
    author = "Dixon, Lance J.",
    title = "{Calculating scattering amplitudes efficiently}",
    booktitle = "{(TASI 95): QCD and Beyond}",
    eprint = "hep-ph/9601359",
    archivePrefix = "arXiv",
    reportNumber = "SLAC-PUB-7106",
    pages = "539--584",
    month = "1",
    year = "1996"
}

@article{Guevara:2018wpp,
    author = "Guevara, Alfredo and Ochirov, Alexander and Vines, Justin",
    title = "{Scattering of Spinning Black Holes from Exponentiated Soft Factors}",
    eprint = "1812.06895",
    archivePrefix = "arXiv",
    primaryClass = "hep-th",
    doi = "10.1007/JHEP09(2019)056",
    journal = "JHEP",
    volume = "09",
    pages = "056",
    year = "2019"
}

@article{Chung:2018kqs,
    author = "Chung, Ming-Zhi and Huang, Yu-Tin and Kim, Jung-Wook and Lee, Sangmin",
    title = "{The simplest massive S-matrix: from minimal coupling to Black Holes}",
    eprint = "1812.08752",
    archivePrefix = "arXiv",
    primaryClass = "hep-th",
    reportNumber = "NCTS-TH/1817",
    doi = "10.1007/JHEP04(2019)156",
    journal = "JHEP",
    volume = "04",
    pages = "156",
    year = "2019"
}

@article{Maybee:2019jus,
    author = "Maybee, Ben and O'Connell, Donal and Vines, Justin",
    title = "{Observables and amplitudes for spinning particles and black holes}",
    eprint = "1906.09260",
    archivePrefix = "arXiv",
    primaryClass = "hep-th",
    doi = "10.1007/JHEP12(2019)156",
    journal = "JHEP",
    volume = "12",
    pages = "156",
    year = "2019"
}

@article{Balkin:2021dko,
    author = "Balkin, Reuven and Durieux, Gauthier and Kitahara, Teppei and Shadmi, Yael and Weiss, Yaniv",
    title = "{On-shell Higgsing for EFTs}",
    eprint = "2112.09688",
    archivePrefix = "arXiv",
    primaryClass = "hep-ph",
    reportNumber = "CERN-TH-2021-212",
    doi = "10.1007/JHEP03(2022)129",
    journal = "JHEP",
    volume = "03",
    pages = "129",
    year = "2022"
}

@article{Liu:2022alx,
    author = "Liu, Da and Yin, Zhewei",
    title = "{Gauge invariance from on-shell massive amplitudes and tree-level unitarity}",
    eprint = "2204.13119",
    archivePrefix = "arXiv",
    primaryClass = "hep-th",
    reportNumber = "UUITP-23/22",
    doi = "10.1103/PhysRevD.106.076003",
    journal = "Phys. Rev. D",
    volume = "106",
    number = "7",
    pages = "076003",
    year = "2022"
}

@article{Bachu:2019ehv,
    author = "Bachu, Brad and Yelleshpur, Akshay",
    title = "{On-Shell Electroweak Sector and the Higgs Mechanism}",
    eprint = "1912.04334",
    archivePrefix = "arXiv",
    primaryClass = "hep-th",
    doi = "10.1007/JHEP08(2020)039",
    journal = "JHEP",
    volume = "08",
    pages = "039",
    year = "2020"
}

@article{Travaglini:2022uwo,
    author = "Travaglini, Gabriele and others",
    title = "{The SAGEX review on scattering amplitudes}",
    eprint = "2203.13011",
    archivePrefix = "arXiv",
    primaryClass = "hep-th",
    reportNumber = "SAGEX-22-01",
    doi = "10.1088/1751-8121/ac8380",
    journal = "J. Phys. A",
    volume = "55",
    number = "44",
    pages = "443001",
    year = "2022"
}

@article{Dittmaier:1998nn,
    author = "Dittmaier, Stefan",
    title = "{Weyl-van der Waerden formalism for helicity amplitudes of massive particles}",
    eprint = "hep-ph/9805445",
    archivePrefix = "arXiv",
    reportNumber = "CERN-TH-98-143",
    doi = "10.1103/PhysRevD.59.016007",
    journal = "Phys. Rev. D",
    volume = "59",
    pages = "016007",
    year = "1998"
}

@article{Parke:1986gb,
    author = "Parke, Stephen J. and Taylor, T. R.",
    title = "{An Amplitude for $n$ Gluon Scattering}",
    reportNumber = "FERMILAB-PUB-86-042-T",
    doi = "10.1103/PhysRevLett.56.2459",
    journal = "Phys. Rev. Lett.",
    volume = "56",
    pages = "2459",
    year = "1986"
}

@article{Bern:1996je,
    author = "Bern, Zvi and Dixon, Lance J. and Kosower, David A.",
    title = "{Progress in one loop QCD computations}",
    eprint = "hep-ph/9602280",
    archivePrefix = "arXiv",
    reportNumber = "SLAC-PUB-7111, UCLA-96-TEP-5, SACLAY-SPH-T-96-10",
    doi = "10.1146/annurev.nucl.46.1.109",
    journal = "Ann. Rev. Nucl. Part. Sci.",
    volume = "46",
    pages = "109--148",
    year = "1996"
}

@article{Durieux:2019eor,
    author = "Durieux, Gauthier and Kitahara, Teppei and Shadmi, Yael and Weiss, Yaniv",
    title = "{The electroweak effective field theory from on-shell amplitudes}",
    eprint = "1909.10551",
    archivePrefix = "arXiv",
    primaryClass = "hep-ph",
    doi = "10.1007/JHEP01(2020)119",
    journal = "JHEP",
    volume = "01",
    pages = "119",
    year = "2020"
}

@article{AccettulliHuber:2021uoa,
    author = "Accettulli Huber, Manuel and De Angelis, Stefano",
    title = "{Standard Model EFTs via on-shell methods}",
    eprint = "2108.03669",
    archivePrefix = "arXiv",
    primaryClass = "hep-th",
    reportNumber = "QMUL-PH-21-32, SAGEX-21-17-E",
    doi = "10.1007/JHEP11(2021)221",
    journal = "JHEP",
    volume = "11",
    pages = "221",
    year = "2021"
}

@article{Ni:2026wiz,
    author = "Ni, Yu-Han and Wu, Chao and Yu, Jiang-Hao",
    title = "{Massless-Massive Amplitude Correspondence I: Helicity-chirality Matching and On-shell Higgsing}",
    eprint = "2601.10620",
    archivePrefix = "arXiv",
    primaryClass = "hep-ph",
    month = "1",
    year = "2026"
}

@article{Ni:2026mia,
    author = "Ni, Yu-Han and Wu, Chao and Yu, Jiang-Hao",
    title = "{Massless-Massive Amplitude Correspondence II: Constructive Massive Amplitudes in Standard Model}",
    eprint = "2601.10622",
    archivePrefix = "arXiv",
    primaryClass = "hep-ph",
    month = "1",
    year = "2026"
}

@article{Ni:2024yrr,
    author = "Ni, Yu-Han and Wang, Yi-Ning and Wu, Chao and Yu, Jiang-Hao",
    title = "{Extended Poincare Symmetry Dictates Massive Scattering Amplitudes}",
    eprint = "2412.03762",
    archivePrefix = "arXiv",
    primaryClass = "hep-ph",
    month = "12",
    year = "2024"
}

@article{Skyrme:1961vq,
    author = "Skyrme, T. H. R.",
    title = "{A Nonlinear field theory}",
    doi = "10.1098/rspa.1961.0018",
    journal = "Proc. Roy. Soc. Lond. A",
    volume = "260",
    pages = "127--138",
    year = "1961"
}

@article{Holzwarth:1985rb,
    author = "Holzwarth, G. and Schwesinger, B.",
    title = "{Baryons in the Skyrme Model}",
    reportNumber = "PRINT-86-0159 (SIEGEN)",
    doi = "10.1088/0034-4885/49/8/001",
    journal = "Rept. Prog. Phys.",
    volume = "49",
    pages = "825",
    year = "1986"
}

@article{Ni:2025xkg,
    author = "Ni, Yu-Han and Wang, Yi-Ning and Wu, Chao and Yu, Jiang-Hao",
    title = "{Massive Helicity-Chirality Spinor Formalism from Massless Amplitudes with On-shell Mass Insertion}",
    eprint = "2501.09062",
    archivePrefix = "arXiv",
    primaryClass = "hep-ph",
    month = "1",
    year = "2025"
}

@book{Rychkov:2016iqz,
    author = "Rychkov, Slava",
    title = "{EPFL Lectures on Conformal Field Theory in D{\ensuremath{>}}= 3 Dimensions}",
    eprint = "1601.05000",
    archivePrefix = "arXiv",
    primaryClass = "hep-th",
    reportNumber = "CERN-TH-2016-012",
    doi = "10.1007/978-3-319-43626-5",
    isbn = "978-3-319-43625-8, 978-3-319-43626-5",
    series = "SpringerBriefs in Physics",
    month = "1",
    year = "2016"
}

@book{Gillioz:2022yze,
    author = "Gillioz, Marc",
    title = "{Conformal field theory for particle physicists}",
    eprint = "2207.09474",
    archivePrefix = "arXiv",
    primaryClass = "hep-th",
    doi = "10.1007/978-3-031-27086-4",
    isbn = "978-3-031-27085-7, 978-3-031-27086-4",
    publisher = "Springer",
    series = "SpringerBriefs in Physics",
    year = "2023"
}

@article{Ni:2026inpre,
    author = "Ni, Yu-Han and Wang, Yi-Ning and Wu, Chao and Yu, Jiang-Hao",
    title = "{An Additional U(1) Symmetry for Massive Scattering Amplitudes}",
    note = {arXiv:2610.xxxxx}
}

@article{Sun:2022ssa,
    author = "Sun, Hao and Xiao, Ming-Lei and Yu, Jiang-Hao",
    title = "{Complete NLO operators in the Higgs effective field theory}",
    eprint = "2206.07722",
    archivePrefix = "arXiv",
    primaryClass = "hep-ph",
    doi = "10.1007/JHEP05(2023)043",
    journal = "JHEP",
    volume = "05",
    pages = "043",
    year = "2023"
}

@article{Sun:2022snw,
    author = "Sun, Hao and Xiao, Ming-Lei and Yu, Jiang-Hao",
    title = "{Complete NNLO operator bases in Higgs effective field theory}",
    eprint = "2210.14939",
    archivePrefix = "arXiv",
    primaryClass = "hep-ph",
    doi = "10.1007/JHEP04(2023)086",
    journal = "JHEP",
    volume = "04",
    pages = "086",
    year = "2023"
}

@article{Song:2023lxf,
    author = "Song, Huayang and Sun, Hao and Yu, Jiang-Hao",
    title = "{Effective field theories of axion, ALP and dark photon}",
    eprint = "2305.16770",
    archivePrefix = "arXiv",
    primaryClass = "hep-ph",
    doi = "10.1007/JHEP01(2024)161",
    journal = "JHEP",
    volume = "01",
    pages = "161",
    year = "2024"
}

@article{Song:2023jqm,
    author = "Song, Huayang and Sun, Hao and Yu, Jiang-Hao",
    title = "{Complete EFT operator bases for dark matter and weakly-interacting light particle}",
    eprint = "2306.05999",
    archivePrefix = "arXiv",
    primaryClass = "hep-ph",
    doi = "10.1007/JHEP05(2024)103",
    journal = "JHEP",
    volume = "05",
    pages = "103",
    year = "2024"
}

@article{Schwinn:2007ee,
    author = "Schwinn, Christian and Weinzierl, Stefan",
    title = "{On-shell recursion relations for all Born QCD amplitudes}",
    eprint = "hep-ph/0703021",
    archivePrefix = "arXiv",
    reportNumber = "MZ-TH-07-02, PITHA-07-01",
    doi = "10.1088/1126-6708/2007/04/072",
    journal = "JHEP",
    volume = "04",
    pages = "072",
    year = "2007"
}
